\documentclass{article}
\usepackage[preprint]{neurips_2021}

\usepackage{amsthm}
\usepackage{graphicx}
\usepackage{dirtytalk}
\usepackage{tabularx}
\usepackage{amsmath}
\usepackage{multirow}
\usepackage{zed-csp}

\usepackage[utf8]{inputenc} 
\usepackage[T1]{fontenc}    
\usepackage[hyphens]{url}            
\usepackage[colorlinks=false,hidelinks]{hyperref}
\usepackage{booktabs}       
\usepackage{amsfonts}       
\usepackage{nicefrac}       
\usepackage{microtype}      
\usepackage{xcolor}         

\setcitestyle{numbers,square}

\title{A Benchmark of JSON-compatible Binary Serialization Specifications}

\author{
  Juan Cruz~Viotti\thanks{\url{https://www.jviotti.com}} \\
  Department of Computer Science \\
  University of Oxford \\
  Oxford, GB OX1 3QD \\
  \texttt{juancruz.viotti@kellogg.ox.ac.uk} \\
  \and
  Mital~Kinderkhedia \\
  Department of Computer Science \\
  University of Oxford \\
  Oxford, GB OX1 3QD \\
  \texttt{mital.kinderkhedia@cs.ox.ac.uk} \\
}

\newtheorem{definition}{Definition}
\newcommand{\benchmarkconclusionrow}{2cm}

\newcommand{\We}{We }
\newcommand{\we}{we }
\newcommand{\our}{our }
\newcommand{\Our}{Our }

\newcommand{\Weare}{We are }

\begin{document}

\maketitle

\begin{abstract}
In this paper, \we present a comprehensive benchmark of JSON-compatible binary
serialization specifications using the SchemaStore open-source test suite
collection of over 400 JSON documents matching their respective schemas and
representative of their use across industries. \We benchmark a set of
schema-driven (ASN.1, Apache Avro, Microsoft Bond, Cap'n Proto, FlatBuffers,
Protocol Buffers, and Apache Thrift) and schema-less (BSON, CBOR, FlexBuffers,
MessagePack, Smile, and UBJSON) JSON-compatible binary serialization
specifications.  Existing literature on benchmarking JSON-compatible binary
serialization specifications demonstrates extensive gaps when it comes to
binary serialization specifications coverage, reproducibility and
representativity, the role of data compression in binary serialization and the
choice and use of obsolete versions of binary serialization specifications.
\We believe \our work is the first of its kind to introduce a tiered taxonomy
for JSON documents consisting of 36 categories classified as Tier 1, Tier 2 and
Tier 3 as a common basis to class JSON documents based on their size, type of
content, characteristics of their structure and redundancy criteria.  \We built
and published a free-to-use online tool to automatically categorize JSON
documents according to \our taxonomy that generates related summary statistics.
In the interest of fairness and transparency, \we adhere to reproducible
software development standards and publicly host the benchmark software and
results on GitHub.  \Our findings provide a number of conclusions: sequential
binary serialization specifications are typically more space-efficient than
pointer-based binary serialization specifications independently of whether they
are schema-less or schema-driven; in comparison to compressed JSON, both
compressed and uncompressed schema-less binary serialization specifications
result in negative median and average size reductions.  Through \our analysis,
\we find that both compressed and uncompressed schema-driven binary
serialization specifications result in positive median and average reduction.
Furthermore, compressed sequential schema-driven binary serialization
specifications are strictly superior to compressed JSON in all the cases from
the input data.

\end{abstract}

\section{Introduction}
\label{sec:benchmark-introduction}

\cite{viotti2022survey} discusses the relevance of the JSON \cite{ECMA-404}
textual schema-less serialization specification in the context of web services,
its history, characteristics, advantages and disadvantages of 13
JSON-compatible schema-driven and schema-less binary serialization
specifications: ASN.1 \cite{asn1}, Apache Avro \cite{avro}, Microsoft Bond
\cite{microsoft-bond}, BSON \cite{bson}, Cap'n Proto \cite{capnproto}, CBOR
\cite{RFC7049}, FlatBuffers \cite{flatbuffers}, FlexBuffers \cite{flexbuffers},
MessagePack \cite{messagepack}, Protocol Buffers \cite{protocolbuffers}, Smile
\cite{smile}, Apache Thrift \cite{slee2007thrift}, and UBJSON \cite{ubjson}.

One of the conclusions found in \cite{viotti2022survey} is that JSON
\cite{ECMA-404} is neither considered a runtime-efficient nor a space-efficient
serialization specification. In terms of runtime-efficiency specification
comparison, \cite{VanuraJ.2018PeoJ} state that serialization and
deserialization speeds vary widely amongst programming languages and data
specification implementations and thus is not comparable. For example, a binary
specification may outperform a textual specification in one programming
language, but the opposite could be true for the same specifications in another
programming language. \We consider runtime-efficiency to be a consequence of
the implementation in the specific programming language rather than a property
of the serialization specification. \cite{krashinsky2003efficient} argues that
network communication is time-expensive and that the network communication
bottleneck \emph{makes computation essentially free in comparison}. It also
argues that the computation overhead of making HTTP/1.1 \cite{RFC2616} payloads
space-efficient using techniques such as data compression can be considered
minimal compared to the time-overhead of a low bandwidth network connection,
which are still common according to \cite{7054230}. For these reasons, \we
focus on studying the space-efficiency characteristics of binary serialization
specifications.

JSON \cite{ECMA-404} documents are typically not sent over the internet
uncompressed. \cite{7054230} found that 92.2\% of the most popular websites
according to Alexa top 500 websites use HTTP/1.1 \cite{RFC2616} compression
formats such as GZIP \cite{RFC1952}. For this reason, \we take into account the
impact of data compression formats as \we investigate space-efficiency
characteristics for the selection of binary serialization specifications.

\subsection{Space-efficiency Benchmark Study}
\ifx\thesis\undefined
Given the JSON-compatible schema-less and schema-driven binary serialization
specifications studied in \cite{viotti2022survey}, this benchmark aims to answer the following
set of research questions:
\else
Given the JSON-compatible schema-less and schema-driven binary serialization
specifications studied in \autoref{sec:thesis-chapter-methodology-survey}, this
benchmark aims to answer the following set of research questions:
\fi

\begin{itemize}

\item \textbf{Q1}: How do JSON-compatible schema-less binary serialization
  specifications compare to JSON in terms of space-efficiency?

\item \textbf{Q2}: How do JSON-compatible schema-driven binary serialization
  specifications compare to JSON and JSON-compatible schema-less binary serialization
    specifications in terms of space-efficiency?

\item \textbf{Q3}: How do JSON-compatible sequential binary serialization
  specifications compare to JSON-compatible pointer-based binary serialization specifications
    in terms of space-efficiency?

\item \textbf{Q4}: How does compressed JSON compares to uncompressed and
  compressed JSON-compatible binary serialization specifications?

\end{itemize}

\subsection{Contributions}
This paper presents and discusses a space-efficiency benchmark of the 13
JSON-compatible binary serialization specifications studied in
\cite{viotti2022survey} using a dataset of 27 methodically JSON documents. \We
believe this work is the first of its kind to produce a comprehensive,
reproducible, extensible and open-source space-efficiency benchmark of
JSON-compatible binary serialization specifications that considers a large and
representative input dataset of real-world JSON documents across industries and
takes data compression into consideration. While producing this benchmark
study, \we identified a lack of an industry-standard automated software to
define JSON-compatible space-efficiency benchmarks. As a solution, \we designed
and implemented an extensible, automated and deterministic benchmark platform
to declare JSON input documents, declare JSON-compatible serialization
specifications written in arbitrary programming languages, declare data
compression formats, extract raw and aggregate statistical resulting data and
generate bar plots and box plots to visualize the results. The benchmark
software is publicly-available on GitHub
\footnote{\url{https://github.com/jviotti/binary-json-size-benchmark}} under
the Apache-2.0 software license. The benchmark runs on the cloud using GitHub
Actions \footnote{\url{https://github.com/features/actions}} and the results
are automatically published to
\url{https://www.jviotti.com/binary-json-size-benchmark/}.

Through the process of conducting a literature review of space-efficiency
benchmarks involving JSON-compatible serialization specifications, \we
identified a lack of a methodical approach for selecting representative sets of
input JSON documents for benchmarking purposes. \We believe this work is the
first of its kind to introduce a formal tiered taxonomy for JSON documents (see
\autoref{sec:taxonomy}) consisting of 36 categories as a common basis to class
JSON documents based on their size, type of content, characteristics of their
structure and redundancy criteria. As part of the taxonomy definition, \we
developed a publicly-available companion web application called \emph{JSON
Stats} to automatically categorize JSON documents according to the taxonomy.
This online tool is available at \url{https://www.jsonbinpack.org/stats/} and
its source code is publicly-available on GitHub
\footnote{\url{https://github.com/jviotti/jsonbinpack}} under the Apache-2.0
software license
\footnote{\url{https://www.apache.org/licenses/LICENSE-2.0.html}}. Refer to
\autoref{fig:json-stats-screenshot} for a screenshot of the tool in action.

\subsection{Paper Organization}
This paper is organized as follows: in \autoref{sec:benchmark-introduction},
\we motivate the need for space-efficiency and outline the proposed benchmark
study. In \autoref{sec:benchmark-literature}, \we summarize existing literature
on space-efficiency benchmark involving the selection of binary serialization
specifications: ASN.1 \cite{asn1}, Apache Avro \cite{avro}, Microsoft Bond
\cite{microsoft-bond}, Cap'n Proto \cite{capnproto}, FlatBuffers
\cite{flatbuffers}, Protocol Buffers \cite{protocolbuffers}, and Apache Thrift
\cite{slee2007thrift}, BSON \cite{bson}, CBOR \cite{RFC7049}, FlexBuffers
\cite{flexbuffers}, MessagePack \cite{messagepack}, Smile \cite{smile}, and
UBJSON \cite{ubjson}. \We found several aspects of the existing literature
lacking merit when it comes to serialization specifications coverage,
reproducibility and representativity, the role of data compression and given
obsolete versions of certain binary serialization specifications. In
\autoref{sec:schemastore}, \we introduce the SchemaStore dataset, an Apache-2.0
licensed collection of JSON Schema \cite{jsonschema-core-2020} documents and
make extensive use of its test suite which consists of over 400 real-world JSON
\cite{ECMA-404} documents matching their respective schemas across various
industries. In \autoref{sec:taxonomy}, \we introduce a tiered taxonomy for JSON
documents based on the size, dominant content, structure and redundancy
characteristics across three main categories: \emph{Tier 1 Minified $<$ 100
bytes}, \emph{Tier 2 Minified $\geq$ 100 $<$ 1000 bytes}, and \emph{Tier 3
Minified $\geq$ 1000 bytes}. \We built and published a free-to-use online tool
to automatically categorize JSON documents according to the taxonomy that
generates related summary statistics. In \autoref{sec:benchmark-methodology},
\we introduce \our methodology to extend the body of literature and demonstrate
what needs to be done to produce a fair benchmark. In
\autoref{sec:benchmark-data}, \we provide comprehensive results along with
their corresponding plots and summary statistics for 27 test cases.  In
\autoref{sec:benchmark-reproducibility}, \we demonstrate all the levels \we
have adhered to in the interest of fairness and reproducibility. In
\autoref{sec:benchmark-conclusions}, \we engage in a reflective discussion
detailing the conclusions of \our benchmark study.

\section{Related Literature}
\label{sec:benchmark-literature}

The existing space-efficiency benchmarks involving the binary serialization
specifications discussed in \cite{viotti2022survey} are summarized in
\autoref{table:benchmark-related-work}. As demonstrated in
\autoref{table:benchmark-related-work}, schema-less binary data serialization
specifications such as BSON \cite{bson}, CBOR \cite{RFC7049}, MessagePack
\cite{messagepack}, and Smile \cite{smile} tend to produce smaller messages
than JSON (up to 63\% size reduction compared to JSON). However, there are some
exceptions. For instance, \cite{7765670} and \cite{petersen2017smart} found
that BSON \cite{bson} and CBOR \cite{RFC7049} tend to produce larger
bit-strings than JSON \cite{ECMA-404} for a subset of their input data (up to
32\% larger than JSON).

In comparison to schema-less serialization specifications, as demonstrated in
\autoref{table:benchmark-related-work}, schema-driven serialization
specifications such as Protocol Buffers \cite{protocolbuffers}, Apache Thrift
\cite{slee2007thrift}, Apache Avro \cite{avro} tend to produce bit-strings that
are up to 95\% smaller than JSON. However, in this case, there are also
exceptions. For instance, \cite{petersen2017smart} and \cite{VanuraJ.2018PeoJ}
found the MessagePack \cite{messagepack} schema-less binary serialization
specification to be space-efficient in comparison to Protocol Buffers
\cite{protocolbuffers} and Apache Avro \cite{avro} (up to 23\% size reduction).
Similarly, \cite{hamerski2018evaluating} and \cite{10589/150617} found
MessagePack \cite{messagepack} to be space-efficient in comparison to Protocol
Buffers \cite{protocolbuffers} and FlatBuffers \cite{flatbuffers} for certain
cases.

Data compression is another approach to achieve space efficiency.
\cite{10.1145/2016716.2016718} conclude that compressed JSON is space-efficient
in comparison to both compressed and uncompressed Protocol Buffers.
\cite{petersen2017smart} conclude that compressed JSON is space-efficient in
comparison to Protocol Buffers, Smile \cite{smile}, compressed and uncompressed
CBOR \cite{RFC7049}, and compressed and uncompressed BSON \cite{bson}.
However, Apache Avro \cite{avro}, MessagePack \cite{messagepack}, compressed
Protocol Buffers, and compressed Smile are space-efficient in comparison to
compressed JSON.

\subsection{Shortcomings}

\We found several aspects of the existing literature to be insufficient,
leading to gaps in the JSON-compatible binary serialization space-efficiency
benchmark literature for the following reasons: \\

\textbf{Coverage of Serialization Specifications.} The binary serialization
specifications covered by existing benchmarks are Protocol Buffers
\cite{protocolbuffers}, MessagePack \cite{messagepack}, FlatBuffers
\cite{flatbuffers}, and to a lesser extent: Apache Avro \cite{avro}, Apache
Thrift \cite{slee2007thrift}, BSON \cite{bson}, CBOR \cite{RFC7049} and Smile
\cite{smile}. \Our previous work \cite{viotti2022survey} also discusses ASN.1 \cite{asn1},
Microsoft Bond \cite{microsoft-bond}, Cap'n Proto \cite{capnproto}, FlexBuffers
\cite{flexbuffers}, and UBJSON \cite{ubjson}. To the best of \our knowledge,
these have not been considered by existing space-efficiency benchmark
literature.  \\

\textbf{Reproducibility and Representativity.} Neither
\cite{10.1145/2016716.2016718}, \cite{7765670}, \cite{petersen2017smart},
\cite{sahlmann2018binary}, \cite{8876986}, nor \cite{9142787} disclose the JSON
\cite{ECMA-404} documents or schema definitions used to arrive at the result of
their space-efficiency benchmarks. Therefore, it is not possible to corroborate
their findings or contextualize their results. The publications that disclose
the input JSON \cite{ECMA-404} documents are \cite{MaedaK2012Peoo} and
\cite{Kyurkchiev-msgpack-json}.  However, they are limited in scope as they
consider a single JSON document in each of their papers. Other publications
disclose schema definitions of varying formality that describe the JSON
documents used as part of their benchmarks. Of those,
\cite{microblogging-protobuf} and \cite{8977050} are concerned with one type of
JSON document, \cite{SumarayAudie2012Acod} and \cite{6784954} are concerned
with two types of JSON documents, and \cite{hamerski2018evaluating} and
\cite{10589/150617} are concerned with three types of JSON documents.
Therefore, \we consider the results from these publications to be either not
reproducible or not representative of the variety of JSON documents that are
widely used in practice across different industries.  \\

\textbf{Data Compression.} \We found two publications that take data
compression into account: \cite{10.1145/2016716.2016718} and
\cite{petersen2017smart}.  However, they are limited in scope as these papers
discuss only the GZIP \cite{RFC1952} data compression format and there is no
mention of the implementation used and the compression level that GZIP is
configured with.  \\

\textbf{Out-of-date.} Some of the existing benchmarks measure obsolete versions
of certain binary serialization specifications. For example, the Protocol
Buffers \cite{protocolbuffers} version 3 was first released in 2014
\footnote{\url{https://github.com/protocolbuffers/protobuf/releases/tag/v3.0.0-alpha-1}}.
However, there are a number of benchmark publications released before that year
that discuss the now-obsolete Protocol Buffers version 2
\cite{10.1145/2016716.2016718} \cite{microblogging-protobuf}
\cite{MaedaK2012Peoo} \cite{SumarayAudie2012Acod} \cite{6784954}.

\begin{table*}[hb!]
\caption{A list of space-efficiency benchmark publications that involve JSON \cite{ECMA-404} and/or a subset of the binary serialization specifications dicussed in \cite{viotti2022survey}. The third column summarises the benchmark conclusions. In this table, a serialization specification is \emph{greater than} another serialization specification if it produced larger bit-strings in the respective publication findings.}
\label{table:benchmark-related-work}
\begin{tabularx}{\linewidth}{X|l|X}
  \toprule
  \textbf{Publication} & \textbf{Year} & \textbf{Conclusion} \\
  \midrule

  {Impacts of data interchange formats on energy consumption and performance in smartphones \cite{10.1145/2016716.2016718}} & {2011} & {JSON $>$ Protocol Buffers $>$ Protocol Buffers with GZIP $>$ JSON with GZIP} \\ \hline
  {Evaluation of Protocol Buffers as Data Serialization Format for Microblogging Communication \cite{microblogging-protobuf}} & {2011} & {JSON $>$ Protocol Buffers} \\ \hline
  {Performance evaluation of object serialization libraries in XML, JSON and binary formats \cite{MaedaK2012Peoo}} & {2012} & {JSON $>$ Apache Thrift $>$ Protocol Buffers $>$ Apache Avro} \\ \hline
  {A comparison of data serialization formats for optimal efficiency on a mobile platform \cite{SumarayAudie2012Acod}} & {2012} & {JSON $>$ Apache Thrift $>$ Protocol Buffers} \\ \hline
  {Google protocol buffers research and application in online game \cite{6784954}} & {2013} & {JSON $>$ Protocol Buffers} \\ \hline
  {Integrating a system for symbol programming of real processes with a cloud service \cite{Kyurkchiev-msgpack-json}} & {2015} & {JSON $>$ MessagePack} \\ \hline
  {Performance evaluation of using Protocol Buffers in the Internet of Things communication \cite{7765670}} & {2016} & {\textbf{In most cases}: JSON $>$ BSON $>$ Protocol Buffers. \textbf{However, in some cases}: BSON $>$ JSON $>$ Protocol Buffers} \\ \hline
  {Smart grid serialization comparison: Comparison of serialization for distributed control in the context of the Internet of Things \cite{petersen2017smart}} & {2017} & {BSON $>$ CBOR $>$ JSON $>$ BSON with GZIP $>$ Smile $>$ Protocol Buffers $>$ CBOR with GZIP $>$ JSON with GZIP $>$ Apache Avro $>$ Protocol Buffers with GZIP $>$ Smile with GZIP $>$ MessagePack $>$ Apache Avro with GZIP $>$ MessagePack with GZIP} \\ \hline
  {Binary Representation of Device Descriptions: CBOR versus RDF HDT \cite{sahlmann2018binary}} & {2018} & {JSON $>$ CBOR} \\ \hline
   {Evaluating Serialization for a Publish-Subscribe Based Middleware for MPSoCs \cite{hamerski2018evaluating}} & {2018} & {FlatBuffers $>$ Protocol Buffers $>$ MessagePack} \\ \hline
  {Performance Evaluation of Java, JavaScript and PHP Serialization Libraries for XML, JSON and Binary Formats \cite{VanuraJ.2018PeoJ}} & {2018} & {JSON $>$ MessagePack $>$ Protocol Buffers $>$ Apache Avro} \\ \hline
  {Analytical assessment of binary data serialization techniques in IoT context (evaluating protocol buffers, flat buffers, message pack, and BSON for sensor nodes) \cite{10589/150617}} & {2019} & {\textbf{For numeric and mixed data}: BSON $>$ FlatBuffers $>$ MessagePack $>$ Protocol Buffers. \textbf{For textual data}: FlatBuffers $>$ BSON $>$ MessagePack $>$ Protocol Buffers} \\ \hline
  {Enabling Model-Driven Software Development Tools for the Internet of Things \cite{8876986}} & {2019} & {JSON $>$ FlatBuffers} \\ \hline
  {Flatbuffers Implementation on MQTT Publish/Subscribe Communication as Data Delivery Format \cite{8977050}} & {2019} & {JSON $>$ FlatBuffers} \\ \hline
  {Performance Comparison of Messaging Protocols and Serialization Formats for Digital Twins in IoV \cite{9142787}} & {2020} & {JSON $>$ FlatBuffers $>$ Protocol Buffers} \\

  \bottomrule
\end{tabularx}
\end{table*}

\clearpage
\section{The SchemaStore Dataset}
\label{sec:schemastore}

SchemaStore \footnote{\url{https://www.schemastore.org}} is an
Apache-2.0-licensed
\footnote{\url{http://www.apache.org/licenses/LICENSE-2.0.html}} open-source
collection of over 300 JSON Schema \cite{jsonschema-core} documents which
describe popular JSON-based \cite{ECMA-404} formats such as CityJSON
\cite{4aad07f4-8f64-46b1-aad3-3d4abe36c5bf} and JSON Patch \cite{RFC6902}. The
SchemaStore API can be integrated with code editors to offer auto-completion
and validation when writing JSON documents. The SchemaStore project was started
by Mads Kristensen \footnote{\url{https://github.com/madskristensen}} in 2014
while working as a Senior Program Manager focused on the Visual Studio IDE at
Microsoft. For the purpose of benchmarking, \we will make use of SchemaStore's
extensive test suite which consists of over 400 real-world JSON documents
matching the respective schemas
\footnote{\url{https://github.com/SchemaStore/schemastore/tree/master/src/test}}.
This paper refers to commit hash
\texttt{0b6bd2a08005e6f7a65a68acaf3064d6e2670872} of the SchemaStore repository
hosted on GitHub \footnote{\url{https://github.com/SchemaStore/schemastore}}.
\We believe that the SchemaStore test suite is a good representation of the set
of JSON documents used across industries.

\begin{figure}[ht!]
  \frame{\includegraphics[width=\linewidth]{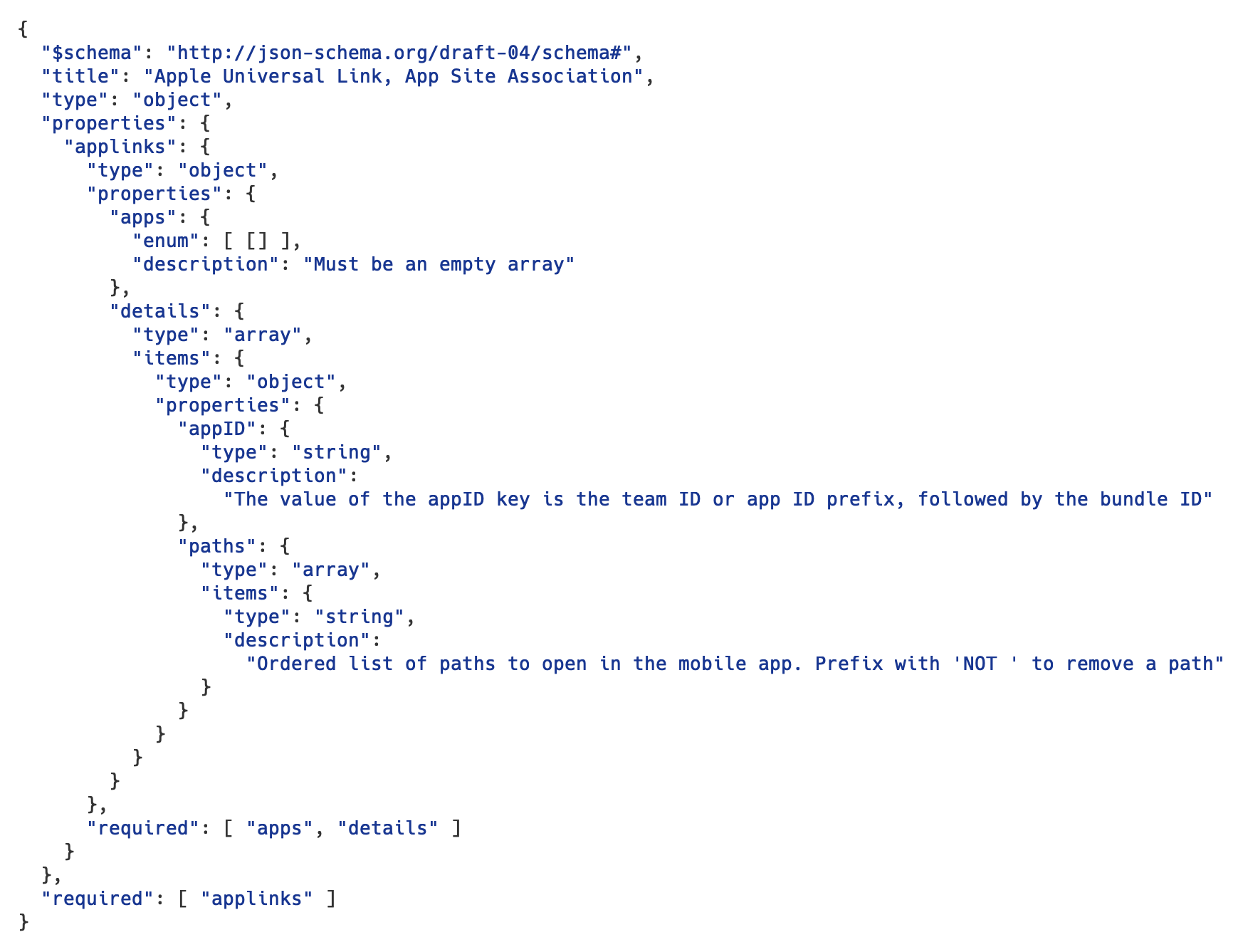}}
  \caption{An example JSON Schema Draft 4 \cite{jsonschema-core-draft4}
  document from SchemaStore
  \protect\footnote{\url{https://github.com/SchemaStore/schemastore/blob/0b6bd2a08005e6f7a65a68acaf3064d6e2670872/src/schemas/json/apple-app-site-association.json}}
  that describes an Apple Associated Domain file
  \protect\footnote{\url{https://developer.apple.com/documentation/safariservices/supporting_associated_domains}} to
  associate an iOS app and a website.} \label{fig:schemastore-schema-example}
\end{figure}

\begin{figure}[ht!]
  \frame{\includegraphics[width=\linewidth]{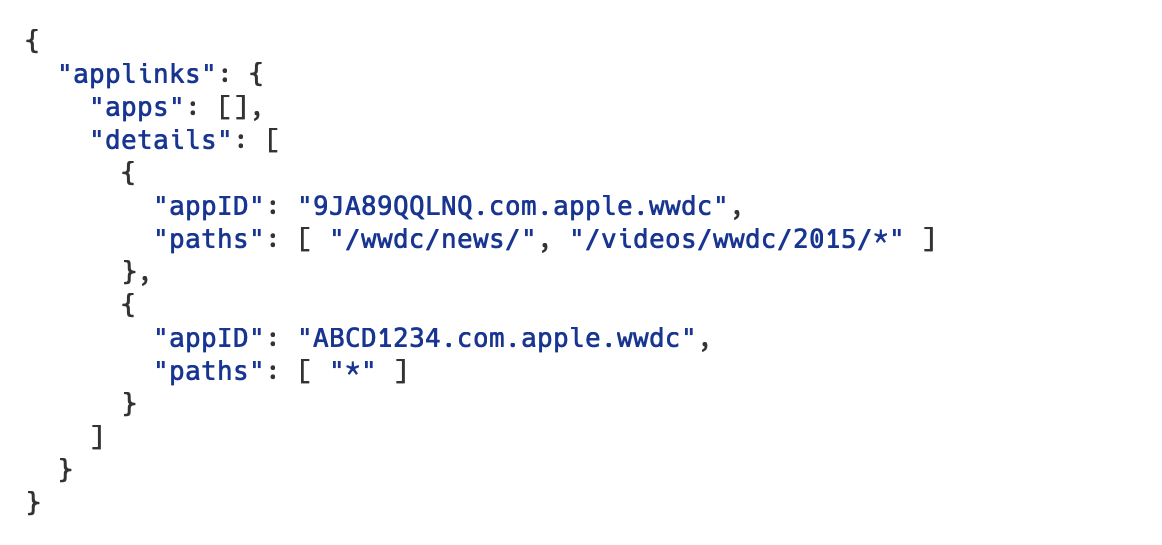}}
  \caption{An example JSON \cite{ECMA-404} document that matches the schema
  definition from \autoref{fig:schemastore-schema-example} taken from
  SchemaStore's test suite
\protect\footnote{\url{https://github.com/SchemaStore/schemastore/blob/0b6bd2a08005e6f7a65a68acaf3064d6e2670872/src/test/apple-app-site-association/apple-app-site-association_getting-started.json}}.}
\label{fig:schemastore-instance-example} \end{figure}

\clearpage
\section{A Taxonomy of JSON Documents}
\label{sec:taxonomy}

Serializing two data structures that match the same schema definition but
consist of different values is likely to result in similar byte sizes.
However, serializating two data structures with the same values but different
structures may produce diverse results, even when utilising the same
serialization specification. Therefore, \we conclude that the structure and the
type of content affects the size of the serialized bit-strings more than the
actual values.  Under this assumption, to produce a representative size
benchmark, it is essential to measure binary serialization specifications using
a set of JSON \cite{ECMA-404} documents that differ in structure, type of
content and size. 

To solve the input data selection problem, it is required to have a process to
categorize JSON \cite{ECMA-404} documents depending on such characteristics.
In this way, \we present a taxonomy consisting of 36 categories listed in
\autoref{table:json-taxonomy}. The taxonomy qualifies JSON documents based on
their size, type of content, nesting, structural, and redundancy
characteristics. While most JSON documents in practice are objects or arrays,
this taxonomy is also applicable to JSON documents consisting of single scalar
values and strings. \We hope that this taxonomy forms a common basis to talk
about JSON documents in a high-level manner beyond the benchmarking problem.

\begin{table*}[hb!]

\caption{There are 36 categories defined in \our JSON documents taxonomy. The
  second column contains acronyms for each category name.  In terms of size, a
  JSON document can either be \emph{Tier 1 Minified $<$ 100 bytes}, \emph{Tier
  2 Minified $\geq$ 100 $<$ 1000 bytes}, or \emph{Tier 3 Minified $\geq$ 1000
  bytes}.  In terms of content, a JSON document can either be \emph{numeric},
  \emph{textual}, or \emph{boolean}. Finally, in terms of structure, a JSON
  document can either be \emph{flat} or \emph{nested}.}

\label{table:json-taxonomy}
\begin{tabularx}{\linewidth}{lllX|l}
  \toprule
  \multicolumn{4}{l|}{\textbf{Category}} & \textbf{Acronym} \\
  \midrule

Tier 1 Minified $<$ 100 bytes             & Numeric & Redundant & Flat           & \texttt{Tier 1 NRF} \\ \hline
Tier 1 Minified $<$ 100 bytes             & Numeric & Redundant & Nested         & \texttt{Tier 1 NRN} \\ \hline
Tier 1 Minified $<$ 100 bytes             & Numeric & Non-Redundant & Flat       & \texttt{Tier 1 NNF} \\ \hline
Tier 1 Minified $<$ 100 bytes             & Numeric & Non-Redundant & Nested     & \texttt{Tier 1 NNN} \\ \hline
Tier 1 Minified $<$ 100 bytes             & Textual & Redundant & Flat           & \texttt{Tier 1 TRF} \\ \hline
Tier 1 Minified $<$ 100 bytes             & Textual & Redundant & Nested         & \texttt{Tier 1 TRN} \\ \hline
Tier 1 Minified $<$ 100 bytes             & Textual & Non-Redundant & Flat       & \texttt{Tier 1 TNF} \\ \hline
Tier 1 Minified $<$ 100 bytes             & Textual & Non-Redundant & Nested     & \texttt{Tier 1 TNN} \\ \hline
Tier 1 Minified $<$ 100 bytes             & Boolean & Redundant & Flat           & \texttt{Tier 1 BRF} \\ \hline
Tier 1 Minified $<$ 100 bytes             & Boolean & Redundant & Nested         & \texttt{Tier 1 BRN} \\ \hline
Tier 1 Minified $<$ 100 bytes             & Boolean & Non-Redundant & Flat       & \texttt{Tier 1 BNF} \\ \hline
Tier 1 Minified $<$ 100 bytes             & Boolean & Non-Redundant & Nested     & \texttt{Tier 1 BNN} \\ \hline \hline
Tier 2 Minified $\geq$ 100 $<$ 1000 bytes & Numeric & Redundant & Flat           & \texttt{Tier 2 NRF} \\ \hline
Tier 2 Minified $\geq$ 100 $<$ 1000 bytes & Numeric & Redundant & Nested         & \texttt{Tier 2 NRN} \\ \hline
Tier 2 Minified $\geq$ 100 $<$ 1000 bytes & Numeric & Non-Redundant & Flat       & \texttt{Tier 2 NNF} \\ \hline
Tier 2 Minified $\geq$ 100 $<$ 1000 bytes & Numeric & Non-Redundant & Nested     & \texttt{Tier 2 NNN} \\ \hline
Tier 2 Minified $\geq$ 100 $<$ 1000 bytes & Textual & Redundant & Flat           & \texttt{Tier 2 TRF} \\ \hline
Tier 2 Minified $\geq$ 100 $<$ 1000 bytes & Textual & Redundant & Nested         & \texttt{Tier 2 TRN} \\ \hline
Tier 2 Minified $\geq$ 100 $<$ 1000 bytes & Textual & Non-Redundant & Flat       & \texttt{Tier 2 TNF} \\ \hline
Tier 2 Minified $\geq$ 100 $<$ 1000 bytes & Textual & Non-Redundant & Nested     & \texttt{Tier 2 TNN} \\ \hline
Tier 2 Minified $\geq$ 100 $<$ 1000 bytes & Boolean & Redundant & Flat           & \texttt{Tier 2 BRF} \\ \hline
Tier 2 Minified $\geq$ 100 $<$ 1000 bytes & Boolean & Redundant & Nested         & \texttt{Tier 2 BRN} \\ \hline
Tier 2 Minified $\geq$ 100 $<$ 1000 bytes & Boolean & Non-Redundant & Flat       & \texttt{Tier 2 BNF} \\ \hline
Tier 2 Minified $\geq$ 100 $<$ 1000 bytes & Boolean & Non-Redundant & Nested     & \texttt{Tier 2 BNN} \\ \hline \hline
Tier 3 Minified $\geq$ 1000 bytes         & Numeric & Redundant & Flat           & \texttt{Tier 3 NRF} \\ \hline
Tier 3 Minified $\geq$ 1000 bytes         & Numeric & Redundant & Nested         & \texttt{Tier 3 NRN} \\ \hline
Tier 3 Minified $\geq$ 1000 bytes         & Numeric & Non-Redundant & Flat       & \texttt{Tier 3 NNF} \\ \hline
Tier 3 Minified $\geq$ 1000 bytes         & Numeric & Non-Redundant & Nested     & \texttt{Tier 3 NNN} \\ \hline
Tier 3 Minified $\geq$ 1000 bytes         & Textual & Redundant & Flat           & \texttt{Tier 3 TRF} \\ \hline
Tier 3 Minified $\geq$ 1000 bytes         & Textual & Redundant & Nested         & \texttt{Tier 3 TRN} \\ \hline
Tier 3 Minified $\geq$ 1000 bytes         & Textual & Non-Redundant & Flat       & \texttt{Tier 3 TNF} \\ \hline
Tier 3 Minified $\geq$ 1000 bytes         & Textual & Non-Redundant & Nested     & \texttt{Tier 3 TNN} \\ \hline
Tier 3 Minified $\geq$ 1000 bytes         & Boolean & Redundant & Flat           & \texttt{Tier 3 BRF} \\ \hline
Tier 3 Minified $\geq$ 1000 bytes         & Boolean & Redundant & Nested         & \texttt{Tier 3 BRN} \\ \hline
Tier 3 Minified $\geq$ 1000 bytes         & Boolean & Non-Redundant & Flat       & \texttt{Tier 3 BNF} \\ \hline
Tier 3 Minified $\geq$ 1000 bytes         & Boolean & Non-Redundant & Nested     & \texttt{Tier 3 BNN} \\

  \bottomrule
\end{tabularx}
\end{table*}

\subsection{Size}
\label{sec:taxonomy-size}

\begin{figure}[hb!]
  \frame{\includegraphics[width=\linewidth]{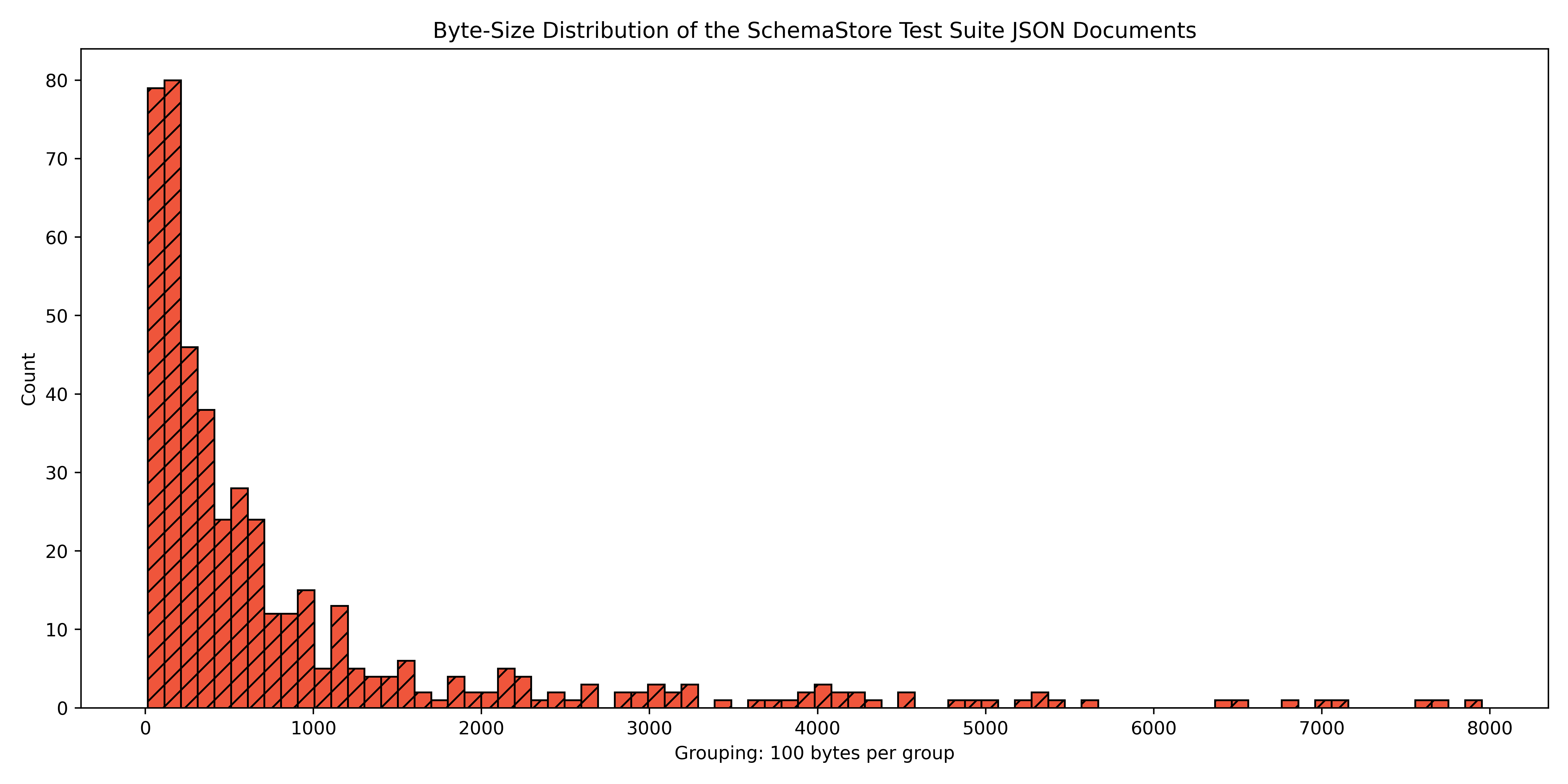}}
  \caption{The byte-size distribution (in 100 byte groups up to 8000 bytes for
  illustration purposes) of the 480 JSON documents present in the SchemaStore
  test suite introduced in \autoref{sec:schemastore}. Most JSON documents weigh
  less than 1000 bytes. The largest JSON document weighs 545392 bytes
  \protect\footnote{\url{https://github.com/SchemaStore/schemastore/blob/0b6bd2a08005e6f7a65a68acaf3064d6e2670872/src/test/sarif/BinSkim.AllRules.sarif.json}}.}
\label{fig:schemastore-byte-size-distribution} \end{figure}

In order to categorize JSON documents in a sensible manner using a small set of
size categories, \we first calculate the byte-size distribution of the JSON
documents in the SchemaStore test suite introduced in
\autoref{sec:schemastore}. The results are illustrated in
\autoref{fig:schemastore-byte-size-distribution}. Based on these results, \we
group JSON documents into three categories:

\begin{itemize}

\item \textbf{Tier 1 Minified $<$ 100 bytes.} A JSON document is in this
  category if its UTF-8 \cite{UnicodeStandard} \emph{minified} form occupies
    less than 100 bytes \cite{viotti2022survey}.

\item \textbf{Tier 2 Minified $\geq$ 100 $<$ 1000 bytes.} A JSON document is in
  this category if its UTF-8 \cite{UnicodeStandard} \emph{minified} form
    occupies 100 bytes or more, but less than 1000 bytes \cite{viotti2022survey}.

\item \textbf{Tier 3 Minified $\geq$ 1000 bytes.} A JSON document is in this
  category if its UTF-8 \cite{UnicodeStandard} \emph{minified} form occupies
    1000 bytes or more \cite{viotti2022survey}.

\begin{figure}[hb!]
  \frame{\includegraphics[width=\linewidth]{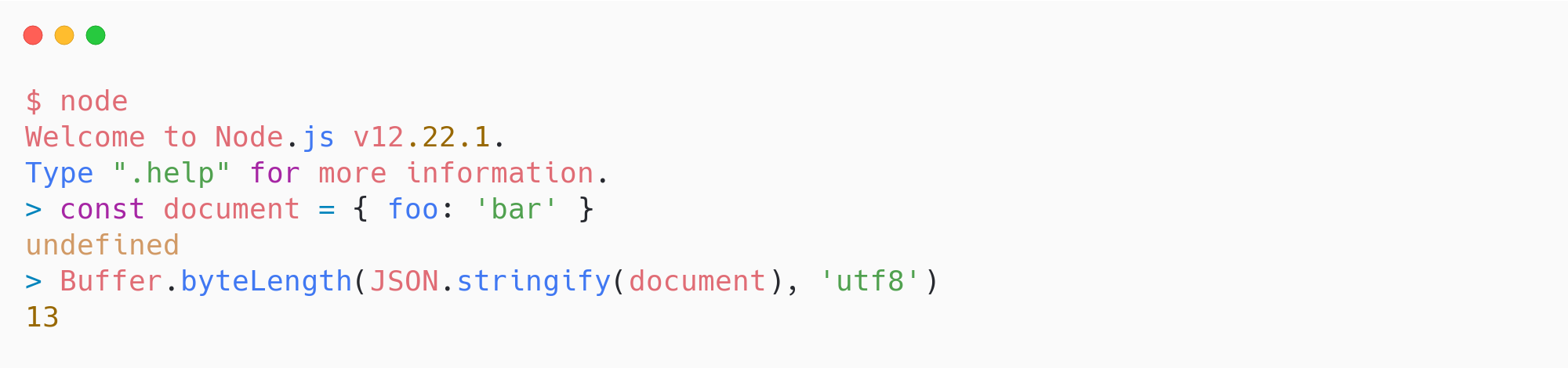}}

  \caption{ The UTF-8 \cite{UnicodeStandard} byte-size of a JSON document in
  \emph{minified} form can be determined using a Node.js
  \protect\footnote{\url{https://nodejs.org/}} interactive REPL session by
  combining the \texttt{JSON.stringify} and the \texttt{Buffer.byteLength}
  functions as demonstrated in this figure. In this example, \we determine the
  size of the JSON document \texttt{\{ "foo": "bar" \}} to be 13 bytes.}

\label{fig:node-json-byte-size} \end{figure}

\end{itemize}

\subsection{Content Type}
\label{sec:taxonomy-content-type}

The taxonomy categorises a JSON document based on the data types that dominate
its content. \We calculate this characteristic based on the number of values of
a certain data type that a JSON document contains and the byte-size that these
data values occupy in the serialized bit-string. \We take both of these
measures into account as serializing many small instances result in more
metadata overhead than serializing a few large instances for a given type.

The JSON \cite{ECMA-404} serialization specification supports the following
data types: \emph{object}, \emph{array}, \emph{boolean}, \emph{string},
\emph{number}, and \emph{null}. \We consider objects and arrays to represent
\emph{structural} values, strings to represent \emph{textual} values, and
numbers to represent \emph{numeric} values. For simplicity, \we consider
\emph{true}, \emph{false}, and \emph{null} to represent \emph{boolean} values
as in three-valued logic \cite{putnam1957three}. \We use this data type
categorization to define the \emph{textual weight}, \emph{numeric weight}, and
\emph{boolean weight} for a given JSON document.

The weight metrics for a JSON document are based on a common formula where $C$
is the total number of data values in the JSON document and $S$ is the total
byte-size of the JSON document in \emph{minified} form \cite{viotti2022survey}:

\begin{equation}
\frac{\frac{K \times 100}{C} \times \frac{B \times 100}{S}}{100}
\end{equation}

\begin{itemize}

  \item \textbf{Textual Weight.} In this case, $K$ is the number of string
    values in the JSON document and $B$ is the cummulative byte-size occupied
    by the string values in the JSON document. The double quotes surrouding
    string values are considered part of the byte-size occupied by a string.
    Therefore, a string value encoded in a UTF-8 \cite{UnicodeStandard} JSON
    document occupies at least $2 + N$ bytes where $N$ corresponds to number of
    code-points in the string.

  \item \textbf{Numeric Weight.} In this case, $K$ is the number of numeric
    values in the JSON document and $B$ is the cummulative byte-size occupied
    by the numeric values in the JSON document. Each numeric digit and
    auxiliary characters such as the minus sign ($-$) and the period (.) for
    representing real numbers count towards the byte-size of the numeric value.

  \item \textbf{Boolean Weight.} In this case, $K$ is the number of boolean
    values in the JSON document and $B$ is the cummulative byte-size occupied
    by the boolean values in the JSON document. The UTF-8
    \cite{UnicodeStandard} JSON encoding represents \emph{true} using 4 bytes,
    \emph{false} using 5 bytes, and \emph{null} using 4 bytes.

\end{itemize}

\We rely on the previous weight definitions to provide the content type
taxonomy for JSON documents based on whether they are \emph{textual},
\emph{numeric}, or \emph{boolean}. Given an input JSON document, consider
$W_t$, $W_n$, and $W_b$ to represent its textual, numeric and boolean weights,
respectively:

\begin{itemize}

\item \textbf{Textual.} A JSON document is textual if $W_t \geq W_n \geq W_b$.
\item \textbf{Numeric.} A JSON document is numeric if $W_n \geq W_t \geq W_b$.
\item \textbf{Boolean.} A JSON document is numeric if $W_b \geq W_t \geq W_n$.

\end{itemize}

If two or more of the content type weight values are equal and greater than the
rest, such JSON document is considered to hold more than one type of content
qualifier. For example, if $W_t = W_n$ and $W_t > W_b$, then the JSON document
is equally considered \emph{textual} and \emph{numeric}.

\begin{figure}[hb!]
\frame{\includegraphics[width=\linewidth]{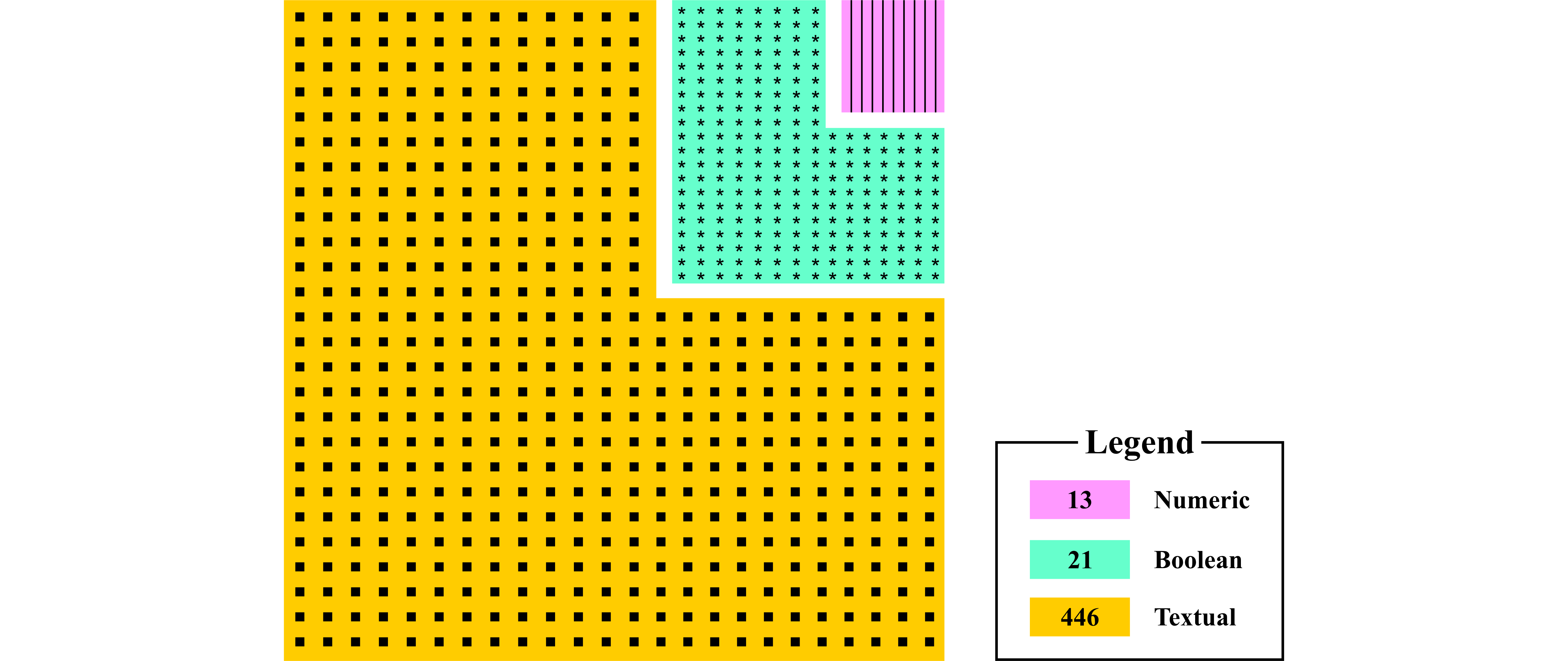}} \caption{Out of
the 480 JSON documents in the SchemaStore test suite introduced in
\autoref{sec:schemastore}, 446 are textual, 21 are boolean, and 13 are
numeric.} \label{fig:schemastore-content-types} \end{figure}

The results of executing this aspect of the taxonomy on the SchemaStore test
suite introduced in \autoref{sec:schemastore} are shown in
\autoref{fig:schemastore-content-types}.

\subsection{Redundancy}
\label{sec:taxonomy-redundancy}

The taxonomy measures redundancy as the percentage of values in a given JSON
document that are duplicated taking scalar and composite data types into
account.

In comparison to schema-less serialization formats, schema-driven serialization
formats make use of schema definitions to avoid encoding object keys. This
taxonomy is designed to aid in categorizing JSON documents based on
characteristics that impact data serialization. For these reasons, the number
of duplicated object keys in the redundancy metric is irrelevant for the
schema-driven subset of the selection of binary serializion formats and is not
taken into account.

Let $JSON$ be the set of JSON documents as defined in the data model introduced
by \cite{10.1145/3034786.3056120} with the exception that the $[\**]$ object
operator results in a \emph{sequence}, instead of a set, of values in the given
JSON object. Consider a new $[\&]$ operator defined using the Z formal
specification notation \cite{ISO13568-2002} that results in the flattened
sequence of atomic and compositional structure values of the given JSON
document:

\begin{axdef}
\_ \thinspace [\&] : JSON \fun \seq JSON
\where
\forall J : JSON @ \\
\;\;\; J[\&] = \langle J \rangle \cat J[\**][0][\&] \cat ... \cat J[\**][\#J][\&] \hspace{3em} \text{if $J$ is an object} \\
\;\;\; J[\&] = \langle J \rangle \cat J_0[\&] \cat J'[\&] \hspace{9.9em} \text{if $J$ is an array} \\
\;\;\; J[\&] = \langle J \rangle \hspace{17.0em} \text{otherwise}
\end{axdef}

Using this operator, $R_J$ is defined as the percentage of duplicate values in
the JSON document $J$:

\begin{equation}
R_J = \frac{(\# J[\&] - \# \{ v | v \: in \: J[\&] \}) \times 100}{\# J[\&]}
\end{equation}

\begin{figure}[hb!]
  \frame{\includegraphics[width=\linewidth]{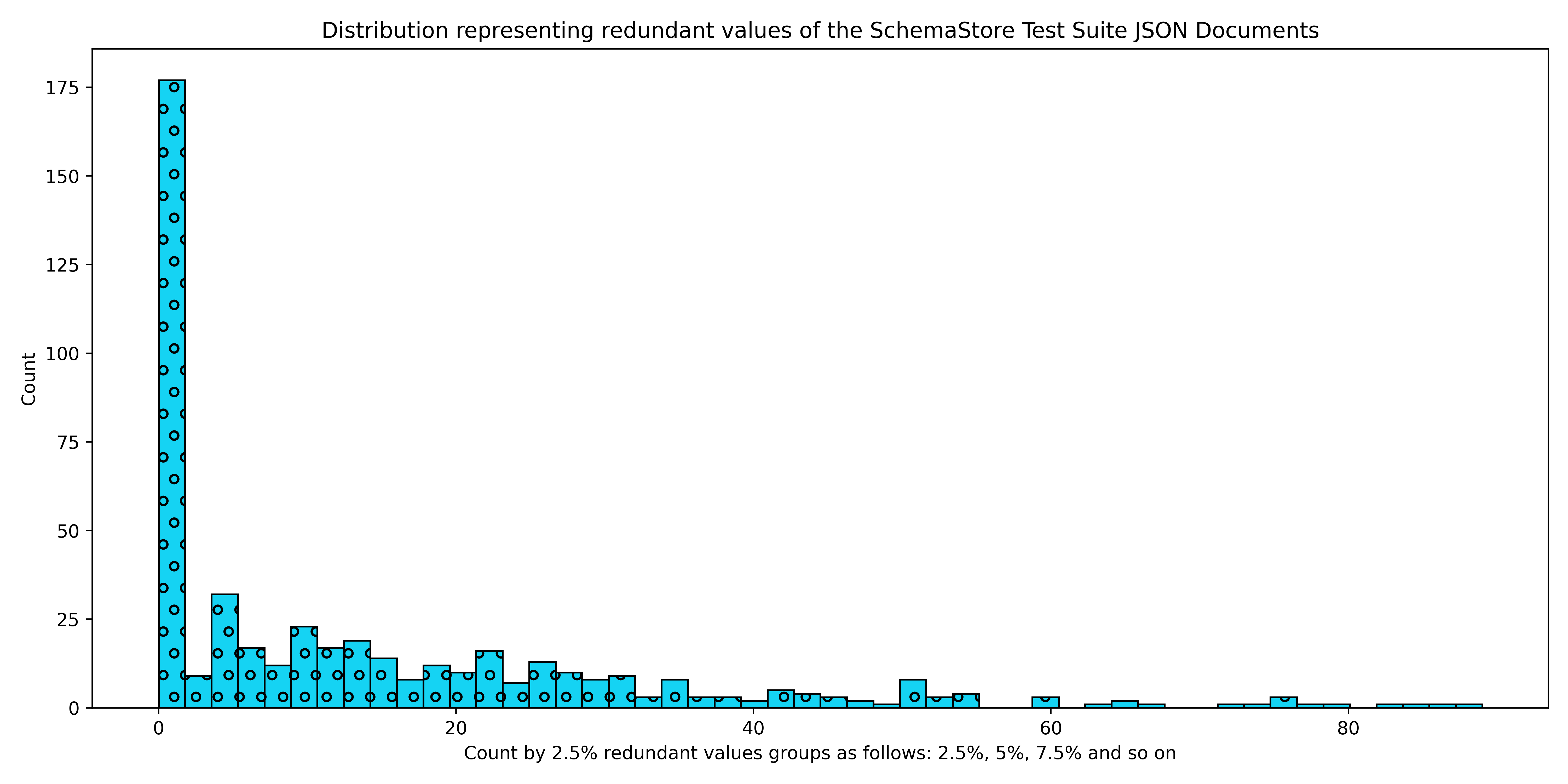}}
  \caption{Distribution representing redundant values. First, \we calculate the
  percentage of redundant values of the 480 JSON documents present in the
  SchemaStore test suite introduced in \autoref{sec:schemastore} and second,
  \we count them by 2.5\% redundant values groups as follows: 2.5\%, 5\%, 7.5\%
  and so on. Most JSON documents are strictly non-redundant.  However there are
  instances of almost every 2.5\% redundancy groups in the plot. The most
  redundant JSON document has a value redundancy of 88.8\%
  \protect\footnote{\url{https://github.com/SchemaStore/schemastore/blob/0b6bd2a08005e6f7a65a68acaf3064d6e2670872/src/test/csslintrc/WebAnalyzer.json}}.}
\label{fig:schemastore-values-redundancy-distribution} \end{figure}

In order to categorize JSON documents in a sensible manner, the taxonomy
distinguishes between \emph{redundant} JSON documents and \emph{non-redundant}
JSON documents. The redundancy distribution of the JSON documents in the
SchemaStore test suite introduced in \autoref{sec:schemastore} is computed in
\autoref{fig:schemastore-values-redundancy-distribution}. Using these results,
this taxonomy aspect is defined as follows:

\begin{itemize}

\item \textbf{Redundant.} A JSON document $J$ is redundant if $R_J \geq 25\%$
\item \textbf{Non-Redundant.} A JSON document $J$ is redundant if $R_J < 25\%$

\end{itemize}

\subsection{Structure}
\label{sec:taxonomy-structure}

\begin{figure}[hb!]
\frame{\includegraphics[width=\linewidth]{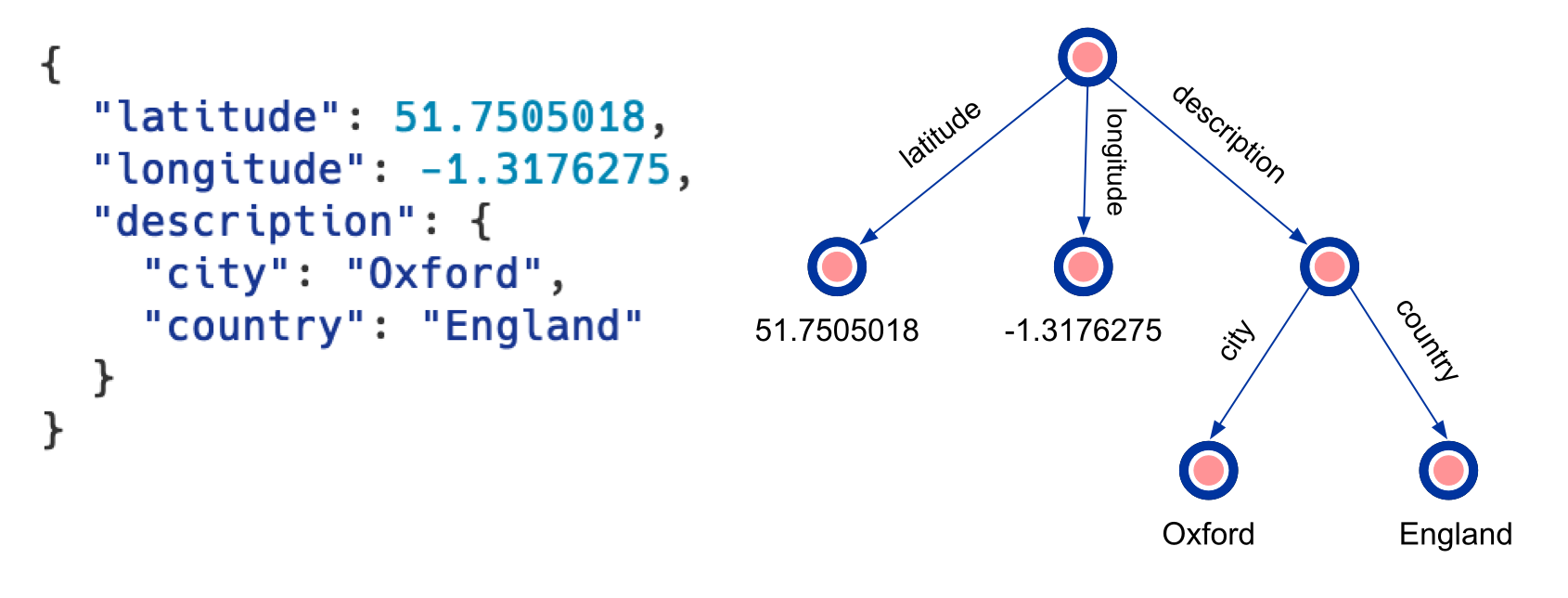}} \caption{An
example JSON document and its corresponding connected acyclic undirected graph
representation.} \label{fig:json-tree} \end{figure}

\cite{10.1145/3034786.3056120} propose that connected acyclic undirected graphs
which resembles a tree structure are a natural representation for JSON
documents as exemplified in \autoref{fig:json-tree}. \We use the following
definitions that define two features associated with the tree: \emph{height}
and \emph{level}.

\begin{definition} The height of a node is the number of edges on the longest
downward path between that node and a leaf.  The height of a tree is the height
of its root.  \end{definition}

\begin{definition} The level of a node is defined by 1 + the number of
connections between the node and the root. The level is depth + 1.
\end{definition}

Using the definition of height, \we extrapolate that the height of the tree
determines the height of a given JSON document. Using the definition of level,
\we extrapolate that the \emph{size} of a level in the tree equals the sum of
the byte-size of every textual, numeric, and boolean values whose nodes have
the corresponding level. Therefore, the \emph{largest level} is the level with
the highest size without taking into account the subtree at depth 0.

The \emph{nesting weight} of a JSON document $J$, referred to as $N_J$, is
defined as the product of its height and largest level minus 1. \We do not
consider the byte-size overhead introduced by compositional structures (object
and array) in the JSON document as \we found that it is highly correlated to
its nesting characteristics.

\begin{equation}
N_J = \text{height} \times \text{largest level} - 1
\end{equation}

\begin{figure}[ht!]
  \frame{\includegraphics[width=\linewidth]{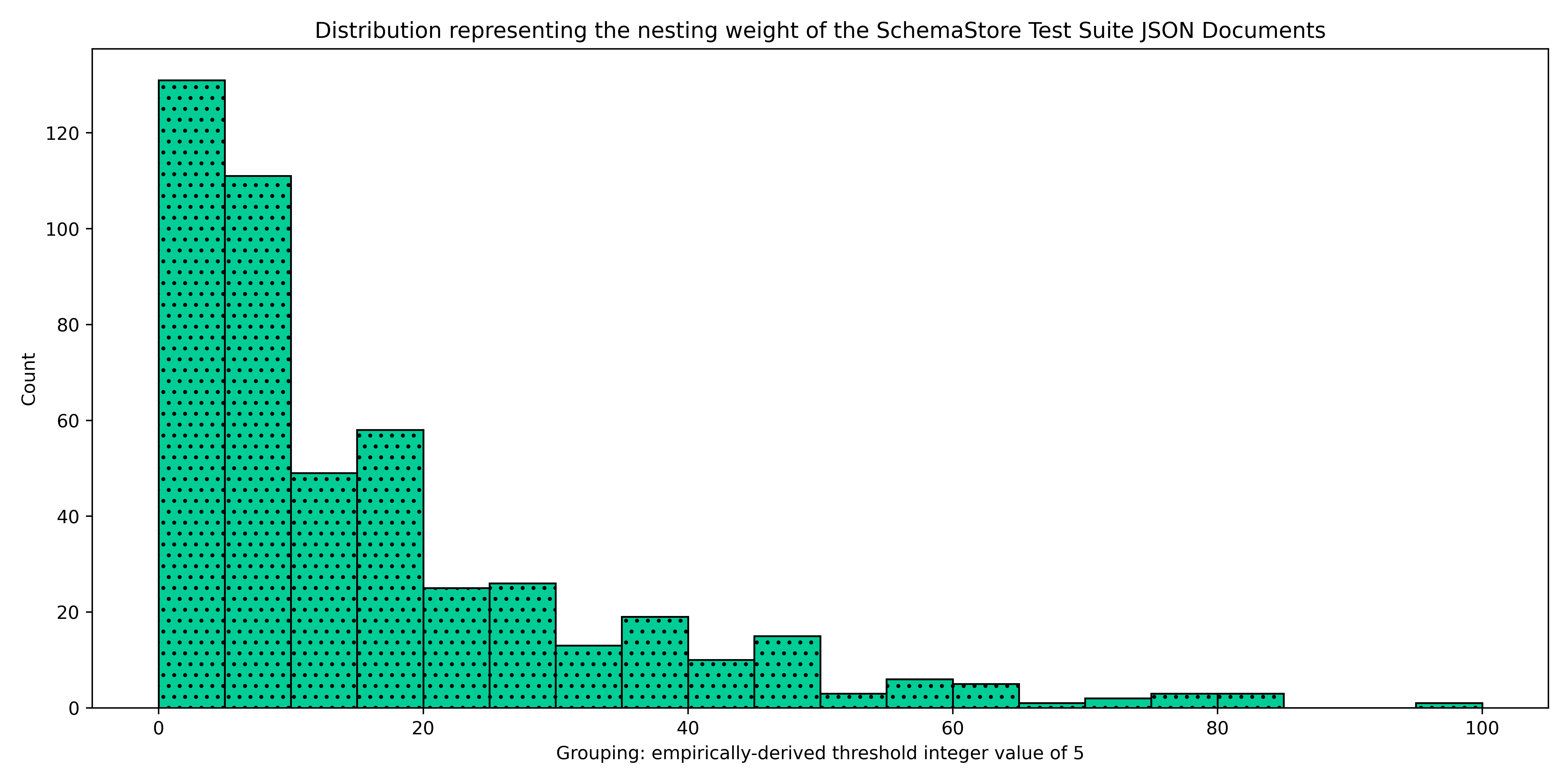}}
  \caption{The nesting weight distribution of the 480 JSON documents present in
  the SchemaStore test suite introduced in \autoref{sec:schemastore} grouped by
  the empirically-derived threshold 5. Most JSON documents have a nesting
  weight of under 20. However, there are JSON documents with a nesting weight
  of up to 100
\protect\footnote{\url{https://github.com/SchemaStore/schemastore/blob/0b6bd2a08005e6f7a65a68acaf3064d6e2670872/src/test/cloudify/utilities-cloudinit-simple.json}}.}
\label{fig:schemastore-nesting-weight-distribution} \end{figure}

In order to categorize JSON documents in a sensible manner, the taxonomy
distinguishes between \emph{flat} JSON documents and \emph{nested} JSON
documents. The nesting weight distribution of the JSON documents in the
SchemaStore test suite introduced in \autoref{sec:schemastore} is computed in
\autoref{fig:schemastore-nesting-weight-distribution}. Using these results,
this taxonomy aspect is defined as follows:

\begin{itemize}

\item \textbf{Flat.} A JSON document $J$ is \emph{flat} if $N_J$ less than the
  empirically-derived threshold integer value 10.

\item \textbf{Nested.} A JSON document $J$ is \emph{nested} if $N_J$ is greater
  than or equal to the empirically-derived threshold integer value 10.

\end{itemize}

\subsection{Demonstration}

\begin{figure}[ht!]
  \frame{\includegraphics[width=\linewidth]{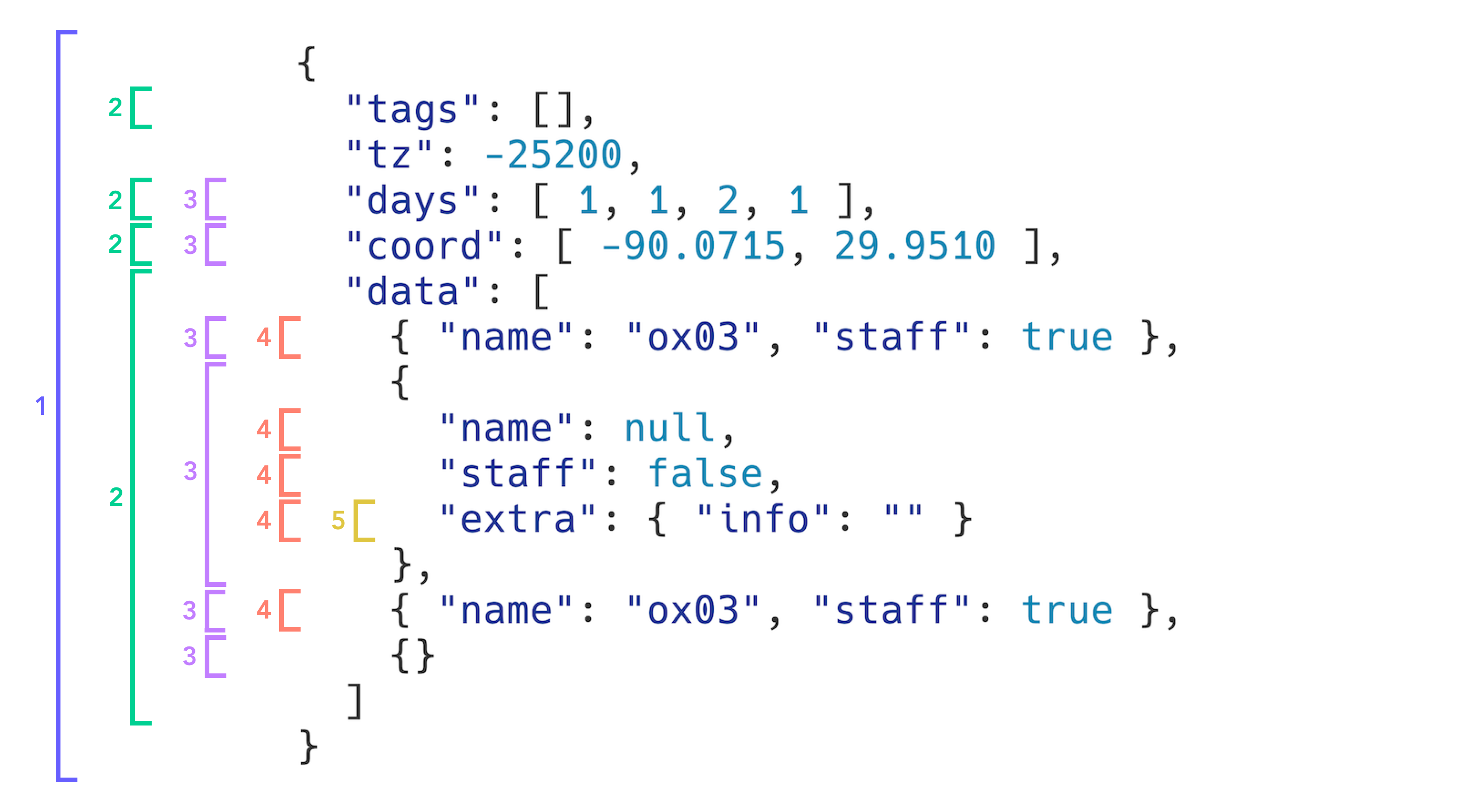}}
  \caption{An example \emph{Tier 2 Minified $\geq$ 100 $<$ 1000 bytes, numeric,
  non-redundant, and nested} JSON document taken from \our previous work
\cite{viotti2022survey}. The annotations at the left highlight each level in the JSON
document. The height of this document is $(5 - 1 = 4)$ (the highest level minus
1).} \label{fig:taxonomy-example} \end{figure}

To demonstrate \our conceptual taxonomy, \we apply it to the JSON document
listed in \autoref{fig:taxonomy-example}.
\autoref{table:taxonomy-example-breakdown} provides a breakdown of this JSON
document including information such as the number of edges and byte size of
every valid JSON Pointer path \cite{RFC6901}.  This document is a \emph{Tier 2
Minified $\geq$ 100 $<$ 1000 bytes}, \emph{numeric}, \emph{non-redundant}, and
\emph{nested} (SNNN according to \autoref{table:json-taxonomy}) JSON document:

\begin{itemize}

  \item \textbf{Tier 2 Minified $\geq$ 100 $<$ 1000 bytes.} The size of the
    JSON document is 184 bytes.  184 is greater than 100 but less than 1000,
    therefore the JSON document from \autoref{fig:taxonomy-example} is
    \emph{Tier 2 Minified $\geq$ 100 $<$ 1000 bytes} according to the taxonomy.

  \item \textbf{Numeric.} \autoref{table:taxonomy-example-breakdown} shows that
    the JSON document has 24 values corresponding to its set of valid JSON
    Pointer \cite{RFC6901} paths. Of those, 7 (29.16\%) are numeric, 3 (12.5\%)
    are textual, and 4 (16.66\%) are boolean. Out of the 184 total bytes from
    the JSON document, 24 bytes (13.04\%) correspond to numeric values, 14
    bytes (7.60\%) correspond to textual values, and 17 bytes (9.23\%)
    correspond to boolean values. The numeric weight is $29.16 \times 13.04 /
    100 = 3.80$, the textual weight is $12.5 \times 7.60 / 100 = 0.95$, and the
    boolean weight is $16.66 \times 9.23 / 100 = 1.53$. $3.80$ is greater than
    $0.95$ and $1.53$, therefore the JSON document from
    \autoref{fig:taxonomy-example} is \emph{numeric} according to the taxonomy.

  \item \textbf{Non-Redundant.} The JSON document consists of 24 values. Out of
    those, the numeric value 1 appears in the JSON Pointer \cite{RFC6901} paths
    \texttt{/days/0}, \texttt{/days/1}, and \texttt{/days/2}. The textual value
    \emph{ox03} appears at \texttt{/data/0/name} and \texttt{/data/2/name}.
    Similarly, the boolean value \emph{true} appears at \texttt{/data/0/staff}
    and \texttt{/data/2/staff}. Furthermore, the objects \texttt{/data/0} and
    \texttt{/data/2} are equal. Therefore, only 19 out of the 24 values in the
    JSON document are unique. \We conclude that only 5 (20.83\%) of its values
    are redundant, so the JSON document from \autoref{fig:taxonomy-example} is
    \emph{non-redundant} according to the taxonomy.

  \item \textbf{Nested.} The height is 4, awarded to the pointer
    \texttt{/data/1/extra/info}. \We calculate the byte-size of each level by
    adding the byte-size of each non-structural value in such level. Level 2
    occupies 6 bytes, level 3 occupies 18 bytes, level 4 occupies 29 bytes, and
    level 5 occupies 2 bytes, so level 4 is the largest level. The nesting
    weight of the JSON document is $4 \times (4 - 1) = 12$ (the height
    multiplied by the largest level minus 1). 12 is greater than 10, therefore
    the JSON document from \autoref{fig:taxonomy-example} is \emph{nested}
    according to the taxonomy.

\end{itemize}

\begin{table*}[ht!]

\caption{A breakdown of the JSON document from \autoref{fig:taxonomy-example}
  in terms of its valid JSON Pointer \cite{RFC6901} paths, value type, level
  , byte-size, and redundancy.}

\label{table:taxonomy-example-breakdown}
\begin{tabularx}{\linewidth}{l|X|X|X|X}
  \toprule
  \textbf{JSON Pointer} & \textbf{Type} & \textbf{Level} & \textbf{Byte-size} & \textbf{Same As} \\
  \midrule

\texttt{/}                  & Structural & 1 & 184 &                                    \\ \hline
\texttt{/tags}              & Structural & 2 & 2   &                                    \\ \hline
\texttt{/tz}                & Numeric    & 2 & 6   &                                    \\ \hline
\texttt{/days}              & Structural & 2 & 9   &                                    \\ \hline
\texttt{/days/0}            & Numeric    & 3 & 1   & \texttt{/days/1}, \texttt{/days/3} \\ \hline
\texttt{/days/1}            & Numeric    & 3 & 1   & \texttt{/days/0}, \texttt{/days/3} \\ \hline
\texttt{/days/2}            & Numeric    & 3 & 1   &                                    \\ \hline
\texttt{/days/3}            & Numeric    & 3 & 1   & \texttt{/days/0}, \texttt{/days/1} \\ \hline
\texttt{/coord}             & Structural & 2 & 17  &                                    \\ \hline
\texttt{/coord/0}           & Numeric    & 3 & 8   &                                    \\ \hline
\texttt{/coord/1}           & Numeric    & 3 & 6   &                                    \\ \hline
\texttt{/data}              & Structural & 2 & 110 &                                    \\ \hline
\texttt{/data/0}            & Structural & 3 & 28  & \texttt{/data/2}                   \\ \hline
\texttt{/data/0/name}       & Textual    & 4 & 6   & \texttt{/data/2/name}              \\ \hline
\texttt{/data/0/staff}      & Boolean    & 4 & 4   & \texttt{/data/2/staff}             \\ \hline
\texttt{/data/1}            & Structural & 3 & 47  &                                    \\ \hline
\texttt{/data/1/name}       & Boolean    & 4 & 4   &                                    \\ \hline
\texttt{/data/1/staff}      & Boolean    & 4 & 5   &                                    \\ \hline
\texttt{/data/1/extra}      & Structural & 4 & 11  &                                    \\ \hline
\texttt{/data/1/extra/info} & Textual    & 5 & 2   &                                    \\ \hline
\texttt{/data/2}            & Structural & 3 & 28  & \texttt{/data/0}                   \\ \hline
\texttt{/data/2/name}       & Textual    & 4 & 6   & \texttt{/data/0/name}              \\ \hline
\texttt{/data/2/staff}      & Boolean    & 4 & 4   & \texttt{/data/0/staff}             \\ \hline
\texttt{/data/3}            & Structural & 3 & 2   &  \\

  \bottomrule
\end{tabularx}
\end{table*}

\subsection{JSON Stats Analyzer}
\label{sec:benchmark-json-stats}

\begin{figure}[ht!]
  \frame{\includegraphics[width=\linewidth]{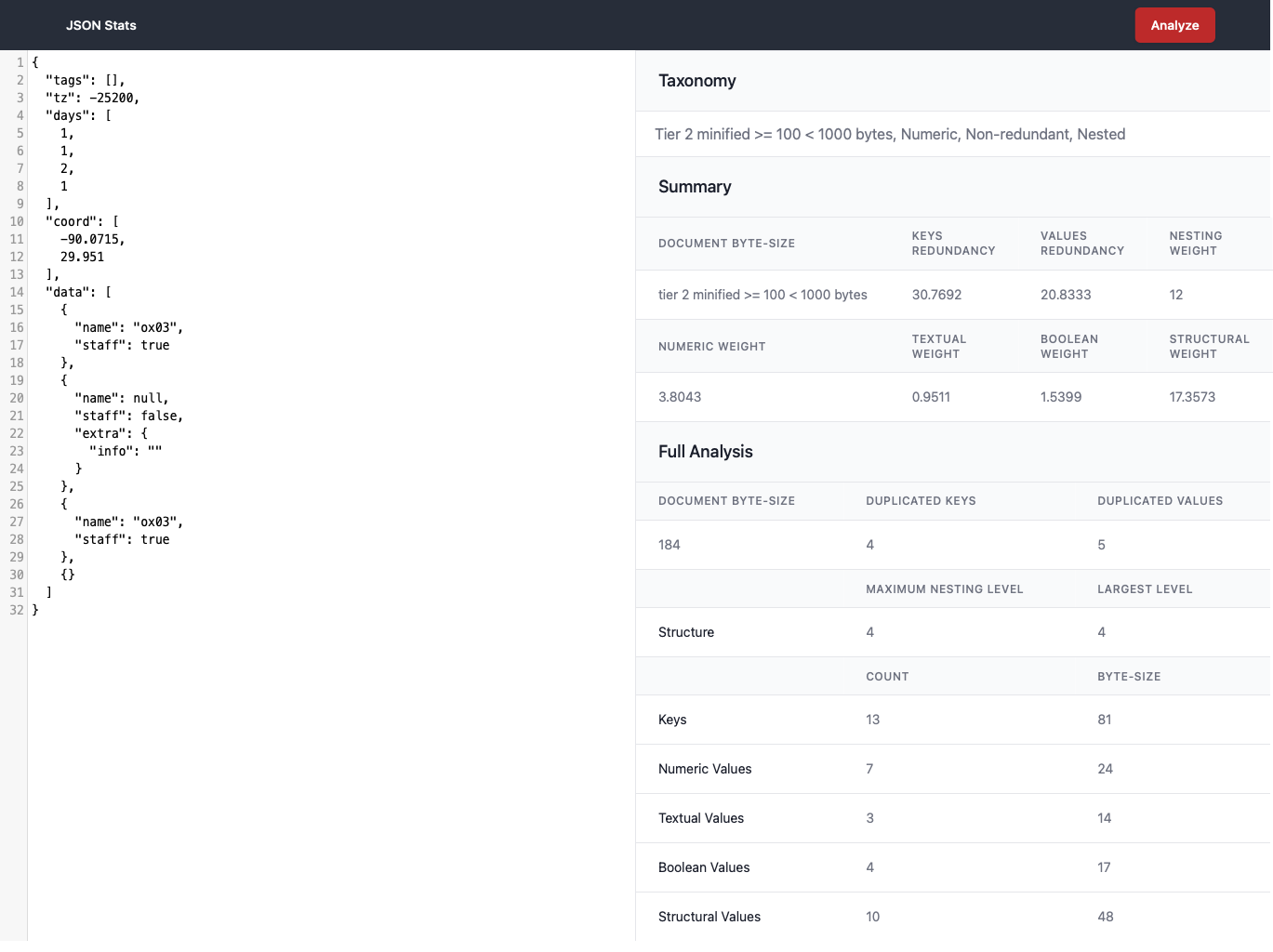}} \caption{A
  screenshot of the online tool published at
  \url{https://www.jsonbinpack.org/stats/} analyzing the JSON document listed
  in \autoref{fig:taxonomy-example}. } \label{fig:json-stats-screenshot}
\end{figure}

\We built and published a free-to-use online tool at
\url{https://www.jsonbinpack.org/stats/} to automatically categorize JSON
documents according to the taxonomy defined in this section and provide summary
statistics.  \autoref{fig:json-stats-screenshot} demonstrates the summary
statistics analyzed for the \emph{Tier 2 Minified $\geq$ 100 $<$ 1000 bytes,
numeric, non-redundant, and nested} JSON document from
\autoref{fig:taxonomy-example}.

The tool is developed using the TypeScript
\footnote{\url{https://www.typescriptlang.org}} programming language, the
CodeMirror \footnote{\url{https://codemirror.net}} open-source embeddable web
editor, and the Tailwind CSS \footnote{\url{https://tailwindcss.com}}
open-source web component framework. The web application is deployed to the
GitHub Pages \footnote{\url{https://pages.github.com}} free static-hosting
service.


\clearpage

\textbf{Discussion.} The document under analysis in
\autoref{fig:json-stats-screenshot} is displayed in the embedded text editor on
the left side of the screen. The analysis results are present on the right side
of the screen and are generated after pressing the red \emph{Analyze} button on
the top right corner. The \emph{Taxonomy} section of the analysis table shows
that the document is a \emph{Tier 2 Minified $\geq$ 100 $<$ 1000 bytes,
numeric, non-redundant, and nested} document according to the taxonomy. The
\emph{Summary} section shows the intermediary results of the size, content
type, redundancy and structural statistics introduced in
\autoref{sec:taxonomy-size}, \autoref{sec:taxonomy-content-type},
\autoref{sec:taxonomy-redundancy} and \autoref{sec:taxonomy-structure},
respectively. The \emph{Full Analysis} section shows all the intermediary
values used throughout every calculation.

\clearpage
\section{Methodology}
\label{sec:benchmark-methodology}

\Our approach to extend the body of literature through a space-efficiency
benchmark of JSON-compatible binary serialization specifications is based on
the following methodology:

\begin{enumerate}

\item \textbf{Input Data.} Select a representative set of real-world JSON
  \cite{ECMA-404} documents across industries according to the taxonomy defined
    in \autoref{sec:taxonomy}.

\item \textbf{Serialization Specifications.} Drawing on research from \cite{viotti2022survey},
  list the set of JSON-compatible schema-less and schema-driven binary
    serialization specifications to be benchmarked along with their respective
    encodings and implementations.

\item \textbf{Compression Formats.} Select a set of popular lossless data
  compression formats along with their respective implementations. These
    compression formats will be used to compress the input JSON \cite{ECMA-404}
    documents and bit-strings generated by the selection of binary
    serialization formats.

\item \textbf{Schema Definitions.} Write schema definitions for each
  combination of input JSON \cite{ECMA-404} document and selected schema-driven
    binary serialization specification.

\item \textbf{Benchmark.} Serialize each JSON \cite{ECMA-404} document using
  the selection of binary serialization specifications.  Then, deserialize the
    bit-strings and compare them to the original JSON \cite{ECMA-404} documents
    to test that there is no accidental loss of information.

\item \textbf{Results.} Measure the byte-size of the JSON \cite{ECMA-404}
  documents and bit-strings generated by each binary serialization
    specification in uncompressed and compressed form using the selection of
    data compression formats.

\item \textbf{Conclusions.} Discuss the results to identify space-efficient
  JSON-compatible binary serialization specifications and the role of data
    compression in increasing space-efficiency of JSON \cite{ECMA-404}
    documents.

\end{enumerate}

\subsection{Input Data}
\label{sec:benchmark-methodology-input-data}

\begin{figure*}[ht!]
  \frame{\includegraphics[width=\linewidth]{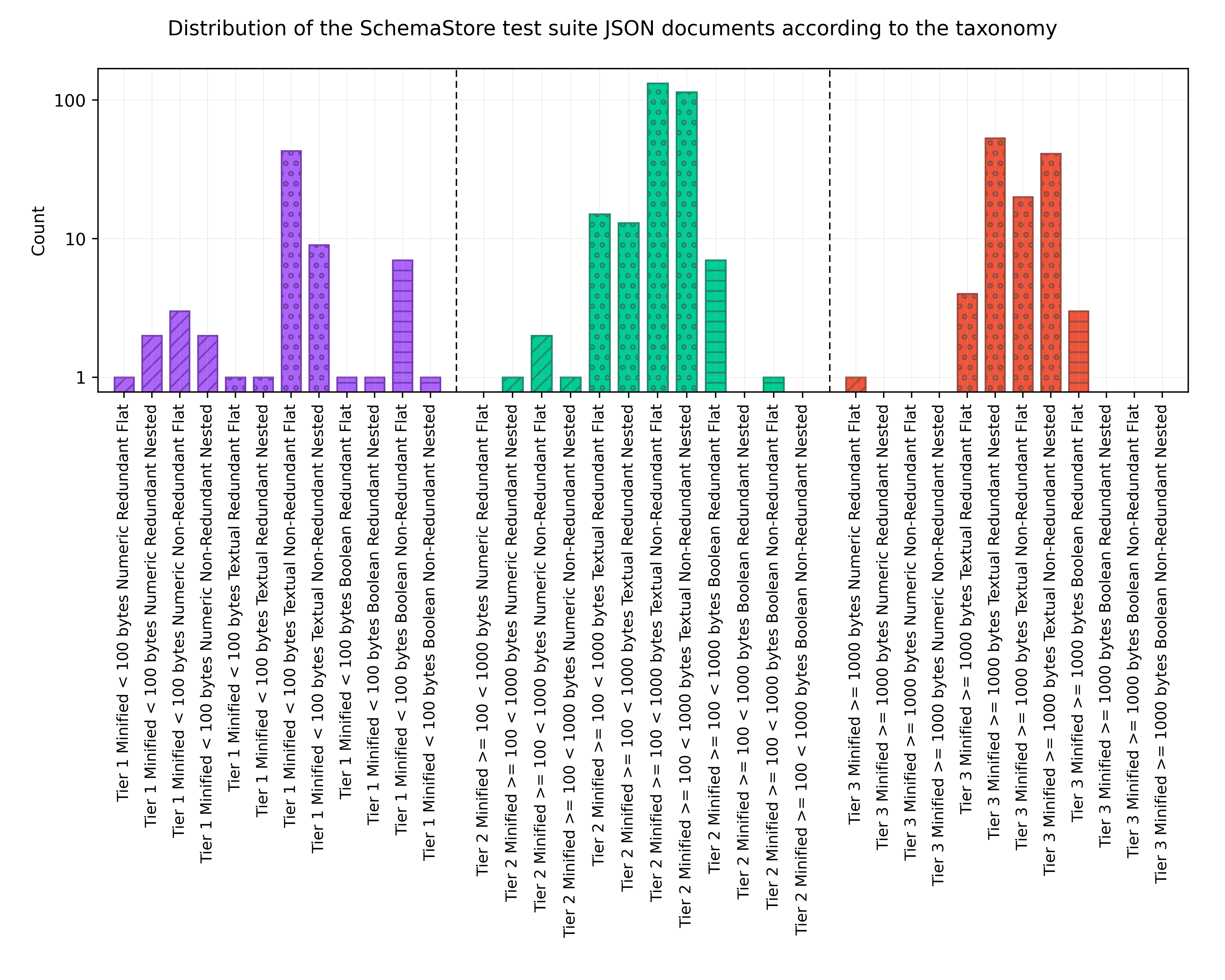}}
  \caption{There are 480 JSON \cite{ECMA-404} documents present in the
  SchemaStore test suite.  These are grouped according to the taxonomy defined
  in \autoref{sec:taxonomy} using a logarithmic y-scale. }
\label{fig:schemastore-taxonomy} \end{figure*}

\autoref{fig:schemastore-taxonomy} categorizes the JSON \cite{ECMA-404}
documents from the SchemaStore test suite introduced in
\autoref{sec:schemastore} according to the taxonomy defined in
\autoref{sec:taxonomy}. The SchemaStore test suite does not contain JSON
\cite{ECMA-404} documents that match 9 out of the 36 categories defined in the
taxonomy, particularly in the \emph{Tier 3 Minified $\geq$ 1000 bytes} size
category which is dominated by \emph{textual} JSON documents.  \We embrace
these results to conclude that the missing categories do not represent
instances of JSON \cite{ECMA-404} documents that are commonly encountered in
practice. The missing categories are the following:

\begin{itemize}

\item Tier 2 Minified $\geq$ 100 $<$ 1000 bytes Numeric Redundant Flat (SNRF)
\item Tier 2 Minified $\geq$ 100 $<$ 1000 bytes Boolean Redundant Nested (SBRN)
\item Tier 2 Minified $\geq$ 100 $<$ 1000 bytes Boolean Non-Redundant Nested (SBNN)
\item Tier 3 Minified $\geq$ 1000 bytes Numeric Redundant Nested (LNRN)
\item Tier 3 Minified $\geq$ 1000 bytes Numeric Non-Redundant Flat (LNNF)
\item Tier 3 Minified $\geq$ 1000 bytes Numeric Non-Redundant Nested (LNNN)
\item Tier 3 Minified $\geq$ 1000 bytes Boolean Redundant Nested (LBRN)
\item Tier 3 Minified $\geq$ 1000 bytes Boolean Non-Redundant Flat (LBNF)
\item Tier 3 Minified $\geq$ 1000 bytes Boolean Non-Redundant Nested (LBNN)

\end{itemize}

\We selected a single JSON \cite{ECMA-404} document from each matching
category. The selection of JSON documents is listed in
\autoref{table:benchmark-documents} and \autoref{table:benchmark-documents-1}.
Some JSON \cite{ECMA-404} documents \we selected from the SchemaStore test
suite, namely \emph{Entry Point Regulation manifest} and \emph{.NET Core
project.json}, include a top level \emph{\$schema} string property that is not
consider in the benchmark. The use of this keyword is a non-standard approach
to make JSON \cite{ECMA-404} documents reference their own JSON Schema
\cite{jsonschema-core-2020} definitions. This keyword is not defined as part of
the formats that these JSON documents represent in the SchemaStore dataset.

\begin{table*}[ht!]

  \caption{The JSON \cite{ECMA-404} documents selected from the SchemaStore
  test suite introduced in \autoref{sec:schemastore} divided by industry. Each
  JSON document matches a different taxonomy category defined in
  \autoref{sec:taxonomy}. The first column consists of a brief description of
  the JSON document. The second column contains the path and link to the test
  case file within the SchemaStore repository. The third column contains the
  taxonomy categories using the acronyms defined in
  \autoref{table:json-taxonomy}. This table is continued in
  \autoref{table:benchmark-documents-1}.}

\label{table:benchmark-documents}
\begin{tabularx}{\linewidth}{X|l|l}
\toprule
\textbf{Description} & \textbf{Test Case Name} & \textbf{Category} \\
\midrule

\multicolumn{3}{c}{\textbf{Continuous Integration / Continuous Deliver (CI/CD)}} \\ \hline

JSON-e templating engine sort example & {\small \href{https://github.com/jviotti/binary-json-size-benchmark/tree/main/benchmark/jsonesort/document.json}{\texttt{jsone/sort.json}}} & TNRF \\ \hline
JSON-e templating engine reverse sort example & {\small \href{https://github.com/jviotti/binary-json-size-benchmark/tree/main/benchmark/jsonereversesort/document.json}{\texttt{jsone/reverse-sort.json}}} & TNRN \\ \hline
CircleCI definition (blank) & {\small \href{https://github.com/jviotti/binary-json-size-benchmark/tree/main/benchmark/circleciblank/document.json}{\texttt{circleciconfig/version-2.0.json}}} & TNNF \\ \hline
CircleCI matrix definition & {\small \href{https://github.com/jviotti/binary-json-size-benchmark/tree/main/benchmark/circlecimatrix/document.json}{\texttt{circleciconfig/matrix-simple.json}}} & TNNN \\ \hline
SAP Cloud SDK Continuous Delivery Toolkit configuration & {\small \href{https://github.com/jviotti/binary-json-size-benchmark/tree/main/benchmark/sapcloudsdkpipeline/document.json}{\texttt{cloud-sdk-pipeline-config-schema/empty.json}}} & TBRF \\ \hline
TravisCI notifications configuration & {\small \href{https://github.com/jviotti/binary-json-size-benchmark/tree/main/benchmark/travisnotifications/document.json}{\texttt{travis/notification-secure.json}}} & STRF \\ \hline
GitHub Workflow Definition & {\small \href{https://github.com/jviotti/binary-json-size-benchmark/tree/main/benchmark/githubworkflow/document.json}{\texttt{github-workflow/919.json}}} & STNN \\ \hline

\multicolumn{3}{c}{\textbf{Software Engineering}} \\ \hline

Grunt.js "clean" task definition & {\small \href{https://github.com/jviotti/binary-json-size-benchmark/tree/main/benchmark/gruntcontribclean/document.json}{\texttt{grunt-clean-task/with-options.json}}} & TTRF \\ \hline
CommitLint configuration & {\small \href{https://github.com/jviotti/binary-json-size-benchmark/tree/main/benchmark/commitlint/document.json}{\texttt{commitlintrc/commitlintrc-test5.json}}} & TTRN \\ \hline
TSLint linter definition (extends only) & {\small \href{https://github.com/jviotti/binary-json-size-benchmark/tree/main/benchmark/tslintextend/document.json}{\texttt{tslint/tslint-test5.json}}} & TTNF \\ \hline
TSLint linter definition (multi-rule) & {\small \href{https://github.com/jviotti/binary-json-size-benchmark/tree/main/benchmark/tslintmulti/document.json}{\texttt{tslint/tslint-test25.json}}} & TBRN \\ \hline
CommitLint configuration (basic) & {\small \href{https://github.com/jviotti/binary-json-size-benchmark/tree/main/benchmark/commitlintbasic/document.json}{\texttt{commitlintrc/commitlintrc-test3.json}}} & TBNF \\ \hline
TSLint linter definition (basic) & {\small \href{https://github.com/jviotti/binary-json-size-benchmark/tree/main/benchmark/tslintbasic/document.json}{\texttt{tslint/tslint-test19.json}}} & TBNN \\ \hline
ESLint configuration document & {\small \href{https://github.com/jviotti/binary-json-size-benchmark/tree/main/benchmark/eslintrc/document.json}{\texttt{eslintrc/WebAnalyzer.json}}} & LNRF \\ \hline
NPM Package.json Linter configuration manifest & {\small \href{https://github.com/jviotti/binary-json-size-benchmark/tree/main/benchmark/packagejsonlintrc/document.json}{\texttt{npmpackagejsonlintrc/npmpackagejsonlintrc-test.json}}} & LTRF \\ \hline
.NET Core project.json & {\small \href{https://github.com/jviotti/binary-json-size-benchmark/tree/main/benchmark/netcoreproject/document.json}{\texttt{project/EF-project.json}}} & LTRN \\ \hline
NPM Package.json example manifest & {\small \href{https://github.com/jviotti/binary-json-size-benchmark/tree/main/benchmark/packagejson/document.json}{\texttt{package/package-test.json}}} & LTNF \\

\bottomrule
\end{tabularx}
\end{table*}

\begin{table*}[ht!]

\caption{Continuation of \autoref{table:benchmark-documents}.}

\label{table:benchmark-documents-1}
\begin{tabularx}{\linewidth}{X|l|l}
\toprule
\textbf{Description} & \textbf{Test Case Name} & \textbf{Category} \\
\midrule

\multicolumn{3}{c}{\textbf{Web}} \\ \hline

ImageOptimizer Azure Webjob configuration & {\small \href{https://github.com/jviotti/binary-json-size-benchmark/tree/main/benchmark/imageoptimizerwebjob/document.json}{\texttt{imageoptimizer/default.json}}} & TTNN \\ \hline
Entry Point Regulation manifest & {\small \href{https://github.com/jviotti/binary-json-size-benchmark/tree/main/benchmark/epr/document.json}{\texttt{epr-manifest/official-example.json}}} & STRN \\ \hline
ECMAScript module loader definition & {\small \href{https://github.com/jviotti/binary-json-size-benchmark/tree/main/benchmark/esmrc/document.json}{\texttt{esmrc/.esmrc\_.json}}} & SBNF \\ \hline
Nightwatch.js Test Framework Configuration & {\small \href{https://github.com/jviotti/binary-json-size-benchmark/tree/main/benchmark/nightwatch/document.json}{\texttt{nightwatch/default.json}}} & LBRF \\ \hline

\multicolumn{3}{c}{\textbf{Geospatial}} \\ \hline

GeoJSON example JSON document & {\small \href{https://github.com/jviotti/binary-json-size-benchmark/tree/main/benchmark/geojson/document.json}{\texttt{geojson/multi-polygon.json}}} & SNRN \\ \hline

\multicolumn{3}{c}{\textbf{Weather}} \\ \hline

OpenWeatherMap API example JSON document & {\small \href{https://github.com/jviotti/binary-json-size-benchmark/tree/main/benchmark/openweathermap/document.json}{\texttt{openweather.current/example.json}}} & SNNF \\ \hline
OpenWeather Road Risk API example & {\small \href{https://github.com/jviotti/binary-json-size-benchmark/tree/main/benchmark/openweatherroadrisk/document.json}{\texttt{openweather.roadrisk/example.json}}} & SNNN \\ \hline

\multicolumn{3}{c}{\textbf{Publishing}} \\ \hline

JSON Feed example document & {\small \href{https://github.com/jviotti/binary-json-size-benchmark/tree/main/benchmark/jsonfeed/document.json}{\texttt{feed/microblog.json}}} & STNF \\ \hline

\multicolumn{3}{c}{\textbf{Open-Source}} \\ \hline

GitHub FUNDING sponsorship definition (empty) & {\small \href{https://github.com/jviotti/binary-json-size-benchmark/tree/main/benchmark/githubfundingblank/document.json}{\texttt{github-funding/ebookfoundation.json}}} & SBRF \\ \hline

\multicolumn{3}{c}{\textbf{Recruitment}} \\ \hline

JSON Resume & {\small \href{https://github.com/jviotti/binary-json-size-benchmark/tree/main/benchmark/jsonresume/document.json}{\texttt{resume/richardhendriks.json}}} & LTNN \\
\bottomrule
\end{tabularx}
\end{table*}

\clearpage

\subsection{Serialization Specifications}
\label{sec:benchmark-specifications}

The selection of schema-driven and schema-less JSON-compatible binary
serialization specifications is listed in
\autoref{table:benchmark-specifications-schema-driven} and
\autoref{table:benchmark-specifications-schema-less}. In comparison to \our
previous work \cite{viotti2022survey}, \we use ASN-1Step 10.0.2 instead of 10.0.1, Microsoft
Bond \cite{microsoft-bond} 9.0.4 instead of 9.0.3, and Protocol Buffers
\cite{protocolbuffers} 3.15.3 instead of 3.13.0. None of these version upgrades
involve changes to the encodings.  Furthermore, \we replaced the third-party
BSON \cite{bson} Python implementation used in \cite{viotti2022survey} with the Node.js
official MongoDB implementation. \We also replaced the Smile \cite{smile}
Python implementation used in \cite{viotti2022survey} with a Clojure implementation as \we
identified issues in the former implementation with respects to floating-point
numbers. For example, encoding the floating-point number 282.55 results in
282.549988 when using \texttt{pysmile} v0.2.

Finally, both the binary and the packed encoding provided by Cap'n Proto
\cite{capnproto} are considered. As described in \cite{viotti2022survey}, the packed encoding
consists of a basic data compression format officially supported as a separate
encoding. These encodings are separately considered to understand the impact of
general-purpose data compression on the uncompressed Cap'n Proto
\cite{capnproto} variant.

\begin{table}[hb!]
\caption{The selection of schema-driven JSON-compatible binary serialization specifications based on \our previous work \cite{viotti2022survey}.}
\label{table:benchmark-specifications-schema-driven}
\begin{tabularx}{\linewidth}{l|X|X|l}
\toprule
\textbf{Specification} & \textbf{Implementation} & \textbf{Encoding} & \textbf{License} \\
\midrule

ASN.1            & OSS ASN-1Step Version 10.0.2               & PER Unaligned \cite{asn1-per}                 & Proprietary \\ \hline
Apache Avro      & Python \texttt{avro} (pip) 1.10.0          & Binary Encoding \footnotemark with no framing & Apache-2.0 \\ \hline
Microsoft Bond   & C++ library 9.0.4                          & Compact Binary v1 \footnotemark               & MIT \\ \hline
Cap'n Proto      & \texttt{capnp} command-line tool 0.8.0     & Binary Encoding \footnotemark                 & MIT \\ \hline
Cap'n Proto      & \texttt{capnp} command-line tool 0.8.0     & Packed Encoding \footnotemark                 & MIT \\ \hline
FlatBuffers      & \texttt{flatc} command-line tool 1.12.0    & Binary Wire Format \footnotemark              & Apache-2.0 \\ \hline
Protocol Buffers & Python \texttt{protobuf} (pip) 3.15.3      & Binary Wire Format \footnotemark              & 3-Clause BSD \\ \hline
Apache Thrift    & Python \texttt{thrift} (pip) 0.13.0        & Compact Protocol \footnotemark                & Apache-2.0 \\

\bottomrule
\end{tabularx}
\end{table}

\footnotetext[\numexpr\thefootnote-6]{\url{https://avro.apache.org/docs/current/spec.html\#binary\_encoding}}
\footnotetext[\numexpr\thefootnote-5]{\url{https://microsoft.github.io/bond/reference/cpp/compact\_\_binary\_8h\_source.html}}
\footnotetext[\numexpr\thefootnote-4]{\url{https://capnproto.org/encoding.html\#packing}}
\footnotetext[\numexpr\thefootnote-3]{\url{https://capnproto.org/encoding.html}}
\footnotetext[\numexpr\thefootnote-2]{\url{https://google.github.io/flatbuffers/flatbuffers\_internals.html}}
\footnotetext[\numexpr\thefootnote-1]{\url{https://developers.google.com/protocol-buffers/docs/encoding}}
\footnotetext[\numexpr\thefootnote]{\url{https://github.com/apache/thrift/blob/master/doc/specs/thrift-compact-protocol.md}}

\begin{table}[hb!]
\caption{The selection of schema-less JSON-compatible binary serialization specifications based on \our previous work \cite{viotti2022survey}.}
\label{table:benchmark-specifications-schema-less}
\begin{tabularx}{\linewidth}{l|X|l}
\toprule
\textbf{Specification} & \textbf{Implementation} & \textbf{License} \\
\midrule
BSON        & Node.js \texttt{bson} (npm) 4.2.2                                      & Apache-2.0 \\ \hline
CBOR        & Python \texttt{cbor2} (pip) 5.1.2                                      & MIT \\ \hline
FlexBuffers & \texttt{flatc} command-line tool 1.12.0                                & Apache-2.0 \\ \hline
MessagePack & \texttt{json2msgpack} command-line tool 0.6 with \texttt{MPack} 0.9dev & MIT \\ \hline
Smile       & Clojure \texttt{cheshire} 5.10.0                                       & MIT \\ \hline
UBJSON      & Python \texttt{py-ubjson} (pip) 0.16.1                                 & Apache-2.0 \\
\bottomrule
\end{tabularx}
\end{table}

\ifx\thesis\undefined
\subsection{Schema Definitions}
\label{sec:schema-definitions}

For brevity, this paper does not include the schema definitions for each input
JSON document listed in \autoref{table:benchmark-documents} and
\autoref{table:benchmark-documents-1} for every selected serialization
specifications listed in
\autoref{table:benchmark-specifications-schema-driven}. The schema definitions
can be found on the GitHub repository implemented as part of the benchmark
study \footnote{\url{https://github.com/jviotti/binary-json-size-benchmark}}.
Direct links to the corresponding schema definitions are provided along with
the benchmark results for each JSON input document.  \fi

\subsection{Fair Benchmarking}
\label{sec:benchmark-testing}

In order to produce a fair benchmark, the resulting bit-strings are ensured to
be lossless encodings of the respective input JSON \cite{ECMA-404} documents.
For some binary serialization specifications such as Cap'n Proto
\cite{capnproto}, providing a schema that only describes a subset of the input
data will result in only such subset being serialized and the remaining of the
input data being silently discarded. In other cases, a serialization
specification may silently coerce an input data type to match the schema
definition even at the expense of loss of information.  For example, Protocol
Buffers \cite{protocolbuffers} may forcefully cast to IEEE 764 32-bit
floating-point encoding \cite{8766229} if requested by the schema even if the
input real number can only be represented without loss of precision by using
the IEEE 764 \emph{64-bit} floating-point encoding \cite{8766229}.

The implemented benchmark program prevents such accidental mistakes by
automatically ensuring that for each combination of serialization specification
listed in \autoref{sec:benchmark-specifications} and input JSON document listed
in \autoref{table:benchmark-documents} and
\autoref{table:benchmark-documents-1}, the produced bit-strings encode the same
information as the respective input JSON document.  The automated test consists
in serializing the input JSON document using a given binary serialization
specification, deserializing the resulting bit-string and asserting that the
original JSON document is strictly equal to the deserialized JSON document.

\subsection{Compression Formats}
\label{sec:benchmark-compression-formats}

\begin{figure*}[hb!]
  \frame{\includegraphics[width=\linewidth]{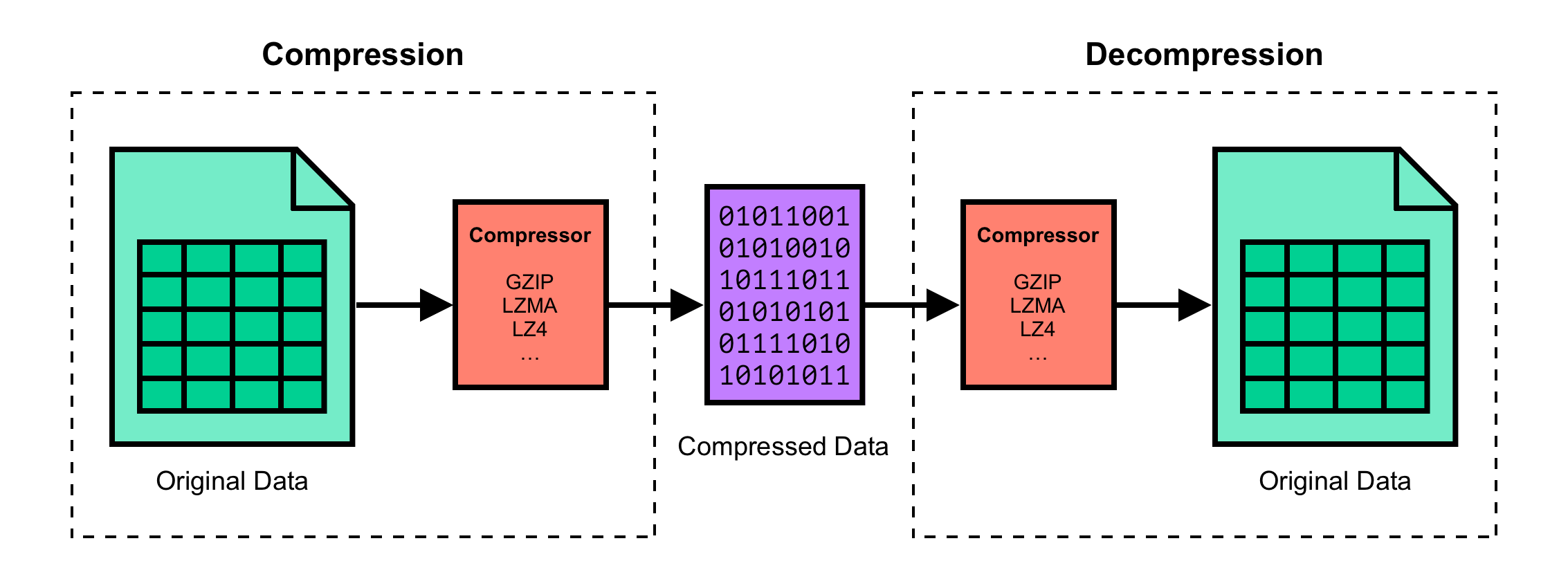}}
  \caption{A general-purpose lossless compressor provides a \emph{compression}
  and a \emph{decompression} process. Compression consists in transforming the
  original data into a compressed representation of the same information.
  Decompression reverses the compression process to obtain the unmodified
  original data from its compressed representation.}
\label{fig:benchmark-lossless-compression} \end{figure*}

\We selected the following data compression formats: GZIP (GNU ZIP)
\cite{RFC1952}, LZ4 \footnote{\url{https://lz4.github.io/lz4/}}, and
Lempel-Ziv-Markov Chain Algorithm (LZMA), a set of compressors commonly used in
the context of web services. These three formats are derived from the LZ77
(Lempel-Ziv) \cite{1055714} dictionary-based coding scheme and are considered
general-purpose lossless compressors \cite{JAYASANKAR2021119}. The LZ77
\cite{1055714} algorithm operates by deduplicating multiple occurrences of the
same data pattern within certain distance \cite{JAYASANKAR2021119}
\cite{10.1145/348751.348754}, a technique also discussed in more detail in
\cite{10.1145/322344.322346}.  According to \cite{10.1145/348751.348754}, the
compression ratios on textual data when using Lempel-Ziv-derived algorithm
ranges between 30\% and 40\% in practice.

\begin{table}[ht!]
\caption{A high-level view of the differences and similarities between the GZIP, LZ4 and LZMA lossless compression formats.}
\label{table:benchmark-compression-formats-comparison}
\begin{tabularx}{\linewidth}{l|X|X|X}
\toprule
\textbf{Compression Format} & \textbf{GZIP} & \textbf{LZ4} & \textbf{LZMA} \\
\midrule
\textbf{Differences}  & Efficient and constant memory usage & High compression and decompression speed & Better compression on large files \\ \hline
\textbf{Similarities} & \multicolumn{3}{l}{Based on LZ77 (Lempel-Ziv) \cite{1055714}} \\
\bottomrule
\end{tabularx}
\end{table}

\textbf{GZIP (GNU ZIP)} \cite{RFC1952} is an open-source compressor based on a
mixture of the LZ77 \cite{1055714} and the Huffman \cite{huffman1952method}
coding schemes. GZIP was developed as part of the GNU Project
\footnote{\url{https://gnu.org}} and released in 1992 as a replacement for the
UNIX \texttt{compress} \footnote{\url{https://ncompress.sourceforge.io}}
program. GZIP it is the most widely-used compression format for HTTP/1.1
\cite{RFC7231}.  \cite{10.1145/1718487.1718536} study the problem of
compressing large amounts of textual and highly-redundant data such as HTML
\cite{HTML5} documents and use GZIP as the reference compressor due to its
popularity. Their findings show that GZIP has been designed to have a small
memory footprint and operate in constant space complexity
\cite{10.1145/348751.348754}.  These characteristics are possible given that
GZIP splits the input data into small blocks of less than 1 MB and compresses
each block separately. As a drawback, this approach limits GZIP ability to
detect redundancy across blocks and reduces its space-efficiency on larger
input data.  \We consider this drawback to be irrelevant for this benchmark as
the largest JSON \cite{RFC8259} document present in the SchemaStore test suite
introduced in \autoref{sec:schemastore} weights \~0.5 MB as discussed in
\autoref{fig:schemastore-byte-size-distribution}.

\textbf{LZ4} is an open-source compressor developed by Yann Collet
\footnote{\url{https://github.com/Cyan4973}} while working at Facebook. LZ4 is
a derivative of LZ77 \cite{1055714}. LZ4 focuses on improving compression and
decompression speed by using a hash table data structure for storing reference
addresses. The hash table provides constant $\mathcal{O}(1)$ instead of linear
$\mathcal{O}(n)$ complexity for match detection \cite{7440278}. LZ4 is a core
part of the Zstandard data compression mechanism \cite{RFC8878}.  Zstandard is
one of the eight compressors, along with GZIP \cite{RFC1952}, that are part of
the IANA HTTP Content Coding Registry
\footnote{\url{https://www.iana.org/assignments/http-parameters/http-parameters.xhtml\#content-coding}}.
LZ4 is also the recommended data compression format for Cap'n Proto
\cite{capnproto} bit-strings
\footnote{\url{https://capnproto.org/encoding.html\#compression}}.

\textbf{LZMA} is an open-source compressor developed as part of the 7-Zip
project \footnote{\url{https://www.7-zip.org}}. LZMA offers high compression
ratios as observed by \cite{parekar2014lossless} and
\cite{10.1145/1718487.1718536}. LZMA is a dictionary-based compressor based on
LZ77 \cite{1055714} with support for dictionaries of up to 4 GB in size. As a
result, LZMA can detect redundancy across large portions of the input data. In
comparison to GZIP \cite{RFC1952}, \cite{10.1145/1718487.1718536} found LZMA to
be space-efficient when taking large files as input. Applying LZMA on their 50
GB and a 440 GB collection of web pages resulted in 4.85\% and 6.15\%
compression ratios compared to 20.35\% and 18.69\% compression ratios in the
case of GZIP. LZMA support is implemented in the Opera web browser
\footnote{\url{https://blogs.opera.com/desktop/changelog-for-31/}}. Official
builds of the Firefox web browser do not include LZMA support.  However, there
exist non-official patches
\footnote{\url{https://wiki.mozilla.org/LZMA2_Compression}} that can be used to
produce a build from source that includes LZMA support.

These data compression formats support multiple compression levels. We are
interested in examining the impact of data compression in the best possible
case, so \we choose the highest recommended compression level supported by each
format.  The implementations, versions and compression levels used for this
benchmark are listed in \autoref{table:benchmark-compression-formats}.

\begin{table}[ht!]
\caption{The selection of lossless data compression formats.}
\label{table:benchmark-compression-formats}
\begin{tabularx}{\linewidth}{l|X|l|X}
\toprule
\textbf{Format} & \textbf{Implementation} & \textbf{Compression Level} & \textbf{License} \\
\midrule
  GZIP  & Apple \texttt{gzip} 321.40.3 (based on FreeBSD \texttt{gzip} 20150113)   & 9 \protect\footnote{\url{https://www.unix.com/man-page/freebsd/0/gzip/}} & 2-Clause BSD \\ \hline
  LZ4   & \texttt{lz4} command-line tool v1.9.3                                    & 9 \protect\footnote{\url{https://man.archlinux.org/man/lz4.1}} & Mixed 2-Clause BSD and GPLv2 \\ \hline
  LZMA  & \texttt{xz} (XZ Utils) 5.2.5 with \texttt{liblzma} 5.2.5                 & 9 \protect\footnote{\url{http://manpages.org/xz}} & Mixed Public Domain and GNU LGPLv2.1+ \\
\bottomrule
\end{tabularx}
\end{table}

\subsection{System Specification}

The implementations of the selected serialization specifications were executed
on a MacBook Pro 13" Dual-Core Intel Core i5 2.9 GHz with 2 cores and 16 GB of
memory (model identifier \texttt{MacBookPro12,1}) running macOS Big Sur 11.2.3,
Xcode 12.4 (12D4e), clang 1200.0.32.29, GNU Make 3.81, Matplotlib 3.4.2, Awk
version 20200816, Python 3.9.2, Node.js 15.11.0, and Clojure 1.10.2.796.

\section{Benchmark}

\begin{figure*}[ht!]
\frame{\includegraphics[width=\linewidth]{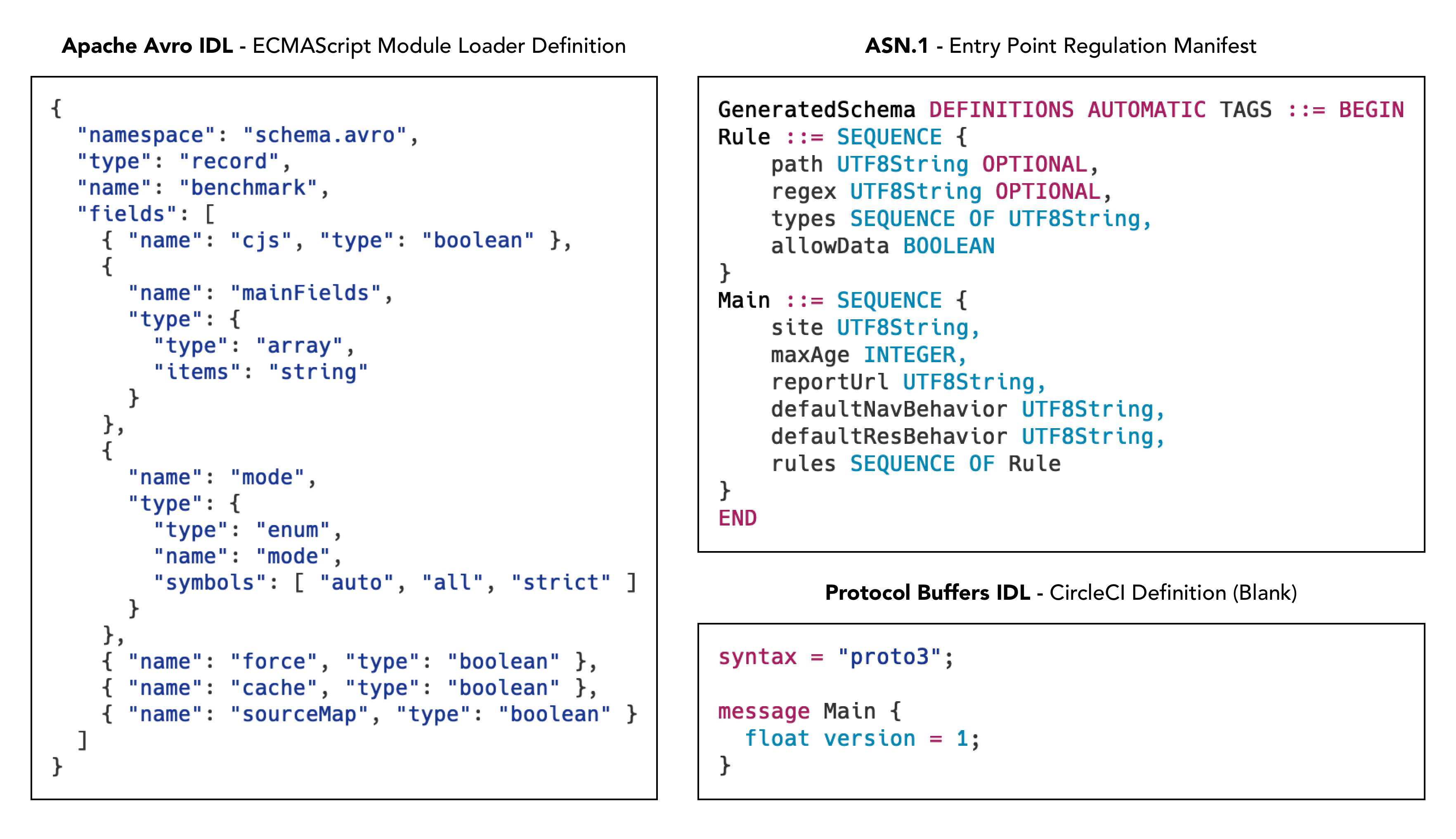}}
\caption{Examples of schema definitions written for 3 different JSON documents
using 3 different interface definition languages: an Apache Avro Interface
Definition Language (IDL) \cite{avro} schema definition for the JSON document
presented in \autoref{sec:benchmark-esmrc} (left), an ASN.1 \cite{asn1} schema
definition for the JSON document presented in \autoref{sec:benchmark-epr} (top
right) and a Protocol Buffers Interface Definition Language (IDL)
\cite{protocolbuffers} schema definition for the JSON document presented in
\autoref{sec:benchmark-circleciblank} (bottom right).}
\label{fig:benchmark-schemas-example} \end{figure*}

In this section, \we present bechmark results for 27 JSON \cite{ECMA-404}
document examples as introduced in \autoref{table:benchmark-documents} and
\autoref{table:benchmark-documents-1}.  \We wrote schema definitions for each
of the JSON documents for each of the 8 schema-driven binary serialization
specifications introduced in
\autoref{table:benchmark-specifications-schema-driven} using their respective
interface definition languages as exemplified in
\autoref{fig:benchmark-schemas-example}.

In \cite{viotti2022survey}, \we discussed that schema-driven serialization
specifications typically define custom schema languages that are not usable by
any other schema-driven serialization specifications rather than relying on
standardized schema languages such as JSON Schema \cite{jsonschema-core-2020}.
The selection of schema-driven binary serialization specifications implement
schema languages that are considered high-level as they abstractly describe
data structures and depend on the serialization specification implementation to
provide the corresponding encoding rules. While schema definitions are
typically written by hand, sometimes schemas are auto-generated from other
schema languages
\footnote{\url{https://github.com/okdistribute/jsonschema-protobuf}} or
inferred from the data \cite{8424731} \cite{mci/Klettke2015}
\cite{baazizi2019parametric} \cite{CANOVASIZQUIERDO201652}
\cite{10.1145/2187980.2188227} \cite{10.1007/978-3-319-61482-3_16}
\cite{10.1007/978-3-642-39200-9_8} \cite{svoboda2020json}.

\clearpage

\label{sec:benchmark-data}

\subsection{JSON-e Templating Engine Sort Example}
\label{sec:benchmark-jsonesort}

\begin{figure*}[ht!]
\frame{\includegraphics[width=\linewidth]{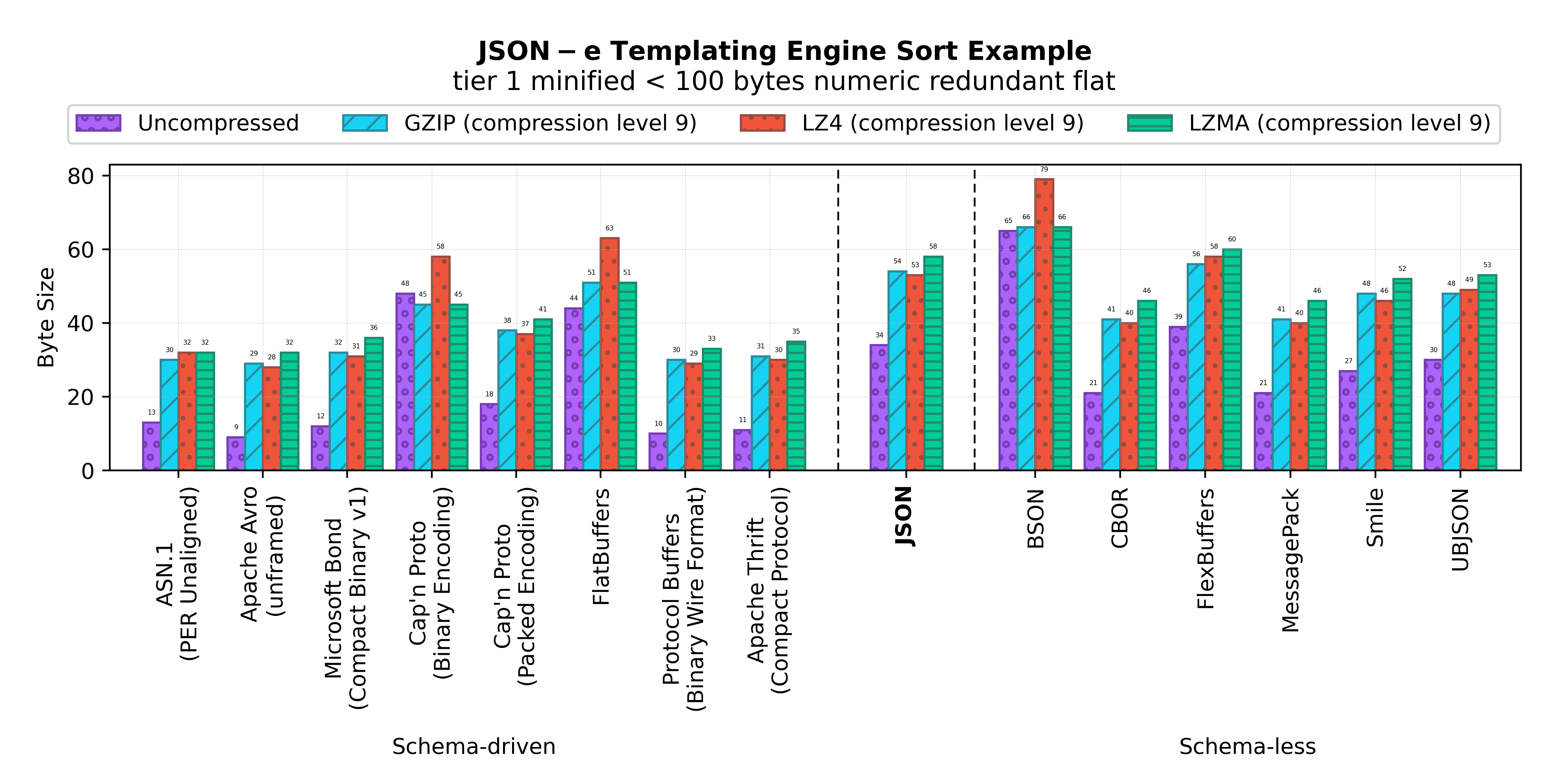}}
\caption{
The benchmark results for the JSON-e Templating Engine Sort Example test case listed in \autoref{table:benchmark-documents} and \autoref{table:benchmark-documents-1}.
}
\label{fig:benchmark-jsonesort}
\end{figure*}

JSON-e \footnote{\url{https://github.com/taskcluster/json-e}} is an open-source
JSON-based templating engine created by Mozilla as part of the TaskCluster
\footnote{\url{https://taskcluster.net}} project, the open-source task
execution framework that supports Mozilla's continuous integration and release
processes. In \autoref{fig:benchmark-jsonesort}, \we demonstrate a \textbf{Tier
1 minified $<$ 100 bytes numeric redundant flat} (Tier 1 NRF from
\autoref{table:json-taxonomy}) JSON document that consists of an example JSON-e
template definition to sort an array of numbers.

The smallest bit-string is produced by Apache Avro \cite{avro} (9 bytes),
closely followed by Protocol Buffers \cite{protocolbuffers} (10 bytes) and
Apache Thrift \cite{slee2007thrift} (11 bytes). The binary serialization
specifications that produced the smallest bit-strings are schema-driven and sequential
\cite{viotti2022survey}. Conversely, the largest bit-string is produced by BSON \cite{bson}
(65 bytes), followed by Cap'n Proto Binary Encoding \cite{capnproto} (48 bytes)
and FlatBuffers \cite{flatbuffers} (44 bytes). With the exception of BSON, the
binary serialization specifications that produced the largest bit-strings are
schema-driven and pointer-based \cite{viotti2022survey}.  In comparison to JSON
\cite{ECMA-404} (34 bytes), binary serialization achieves a \textbf{3.7x} size
reduction in the best case for this input document.  However, 4 out of the 14
JSON-compatible binary serialization specifications listed in
\autoref{table:benchmark-specifications-schema-driven} and
\autoref{table:benchmark-specifications-schema-less} result in bit-strings that are
larger than JSON: Cap'n Proto Binary Encoding \cite{capnproto}, FlatBuffers
\cite{flatbuffers}, BSON \cite{bson} and FlexBuffers \cite{flexbuffers}. These
binary serialization specifications are either schema-less or schema-driven and
pointer-based.

For this Tier 1 NRF document, the best performing schema-driven serialization
specification achieves a \textbf{2.3x} size reduction compared to the best performing
schema-less serialization specification: CBOR \cite{RFC7049} and MessagePack
\cite{messagepack} (21 bytes).  As shown in
\autoref{table:benchmark-stats-jsonesort}, uncompressed schema-driven specifications
provide smaller \emph{average} and \emph{median} bit-strings than uncompressed
schema-less specifications. Additionally, as highlighted by the \emph{range} and
\emph{standard deviation}, uncompressed schema-less specifications exhibit higher size
reduction variability given that BSON \cite{bson} produces a notably large
bit-string. With the exception of the pointer-based binary serialization
specifications Cap'n Proto Binary Encoding \cite{capnproto} and FlatBuffers
\cite{flatbuffers}, the selection of schema-driven serialization specifications listed
in \autoref{table:benchmark-specifications-schema-driven} produce bit-strings that are
equal to or smaller than their schema-less counterparts listed in
\autoref{table:benchmark-specifications-schema-less}.  The best performing sequential
serialization specification achieves a \textbf{2x} size reduction compared to the best
performing pointer-based serialization specification: Cap'n Proto Packed Encoding
\cite{capnproto} (18 bytes).

The compression formats listed in
\autoref{sec:benchmark-compression-formats} result in positive gains for
the bit-string produced by Cap'n Proto Binary Encoding \cite{capnproto}. The
best performing uncompressed binary serialization specification achieves a
\textbf{5.8x} size reduction compared to the best performing compression format
for JSON: LZ4 (53 bytes).

\begin{table*}[hb!]
\caption{A byte-size statistical analysis of the benchmark results shown in \autoref{fig:benchmark-jsonesort} divided by schema-driven and schema-less specifications.}
\label{table:benchmark-stats-jsonesort}
\begin{tabularx}{\linewidth}{X|l|l|l|l|l|l|l|l}
\toprule
\multirow{2}{*}{\textbf{Category}} &
\multicolumn{4}{c|}{\textbf{Schema-driven}} &
\multicolumn{4}{c}{\textbf{Schema-less}} \\
\cline{2-9}
& \small\textbf{Average} & \small\textbf{Median} & \small\textbf{Range} & \small\textbf{Std.dev} & \small\textbf{Average} & \small\textbf{Median} & \small\textbf{Range} & \small\textbf{Std.dev} \\
\midrule
Uncompressed & \small{20.6} & \small{12.5} & \small{39} & \small{14.9} & \small{33.8} & \small{28.5} & \small{44} & \small{15.2} \\ \hline
GZIP (compression level 9) & \small{35.8} & \small{31.5} & \small{22} & \small{7.7} & \small{50} & \small{48} & \small{25} & \small{8.8} \\ \hline
LZ4 (compression level 9) & \small{38.5} & \small{31.5} & \small{35} & \small{13.0} & \small{52} & \small{47.5} & \small{39} & \small{13.5} \\ \hline
LZMA (compression level 9) & \small{38.1} & \small{35.5} & \small{19} & \small{6.5} & \small{53.8} & \small{52.5} & \small{20} & \small{7.2} \\
\bottomrule
\end{tabularx}
\end{table*}

\begin{table*}[hb!]
\caption{The benchmark raw data results and schemas for the plot in \autoref{fig:benchmark-jsonesort}.}
\label{table:benchmark-jsonesort}
\begin{tabularx}{\linewidth}{X|l|l|l|l|l}
\toprule
\textbf{Serialization Format} & \textbf{Schema} & \textbf{Uncompressed} & \textbf{GZIP} & \textbf{LZ4} & \textbf{LZMA} \\
\midrule
ASN.1 (PER Unaligned) & \href{https://github.com/jviotti/binary-json-size-benchmark/blob/main/benchmark/jsonesort/asn1/schema.asn}{\small{\texttt{schema.asn}}} & 13 & 30 & 32 & 32 \\ \hline
Apache Avro (unframed) & \href{https://github.com/jviotti/binary-json-size-benchmark/blob/main/benchmark/jsonesort/avro/schema.json}{\small{\texttt{schema.json}}} & 9 & 29 & 28 & 32 \\ \hline
Microsoft Bond (Compact Binary v1) & \href{https://github.com/jviotti/binary-json-size-benchmark/blob/main/benchmark/jsonesort/bond/schema.bond}{\small{\texttt{schema.bond}}} & 12 & 32 & 31 & 36 \\ \hline
Cap'n Proto (Binary Encoding) & \href{https://github.com/jviotti/binary-json-size-benchmark/blob/main/benchmark/jsonesort/capnproto/schema.capnp}{\small{\texttt{schema.capnp}}} & 48 & 45 & 58 & 45 \\ \hline
Cap'n Proto (Packed Encoding) & \href{https://github.com/jviotti/binary-json-size-benchmark/blob/main/benchmark/jsonesort/capnproto-packed/schema.capnp}{\small{\texttt{schema.capnp}}} & 18 & 38 & 37 & 41 \\ \hline
FlatBuffers & \href{https://github.com/jviotti/binary-json-size-benchmark/blob/main/benchmark/jsonesort/flatbuffers/schema.fbs}{\small{\texttt{schema.fbs}}} & 44 & 51 & 63 & 51 \\ \hline
Protocol Buffers (Binary Wire Format) & \href{https://github.com/jviotti/binary-json-size-benchmark/blob/main/benchmark/jsonesort/protobuf/schema.proto}{\small{\texttt{schema.proto}}} & 10 & 30 & 29 & 33 \\ \hline
Apache Thrift (Compact Protocol) & \href{https://github.com/jviotti/binary-json-size-benchmark/blob/main/benchmark/jsonesort/thrift/schema.thrift}{\small{\texttt{schema.thrift}}} & 11 & 31 & 30 & 35 \\ \hline
\hline \textbf{JSON} & - & 34 & 54 & 53 & 58 \\ \hline \hline
BSON & - & 65 & 66 & 79 & 66 \\ \hline
CBOR & - & 21 & 41 & 40 & 46 \\ \hline
FlexBuffers & - & 39 & 56 & 58 & 60 \\ \hline
MessagePack & - & 21 & 41 & 40 & 46 \\ \hline
Smile & - & 27 & 48 & 46 & 52 \\ \hline
UBJSON & - & 30 & 48 & 49 & 53 \\
\bottomrule
\end{tabularx}
\end{table*}

\begin{figure*}[ht!]
\frame{\includegraphics[width=\linewidth]{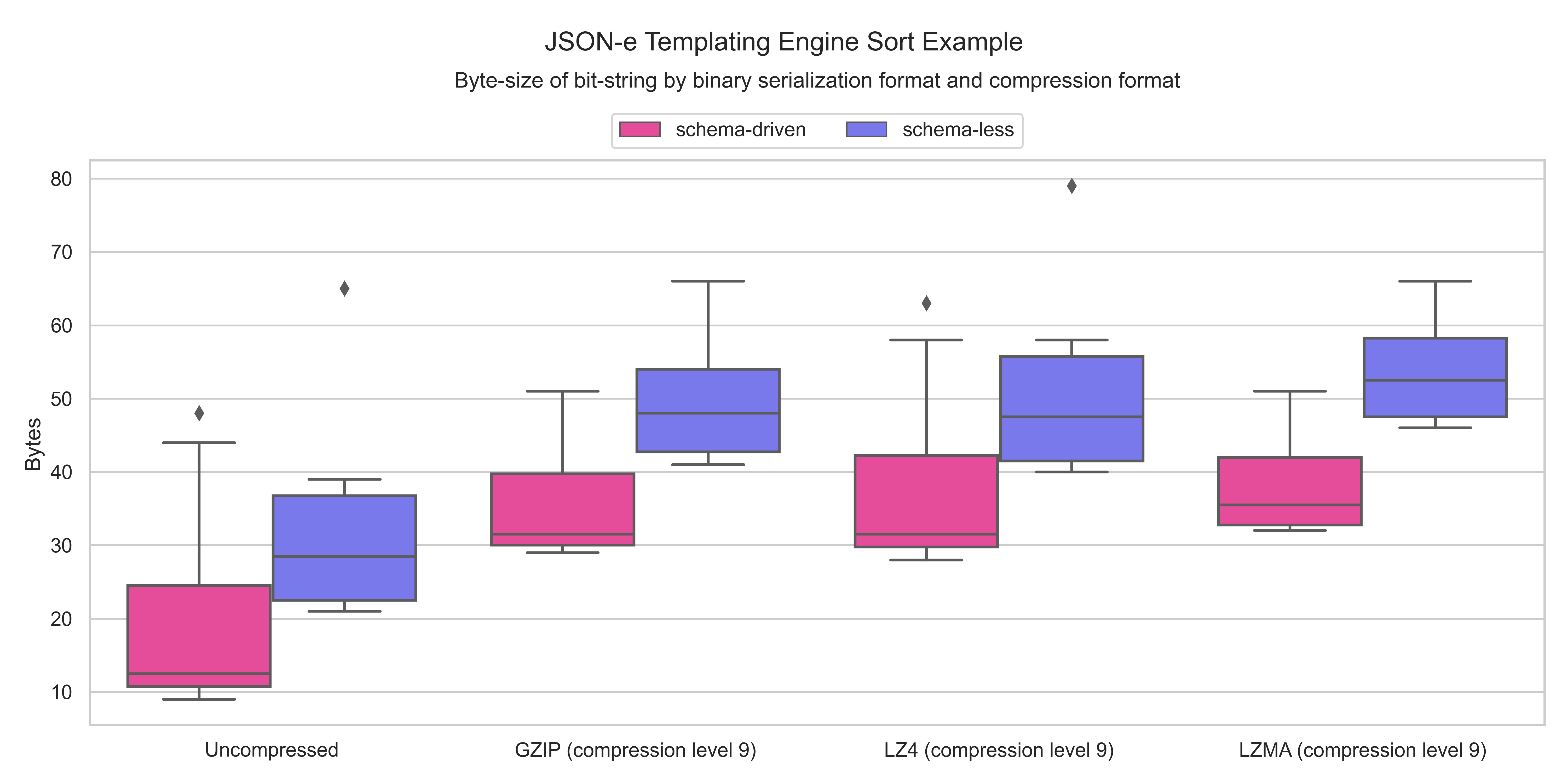}}
\caption{
Box plot of the statistical results in \autoref{table:benchmark-stats-jsonesort}.
}
\label{fig:benchmark-jsonesort-boxplot}
\end{figure*}

In \autoref{fig:benchmark-jsonesort-boxplot}, we observe the medians for
uncompressed schema-driven binary serialization specifications to be smaller in
comparison to uncompressed schema-less binary serialization specifications.  The range
between the upper and lower whiskers of uncompressed schema-less binary
serialization specifications is smaller than the range between the upper and lower
whiskers of uncompressed schema-driven binary serialization specifications.  However,
the inter-quartile range of both both uncompressed schema-driven and
schema-less binary serialization specifications is similar.  Additionally, their
respective quartiles overlap.


In terms of compression, GZIP and LZ4 result in the lower medians for
schema-driven binary serialization specifications while LZ4 results in the
lower median for schema-less binary serialization specifications.  However,
compression is not space-efficient in terms of the median for both
schema-driven and schema-less binary serialization specifications.
Additionally, the use of LZ4 for both schema-driven binary serialization
specifications and schema-less binary serialization specifications exhibits
upper outliers.  While compression does not contribute to space-efficiency, it
reduces the range between the upper and lower whiskers and inter-quartile range
for schema-driven binary serialization specifications and it reduces the
inter-quartile range for schema-less binary serialization specifications.  In
particular, the compression format with the smaller range between the upper and
lower whiskers for schema-driven binary serialization specifications is LZMA,
the compression formats with the smaller inter-quartile range for schema-driven
binary serialization specifications are GZIP and LZMA, the compression format
with the smaller range between the upper and lower whiskers for schema-less
binary serialization specifications is LZ4, and the compression formats with
the smaller inter-quartile range for schema-less binary serialization
specifications are GZIP and LZMA.


Overall, \we conclude that uncompressed schema-driven binary serialization
specifications are space-efficient in comparison to uncompressed schema-less
binary serialization specifications and that compression does not contribute to
space-efficiency in comparison to both uncompressed schema-driven and
schema-less binary serialization specifications.

\clearpage

\subsection{JSON-e Templating Engine Reverse Sort Example}
\label{sec:benchmark-jsonereversesort}

\begin{figure*}[ht!]
\frame{\includegraphics[width=\linewidth]{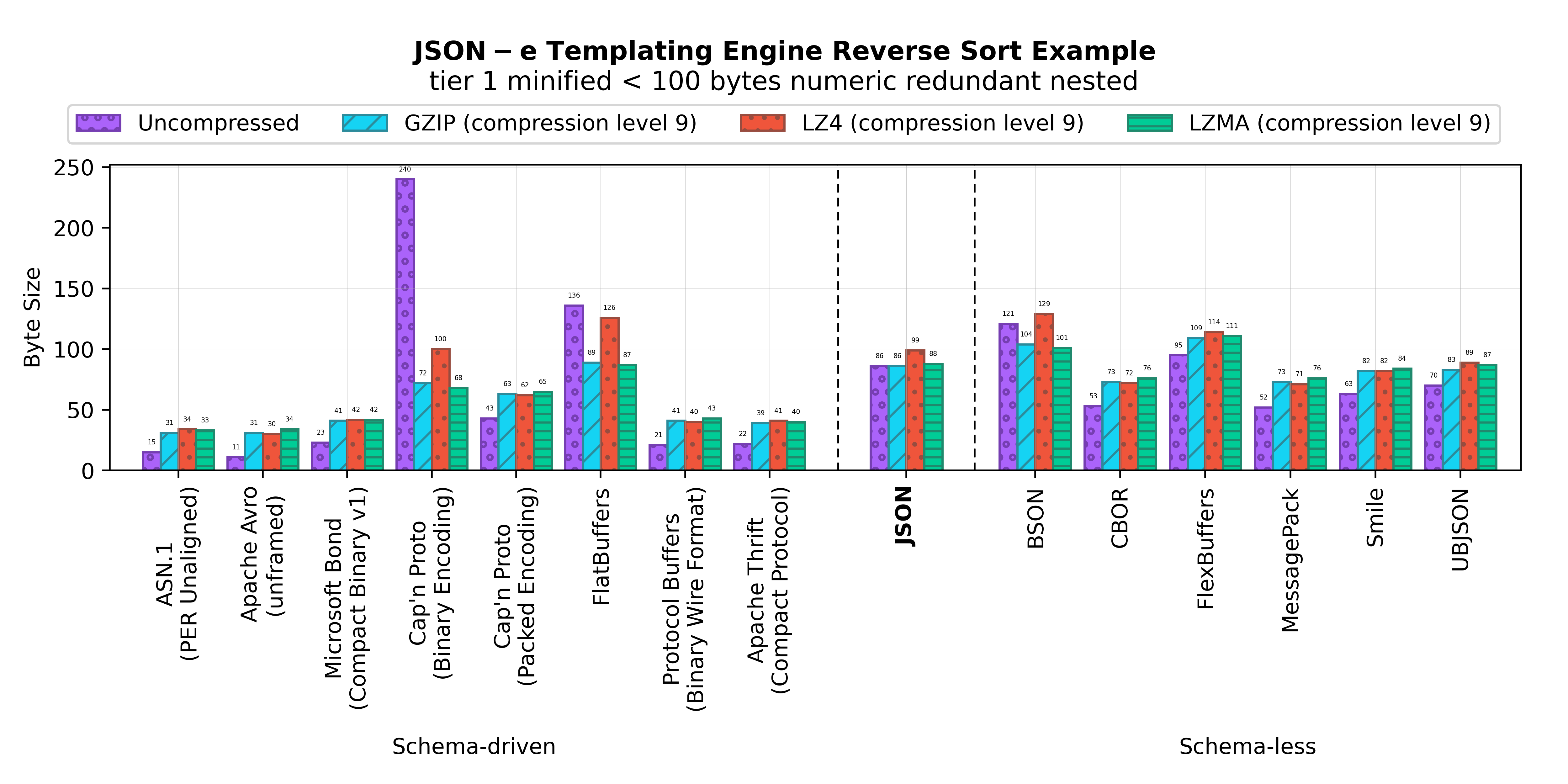}}
\caption{
The benchmark results for the JSON-e Templating Engine Reverse Sort Example test case listed in \autoref{table:benchmark-documents} and \autoref{table:benchmark-documents-1}.
}
\label{fig:benchmark-jsonereversesort}
\end{figure*}

JSON-e \footnote{\url{https://github.com/taskcluster/json-e}} is an open-source
JSON-based templating engine created by Mozilla as part of the TaskCluster
\footnote{\url{https://taskcluster.net}} project, the open-source task
execution framework that supports Mozilla's continuous integration and release
processes. In \autoref{fig:benchmark-jsonereversesort}, \we demonstrate a
\textbf{Tier 1 minified $<$ 100 bytes numeric redundant nested} (Tier 1 NRN
from \autoref{table:json-taxonomy}) JSON document that consists of an example
JSON-e template definition to sort and reverse an array of numbers.

The smallest bit-string is produced by Apache Avro \cite{avro} (11 bytes),
followed by ASN.1 PER Unaligned \cite{asn1-per} (15 bytes) and Protocol Buffers
\cite{protocolbuffers} (21 bytes). The binary serialization specifications that
produced the smallest bit-strings are schema-driven and sequential \cite{viotti2022survey}.
Conversely, the largest bit-string is produced by Cap'n Proto Binary Encoding
\cite{capnproto} (240 bytes), followed by FlatBuffers \cite{flatbuffers} (136
bytes) and BSON \cite{bson} (121 bytes).  With the exception of BSON, the
binary serialization specifications that produced the largest bit-strings are
schema-driven and pointer-based \cite{viotti2022survey}.  In comparison to JSON
\cite{ECMA-404} (86 bytes), binary serialization achieves a \textbf{7.8x} size
reduction in the best case for this input document.  However, 4 out of the 14
JSON-compatible binary serialization specifications listed in
\autoref{table:benchmark-specifications-schema-driven} and
\autoref{table:benchmark-specifications-schema-less} result in bit-strings that are
larger than JSON: Cap'n Proto Binary Encoding \cite{capnproto}, FlatBuffers
\cite{flatbuffers}, BSON \cite{bson} and FlexBuffers \cite{flexbuffers}. These
binary serialization specifications are either schema-less or schema-driven and
pointer-based.

For this Tier 1 NRN document, the best performing schema-driven serialization
specification achieves a \textbf{4.7x} size reduction compared to the best performing
schema-less serialization specification: CBOR \cite{RFC7049} and MessagePack
\cite{messagepack} (52 bytes).  As shown in
\autoref{table:benchmark-stats-jsonereversesort}, uncompressed schema-driven
specifications provide smaller \emph{average} and \emph{median} bit-strings than
uncompressed schema-less specifications. However, as highlighted by the \emph{range}
and \emph{standard deviation}, uncompressed schema-driven specifications exhibit
higher size reduction variability depending on the expressiveness of the schema
language (i.e. how the language constructs allow you to model the data) and the
size optimizations devised by its authors. With the exception of the
pointer-based binary serialization specifications Cap'n Proto Binary Encoding
\cite{capnproto} and FlatBuffers \cite{flatbuffers}, the selection of
schema-driven serialization specifications listed in
\autoref{table:benchmark-specifications-schema-driven} produce bit-strings that are
equal to or smaller than their schema-less counterparts listed in
\autoref{table:benchmark-specifications-schema-less}.  The best performing sequential
serialization specification achieves a \textbf{3.9x} size reduction compared to the
best performing pointer-based serialization specification: Cap'n Proto Packed Encoding
\cite{capnproto} (43 bytes).

The compression formats listed in
\autoref{sec:benchmark-compression-formats} result in positive gains for
the bit-strings produced by Cap'n Proto Binary Encoding \cite{capnproto},
FlatBuffers \cite{flatbuffers} and BSON \cite{bson}.  The best performing
uncompressed binary serialization specification achieves a \textbf{7.8x} size
reduction compared to the best performing compression format for JSON: GZIP
\cite{RFC1952} (86 bytes).

\begin{table*}[hb!]
\caption{A byte-size statistical analysis of the benchmark results shown in \autoref{fig:benchmark-jsonereversesort} divided by schema-driven and schema-less specifications.}
\label{table:benchmark-stats-jsonereversesort}
\begin{tabularx}{\linewidth}{X|l|l|l|l|l|l|l|l}
\toprule
\multirow{2}{*}{\textbf{Category}} &
\multicolumn{4}{c|}{\textbf{Schema-driven}} &
\multicolumn{4}{c}{\textbf{Schema-less}} \\
\cline{2-9}
& \small\textbf{Average} & \small\textbf{Median} & \small\textbf{Range} & \small\textbf{Std.dev} & \small\textbf{Average} & \small\textbf{Median} & \small\textbf{Range} & \small\textbf{Std.dev} \\
\midrule
Uncompressed & \small{63.9} & \small{22.5} & \small{229} & \small{76.7} & \small{75.7} & \small{66.5} & \small{69} & \small{24.8} \\ \hline
GZIP (compression level 9) & \small{50.9} & \small{41} & \small{58} & \small{19.9} & \small{87.3} & \small{82.5} & \small{36} & \small{14.2} \\ \hline
LZ4 (compression level 9) & \small{59.4} & \small{41.5} & \small{96} & \small{32.8} & \small{92.8} & \small{85.5} & \small{58} & \small{21.6} \\ \hline
LZMA (compression level 9) & \small{51.5} & \small{42.5} & \small{54} & \small{18.2} & \small{89.2} & \small{85.5} & \small{35} & \small{12.9} \\
\bottomrule
\end{tabularx}
\end{table*}

\begin{table*}[hb!]
\caption{The benchmark raw data results and schemas for the plot in \autoref{fig:benchmark-jsonereversesort}.}
\label{table:benchmark-jsonereversesort}
\begin{tabularx}{\linewidth}{X|l|l|l|l|l}
\toprule
\textbf{Serialization Format} & \textbf{Schema} & \textbf{Uncompressed} & \textbf{GZIP} & \textbf{LZ4} & \textbf{LZMA} \\
\midrule
ASN.1 (PER Unaligned) & \href{https://github.com/jviotti/binary-json-size-benchmark/blob/main/benchmark/jsonereversesort/asn1/schema.asn}{\small{\texttt{schema.asn}}} & 15 & 31 & 34 & 33 \\ \hline
Apache Avro (unframed) & \href{https://github.com/jviotti/binary-json-size-benchmark/blob/main/benchmark/jsonereversesort/avro/schema.json}{\small{\texttt{schema.json}}} & 11 & 31 & 30 & 34 \\ \hline
Microsoft Bond (Compact Binary v1) & \href{https://github.com/jviotti/binary-json-size-benchmark/blob/main/benchmark/jsonereversesort/bond/schema.bond}{\small{\texttt{schema.bond}}} & 23 & 41 & 42 & 42 \\ \hline
Cap'n Proto (Binary Encoding) & \href{https://github.com/jviotti/binary-json-size-benchmark/blob/main/benchmark/jsonereversesort/capnproto/schema.capnp}{\small{\texttt{schema.capnp}}} & 240 & 72 & 100 & 68 \\ \hline
Cap'n Proto (Packed Encoding) & \href{https://github.com/jviotti/binary-json-size-benchmark/blob/main/benchmark/jsonereversesort/capnproto-packed/schema.capnp}{\small{\texttt{schema.capnp}}} & 43 & 63 & 62 & 65 \\ \hline
FlatBuffers & \href{https://github.com/jviotti/binary-json-size-benchmark/blob/main/benchmark/jsonereversesort/flatbuffers/schema.fbs}{\small{\texttt{schema.fbs}}} & 136 & 89 & 126 & 87 \\ \hline
Protocol Buffers (Binary Wire Format) & \href{https://github.com/jviotti/binary-json-size-benchmark/blob/main/benchmark/jsonereversesort/protobuf/schema.proto}{\small{\texttt{schema.proto}}} & 21 & 41 & 40 & 43 \\ \hline
Apache Thrift (Compact Protocol) & \href{https://github.com/jviotti/binary-json-size-benchmark/blob/main/benchmark/jsonereversesort/thrift/schema.thrift}{\small{\texttt{schema.thrift}}} & 22 & 39 & 41 & 40 \\ \hline
\hline \textbf{JSON} & - & 86 & 86 & 99 & 88 \\ \hline \hline
BSON & - & 121 & 104 & 129 & 101 \\ \hline
CBOR & - & 53 & 73 & 72 & 76 \\ \hline
FlexBuffers & - & 95 & 109 & 114 & 111 \\ \hline
MessagePack & - & 52 & 73 & 71 & 76 \\ \hline
Smile & - & 63 & 82 & 82 & 84 \\ \hline
UBJSON & - & 70 & 83 & 89 & 87 \\
\bottomrule
\end{tabularx}
\end{table*}

\begin{figure*}[ht!]
\frame{\includegraphics[width=\linewidth]{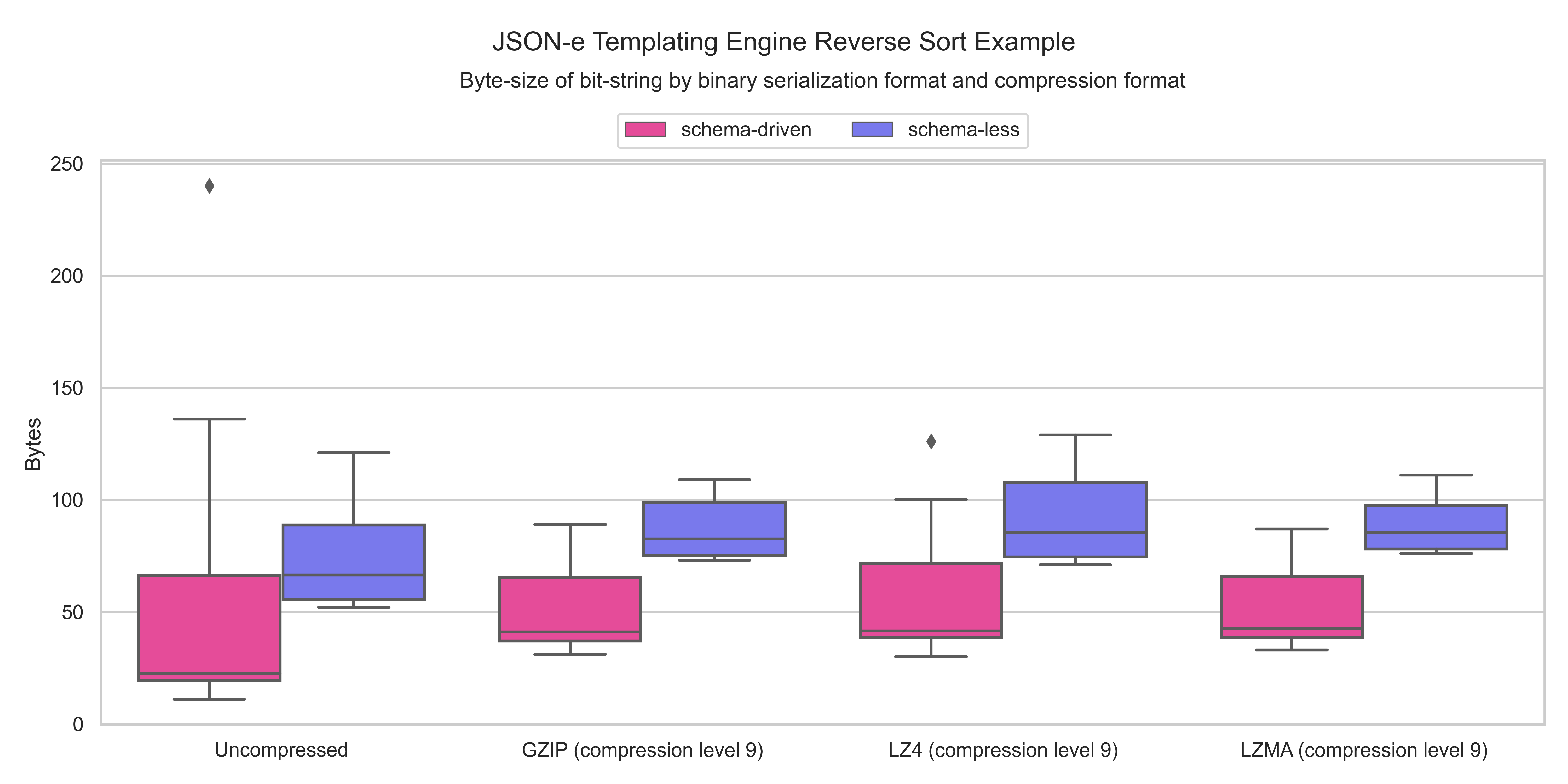}}
\caption{
Box plot of the statistical results in \autoref{table:benchmark-stats-jsonereversesort}.
}
\label{fig:benchmark-jsonereversesort-boxplot}
\end{figure*}

In \autoref{fig:benchmark-jsonereversesort-boxplot}, we observe the medians for
uncompressed schema-driven binary serialization specifications to be smaller in
comparison to uncompressed schema-less binary serialization specifications.  The range
between the upper and lower whiskers and the inter-quartile range of
uncompressed schema-less binary serialization specifications is smaller than the range
between the upper and lower whiskers and the inter-quartile range of
uncompressed schema-driven binary serialization specifications. However, their
respective quartiles overlap.


In terms of compression, GZIP, LZ4 and LZMA resuls in similar medians for
schema-driven binary serialization specifications while GZIP results in the
lower median for schema-less binary serialization specifications.  However,
compression is not space-efficient in terms of the median for both
schema-driven and schema-less binary serialization specifications.
Additionally, the use of LZ4 for schema-driven binary serialization
specifications exhibits upper outliers.  While compression does not contribute
to space-efficiency, it reduces the range between the upper and lower whiskers
and inter-quartile range for both schema-driven and schema-less binary
serialization specifications.  In particular, the compression formats with the
smaller range between the upper and lower whiskers for schema-driven binary
serialization specifications are GZIP and LZMA, the compression formats with
the smaller inter-quartile range for schema-driven binary serialization
specifications are GZIP and LZMA, the compression formats with the smaller
range between the upper and lower whiskers for schema-less binary serialization
specifications are GZIP and LZMA, and the compression format with the smaller
inter-quartile range for schema-less binary serialization specifications is
LZMA.


Overall, \we conclude that uncompressed schema-driven binary serialization
specifications are space-efficient in comparison to uncompressed schema-less binary
serialization specifications and that compression does not contribute to
space-efficiency in comparison to both uncompressed schema-driven and
schema-less binary serialization specifications.

\clearpage

\subsection{CircleCI Definition (Blank)}
\label{sec:benchmark-circleciblank}

\begin{figure*}[ht!]
\frame{\includegraphics[width=\linewidth]{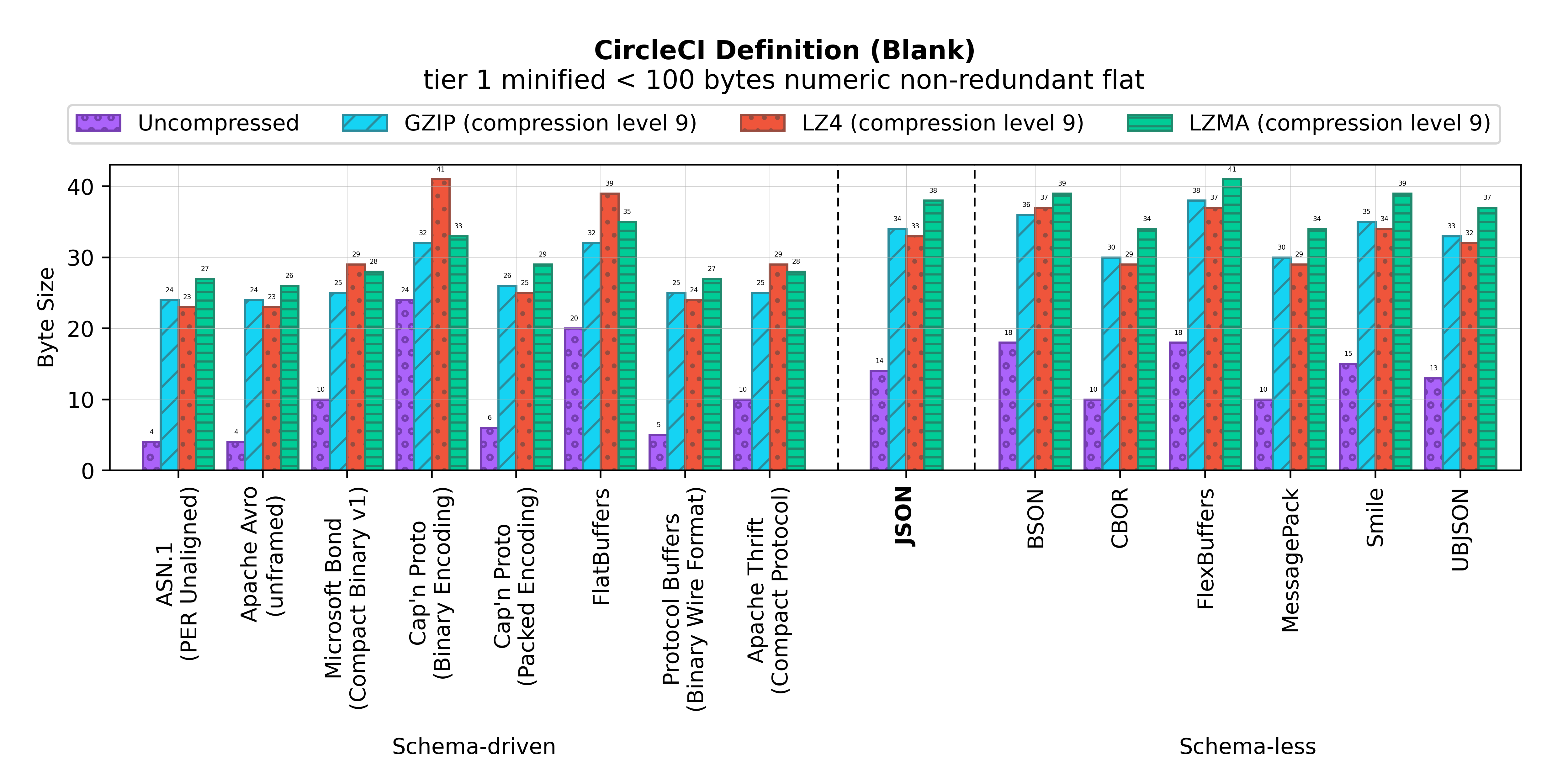}}
\caption{
The benchmark results for the CircleCI Definition (Blank) test case listed in \autoref{table:benchmark-documents} and \autoref{table:benchmark-documents-1}.
}
\label{fig:benchmark-circleciblank}
\end{figure*}

CircleCI \footnote{\url{https://circleci.com}} is a commercial cloud-provider
of continuous integration and deployment pipelines used by a wide range of
companies in the software development industry such as Facebook, Spotify, and
Heroku \footnote{\url{https://circleci.com/customers/}}. In
\autoref{fig:benchmark-circleciblank}, \we demonstrate a \textbf{Tier 1
minified $<$ 100 bytes numeric non-redundant flat} (Tier 1 NNF from
\autoref{table:json-taxonomy}) JSON document that represents a simple pipeline
configuration file for CircleCI that declares the desired CircleCI version
without defining any workflows.

The smallest bit-string is produced by both ASN.1 PER Unaligned \cite{asn1-per}
and Apache Avro \cite{avro} (4 bytes), closely followed by Protocol Buffers
\cite{protocolbuffers} (5 bytes) and Cap'n Proto Packed Encoding
\cite{capnproto} (6 bytes). These serialization specifications are
schema-driven and with the exception of Cap'n Proto Packed Encoding, which
occupies the fourth place, they are also sequential \cite{viotti2022survey}. Conversely, the
largest bit-string is produced by Cap'n Proto Binary Encoding \cite{capnproto}
(24 bytes), followed by FlatBuffers \cite{flatbuffers} (20 bytes) and both BSON
\cite{bson} and FlexBuffers \cite{flexbuffers} (18 bytes).  With the exception
of BSON, the binary serialization specifications that produced the largest
bit-strings are pointer-based \cite{viotti2022survey}. In comparison to JSON \cite{ECMA-404}
(14 bytes), binary serialization achieves a \textbf{3.5x} size reduction in the
best case for this input document. However, 5 out of the 14 JSON-compatible
binary serialization specifications listed in
\autoref{table:benchmark-specifications-schema-driven} and
\autoref{table:benchmark-specifications-schema-less} result in bit-strings that are
larger than JSON: Cap'n Proto Binary Encoding \cite{capnproto}, FlatBuffers
\cite{flatbuffers}, BSON \cite{bson}, FlexBuffers \cite{flexbuffers} and Smile
\cite{smile}. These binary serialization specifications are either schema-less
or schema-driven and pointer-based.

For this Tier 1 NNF document, the best performing schema-driven serialization
specification achieves a \textbf{2.5x} size reduction compared to the best
performing schema-less serialization specification: CBOR \cite{RFC7049} and
MessagePack \cite{messagepack} (10 bytes). As shown in
\autoref{table:benchmark-stats-circleciblank}, uncompressed schema-driven
specifications provide smaller \emph{average} and \emph{median} bit-strings
than uncompressed schema-less specifications. However, as highlighted by the
\emph{range} and \emph{standard deviation}, uncompressed schema-driven
specifications exhibit higher size reduction variability depending on the
schema language capabilities and implemented size optimizations. With the
exception of the pointer-based binary serialization specifications Cap'n Proto
Binary Encoding \cite{capnproto} and FlatBuffers \cite{flatbuffers}, the
selection of schema-driven serialization specifications listed in
\autoref{table:benchmark-specifications-schema-driven} produce bit-strings that are
equal to or smaller than their schema-less counterparts listed in
\autoref{table:benchmark-specifications-schema-less}.  The best performing sequential
serialization specification only achieves a \textbf{1.5x} size reduction
compared to the best performing pointer-based serialization specification:
Cap'n Proto Packed Encoding \cite{capnproto} (6 bytes).

The compression formats listed in
\autoref{sec:benchmark-compression-formats} do not result in positive
gains for any case due to the overhead of encoding the dictionary data
structures and the low redundancy of the input data. The best performing
uncompressed binary serialization specification achieves a \textbf{8.2x} size
reduction compared to the best performing compression format for JSON: LZ4 (33
bytes).

\begin{table*}[hb!]
\caption{A byte-size statistical analysis of the benchmark results shown in \autoref{fig:benchmark-circleciblank} divided by schema-driven and schema-less specifications.}
\label{table:benchmark-stats-circleciblank}
\begin{tabularx}{\linewidth}{X|l|l|l|l|l|l|l|l}
\toprule
\multirow{2}{*}{\textbf{Category}} &
\multicolumn{4}{c|}{\textbf{Schema-driven}} &
\multicolumn{4}{c}{\textbf{Schema-less}} \\
\cline{2-9}
& \small\textbf{Average} & \small\textbf{Median} & \small\textbf{Range} & \small\textbf{Std.dev} & \small\textbf{Average} & \small\textbf{Median} & \small\textbf{Range} & \small\textbf{Std.dev} \\
\midrule
Uncompressed & \small{10.4} & \small{8} & \small{20} & \small{7.1} & \small{14} & \small{14} & \small{8} & \small{3.3} \\ \hline
GZIP (compression level 9) & \small{26.6} & \small{25} & \small{8} & \small{3.2} & \small{33.7} & \small{34} & \small{8} & \small{3.0} \\ \hline
LZ4 (compression level 9) & \small{29.1} & \small{27} & \small{18} & \small{6.7} & \small{33} & \small{33} & \small{8} & \small{3.3} \\ \hline
LZMA (compression level 9) & \small{29.1} & \small{28} & \small{9} & \small{3.0} & \small{37.3} & \small{38} & \small{7} & \small{2.6} \\
\bottomrule
\end{tabularx}
\end{table*}

\begin{table*}[hb!]
\caption{The benchmark raw data results and schemas for the plot in \autoref{fig:benchmark-circleciblank}.}
\label{table:benchmark-circleciblank}
\begin{tabularx}{\linewidth}{X|l|l|l|l|l}
\toprule
\textbf{Serialization Format} & \textbf{Schema} & \textbf{Uncompressed} & \textbf{GZIP} & \textbf{LZ4} & \textbf{LZMA} \\
\midrule
ASN.1 (PER Unaligned) & \href{https://github.com/jviotti/binary-json-size-benchmark/blob/main/benchmark/circleciblank/asn1/schema.asn}{\small{\texttt{schema.asn}}} & 4 & 24 & 23 & 27 \\ \hline
Apache Avro (unframed) & \href{https://github.com/jviotti/binary-json-size-benchmark/blob/main/benchmark/circleciblank/avro/schema.json}{\small{\texttt{schema.json}}} & 4 & 24 & 23 & 26 \\ \hline
Microsoft Bond (Compact Binary v1) & \href{https://github.com/jviotti/binary-json-size-benchmark/blob/main/benchmark/circleciblank/bond/schema.bond}{\small{\texttt{schema.bond}}} & 10 & 25 & 29 & 28 \\ \hline
Cap'n Proto (Binary Encoding) & \href{https://github.com/jviotti/binary-json-size-benchmark/blob/main/benchmark/circleciblank/capnproto/schema.capnp}{\small{\texttt{schema.capnp}}} & 24 & 32 & 41 & 33 \\ \hline
Cap'n Proto (Packed Encoding) & \href{https://github.com/jviotti/binary-json-size-benchmark/blob/main/benchmark/circleciblank/capnproto-packed/schema.capnp}{\small{\texttt{schema.capnp}}} & 6 & 26 & 25 & 29 \\ \hline
FlatBuffers & \href{https://github.com/jviotti/binary-json-size-benchmark/blob/main/benchmark/circleciblank/flatbuffers/schema.fbs}{\small{\texttt{schema.fbs}}} & 20 & 32 & 39 & 35 \\ \hline
Protocol Buffers (Binary Wire Format) & \href{https://github.com/jviotti/binary-json-size-benchmark/blob/main/benchmark/circleciblank/protobuf/schema.proto}{\small{\texttt{schema.proto}}} & 5 & 25 & 24 & 27 \\ \hline
Apache Thrift (Compact Protocol) & \href{https://github.com/jviotti/binary-json-size-benchmark/blob/main/benchmark/circleciblank/thrift/schema.thrift}{\small{\texttt{schema.thrift}}} & 10 & 25 & 29 & 28 \\ \hline
\hline \textbf{JSON} & - & 14 & 34 & 33 & 38 \\ \hline \hline
BSON & - & 18 & 36 & 37 & 39 \\ \hline
CBOR & - & 10 & 30 & 29 & 34 \\ \hline
FlexBuffers & - & 18 & 38 & 37 & 41 \\ \hline
MessagePack & - & 10 & 30 & 29 & 34 \\ \hline
Smile & - & 15 & 35 & 34 & 39 \\ \hline
UBJSON & - & 13 & 33 & 32 & 37 \\
\bottomrule
\end{tabularx}
\end{table*}

\begin{figure*}[ht!]
\frame{\includegraphics[width=\linewidth]{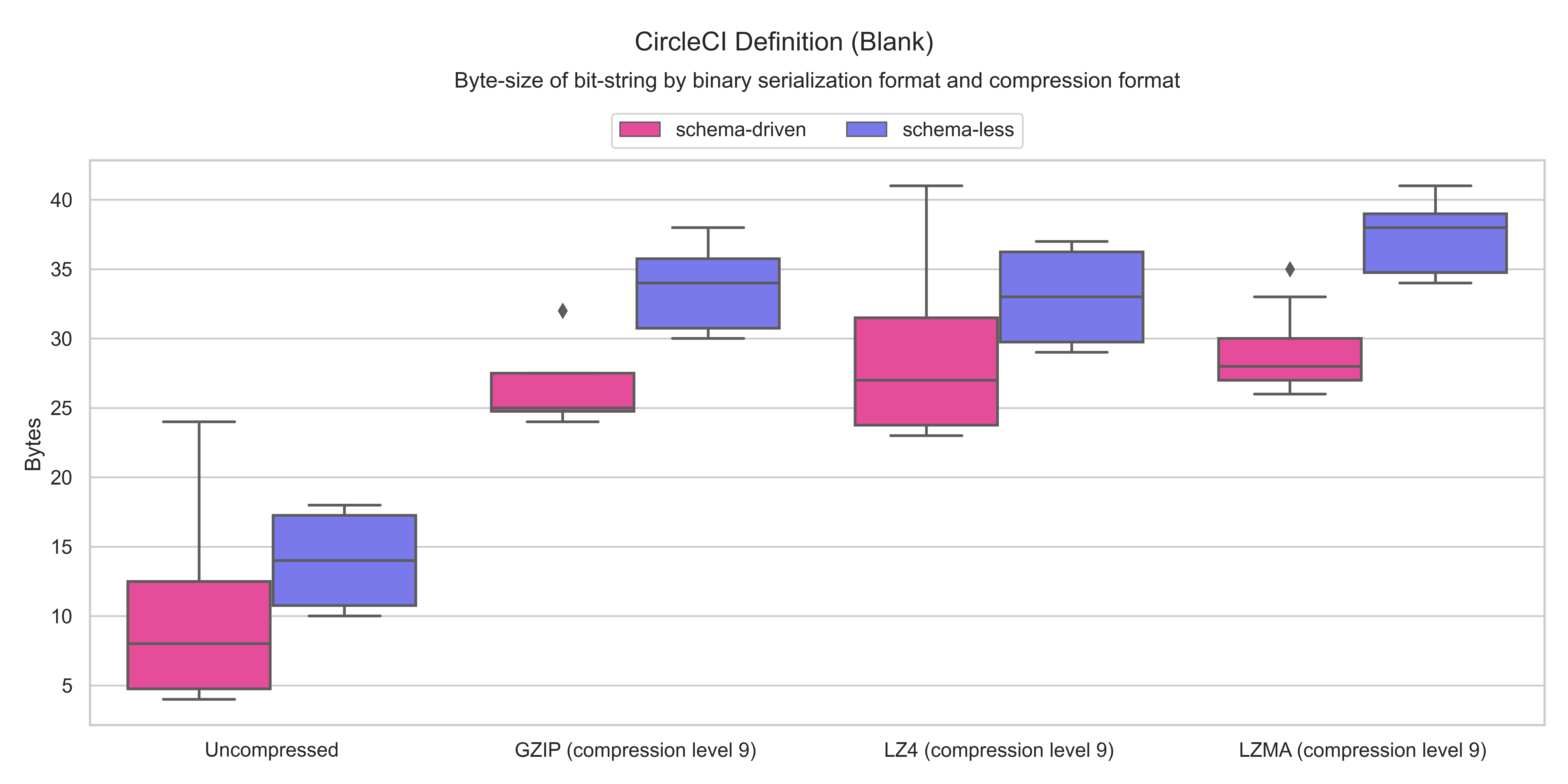}}
\caption{
Box plot of the statistical results in \autoref{table:benchmark-stats-circleciblank}.
}
\label{fig:benchmark-circleciblank-boxplot}
\end{figure*}

In \autoref{fig:benchmark-circleciblank-boxplot}, we observe the medians for
uncompressed schema-driven binary serialization specifications to be smaller in
comparison to uncompressed schema-less binary serialization specifications. The
range between the upper and lower whiskers and the inter-quartile range of
uncompressed schema-less binary serialization specifications is smaller than
the range between the upper and lower whiskers and the inter-quartile range of
uncompressed schema-driven binary serialization specifications. However, their
respective quartiles overlap.


In terms of compression, GZIP results in the lower median for schema-driven
binary serialization specifications while LZ4 results in the lower median for
schema-less binary serialization specifications. However, compression is not
space-efficient in terms of the median for both schema-driven and schema-less
binary serialization specifications. Additionally, the use of GZIP and LZMA for
schema-driven binary serialization specifications exhibit upper outliers. While
compression does not contribute to space-efficiency, it reduces the range
between the upper and lower whiskers and inter-quartile range for schema-driven
binary serialization specifications and it reduces the inter-quartile range for
schema-less binary serialization specifications.  In particular, the
compression format with the smaller range between the upper and lower whiskers
for schema-driven binary serialization specifications is GZIP, the compression
formats with the smaller inter-quartile range for schema-driven binary
serialization specifications are GZIP and LZMA, and the compression format with
the smaller inter-quartile range for schema-less binary serialization
specifications is LZMA.


Overall, \we conclude that uncompressed schema-driven binary serialization
specifications are space-efficient in comparison to uncompressed schema-less
binary serialization specifications and that compression does not contribute to
space-efficiency in comparison to both uncompressed schema-driven and
schema-less binary serialization specifications.

\clearpage

\subsection{CircleCI Matrix Definition}
\label{sec:benchmark-circlecimatrix}

\begin{figure*}[ht!]
\frame{\includegraphics[width=\linewidth]{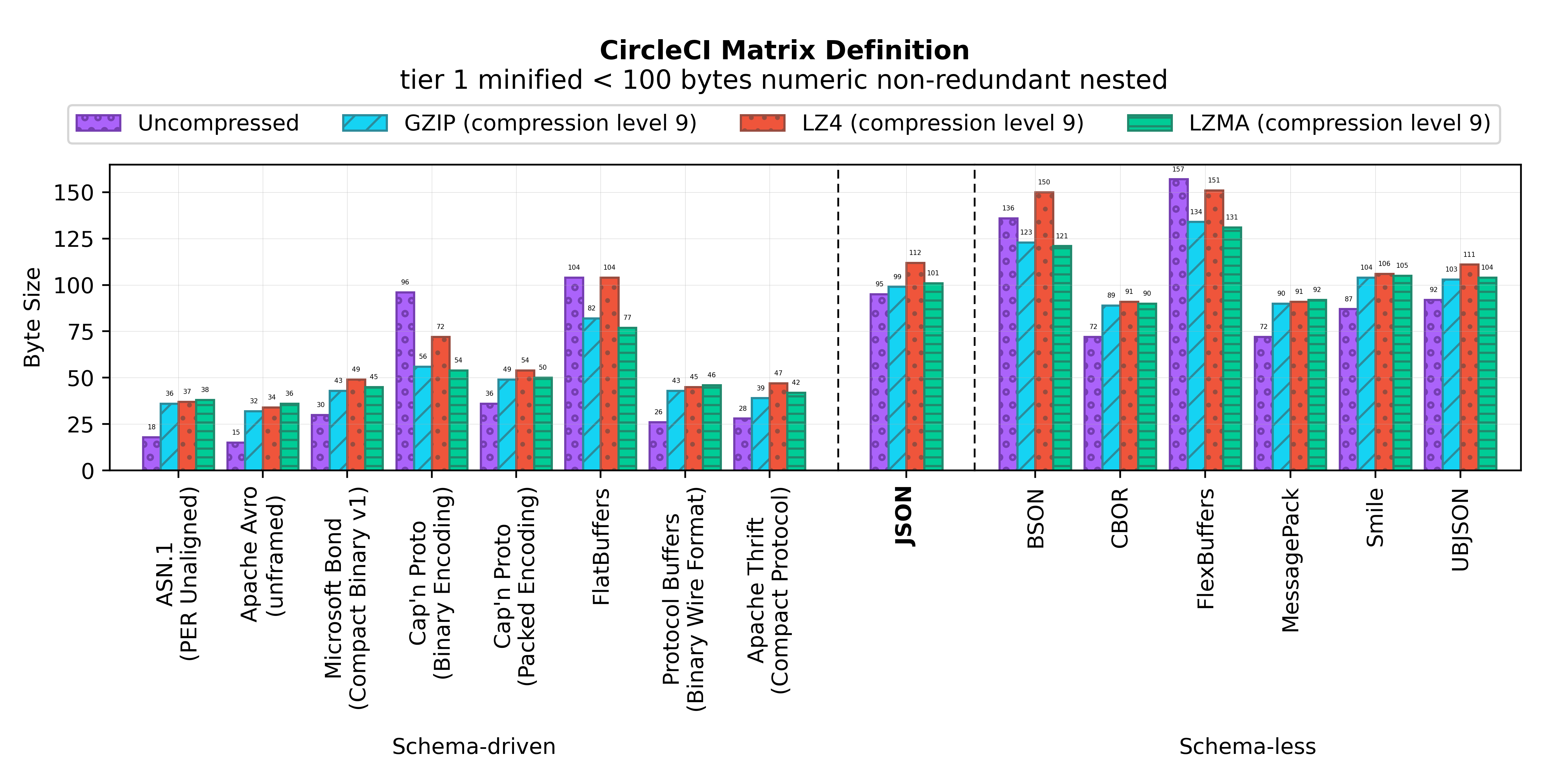}}
\caption{
The benchmark results for the CircleCI Matrix Definition test case listed in \autoref{table:benchmark-documents} and \autoref{table:benchmark-documents-1}.
}
\label{fig:benchmark-circlecimatrix}
\end{figure*}

CircleCI \footnote{\url{https://circleci.com}} is a commercial cloud-provider
of continuous integration and deployment pipelines used by a wide range of
companies in the software development industry such as Facebook, Spotify, and
Heroku \footnote{\url{https://circleci.com/customers/}}. In
\autoref{fig:benchmark-circlecimatrix}, \we demonstrate a \textbf{Tier 1
minified $<$ 100 bytes numeric non-redundant nested} (Tier 1 NNN from
\autoref{table:json-taxonomy}) JSON document that represents a pipeline
configuration file for CircleCI that declares the desired CircleCI version and
defines a workflow that contains a single blank matrix-based job.

The smallest bit-string is produced by Apache Avro \cite{avro} (15 bytes),
followed by ASN.1 PER Unaligned \cite{asn1-per} (18 bytes) and Protocol Buffers
\cite{protocolbuffers} (26 bytes). The binary serialization specifications that
produced the smallest bit-strings are schema-driven and sequential \cite{viotti2022survey}.
Conversely, the largest bit-string is produced by FlexBuffers
\cite{flexbuffers} (157 bytes), BSON \cite{bson} (136 bytes) and FlatBuffers
\cite{flatbuffers} (104 bytes). With the exception of BSON, the binary
serialization specifications that produced the largest bit-strings are pointer-based
\cite{viotti2022survey}.  In comparison to JSON \cite{ECMA-404} (95 bytes), binary
serialization achieves a \textbf{6.3x} size reduction in the best case for this
input document. However, 4 out of the 14 JSON-compatible binary serialization
specifications listed in \autoref{table:benchmark-specifications-schema-driven} and
\autoref{table:benchmark-specifications-schema-less} result in bit-strings that are
larger than JSON: Cap'n Proto Binary Encoding \cite{capnproto}, FlatBuffers
\cite{flatbuffers}, BSON \cite{bson} and FlexBuffers \cite{flexbuffers}. These
binary serialization specifications are either schema-less or schema-driven and
pointer-based.

For this Tier 1 NNN document, the best performing schema-driven serialization
specification achieves a \textbf{4.8x} size reduction compared to the best performing
schema-less serialization specification: CBOR \cite{RFC7049} and MessagePack
\cite{messagepack} (72 bytes).  As shown in
\autoref{table:benchmark-stats-circlecimatrix}, uncompressed schema-driven
specifications provide smaller \emph{average} and \emph{median} bit-strings than
uncompressed schema-less specifications. However, as highlighted by the \emph{range}
and \emph{standard deviation}, uncompressed schema-driven specifications exhibit
higher size reduction variability depending on the schema language capabilities
and implemented size optimizations. With the exception of the pointer-based
binary serialization specifications Cap'n Proto Binary Encoding \cite{capnproto} and
FlatBuffers \cite{flatbuffers}, the selection of schema-driven serialization
specifications listed in \autoref{table:benchmark-specifications-schema-driven} produce
bit-strings that are equal to or smaller than their schema-less counterparts
listed in \autoref{table:benchmark-specifications-schema-less}.  The best performing
sequential serialization specification achieves a \textbf{2.4x} size reduction
compared to the best performing pointer-based serialization specification: Cap'n Proto
Packed Encoding \cite{capnproto} (36 bytes).

The compression formats listed in
\autoref{sec:benchmark-compression-formats} result in positive gains for
the bit-strings produced by Cap'n Proto Binary Encoding \cite{capnproto},
FlatBuffers \cite{flatbuffers}, BSON \cite{bson} and FlexBuffers
\cite{flexbuffers}. The best performing uncompressed binary serialization
specification achieves a \textbf{6.6x} size reduction compared to the best
performing compression format for JSON: GZIP \cite{RFC1952} (99 bytes).

\begin{table*}[hb!]
\caption{A byte-size statistical analysis of the benchmark results shown in \autoref{fig:benchmark-circlecimatrix} divided by schema-driven and schema-less specifications.}
\label{table:benchmark-stats-circlecimatrix}
\begin{tabularx}{\linewidth}{X|l|l|l|l|l|l|l|l}
\toprule
\multirow{2}{*}{\textbf{Category}} &
\multicolumn{4}{c|}{\textbf{Schema-driven}} &
\multicolumn{4}{c}{\textbf{Schema-less}} \\
\cline{2-9}
& \small\textbf{Average} & \small\textbf{Median} & \small\textbf{Range} & \small\textbf{Std.dev} & \small\textbf{Average} & \small\textbf{Median} & \small\textbf{Range} & \small\textbf{Std.dev} \\
\midrule
Uncompressed & \small{44.1} & \small{29} & \small{89} & \small{32.9} & \small{102.7} & \small{89.5} & \small{85} & \small{32.4} \\ \hline
GZIP (compression level 9) & \small{47.5} & \small{43} & \small{50} & \small{14.8} & \small{107.2} & \small{103.5} & \small{45} & \small{16.4} \\ \hline
LZ4 (compression level 9) & \small{55.3} & \small{48} & \small{70} & \small{21.4} & \small{116.7} & \small{108.5} & \small{60} & \small{25.0} \\ \hline
LZMA (compression level 9) & \small{48.5} & \small{45.5} & \small{41} & \small{12.1} & \small{107.2} & \small{104.5} & \small{41} & \small{14.7} \\
\bottomrule
\end{tabularx}
\end{table*}

\begin{table*}[hb!]
\caption{The benchmark raw data results and schemas for the plot in \autoref{fig:benchmark-circlecimatrix}.}
\label{table:benchmark-circlecimatrix}
\begin{tabularx}{\linewidth}{X|l|l|l|l|l}
\toprule
\textbf{Serialization Format} & \textbf{Schema} & \textbf{Uncompressed} & \textbf{GZIP} & \textbf{LZ4} & \textbf{LZMA} \\
\midrule
ASN.1 (PER Unaligned) & \href{https://github.com/jviotti/binary-json-size-benchmark/blob/main/benchmark/circlecimatrix/asn1/schema.asn}{\small{\texttt{schema.asn}}} & 18 & 36 & 37 & 38 \\ \hline
Apache Avro (unframed) & \href{https://github.com/jviotti/binary-json-size-benchmark/blob/main/benchmark/circlecimatrix/avro/schema.json}{\small{\texttt{schema.json}}} & 15 & 32 & 34 & 36 \\ \hline
Microsoft Bond (Compact Binary v1) & \href{https://github.com/jviotti/binary-json-size-benchmark/blob/main/benchmark/circlecimatrix/bond/schema.bond}{\small{\texttt{schema.bond}}} & 30 & 43 & 49 & 45 \\ \hline
Cap'n Proto (Binary Encoding) & \href{https://github.com/jviotti/binary-json-size-benchmark/blob/main/benchmark/circlecimatrix/capnproto/schema.capnp}{\small{\texttt{schema.capnp}}} & 96 & 56 & 72 & 54 \\ \hline
Cap'n Proto (Packed Encoding) & \href{https://github.com/jviotti/binary-json-size-benchmark/blob/main/benchmark/circlecimatrix/capnproto-packed/schema.capnp}{\small{\texttt{schema.capnp}}} & 36 & 49 & 54 & 50 \\ \hline
FlatBuffers & \href{https://github.com/jviotti/binary-json-size-benchmark/blob/main/benchmark/circlecimatrix/flatbuffers/schema.fbs}{\small{\texttt{schema.fbs}}} & 104 & 82 & 104 & 77 \\ \hline
Protocol Buffers (Binary Wire Format) & \href{https://github.com/jviotti/binary-json-size-benchmark/blob/main/benchmark/circlecimatrix/protobuf/schema.proto}{\small{\texttt{schema.proto}}} & 26 & 43 & 45 & 46 \\ \hline
Apache Thrift (Compact Protocol) & \href{https://github.com/jviotti/binary-json-size-benchmark/blob/main/benchmark/circlecimatrix/thrift/schema.thrift}{\small{\texttt{schema.thrift}}} & 28 & 39 & 47 & 42 \\ \hline
\hline \textbf{JSON} & - & 95 & 99 & 112 & 101 \\ \hline \hline
BSON & - & 136 & 123 & 150 & 121 \\ \hline
CBOR & - & 72 & 89 & 91 & 90 \\ \hline
FlexBuffers & - & 157 & 134 & 151 & 131 \\ \hline
MessagePack & - & 72 & 90 & 91 & 92 \\ \hline
Smile & - & 87 & 104 & 106 & 105 \\ \hline
UBJSON & - & 92 & 103 & 111 & 104 \\
\bottomrule
\end{tabularx}
\end{table*}

\begin{figure*}[ht!]
\frame{\includegraphics[width=\linewidth]{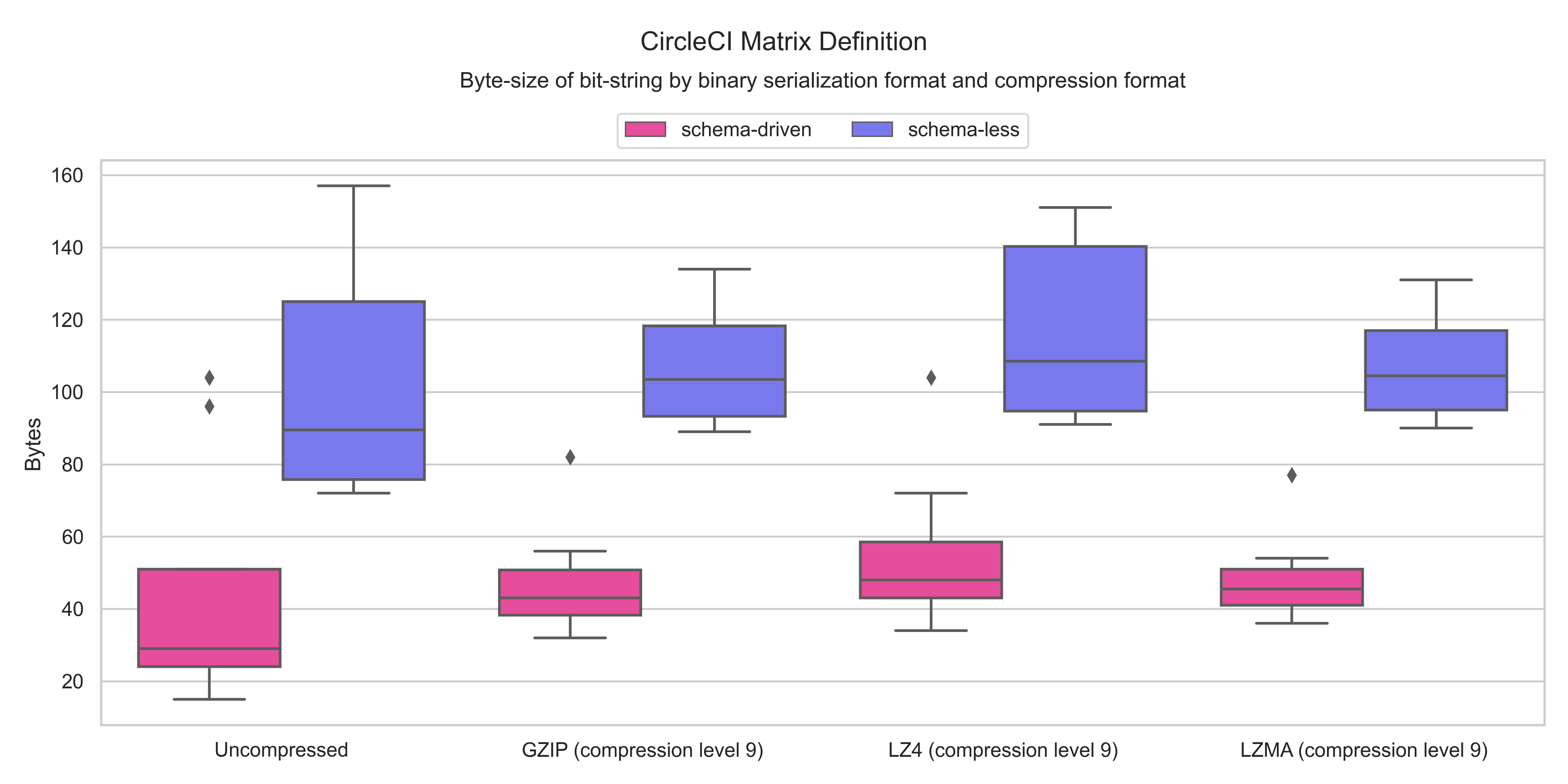}}
\caption{
Box plot of the statistical results in \autoref{table:benchmark-stats-circlecimatrix}.
}
\label{fig:benchmark-circlecimatrix-boxplot}
\end{figure*}

In \autoref{fig:benchmark-circlecimatrix-boxplot}, we observe the medians for
uncompressed schema-driven binary serialization specifications to be smaller in
comparison to uncompressed schema-less binary serialization specifications.  The range
between the upper and lower whiskers and the inter-quartile range of
uncompressed schema-driven binary serialization specifications is smaller than the
range between the upper and lower whiskers and the inter-quartile range of
uncompressed schema-less binary serialization specifications.


In terms of compression, LZMA results in the lower median for schema-driven
binary serialization specifications while GZIP results in the lower median for
schema-less binary serialization specifications. However, compression is not
space-efficient in terms of the median for both schema-driven and schema-less
binary serialization specifications. Additionally, the use of GZIP, LZ4 and
LZMA for schema-driven binary serialization specifications exhibit upper
outliers.  While compression does not contribute to space-efficiency, it
reduces the range between the upper and lower whiskers and inter-quartile range
for both schema-driven and schema-less binary serialization specifications.  In
particular, the compression format with the smaller range between the upper and
lower whiskers and the smaller inter-quartile range for schema-driven binary
serialization specifications is LZMA, the compression formats with the smaller
range between the upper and lower whiskers for schema-less binary serialization
specifications are GZIP and LZMA, and the compression format with the smaller
inter-quartile range for schema-less binary serialization specifications is
LZMA.


Overall, \we conclude that uncompressed schema-driven binary serialization
specifications are space-efficient in comparison to uncompressed schema-less binary
serialization specifications and that compression does not contribute to
space-efficiency in comparison to both uncompressed schema-driven and
schema-less binary serialization specifications.

\clearpage

\subsection{Grunt.js Clean Task Definition}
\label{sec:benchmark-gruntcontribclean}

\begin{figure*}[ht!]
\frame{\includegraphics[width=\linewidth]{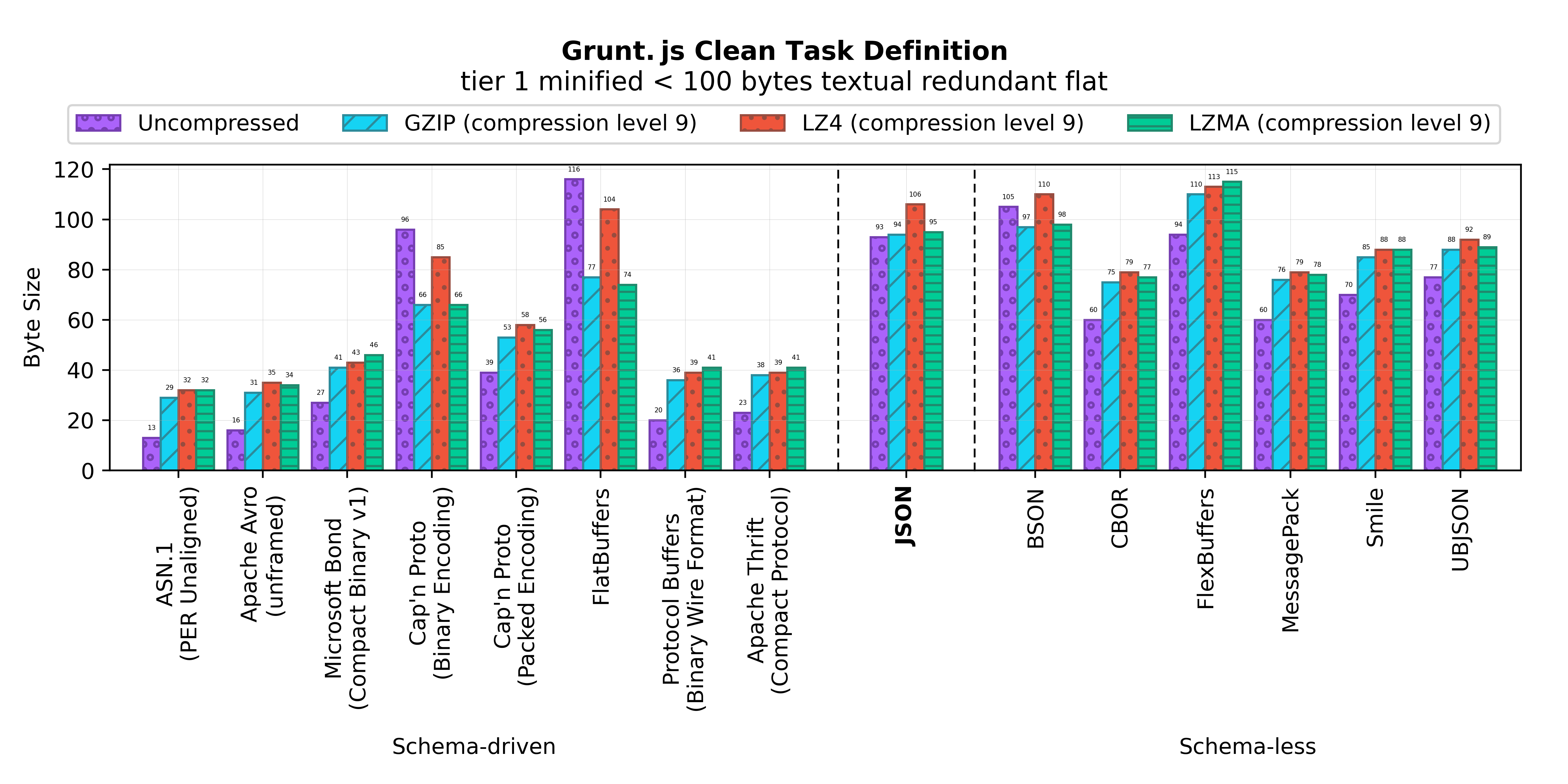}}
\caption{
The benchmark results for the Grunt.js Clean Task Definition test case listed in \autoref{table:benchmark-documents} and \autoref{table:benchmark-documents-1}.
}
\label{fig:benchmark-gruntcontribclean}
\end{figure*}

Grunt.js \footnote{\url{https://gruntjs.com}} is an open-source task runner for
the JavaScript \cite{ECMA-262} programming language used by a wide range of
companies in the software development industry such as Twitter, Adobe, and
Mozilla \footnote{\url{https://gruntjs.com/who-uses-grunt}}. In
\autoref{fig:benchmark-gruntcontribclean}, \we demonstrate a \textbf{Tier 1
minified $<$ 100 bytes textual redundant flat} (Tier 1 TRF from
\autoref{table:json-taxonomy}) JSON document that consists of an example
configuration for a built-in plugin to clear files and folders called
\texttt{grunt-contrib-clean}
\footnote{\url{https://github.com/gruntjs/grunt-contrib-clean}}.

The smallest bit-string is produced by ASN.1 PER Unaligned \cite{asn1-per} (13
bytes), followed by Apache Avro \cite{avro} (16 bytes) and Protocol Buffers
\cite{protocolbuffers} (20 bytes). The binary serialization specifications that
produced the smallest bit-strings are schema-driven and sequential \cite{viotti2022survey}.
Conversely, the largest bit-string is produced by FlatBuffers
\cite{flatbuffers} (116 bytes), followed by BSON \cite{bson} (105 bytes) and
Cap'n Proto Binary Encoding \cite{capnproto} (96 bytes).  With the exception of
BSON, the binary serialization specifications that produced the largest bit-strings
are schema-driven and pointer-based \cite{viotti2022survey}.  In comparison to JSON
\cite{ECMA-404} (93 bytes), binary serialization achieves a \textbf{7.1x} size
reduction in the best case for this input document.  However, 4 out of the 14
JSON-compatible binary serialization specifications listed in
\autoref{table:benchmark-specifications-schema-driven} and
\autoref{table:benchmark-specifications-schema-less} result in bit-strings that are
larger than JSON: Cap'n Proto Binary Encoding \cite{capnproto}, FlatBuffers
\cite{flatbuffers}, BSON \cite{bson} and FlexBuffers \cite{flexbuffers}. These
binary serialization specifications are either schema-less or schema-driven and
pointer-based.

For this Tier 1 TRF document, the best performing schema-driven serialization
specification achieves a \textbf{4.6x} size reduction compared to the best performing
schema-less serialization specification: CBOR \cite{RFC7049} and MessagePack
\cite{messagepack} (60 bytes).  As shown in
\autoref{table:benchmark-stats-gruntcontribclean}, uncompressed schema-driven
specifications provide smaller \emph{average} and \emph{median} bit-strings than
uncompressed schema-less specifications. However, as highlighted by the \emph{range}
and \emph{standard deviation}, uncompressed schema-driven specifications exhibit
higher size reduction variability depending on the expressiveness of the schema
language (i.e. how the language constructs allow you to model the data) and the
size optimizations devised by its authors. With the exception of the
pointer-based binary serialization specifications Cap'n Proto Binary Encoding
\cite{capnproto} and FlatBuffers \cite{flatbuffers}, the selection of
schema-driven serialization specifications listed in
\autoref{table:benchmark-specifications-schema-driven} produce bit-strings that are
equal to or smaller than their schema-less counterparts listed in
\autoref{table:benchmark-specifications-schema-less}.  The best performing sequential
serialization specification achieves a \textbf{3x} size reduction compared to the best
performing pointer-based serialization specification: Cap'n Proto Packed Encoding
\cite{capnproto} (39 bytes).

The compression formats listed in
\autoref{sec:benchmark-compression-formats} result in positive gains for
the bit-strings produced by Cap'n Proto Binary Encoding \cite{capnproto},
FlatBuffers \cite{flatbuffers} and BSON \cite{bson}.  The best performing
uncompressed binary serialization specification achieves a \textbf{7.2x} size
reduction compared to the best performing compression format for JSON: GZIP
\cite{RFC1952} (94 bytes).

\begin{table*}[hb!]
\caption{A byte-size statistical analysis of the benchmark results shown in \autoref{fig:benchmark-gruntcontribclean} divided by schema-driven and schema-less specifications.}
\label{table:benchmark-stats-gruntcontribclean}
\begin{tabularx}{\linewidth}{X|l|l|l|l|l|l|l|l}
\toprule
\multirow{2}{*}{\textbf{Category}} &
\multicolumn{4}{c|}{\textbf{Schema-driven}} &
\multicolumn{4}{c}{\textbf{Schema-less}} \\
\cline{2-9}
& \small\textbf{Average} & \small\textbf{Median} & \small\textbf{Range} & \small\textbf{Std.dev} & \small\textbf{Average} & \small\textbf{Median} & \small\textbf{Range} & \small\textbf{Std.dev} \\
\midrule
Uncompressed & \small{43.8} & \small{25} & \small{103} & \small{37.0} & \small{77.7} & \small{73.5} & \small{45} & \small{16.8} \\ \hline
GZIP (compression level 9) & \small{46.4} & \small{39.5} & \small{48} & \small{16.2} & \small{88.5} & \small{86.5} & \small{35} & \small{12.1} \\ \hline
LZ4 (compression level 9) & \small{54.4} & \small{41} & \small{72} & \small{24.7} & \small{93.5} & \small{90} & \small{34} & \small{13.6} \\ \hline
LZMA (compression level 9) & \small{48.8} & \small{43.5} & \small{42} & \small{14.2} & \small{90.8} & \small{88.5} & \small{38} & \small{12.9} \\
\bottomrule
\end{tabularx}
\end{table*}

\begin{table*}[hb!]
\caption{The benchmark raw data results and schemas for the plot in \autoref{fig:benchmark-gruntcontribclean}.}
\label{table:benchmark-gruntcontribclean}
\begin{tabularx}{\linewidth}{X|l|l|l|l|l}
\toprule
\textbf{Serialization Format} & \textbf{Schema} & \textbf{Uncompressed} & \textbf{GZIP} & \textbf{LZ4} & \textbf{LZMA} \\
\midrule
ASN.1 (PER Unaligned) & \href{https://github.com/jviotti/binary-json-size-benchmark/blob/main/benchmark/gruntcontribclean/asn1/schema.asn}{\small{\texttt{schema.asn}}} & 13 & 29 & 32 & 32 \\ \hline
Apache Avro (unframed) & \href{https://github.com/jviotti/binary-json-size-benchmark/blob/main/benchmark/gruntcontribclean/avro/schema.json}{\small{\texttt{schema.json}}} & 16 & 31 & 35 & 34 \\ \hline
Microsoft Bond (Compact Binary v1) & \href{https://github.com/jviotti/binary-json-size-benchmark/blob/main/benchmark/gruntcontribclean/bond/schema.bond}{\small{\texttt{schema.bond}}} & 27 & 41 & 43 & 46 \\ \hline
Cap'n Proto (Binary Encoding) & \href{https://github.com/jviotti/binary-json-size-benchmark/blob/main/benchmark/gruntcontribclean/capnproto/schema.capnp}{\small{\texttt{schema.capnp}}} & 96 & 66 & 85 & 66 \\ \hline
Cap'n Proto (Packed Encoding) & \href{https://github.com/jviotti/binary-json-size-benchmark/blob/main/benchmark/gruntcontribclean/capnproto-packed/schema.capnp}{\small{\texttt{schema.capnp}}} & 39 & 53 & 58 & 56 \\ \hline
FlatBuffers & \href{https://github.com/jviotti/binary-json-size-benchmark/blob/main/benchmark/gruntcontribclean/flatbuffers/schema.fbs}{\small{\texttt{schema.fbs}}} & 116 & 77 & 104 & 74 \\ \hline
Protocol Buffers (Binary Wire Format) & \href{https://github.com/jviotti/binary-json-size-benchmark/blob/main/benchmark/gruntcontribclean/protobuf/schema.proto}{\small{\texttt{schema.proto}}} & 20 & 36 & 39 & 41 \\ \hline
Apache Thrift (Compact Protocol) & \href{https://github.com/jviotti/binary-json-size-benchmark/blob/main/benchmark/gruntcontribclean/thrift/schema.thrift}{\small{\texttt{schema.thrift}}} & 23 & 38 & 39 & 41 \\ \hline
\hline \textbf{JSON} & - & 93 & 94 & 106 & 95 \\ \hline \hline
BSON & - & 105 & 97 & 110 & 98 \\ \hline
CBOR & - & 60 & 75 & 79 & 77 \\ \hline
FlexBuffers & - & 94 & 110 & 113 & 115 \\ \hline
MessagePack & - & 60 & 76 & 79 & 78 \\ \hline
Smile & - & 70 & 85 & 88 & 88 \\ \hline
UBJSON & - & 77 & 88 & 92 & 89 \\
\bottomrule
\end{tabularx}
\end{table*}

\begin{figure*}[ht!]
\frame{\includegraphics[width=\linewidth]{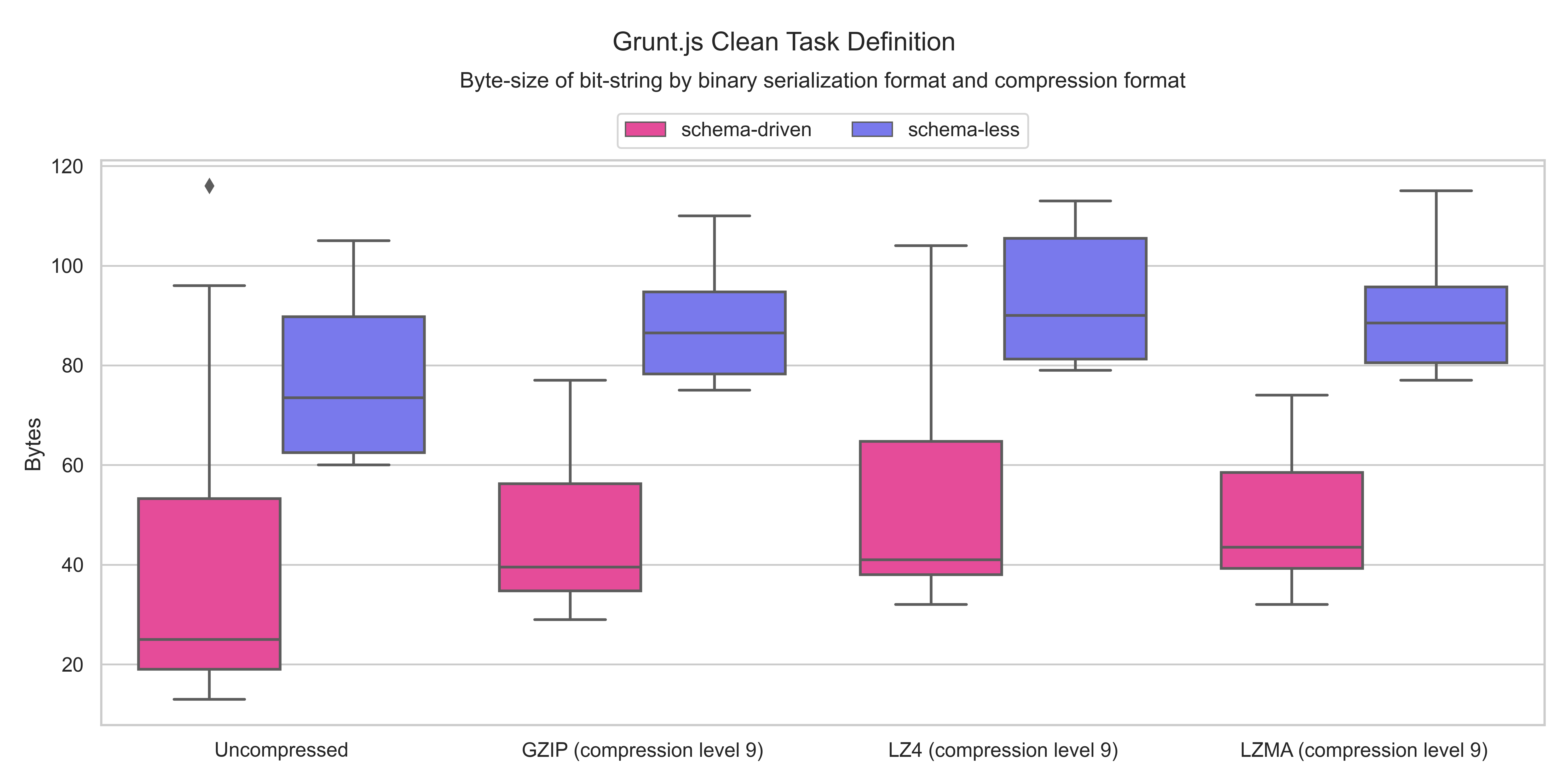}}
\caption{
Box plot of the statistical results in \autoref{table:benchmark-stats-gruntcontribclean}.
}
\label{fig:benchmark-gruntcontribclean-boxplot}
\end{figure*}

In \autoref{fig:benchmark-gruntcontribclean-boxplot}, we observe the medians
for uncompressed schema-driven binary serialization specifications to be smaller in
comparison to uncompressed schema-less binary serialization specifications.  The range
between the upper and lower whiskers and the inter-quartile range of
uncompressed schema-less binary serialization specifications is smaller than the range
between the upper and lower whiskers and the inter-quartile range of
uncompressed schema-driven binary serialization specifications.


In terms of compression, GZIP results in the lower median for both
schema-driven and schema-less binary serialization specifications.  However,
compression is not space-efficient in terms of the median for both
schema-driven and schema-less binary serialization specifications.  While
compression does not contribute to space-efficiency, it reduces the range
between the upper and lower whiskers and inter-quartile range for both
schema-driven and schema-less binary serialization specifications.  In
particular, the compression format with the smaller range between the upper and
lower whiskers for schema-driven binary serialization specifications is LZMA,
the compression formats with the smaller inter-quartile range for schema-driven
binary serialization specifications are GZIP and LZMA, the compression formats
with the smaller range between the upper and lower whiskers for schema-less
binary serialization specifications are GZIP and LZ4, and the compression
formats with the smaller inter-quartile range for schema-less binary
serialization specifications are GZIP and LZMA.


Overall, \we conclude that uncompressed schema-driven binary serialization
specifications are space-efficient in comparison to uncompressed schema-less binary
serialization specifications and that compression does not contribute to
space-efficiency in comparison to both uncompressed schema-driven and
schema-less binary serialization specifications.

\clearpage

\subsection{CommitLint Configuration}
\label{sec:benchmark-commitlint}

\begin{figure*}[ht!]
\frame{\includegraphics[width=\linewidth]{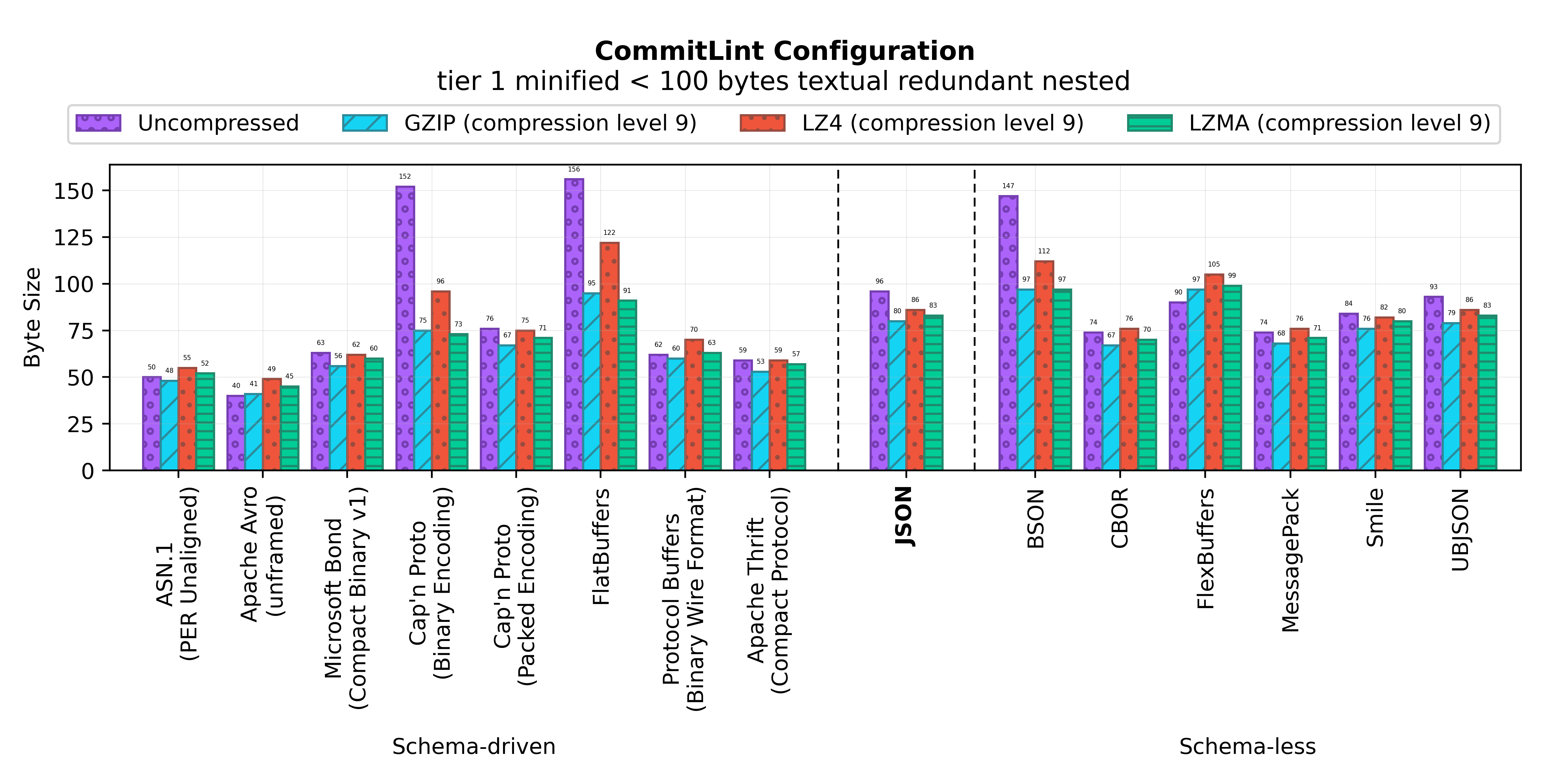}}
\caption{
The benchmark results for the CommitLint Configuration test case listed in \autoref{table:benchmark-documents} and \autoref{table:benchmark-documents-1}.
}
\label{fig:benchmark-commitlint}
\end{figure*}

CommitLint \footnote{\url{https://commitlint.js.org/\#/}} is an open-source
command-line tool to enforce version-control commit conventions in software
engineering projects. CommitLint is a community effort under the Conventional
Changelog \footnote{\url{https://github.com/conventional-changelog}}
organization formed by employees from companies including GitHub
\footnote{\url{https://github.com/zeke}} and Google
\footnote{\url{https://github.com/bcoe}}. In
\autoref{fig:benchmark-commitlint}, \we demonstrate a \textbf{Tier 1 minified
$<$ 100 bytes textual redundant nested} (Tier 1 TRN from
\autoref{table:json-taxonomy}) JSON document that represents a CommitLint
configuration file which declares that the subject and the scope of any commit
must be written in lower-case form.

The smallest bit-string is produced by Apache Avro \cite{avro} (40 bytes),
followed by ASN.1 PER Unaligned \cite{asn1-per} (50 bytes) and Apache Thrift
\cite{slee2007thrift} (59 bytes).  The binary serialization specifications that
produced the smallest bit-strings are schema-driven and sequential \cite{viotti2022survey}.
Conversely, the largest bit-string is produced by FlatBuffers
\cite{flatbuffers} (156 bytes), Cap'n Proto Binary Encoding \cite{capnproto}
(152 bytes) and BSON \cite{bson} (147 bytes). With the exception of BSON, the
binary serialization specifications that produced the largest bit-strings are
pointer-based \cite{viotti2022survey}.  In comparison to JSON \cite{ECMA-404} (96 bytes),
binary serialization achieves a \textbf{2.4x} size reduction in the best case
for this input document. However, 3 out of the 14 JSON-compatible binary
serialization specifications listed in \autoref{table:benchmark-specifications-schema-driven}
and \autoref{table:benchmark-specifications-schema-less} result in bit-strings that
are larger than JSON: Cap'n Proto Binary Encoding \cite{capnproto}, FlatBuffers
\cite{flatbuffers} and BSON \cite{bson}. These binary serialization specifications are
either schema-less or schema-driven and pointer-based.

For this Tier 1 TRN document, the best performing schema-driven serialization
specification only achieves a \textbf{1.8x} size reduction compared to the best
performing schema-less serialization specification: CBOR \cite{RFC7049} and
MessagePack \cite{messagepack} (74 bytes).  As shown in
\autoref{table:benchmark-stats-commitlint}, uncompressed schema-driven specifications
provide smaller \emph{average} and \emph{median} bit-strings than uncompressed
schema-less specifications. However, as highlighted by the \emph{range} and
\emph{standard deviation}, uncompressed schema-driven specifications exhibit higher
size reduction variability depending on the expressiveness of the schema
language (i.e. how the language constructs allow you to model the data) and the
size optimizations devised by its authors. With the exception of the
pointer-based binary serialization specifications Cap'n Proto Binary Encoding
\cite{capnproto}, Cap'n Proto Packed Encoding \cite{capnproto} and FlatBuffers
\cite{flatbuffers}, the selection of schema-driven serialization specifications listed
in \autoref{table:benchmark-specifications-schema-driven} produce bit-strings that are
equal to or smaller than their schema-less counterparts listed in
\autoref{table:benchmark-specifications-schema-less}.  The best performing sequential
serialization specification achieves a \textbf{1.9x} size reduction compared to the
best performing pointer-based serialization specification: Cap'n Proto Packed Encoding
\cite{capnproto} (76 bytes).

The compression formats listed in
\autoref{sec:benchmark-compression-formats} result in positive gains for
all the bit-strings except the ones produced by Apache Avro \cite{avro} and
FlexBuffers \cite{flexbuffers}. The best performing uncompressed binary
serialization specification achieves a \textbf{2x} size reduction compared to
the best performing compression format for JSON: GZIP \cite{RFC1952} (80
bytes).

\begin{table*}[hb!]
\caption{A byte-size statistical analysis of the benchmark results shown in \autoref{fig:benchmark-commitlint} divided by schema-driven and schema-less specifications.}
\label{table:benchmark-stats-commitlint}
\begin{tabularx}{\linewidth}{X|l|l|l|l|l|l|l|l}
\toprule
\multirow{2}{*}{\textbf{Category}} &
\multicolumn{4}{c|}{\textbf{Schema-driven}} &
\multicolumn{4}{c}{\textbf{Schema-less}} \\
\cline{2-9}
& \small\textbf{Average} & \small\textbf{Median} & \small\textbf{Range} & \small\textbf{Std.dev} & \small\textbf{Average} & \small\textbf{Median} & \small\textbf{Range} & \small\textbf{Std.dev} \\
\midrule
Uncompressed & \small{82.3} & \small{62.5} & \small{116} & \small{42.6} & \small{93.7} & \small{87} & \small{73} & \small{24.9} \\ \hline
GZIP (compression level 9) & \small{61.9} & \small{58} & \small{54} & \small{16.0} & \small{80.7} & \small{77.5} & \small{30} & \small{12.3} \\ \hline
LZ4 (compression level 9) & \small{73.5} & \small{66} & \small{73} & \small{22.8} & \small{89.5} & \small{84} & \small{36} & \small{14.0} \\ \hline
LZMA (compression level 9) & \small{64} & \small{61.5} & \small{46} & \small{13.4} & \small{83.3} & \small{81.5} & \small{29} & \small{11.4} \\
\bottomrule
\end{tabularx}
\end{table*}

\begin{table*}[hb!]
\caption{The benchmark raw data results and schemas for the plot in \autoref{fig:benchmark-commitlint}.}
\label{table:benchmark-commitlint}
\begin{tabularx}{\linewidth}{X|l|l|l|l|l}
\toprule
\textbf{Serialization Format} & \textbf{Schema} & \textbf{Uncompressed} & \textbf{GZIP} & \textbf{LZ4} & \textbf{LZMA} \\
\midrule
ASN.1 (PER Unaligned) & \href{https://github.com/jviotti/binary-json-size-benchmark/blob/main/benchmark/commitlint/asn1/schema.asn}{\small{\texttt{schema.asn}}} & 50 & 48 & 55 & 52 \\ \hline
Apache Avro (unframed) & \href{https://github.com/jviotti/binary-json-size-benchmark/blob/main/benchmark/commitlint/avro/schema.json}{\small{\texttt{schema.json}}} & 40 & 41 & 49 & 45 \\ \hline
Microsoft Bond (Compact Binary v1) & \href{https://github.com/jviotti/binary-json-size-benchmark/blob/main/benchmark/commitlint/bond/schema.bond}{\small{\texttt{schema.bond}}} & 63 & 56 & 62 & 60 \\ \hline
Cap'n Proto (Binary Encoding) & \href{https://github.com/jviotti/binary-json-size-benchmark/blob/main/benchmark/commitlint/capnproto/schema.capnp}{\small{\texttt{schema.capnp}}} & 152 & 75 & 96 & 73 \\ \hline
Cap'n Proto (Packed Encoding) & \href{https://github.com/jviotti/binary-json-size-benchmark/blob/main/benchmark/commitlint/capnproto-packed/schema.capnp}{\small{\texttt{schema.capnp}}} & 76 & 67 & 75 & 71 \\ \hline
FlatBuffers & \href{https://github.com/jviotti/binary-json-size-benchmark/blob/main/benchmark/commitlint/flatbuffers/schema.fbs}{\small{\texttt{schema.fbs}}} & 156 & 95 & 122 & 91 \\ \hline
Protocol Buffers (Binary Wire Format) & \href{https://github.com/jviotti/binary-json-size-benchmark/blob/main/benchmark/commitlint/protobuf/schema.proto}{\small{\texttt{schema.proto}}} & 62 & 60 & 70 & 63 \\ \hline
Apache Thrift (Compact Protocol) & \href{https://github.com/jviotti/binary-json-size-benchmark/blob/main/benchmark/commitlint/thrift/schema.thrift}{\small{\texttt{schema.thrift}}} & 59 & 53 & 59 & 57 \\ \hline
\hline \textbf{JSON} & - & 96 & 80 & 86 & 83 \\ \hline \hline
BSON & - & 147 & 97 & 112 & 97 \\ \hline
CBOR & - & 74 & 67 & 76 & 70 \\ \hline
FlexBuffers & - & 90 & 97 & 105 & 99 \\ \hline
MessagePack & - & 74 & 68 & 76 & 71 \\ \hline
Smile & - & 84 & 76 & 82 & 80 \\ \hline
UBJSON & - & 93 & 79 & 86 & 83 \\
\bottomrule
\end{tabularx}
\end{table*}

\begin{figure*}[ht!]
\frame{\includegraphics[width=\linewidth]{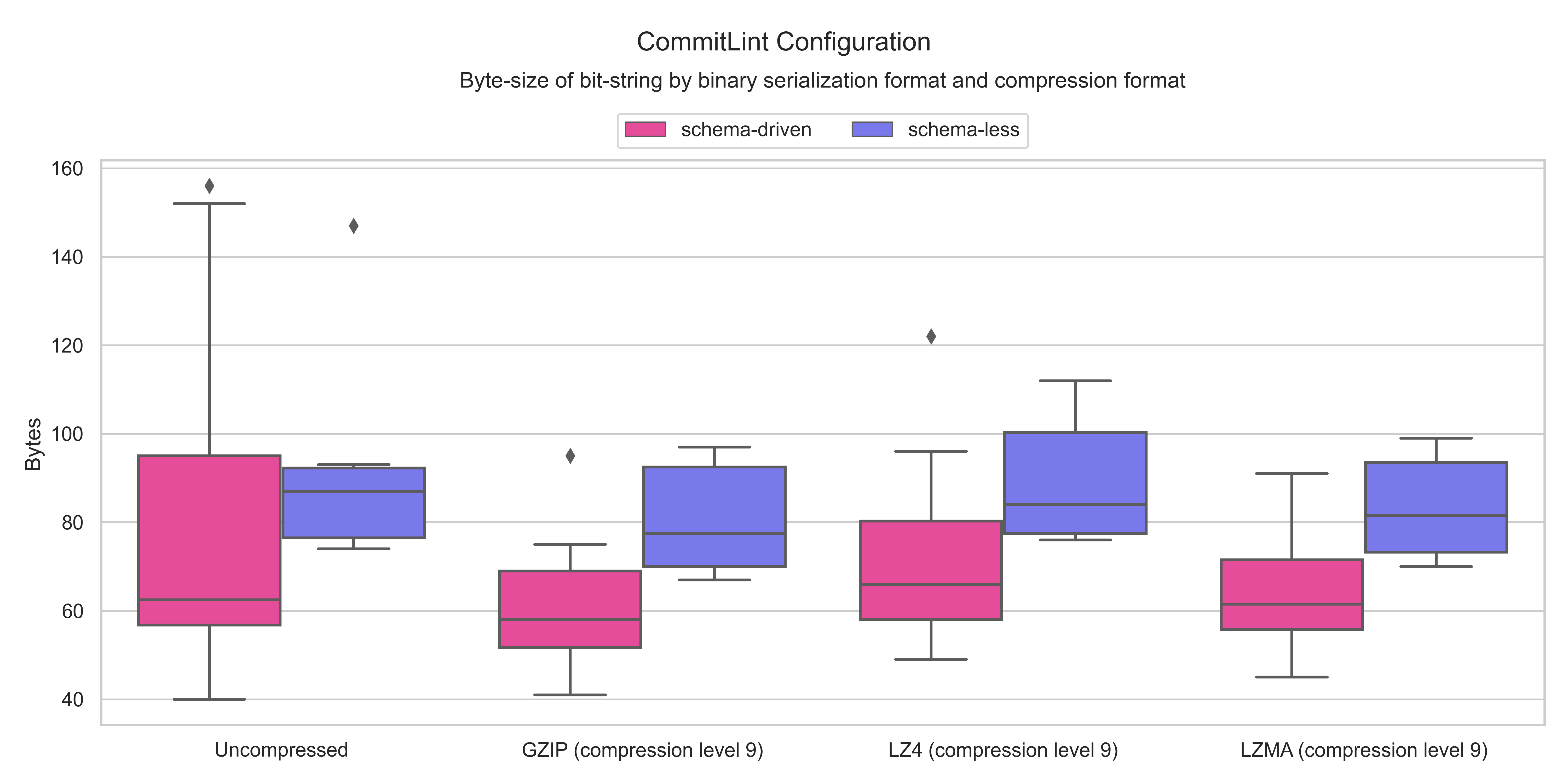}}
\caption{
Box plot of the statistical results in \autoref{table:benchmark-stats-commitlint}.
}
\label{fig:benchmark-commitlint-boxplot}
\end{figure*}

In \autoref{fig:benchmark-commitlint-boxplot}, we observe the medians for
uncompressed schema-driven binary serialization specifications to be smaller in
comparison to uncompressed schema-less binary serialization specifications.  The range
between the upper and lower whiskers and the inter-quartile range of
uncompressed schema-less binary serialization specifications is smaller than the range
between the upper and lower whiskers and the inter-quartile range of
uncompressed schema-driven binary serialization specifications. However, their
respective quartiles overlap.


In terms of compression, GZIP results in the lower median for both
schema-driven and schema-less binary serialization specifications.
Additionally, GZIP and LZMA are space-efficient in terms of the median in
comparison to uncompressed schema-driven binary serialization specifications
and GZIP, LZ4 and LZMA are space-efficient in terms of the median in comparison
to uncompressed schema-less binary serialization specifications. However, the
use of GZIP and LZ4 for schema-driven binary serialization specifications
exhibit upper outliers.  Nevertheless, compression reduces the range between
the upper and lower whiskers and inter-quartile range for schema-driven binary
serialization specifications.  In particular, the compression format with the
smaller range between the upper and lower whiskers for schema-driven binary
serialization specifications is GZIP, the compression format with the smaller
inter-quartile range for schema-driven binary serialization specifications is
LZMA, and the compression formats with the smaller range between the upper and
lower whiskers and the smaller inter-quartile range for schema-less binary
serialization specifications are GZIP and LZMA.


Overall, \we conclude that uncompressed schema-driven binary serialization
specifications are space-efficient in comparison to uncompressed schema-less
binary serialization specifications. GZIP and LZMA are space-efficient in
comparison to uncompressed schema-driven binary serialization specifications
and all the considered compression formats are space-efficient in comparison to
uncompressed schema-less binary serialization specifications.

\clearpage

\subsection{TSLint Linter Definition (Extends Only)}
\label{sec:benchmark-tslintextend}

\begin{figure*}[ht!]
\frame{\includegraphics[width=\linewidth]{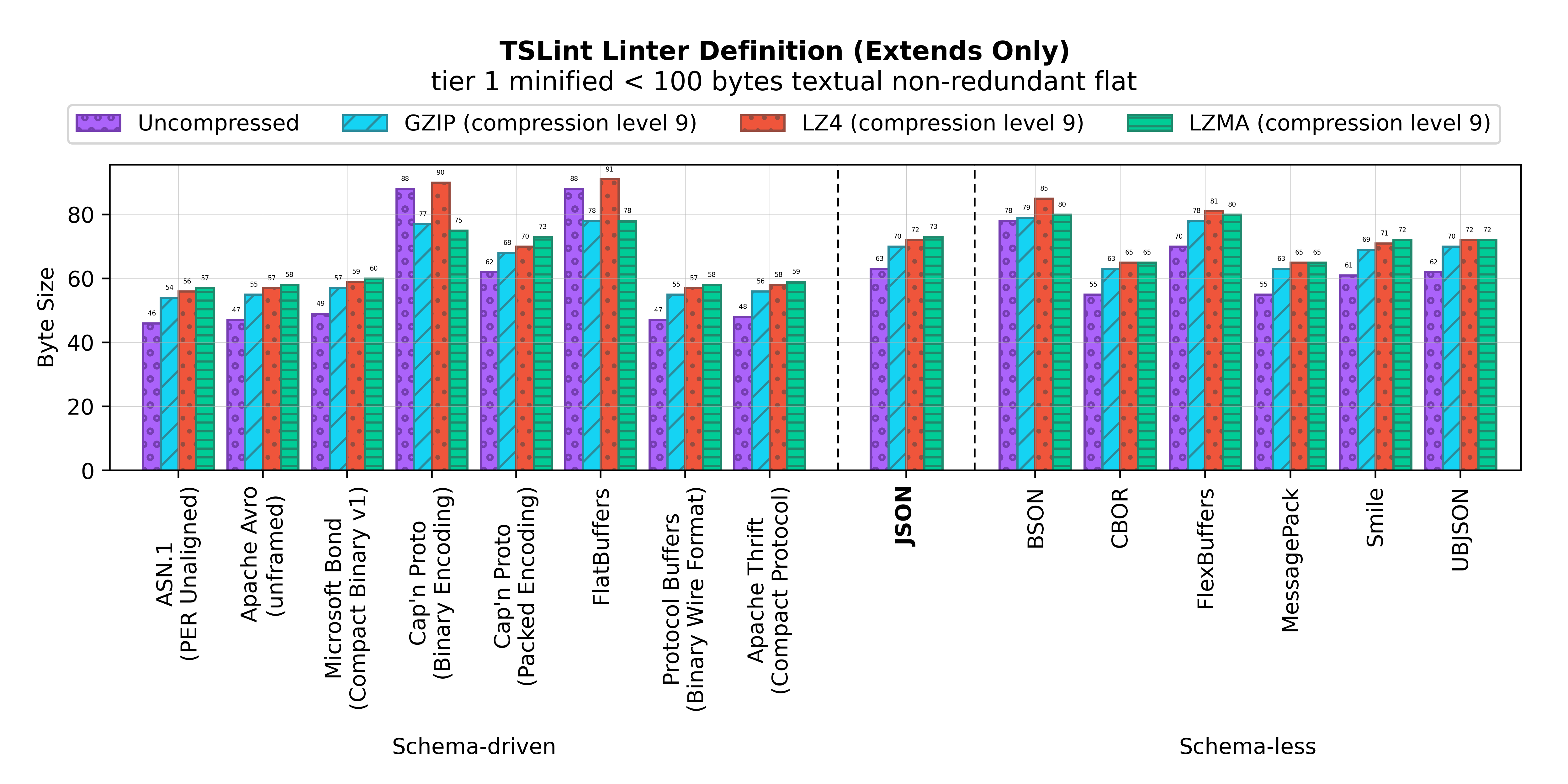}}
\caption{
The benchmark results for the TSLint Linter Definition (Extends Only) test case listed in \autoref{table:benchmark-documents} and \autoref{table:benchmark-documents-1}.
}
\label{fig:benchmark-tslintextend}
\end{figure*}

TSLint \footnote{\url{https://palantir.github.io/tslint}} is now an obsolete
open-source linter for the TypeScript
\footnote{\url{https://www.typescriptlang.org}} programming language. TSLint
was created by the Big Data analytics company Palantir
\footnote{\url{https://www.palantir.com}} and was merged with the ESLint
open-source JavaScript linter in 2019
\footnote{\url{https://github.com/palantir/tslint/issues/4534}}. In
\autoref{fig:benchmark-tslintextend}, \we demonstrate a \textbf{Tier 1 minified
$<$ 100 bytes textual non-redundant flat} (Tier 1 TNF from
\autoref{table:json-taxonomy}) JSON document that consists of a basic TSLint
configuration that only extends a set of existing TSLint configurations.

The smallest bit-string is produced by ASN.1 PER Unaligned \cite{asn1-per} (46
bytes), closely followed by both Apache Avro \cite{avro} and Protocol Buffers
\cite{protocolbuffers} (47 bytes), and Apache Thrift \cite{slee2007thrift} (48
bytes). The binary serialization specifications that produced the smallest bit-strings
are schema-driven and sequential \cite{viotti2022survey}.  Conversely, the largest bit-string
is produced by both Cap'n Proto Binary Encoding \cite{capnproto} and
FlatBuffers \cite{flatbuffers} (88 bytes), followed by BSON \cite{bson} (78
bytes) and FlexBuffers \cite{flexbuffers} (70 bytes). With the exception of
BSON, the binary serialization specifications that produced the largest bit-strings
are pointer-based \cite{viotti2022survey}.  In comparison to JSON \cite{ECMA-404} (63 bytes),
binary serialization only achieves a \textbf{1.3x} size reduction in the best
case for this input document.  Additionally, 4 out of the 14 JSON-compatible
binary serialization specifications listed in
\autoref{table:benchmark-specifications-schema-driven} and
\autoref{table:benchmark-specifications-schema-less} result in bit-strings that are
larger than JSON: Cap'n Proto Binary Encoding \cite{capnproto}, FlatBuffers
\cite{flatbuffers}, BSON \cite{bson} and FlexBuffers \cite{flexbuffers}. These
binary serialization specifications are either schema-less or schema-driven and
pointer-based.

For this Tier 1 TNF document, the best performing schema-driven serialization
specification only achieves a \textbf{1.1x} size reduction compared to the best
performing schema-less serialization specification: CBOR \cite{RFC7049} and
MessagePack \cite{messagepack} (55 bytes).  As shown in
\autoref{table:benchmark-stats-tslintextend}, uncompressed schema-driven
specifications provide smaller \emph{average} and \emph{median} bit-strings than
uncompressed schema-less specifications. However, as highlighted by the \emph{range}
and \emph{standard deviation}, uncompressed schema-driven specifications exhibit
higher size reduction variability depending on the expressiveness of the schema
language (i.e. how the language constructs allow you to model the data) and the
size optimizations devised by its authors.  With the exception of the
pointer-based binary serialization specifications Cap'n Proto Binary Encoding
\cite{capnproto}, Cap'n Proto Packed Encoding \cite{capnproto} and FlatBuffers
\cite{flatbuffers}, the selection of schema-driven serialization specifications listed
in \autoref{table:benchmark-specifications-schema-driven} produce bit-strings that are
equal to or smaller than their schema-less counterparts listed in
\autoref{table:benchmark-specifications-schema-less}.  The best performing sequential
serialization specification only achieves a \textbf{1.3x} size reduction compared to
the best performing pointer-based serialization specification: Cap'n Proto Packed
Encoding \cite{capnproto} (62 bytes).

The compression formats listed in
\autoref{sec:benchmark-compression-formats} result in positive gains for
the bit-strings produced by Cap'n Proto Binary Encoding \cite{capnproto} and
FlatBuffers \cite{flatbuffers}. The best performing uncompressed binary
serialization specification achieves a \textbf{1.5x} size reduction compared to
the best performing compression format for JSON: GZIP \cite{RFC1952} (70
bytes).

\begin{table*}[hb!]
\caption{A byte-size statistical analysis of the benchmark results shown in \autoref{fig:benchmark-tslintextend} divided by schema-driven and schema-less specifications.}
\label{table:benchmark-stats-tslintextend}
\begin{tabularx}{\linewidth}{X|l|l|l|l|l|l|l|l}
\toprule
\multirow{2}{*}{\textbf{Category}} &
\multicolumn{4}{c|}{\textbf{Schema-driven}} &
\multicolumn{4}{c}{\textbf{Schema-less}} \\
\cline{2-9}
& \small\textbf{Average} & \small\textbf{Median} & \small\textbf{Range} & \small\textbf{Std.dev} & \small\textbf{Average} & \small\textbf{Median} & \small\textbf{Range} & \small\textbf{Std.dev} \\
\midrule
Uncompressed & \small{59.4} & \small{48.5} & \small{42} & \small{17.2} & \small{63.5} & \small{61.5} & \small{23} & \small{8.2} \\ \hline
GZIP (compression level 9) & \small{62.5} & \small{56.5} & \small{24} & \small{9.6} & \small{70.3} & \small{69.5} & \small{16} & \small{6.4} \\ \hline
LZ4 (compression level 9) & \small{67.3} & \small{58.5} & \small{35} & \small{14.1} & \small{73.2} & \small{71.5} & \small{20} & \small{7.5} \\ \hline
LZMA (compression level 9) & \small{64.8} & \small{59.5} & \small{21} & \small{8.3} & \small{72.3} & \small{72} & \small{15} & \small{6.1} \\
\bottomrule
\end{tabularx}
\end{table*}

\begin{table*}[hb!]
\caption{The benchmark raw data results and schemas for the plot in \autoref{fig:benchmark-tslintextend}.}
\label{table:benchmark-tslintextend}
\begin{tabularx}{\linewidth}{X|l|l|l|l|l}
\toprule
\textbf{Serialization Format} & \textbf{Schema} & \textbf{Uncompressed} & \textbf{GZIP} & \textbf{LZ4} & \textbf{LZMA} \\
\midrule
ASN.1 (PER Unaligned) & \href{https://github.com/jviotti/binary-json-size-benchmark/blob/main/benchmark/tslintextend/asn1/schema.asn}{\small{\texttt{schema.asn}}} & 46 & 54 & 56 & 57 \\ \hline
Apache Avro (unframed) & \href{https://github.com/jviotti/binary-json-size-benchmark/blob/main/benchmark/tslintextend/avro/schema.json}{\small{\texttt{schema.json}}} & 47 & 55 & 57 & 58 \\ \hline
Microsoft Bond (Compact Binary v1) & \href{https://github.com/jviotti/binary-json-size-benchmark/blob/main/benchmark/tslintextend/bond/schema.bond}{\small{\texttt{schema.bond}}} & 49 & 57 & 59 & 60 \\ \hline
Cap'n Proto (Binary Encoding) & \href{https://github.com/jviotti/binary-json-size-benchmark/blob/main/benchmark/tslintextend/capnproto/schema.capnp}{\small{\texttt{schema.capnp}}} & 88 & 77 & 90 & 75 \\ \hline
Cap'n Proto (Packed Encoding) & \href{https://github.com/jviotti/binary-json-size-benchmark/blob/main/benchmark/tslintextend/capnproto-packed/schema.capnp}{\small{\texttt{schema.capnp}}} & 62 & 68 & 70 & 73 \\ \hline
FlatBuffers & \href{https://github.com/jviotti/binary-json-size-benchmark/blob/main/benchmark/tslintextend/flatbuffers/schema.fbs}{\small{\texttt{schema.fbs}}} & 88 & 78 & 91 & 78 \\ \hline
Protocol Buffers (Binary Wire Format) & \href{https://github.com/jviotti/binary-json-size-benchmark/blob/main/benchmark/tslintextend/protobuf/schema.proto}{\small{\texttt{schema.proto}}} & 47 & 55 & 57 & 58 \\ \hline
Apache Thrift (Compact Protocol) & \href{https://github.com/jviotti/binary-json-size-benchmark/blob/main/benchmark/tslintextend/thrift/schema.thrift}{\small{\texttt{schema.thrift}}} & 48 & 56 & 58 & 59 \\ \hline
\hline \textbf{JSON} & - & 63 & 70 & 72 & 73 \\ \hline \hline
BSON & - & 78 & 79 & 85 & 80 \\ \hline
CBOR & - & 55 & 63 & 65 & 65 \\ \hline
FlexBuffers & - & 70 & 78 & 81 & 80 \\ \hline
MessagePack & - & 55 & 63 & 65 & 65 \\ \hline
Smile & - & 61 & 69 & 71 & 72 \\ \hline
UBJSON & - & 62 & 70 & 72 & 72 \\
\bottomrule
\end{tabularx}
\end{table*}

\begin{figure*}[ht!]
\frame{\includegraphics[width=\linewidth]{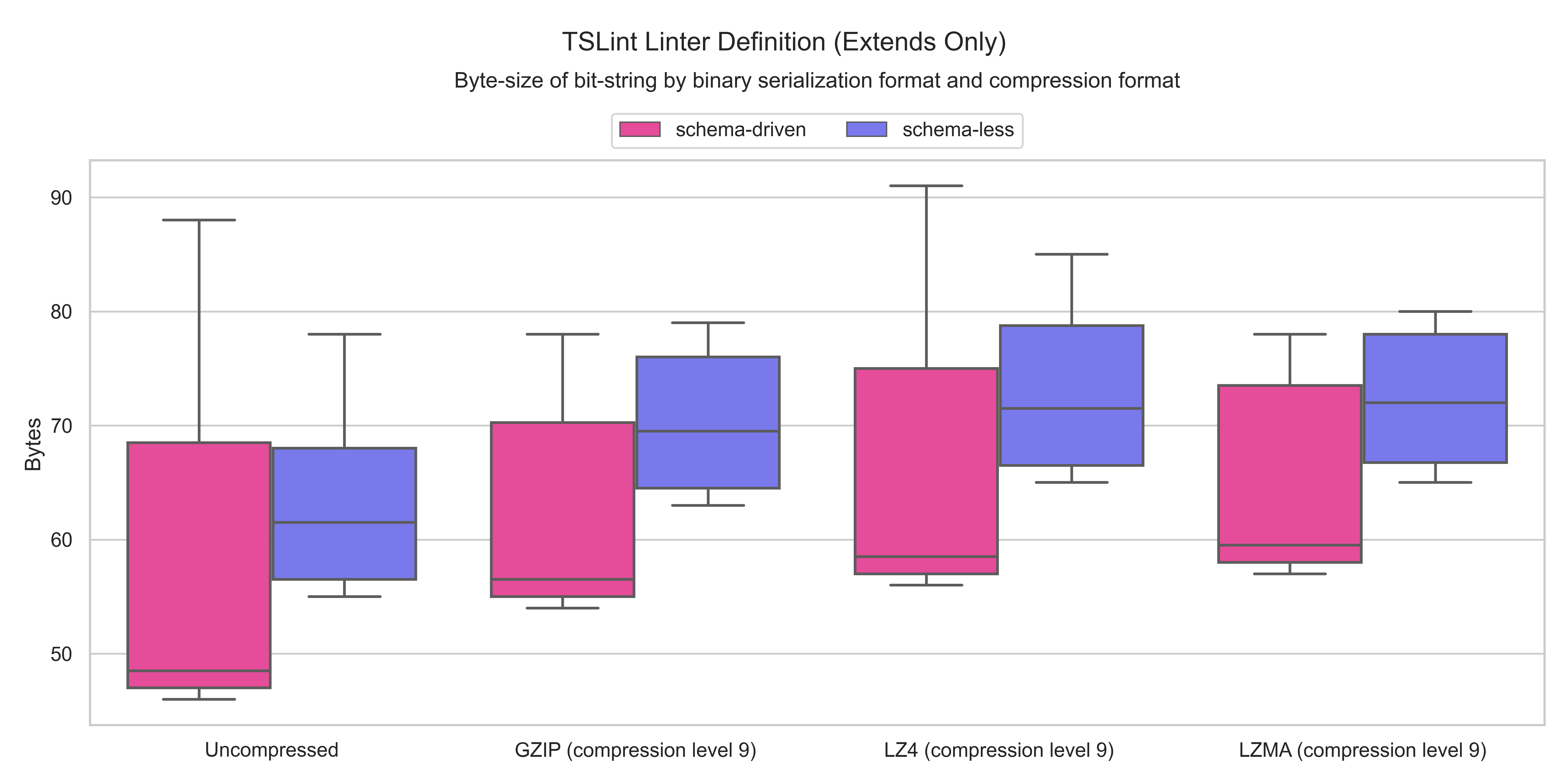}}
\caption{
Box plot of the statistical results in \autoref{table:benchmark-stats-tslintextend}.
}
\label{fig:benchmark-tslintextend-boxplot}
\end{figure*}

In \autoref{fig:benchmark-tslintextend-boxplot}, we observe the medians for
uncompressed schema-driven binary serialization specifications to be smaller in
comparison to uncompressed schema-less binary serialization specifications.  The range
between the upper and lower whiskers and the inter-quartile range of
uncompressed schema-less binary serialization specifications is smaller than the range
between the upper and lower whiskers and the inter-quartile range of
uncompressed schema-driven binary serialization specifications. However, their
respective quartiles overlap.


In terms of compression, GZIP results in the lower median for both
schema-driven and schema-less binary serialization specifications.  However,
compression is not space-efficient in terms of the median for both
schema-driven and schema-less binary serialization specifications.  While
compression does not contribute to space-efficiency, it reduces the range
between the upper and lower whiskers and inter-quartile range for schema-driven
binary serialization specifications.  In particular, the compression format
with the smaller range between the upper and lower whiskers for schema-driven
binary serialization specifications is LZMA, the compression formats with the
smaller inter-quartile range for schema-driven binary serialization
specifications are GZIP and LZMA, and the compression format with the smaller
range between the upper and lower whiskers for schema-less binary serialization
specifications is LZMA.


Overall, \we conclude that uncompressed schema-driven binary serialization
specifications are space-efficient in comparison to uncompressed schema-less
binary serialization specifications and that compression does not contribute to
space-efficiency in comparison to both uncompressed schema-driven and
schema-less binary serialization specifications.

\clearpage

\subsection{ImageOptimizer Azure Webjob Configuration}
\label{sec:benchmark-imageoptimizerwebjob}

\begin{figure*}[ht!]
\frame{\includegraphics[width=\linewidth]{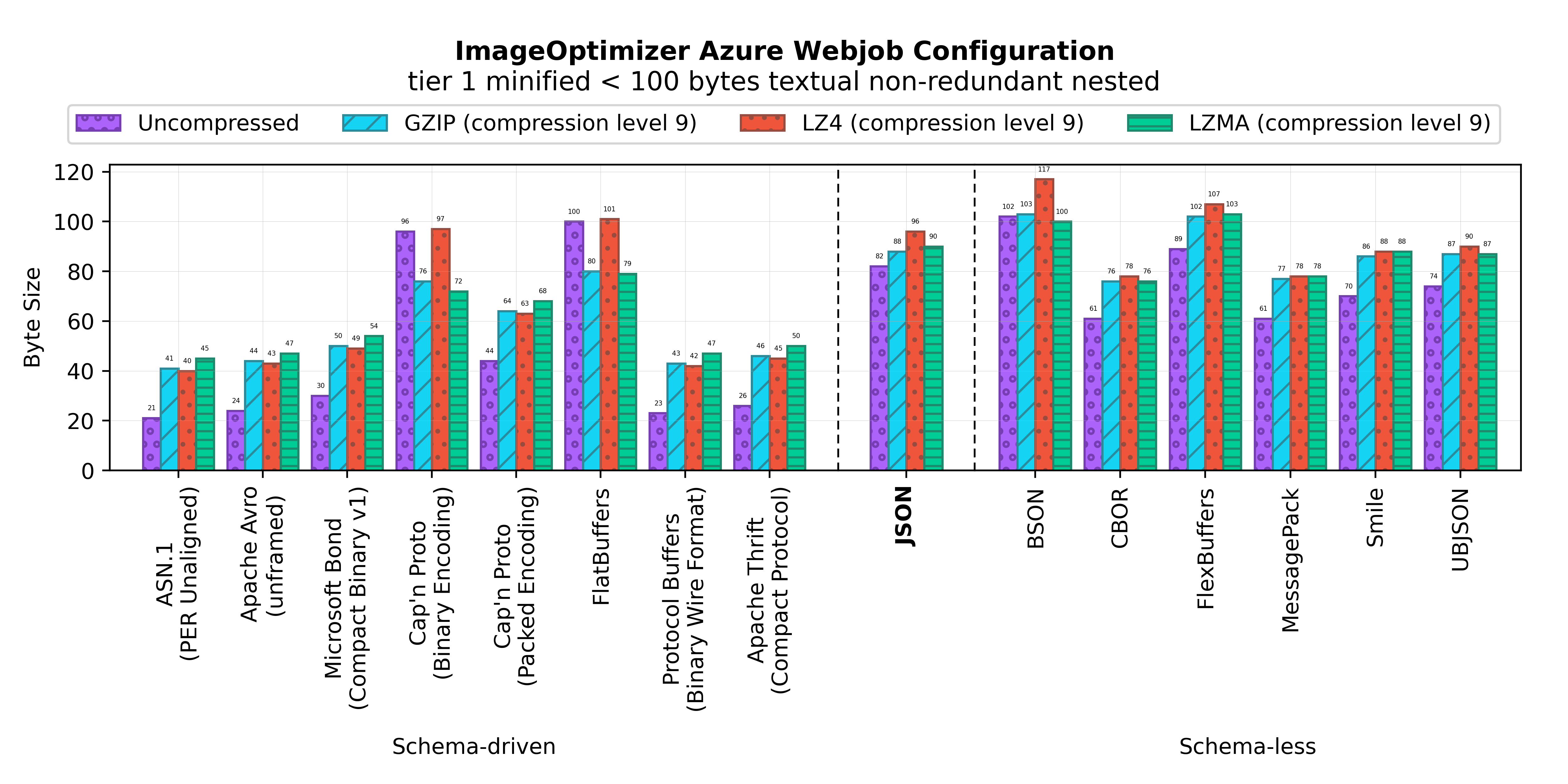}}
\caption{
The benchmark results for the ImageOptimizer Azure Webjob Configuration test case listed in \autoref{table:benchmark-documents} and \autoref{table:benchmark-documents-1}.
}
\label{fig:benchmark-imageoptimizerwebjob}
\end{figure*}

Image Optimizer
\footnote{\url{https://github.com/madskristensen/ImageOptimizerWebJob}} is an
Azure App Services WebJob
\footnote{\url{https://docs.microsoft.com/en-us/azure/app-service/webjobs-create}}
to compress website images used in the web development industry. In
\autoref{fig:benchmark-imageoptimizerwebjob}, \we demonstrate a \textbf{Tier 1
minified $<$ 100 bytes textual non-redundant nested} (Tier 1 TNN from
\autoref{table:json-taxonomy}) JSON document that consists of an Image
Optimizer configuration to perform lossy compression on images inside a
particular folder.

The smallest bit-string is produced by ASN.1 PER Unaligned \cite{asn1-per} (21
bytes), closely followed by Protocol Buffers \cite{protocolbuffers} (23 bytes)
and Apache Avro \cite{avro} (24 bytes).  The binary serialization specifications that
produced the smallest bit-strings are schema-driven and sequential \cite{viotti2022survey}.
Conversely, the largest bit-string is produced by BSON \cite{bson} (102 bytes),
closely followed by FlatBuffers \cite{flatbuffers} (100 bytes) and Cap'n Proto
Binary Encoding \cite{capnproto} (96 bytes).  With the exception of BSON, the
binary serialization specifications that produced the largest bit-strings are
schema-driven and pointer-based \cite{viotti2022survey}.  In comparison to JSON
\cite{ECMA-404} (82 bytes), binary serialization achieves a \textbf{3.9x} size
reduction in the best case for this input document.  However, 4 out of the 14
JSON-compatible binary serialization specifications listed in
\autoref{table:benchmark-specifications-schema-driven} and
\autoref{table:benchmark-specifications-schema-less} result in bit-strings that are
larger than JSON: Cap'n Proto Binary Encoding \cite{capnproto}, FlatBuffers
\cite{flatbuffers}, BSON \cite{bson} and FlexBuffers \cite{flexbuffers}. These
binary serialization specifications are either schema-less or schema-driven and
pointer-based.

For this Tier 1 TNN document, the best performing schema-driven serialization
specification achieves a \textbf{2.9x} size reduction compared to the best performing
schema-less serialization specification: CBOR \cite{RFC7049} and MessagePack
\cite{messagepack} (61 bytes).  As shown in
\autoref{table:benchmark-stats-imageoptimizerwebjob}, uncompressed
schema-driven specifications provide smaller \emph{average} and \emph{median}
bit-strings than uncompressed schema-less specifications. However, as highlighted by
the \emph{range} and \emph{standard deviation}, uncompressed schema-driven
specifications exhibit higher size reduction variability depending on the
expressiveness of the schema language (i.e. how the language constructs allow
you to model the data) and the size optimizations devised by its authors. With
the exception of the pointer-based binary serialization specifications Cap'n Proto
Binary Encoding \cite{capnproto} and FlatBuffers \cite{flatbuffers}, the
selection of schema-driven serialization specifications listed in
\autoref{table:benchmark-specifications-schema-driven} produce bit-strings that are
equal to or smaller than their schema-less counterparts listed in
\autoref{table:benchmark-specifications-schema-less}.  The best performing sequential
serialization specification achieves a \textbf{2x} size reduction compared to the best
performing pointer-based serialization specification: Cap'n Proto Packed Encoding
\cite{capnproto} (44 bytes).

The compression formats listed in
\autoref{sec:benchmark-compression-formats} result in positive gains for
the bit-strings produced by Cap'n Proto Binary Encoding \cite{capnproto} and
FlatBuffers \cite{flatbuffers}. The best performing uncompressed binary
serialization specification achieves a \textbf{4.1x} size reduction compared to
the best performing compression format for JSON: GZIP \cite{RFC1952} (88
bytes).

\begin{table*}[hb!]
\caption{A byte-size statistical analysis of the benchmark results shown in \autoref{fig:benchmark-imageoptimizerwebjob} divided by schema-driven and schema-less specifications.}
\label{table:benchmark-stats-imageoptimizerwebjob}
\begin{tabularx}{\linewidth}{X|l|l|l|l|l|l|l|l}
\toprule
\multirow{2}{*}{\textbf{Category}} &
\multicolumn{4}{c|}{\textbf{Schema-driven}} &
\multicolumn{4}{c}{\textbf{Schema-less}} \\
\cline{2-9}
& \small\textbf{Average} & \small\textbf{Median} & \small\textbf{Range} & \small\textbf{Std.dev} & \small\textbf{Average} & \small\textbf{Median} & \small\textbf{Range} & \small\textbf{Std.dev} \\
\midrule
Uncompressed & \small{45.5} & \small{28} & \small{79} & \small{31.0} & \small{76.2} & \small{72} & \small{41} & \small{14.9} \\ \hline
GZIP (compression level 9) & \small{55.5} & \small{48} & \small{39} & \small{14.6} & \small{88.5} & \small{86.5} & \small{27} & \small{10.7} \\ \hline
LZ4 (compression level 9) & \small{60} & \small{47} & \small{61} & \small{23.5} & \small{93} & \small{89} & \small{39} & \small{14.5} \\ \hline
LZMA (compression level 9) & \small{57.8} & \small{52} & \small{34} & \small{12.4} & \small{88.7} & \small{87.5} & \small{27} & \small{10.1} \\
\bottomrule
\end{tabularx}
\end{table*}

\begin{table*}[hb!]
\caption{The benchmark raw data results and schemas for the plot in \autoref{fig:benchmark-imageoptimizerwebjob}.}
\label{table:benchmark-imageoptimizerwebjob}
\begin{tabularx}{\linewidth}{X|l|l|l|l|l}
\toprule
\textbf{Serialization Format} & \textbf{Schema} & \textbf{Uncompressed} & \textbf{GZIP} & \textbf{LZ4} & \textbf{LZMA} \\
\midrule
ASN.1 (PER Unaligned) & \href{https://github.com/jviotti/binary-json-size-benchmark/blob/main/benchmark/imageoptimizerwebjob/asn1/schema.asn}{\small{\texttt{schema.asn}}} & 21 & 41 & 40 & 45 \\ \hline
Apache Avro (unframed) & \href{https://github.com/jviotti/binary-json-size-benchmark/blob/main/benchmark/imageoptimizerwebjob/avro/schema.json}{\small{\texttt{schema.json}}} & 24 & 44 & 43 & 47 \\ \hline
Microsoft Bond (Compact Binary v1) & \href{https://github.com/jviotti/binary-json-size-benchmark/blob/main/benchmark/imageoptimizerwebjob/bond/schema.bond}{\small{\texttt{schema.bond}}} & 30 & 50 & 49 & 54 \\ \hline
Cap'n Proto (Binary Encoding) & \href{https://github.com/jviotti/binary-json-size-benchmark/blob/main/benchmark/imageoptimizerwebjob/capnproto/schema.capnp}{\small{\texttt{schema.capnp}}} & 96 & 76 & 97 & 72 \\ \hline
Cap'n Proto (Packed Encoding) & \href{https://github.com/jviotti/binary-json-size-benchmark/blob/main/benchmark/imageoptimizerwebjob/capnproto-packed/schema.capnp}{\small{\texttt{schema.capnp}}} & 44 & 64 & 63 & 68 \\ \hline
FlatBuffers & \href{https://github.com/jviotti/binary-json-size-benchmark/blob/main/benchmark/imageoptimizerwebjob/flatbuffers/schema.fbs}{\small{\texttt{schema.fbs}}} & 100 & 80 & 101 & 79 \\ \hline
Protocol Buffers (Binary Wire Format) & \href{https://github.com/jviotti/binary-json-size-benchmark/blob/main/benchmark/imageoptimizerwebjob/protobuf/schema.proto}{\small{\texttt{schema.proto}}} & 23 & 43 & 42 & 47 \\ \hline
Apache Thrift (Compact Protocol) & \href{https://github.com/jviotti/binary-json-size-benchmark/blob/main/benchmark/imageoptimizerwebjob/thrift/schema.thrift}{\small{\texttt{schema.thrift}}} & 26 & 46 & 45 & 50 \\ \hline
\hline \textbf{JSON} & - & 82 & 88 & 96 & 90 \\ \hline \hline
BSON & - & 102 & 103 & 117 & 100 \\ \hline
CBOR & - & 61 & 76 & 78 & 76 \\ \hline
FlexBuffers & - & 89 & 102 & 107 & 103 \\ \hline
MessagePack & - & 61 & 77 & 78 & 78 \\ \hline
Smile & - & 70 & 86 & 88 & 88 \\ \hline
UBJSON & - & 74 & 87 & 90 & 87 \\
\bottomrule
\end{tabularx}
\end{table*}

\begin{figure*}[ht!]
\frame{\includegraphics[width=\linewidth]{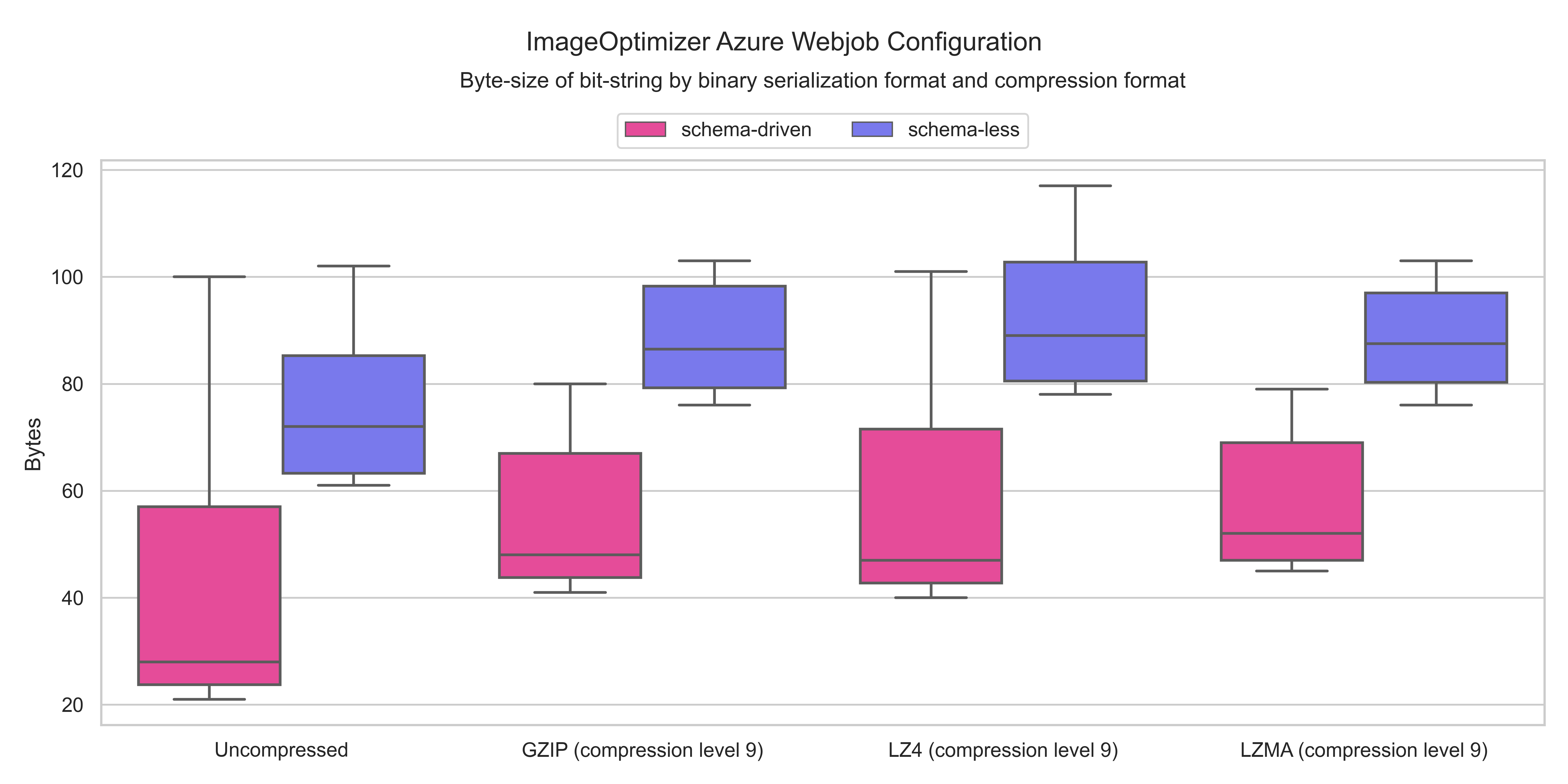}}
\caption{
Box plot of the statistical results in \autoref{table:benchmark-stats-imageoptimizerwebjob}.
}
\label{fig:benchmark-imageoptimizerwebjob-boxplot}
\end{figure*}

In \autoref{fig:benchmark-imageoptimizerwebjob-boxplot}, we observe the medians
for uncompressed schema-driven binary serialization specifications to be smaller in
comparison to uncompressed schema-less binary serialization specifications.  The range
between the upper and lower whiskers and the inter-quartile range of
uncompressed schema-less binary serialization specifications is smaller than the range
between the upper and lower whiskers and the inter-quartile range of
uncompressed schema-driven binary serialization specifications.


In terms of compression, LZ4 results in the lower median for schema-driven
binary serialization specifications while GZIP results in the lower median for
schema-less binary serialization specifications.  However, compression is not
space-efficient in terms of the median for both schema-driven and schema-less
binary serialization specifications.  While compression does not contribute to
space-efficiency, it reduces the range between the upper and lower whiskers and
inter-quartile range for both schema-driven and schema-less binary
serialization specifications.  In particular, the compression format with the
smaller range between the upper and lower whiskers for schema-driven binary
serialization specifications is LZMA, the compression formats with the smaller
inter-quartile range for schema-driven binary serialization specifications are
GZIP and LZMA, the compression format with the smaller range between the upper
and lower whiskers for schema-less binary serialization specifications is GZIP,
and the compression format with the smaller inter-quartile range for
schema-less binary serialization specifications is LZMA.


Overall, \we conclude that uncompressed schema-driven binary serialization
specifications are space-efficient in comparison to uncompressed schema-less binary
serialization specifications and that compression does not contribute to
space-efficiency in comparison to both uncompressed schema-driven and
schema-less binary serialization specifications.

\clearpage

\subsection{SAP Cloud SDK Continuous Delivery Toolkit Configuration}
\label{sec:benchmark-sapcloudsdkpipeline}

\begin{figure*}[ht!]
\frame{\includegraphics[width=\linewidth]{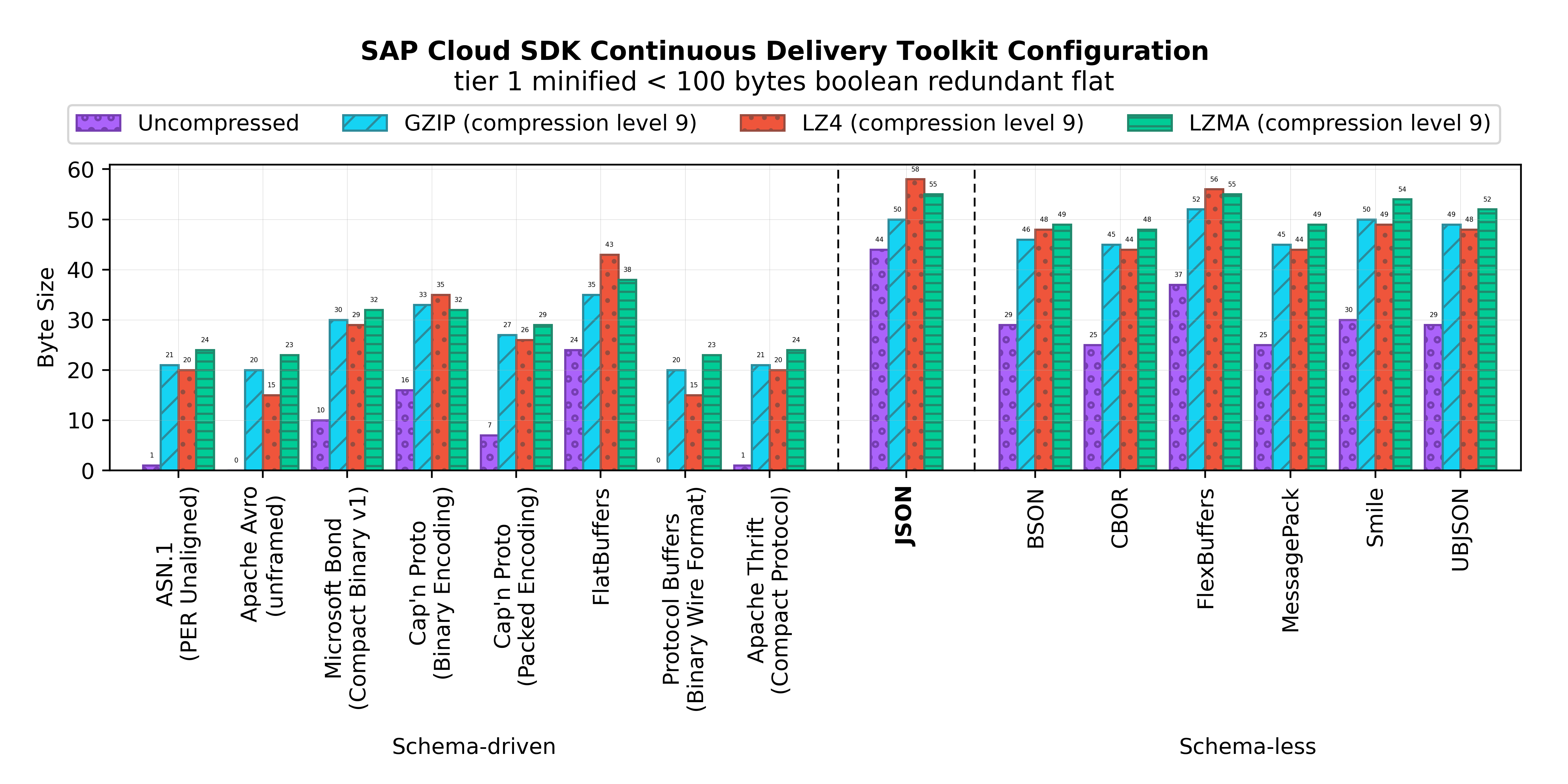}}
\caption{
The benchmark results for the SAP Cloud SDK Continuous Delivery Toolkit Configuration test case listed in \autoref{table:benchmark-documents} and \autoref{table:benchmark-documents-1}.
}
\label{fig:benchmark-sapcloudsdkpipeline}
\end{figure*}

SAP Cloud SDK \footnote{\url{https://sap.github.io/cloud-sdk/}} is a framework
that includes support for continuous integration and delivery pipelines to
develop applications for the SAP
\footnote{\url{https://www.sap.com/index.html}} enterprise resource planning
platform used by industries such as finance, healthcare and retail. In
\autoref{fig:benchmark-sapcloudsdkpipeline}, \we demonstrate a \textbf{Tier 1
minified $<$ 100 bytes boolean redundant flat} (Tier 1 BRF from
\autoref{table:json-taxonomy}) JSON document that defines a blank pipeline with
no declared steps.

The smallest bit-string is produced by both Apache Avro \cite{avro} and
Protocol Buffers \cite{protocolbuffers} (0 bytes), closely followed by both
ASN.1 PER Unaligned \cite{asn1-per} and Apache Thrift \cite{slee2007thrift} (1
byte), and Cap'n Proto Packed Encoding \cite{capnproto} (7 bytes). Apache Avro
and Protocol Buffers achieve a zero byte-size as the input document consists of
a set of \emph{null} values, which these serialization specifications represent by not
encoding the corresponding fields.  The binary serialization specifications that
produced the smallest bit-strings are schema-driven, and with the exception of
Cap'n Proto Packed Encoding, which takes the fifth place, they are also
sequential. Conversely, the largest bit-string is produced by FlexBuffers
\cite{flexbuffers} (37 bytes), followed by Smile \cite{smile} (30 bytes) and
both BSON \cite{bson} and UBJSON \cite{ubjson} (29 bytes). The binary
serialization specifications that produced the largest bit-strings are schema-less and
with the exception of FlexBuffers, they are also sequential \cite{viotti2022survey}.  In
comparison to JSON \cite{ECMA-404} (44 bytes), binary serialization achieves 0
bytes in the best case for this input document, the maximal possible size
reduction. Additionally, none of the 14 JSON-compatible binary serialization
specifications listed in \autoref{table:benchmark-specifications-schema-driven} and
\autoref{table:benchmark-specifications-schema-less} result in bit-strings that are
larger than JSON.

For this Tier 1 BRF document, the best performing schema-less serialization
specifications are CBOR \cite{RFC7049} and MessagePack \cite{messagepack} (25 bytes).
As shown in \autoref{table:benchmark-stats-sapcloudsdkpipeline}, uncompressed
schema-driven specifications provide smaller \emph{average} and \emph{median}
bit-strings than uncompressed schema-less specifications. However, as highlighted by
the \emph{range} and \emph{standard deviation}, uncompressed schema-driven
specifications exhibit higher size reduction variability depending on the
expressiveness of the schema language (i.e. how the language constructs allow
you to model the data) and the size optimizations devised by its authors.  The
entire selection of schema-driven serialization specifications listed in
\autoref{table:benchmark-specifications-schema-driven} produce bit-strings that are
equal to or smaller than their schema-less counterparts listed in
\autoref{table:benchmark-specifications-schema-less}.  The best performing
pointer-based serialization specification is Cap'n Proto Packed Encoding
\cite{capnproto} (7 bytes).

The compression formats listed in
\autoref{sec:benchmark-compression-formats} do not result in positive
gains for any bit-string. The best performing compression format for JSON is
GZIP \cite{RFC1952} (50 bytes).

\begin{table*}[hb!]
\caption{A byte-size statistical analysis of the benchmark results shown in \autoref{fig:benchmark-sapcloudsdkpipeline} divided by schema-driven and schema-less specifications.}
\label{table:benchmark-stats-sapcloudsdkpipeline}
\begin{tabularx}{\linewidth}{X|l|l|l|l|l|l|l|l}
\toprule
\multirow{2}{*}{\textbf{Category}} &
\multicolumn{4}{c|}{\textbf{Schema-driven}} &
\multicolumn{4}{c}{\textbf{Schema-less}} \\
\cline{2-9}
& \small\textbf{Average} & \small\textbf{Median} & \small\textbf{Range} & \small\textbf{Std.dev} & \small\textbf{Average} & \small\textbf{Median} & \small\textbf{Range} & \small\textbf{Std.dev} \\
\midrule
Uncompressed & \small{7.4} & \small{4} & \small{24} & \small{8.3} & \small{29.2} & \small{29} & \small{12} & \small{4.0} \\ \hline
GZIP (compression level 9) & \small{25.9} & \small{24} & \small{15} & \small{5.8} & \small{47.8} & \small{47.5} & \small{7} & \small{2.7} \\ \hline
LZ4 (compression level 9) & \small{25.4} & \small{23} & \small{28} & \small{9.3} & \small{48.2} & \small{48} & \small{12} & \small{4.0} \\ \hline
LZMA (compression level 9) & \small{28.1} & \small{26.5} & \small{15} & \small{5.2} & \small{51.2} & \small{50.5} & \small{7} & \small{2.7} \\
\bottomrule
\end{tabularx}
\end{table*}

\begin{table*}[hb!]
\caption{The benchmark raw data results and schemas for the plot in \autoref{fig:benchmark-sapcloudsdkpipeline}.}
\label{table:benchmark-sapcloudsdkpipeline}
\begin{tabularx}{\linewidth}{X|l|l|l|l|l}
\toprule
\textbf{Serialization Format} & \textbf{Schema} & \textbf{Uncompressed} & \textbf{GZIP} & \textbf{LZ4} & \textbf{LZMA} \\
\midrule
ASN.1 (PER Unaligned) & \href{https://github.com/jviotti/binary-json-size-benchmark/blob/main/benchmark/sapcloudsdkpipeline/asn1/schema.asn}{\small{\texttt{schema.asn}}} & 1 & 21 & 20 & 24 \\ \hline
Apache Avro (unframed) & \href{https://github.com/jviotti/binary-json-size-benchmark/blob/main/benchmark/sapcloudsdkpipeline/avro/schema.json}{\small{\texttt{schema.json}}} & 0 & 20 & 15 & 23 \\ \hline
Microsoft Bond (Compact Binary v1) & \href{https://github.com/jviotti/binary-json-size-benchmark/blob/main/benchmark/sapcloudsdkpipeline/bond/schema.bond}{\small{\texttt{schema.bond}}} & 10 & 30 & 29 & 32 \\ \hline
Cap'n Proto (Binary Encoding) & \href{https://github.com/jviotti/binary-json-size-benchmark/blob/main/benchmark/sapcloudsdkpipeline/capnproto/schema.capnp}{\small{\texttt{schema.capnp}}} & 16 & 33 & 35 & 32 \\ \hline
Cap'n Proto (Packed Encoding) & \href{https://github.com/jviotti/binary-json-size-benchmark/blob/main/benchmark/sapcloudsdkpipeline/capnproto-packed/schema.capnp}{\small{\texttt{schema.capnp}}} & 7 & 27 & 26 & 29 \\ \hline
FlatBuffers & \href{https://github.com/jviotti/binary-json-size-benchmark/blob/main/benchmark/sapcloudsdkpipeline/flatbuffers/schema.fbs}{\small{\texttt{schema.fbs}}} & 24 & 35 & 43 & 38 \\ \hline
Protocol Buffers (Binary Wire Format) & \href{https://github.com/jviotti/binary-json-size-benchmark/blob/main/benchmark/sapcloudsdkpipeline/protobuf/schema.proto}{\small{\texttt{schema.proto}}} & 0 & 20 & 15 & 23 \\ \hline
Apache Thrift (Compact Protocol) & \href{https://github.com/jviotti/binary-json-size-benchmark/blob/main/benchmark/sapcloudsdkpipeline/thrift/schema.thrift}{\small{\texttt{schema.thrift}}} & 1 & 21 & 20 & 24 \\ \hline
\hline \textbf{JSON} & - & 44 & 50 & 58 & 55 \\ \hline \hline
BSON & - & 29 & 46 & 48 & 49 \\ \hline
CBOR & - & 25 & 45 & 44 & 48 \\ \hline
FlexBuffers & - & 37 & 52 & 56 & 55 \\ \hline
MessagePack & - & 25 & 45 & 44 & 49 \\ \hline
Smile & - & 30 & 50 & 49 & 54 \\ \hline
UBJSON & - & 29 & 49 & 48 & 52 \\
\bottomrule
\end{tabularx}
\end{table*}

\begin{figure*}[ht!]
\frame{\includegraphics[width=\linewidth]{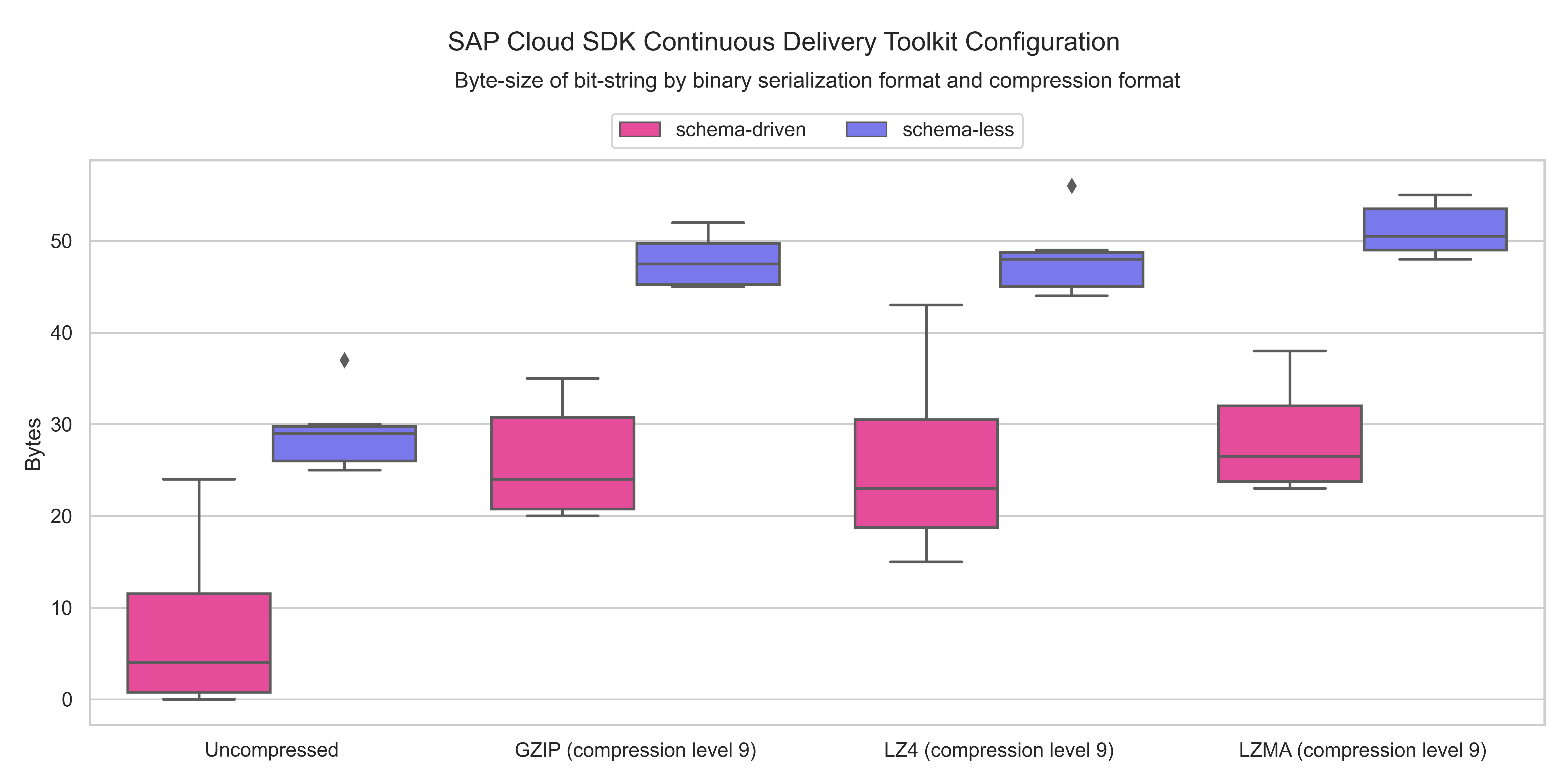}}
\caption{
Box plot of the statistical results in \autoref{table:benchmark-stats-sapcloudsdkpipeline}.
}
\label{fig:benchmark-sapcloudsdkpipeline-boxplot}
\end{figure*}

In \autoref{fig:benchmark-sapcloudsdkpipeline-boxplot}, we observe the medians
for uncompressed schema-driven binary serialization specifications to be smaller in
comparison to uncompressed schema-less binary serialization specifications.  The range
between the upper and lower whiskers and the inter-quartile range of
uncompressed schema-less binary serialization specifications is smaller than the range
between the upper and lower whiskers and the inter-quartile range of
uncompressed schema-driven binary serialization specifications.


In terms of compression, LZ4 results in the lower medians for schema-driven
binary serialization specifications while GZIP results in the lower median for
schema-less binary serialization specifications.  However, compression is not
space-efficient in terms of the median for both schema-driven and schema-less
binary serialization specifications.  Additionally, the use of LZ4 for
schema-less binary serialization specifications exhibits upper outliers.  While
compression does not contribute to space-efficiency, it reduces the range
between the upper and lower whiskers and inter-quartile range for schema-driven
binary serialization specifications.  In particular, the compression formats
with the smaller range between the upper and lower whiskers for schema-driven
binary serialization specifications are GZIP and LZMA, the compression format
with the smaller inter-quartile range for schema-driven binary serialization
specifications is LZMA, and the compression format with the smaller range
between the upper and lower whiskers and the smaller inter-quartile range for
schema-less binary serialization specifications is LZ4.


Overall, \we conclude that uncompressed schema-driven binary serialization
specifications are space-efficient in comparison to uncompressed schema-less binary
serialization specifications and that compression does not contribute to
space-efficiency in comparison to both uncompressed schema-driven and
schema-less binary serialization specifications.

\clearpage

\subsection{TSLint Linter Definition (Multi-rule)}
\label{sec:benchmark-tslintmulti}

\begin{figure*}[ht!]
\frame{\includegraphics[width=\linewidth]{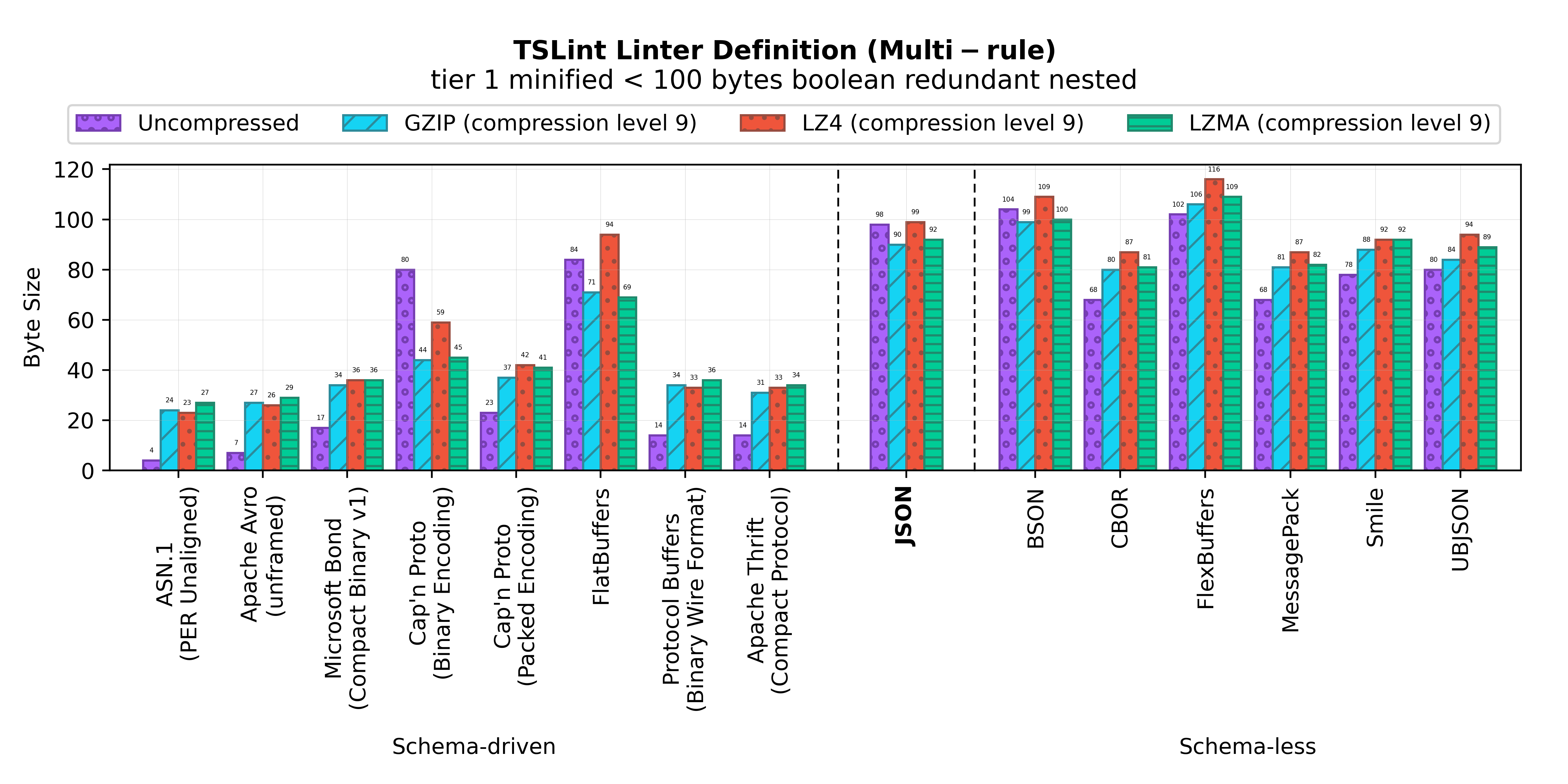}}
\caption{
The benchmark results for the TSLint Linter Definition (Multi-rule) test case listed in \autoref{table:benchmark-documents} and \autoref{table:benchmark-documents-1}.
}
\label{fig:benchmark-tslintmulti}
\end{figure*}

TSLint \footnote{\url{https://palantir.github.io/tslint}} is now an obsolete
open-source linter for the TypeScript
\footnote{\url{https://www.typescriptlang.org}} programming language. TSLint
was created by the Big Data analytics company Palantir
\footnote{\url{https://www.palantir.com}} and was merged with the ESLint
open-source JavaScript linter in 2019
\footnote{\url{https://github.com/palantir/tslint/issues/4534}}. In
\autoref{fig:benchmark-tslintmulti}, \we demonstrate a \textbf{Tier 1 minified
$<$ 100 bytes boolean redundant nested} (Tier 1 BRN from
\autoref{table:json-taxonomy}) JSON document that consists of a TSLint
configuration that enables and configures a set of built-in rules.

The smallest bit-string is produced by ASN.1 PER Unaligned \cite{asn1-per} (4
bytes), followed by Apache Avro \cite{avro} (7 bytes) and both Protocol Buffers
\cite{protocolbuffers} and Apache Thrift \cite{slee2007thrift} (14 bytes). The
binary serialization specifications that produced the smallest bit-strings are
schema-driven and sequential \cite{viotti2022survey}.  Conversely, the largest bit-string is
produced by BSON \cite{bson} (104 bytes), closely followed by FlexBuffers
\cite{flexbuffers} (102 bytes) and FlatBuffers \cite{flatbuffers} (84 bytes).
With the exception of BSON, the binary serialization specifications that produced the
largest bit-strings are pointer-based \cite{viotti2022survey}.  In comparison to JSON
\cite{ECMA-404} (98 bytes), binary serialization achieves a \textbf{24.5x} size
reduction in the best case for this input document. Similar large size
reductions are observed in JSON documents whose content is dominated by
\emph{boolean} and \emph{numeric} values. However, 2 out of the 14
JSON-compatible binary serialization specifications listed in
\autoref{table:benchmark-specifications-schema-driven} and
\autoref{table:benchmark-specifications-schema-less} result in bit-strings that are
larger than JSON: BSON \cite{bson} and FlexBuffers \cite{flexbuffers}. These
binary serialization specifications are schema-less.

For this Tier 1 BRN document, the best performing schema-driven serialization
specification achieves a \textbf{17x} size reduction compared to the best performing
schema-less serialization specification: CBOR \cite{RFC7049} and MessagePack
\cite{messagepack} (68 bytes).  As shown in
\autoref{table:benchmark-stats-tslintmulti}, uncompressed schema-driven specifications
provide smaller \emph{average} and \emph{median} bit-strings than uncompressed
schema-less specifications. However, as highlighted by the \emph{range} and
\emph{standard deviation}, uncompressed schema-driven specifications exhibit higher
size reduction variability depending on the expressiveness of the schema
language (i.e. how the language constructs allow you to model the data) and the
size optimizations devised by its authors.  With the exception of the
pointer-based binary serialization specifications Cap'n Proto Binary Encoding
\cite{capnproto} and FlatBuffers \cite{flatbuffers}, the selection of
schema-driven serialization specifications listed in
\autoref{table:benchmark-specifications-schema-driven} produce bit-strings that are
equal to or smaller than their schema-less counterparts listed in
\autoref{table:benchmark-specifications-schema-less}.  The best performing sequential
serialization specification achieves a \textbf{5.7x} size reduction compared to the
best performing pointer-based serialization specification: Cap'n Proto Packed Encoding
\cite{capnproto} (23 bytes).

The compression formats listed in
\autoref{sec:benchmark-compression-formats} result in positive gains for
the bit-strings produced by Cap'n Proto Binary Encoding \cite{capnproto},
FlatBuffers \cite{flatbuffers}, JSON \cite{ECMA-404} and BSON \cite{bson}. The
best performing uncompressed binary serialization specification achieves a
\textbf{22.5x} size reduction compared to the best performing compression
format for JSON: GZIP \cite{RFC1952} (90 bytes).

\begin{table*}[hb!]
\caption{A byte-size statistical analysis of the benchmark results shown in \autoref{fig:benchmark-tslintmulti} divided by schema-driven and schema-less specifications.}
\label{table:benchmark-stats-tslintmulti}
\begin{tabularx}{\linewidth}{X|l|l|l|l|l|l|l|l}
\toprule
\multirow{2}{*}{\textbf{Category}} &
\multicolumn{4}{c|}{\textbf{Schema-driven}} &
\multicolumn{4}{c}{\textbf{Schema-less}} \\
\cline{2-9}
& \small\textbf{Average} & \small\textbf{Median} & \small\textbf{Range} & \small\textbf{Std.dev} & \small\textbf{Average} & \small\textbf{Median} & \small\textbf{Range} & \small\textbf{Std.dev} \\
\midrule
Uncompressed & \small{30.4} & \small{15.5} & \small{80} & \small{30.3} & \small{83.3} & \small{79} & \small{36} & \small{14.6} \\ \hline
GZIP (compression level 9) & \small{37.8} & \small{34} & \small{47} & \small{13.8} & \small{89.7} & \small{86} & \small{26} & \small{9.6} \\ \hline
LZ4 (compression level 9) & \small{43.3} & \small{34.5} & \small{71} & \small{21.8} & \small{97.5} & \small{93} & \small{29} & \small{11.1} \\ \hline
LZMA (compression level 9) & \small{39.6} & \small{36} & \small{42} & \small{12.4} & \small{92.2} & \small{90.5} & \small{28} & \small{9.9} \\
\bottomrule
\end{tabularx}
\end{table*}

\begin{table*}[hb!]
\caption{The benchmark raw data results and schemas for the plot in \autoref{fig:benchmark-tslintmulti}.}
\label{table:benchmark-tslintmulti}
\begin{tabularx}{\linewidth}{X|l|l|l|l|l}
\toprule
\textbf{Serialization Format} & \textbf{Schema} & \textbf{Uncompressed} & \textbf{GZIP} & \textbf{LZ4} & \textbf{LZMA} \\
\midrule
ASN.1 (PER Unaligned) & \href{https://github.com/jviotti/binary-json-size-benchmark/blob/main/benchmark/tslintmulti/asn1/schema.asn}{\small{\texttt{schema.asn}}} & 4 & 24 & 23 & 27 \\ \hline
Apache Avro (unframed) & \href{https://github.com/jviotti/binary-json-size-benchmark/blob/main/benchmark/tslintmulti/avro/schema.json}{\small{\texttt{schema.json}}} & 7 & 27 & 26 & 29 \\ \hline
Microsoft Bond (Compact Binary v1) & \href{https://github.com/jviotti/binary-json-size-benchmark/blob/main/benchmark/tslintmulti/bond/schema.bond}{\small{\texttt{schema.bond}}} & 17 & 34 & 36 & 36 \\ \hline
Cap'n Proto (Binary Encoding) & \href{https://github.com/jviotti/binary-json-size-benchmark/blob/main/benchmark/tslintmulti/capnproto/schema.capnp}{\small{\texttt{schema.capnp}}} & 80 & 44 & 59 & 45 \\ \hline
Cap'n Proto (Packed Encoding) & \href{https://github.com/jviotti/binary-json-size-benchmark/blob/main/benchmark/tslintmulti/capnproto-packed/schema.capnp}{\small{\texttt{schema.capnp}}} & 23 & 37 & 42 & 41 \\ \hline
FlatBuffers & \href{https://github.com/jviotti/binary-json-size-benchmark/blob/main/benchmark/tslintmulti/flatbuffers/schema.fbs}{\small{\texttt{schema.fbs}}} & 84 & 71 & 94 & 69 \\ \hline
Protocol Buffers (Binary Wire Format) & \href{https://github.com/jviotti/binary-json-size-benchmark/blob/main/benchmark/tslintmulti/protobuf/schema.proto}{\small{\texttt{schema.proto}}} & 14 & 34 & 33 & 36 \\ \hline
Apache Thrift (Compact Protocol) & \href{https://github.com/jviotti/binary-json-size-benchmark/blob/main/benchmark/tslintmulti/thrift/schema.thrift}{\small{\texttt{schema.thrift}}} & 14 & 31 & 33 & 34 \\ \hline
\hline \textbf{JSON} & - & 98 & 90 & 99 & 92 \\ \hline \hline
BSON & - & 104 & 99 & 109 & 100 \\ \hline
CBOR & - & 68 & 80 & 87 & 81 \\ \hline
FlexBuffers & - & 102 & 106 & 116 & 109 \\ \hline
MessagePack & - & 68 & 81 & 87 & 82 \\ \hline
Smile & - & 78 & 88 & 92 & 92 \\ \hline
UBJSON & - & 80 & 84 & 94 & 89 \\
\bottomrule
\end{tabularx}
\end{table*}

\begin{figure*}[ht!]
\frame{\includegraphics[width=\linewidth]{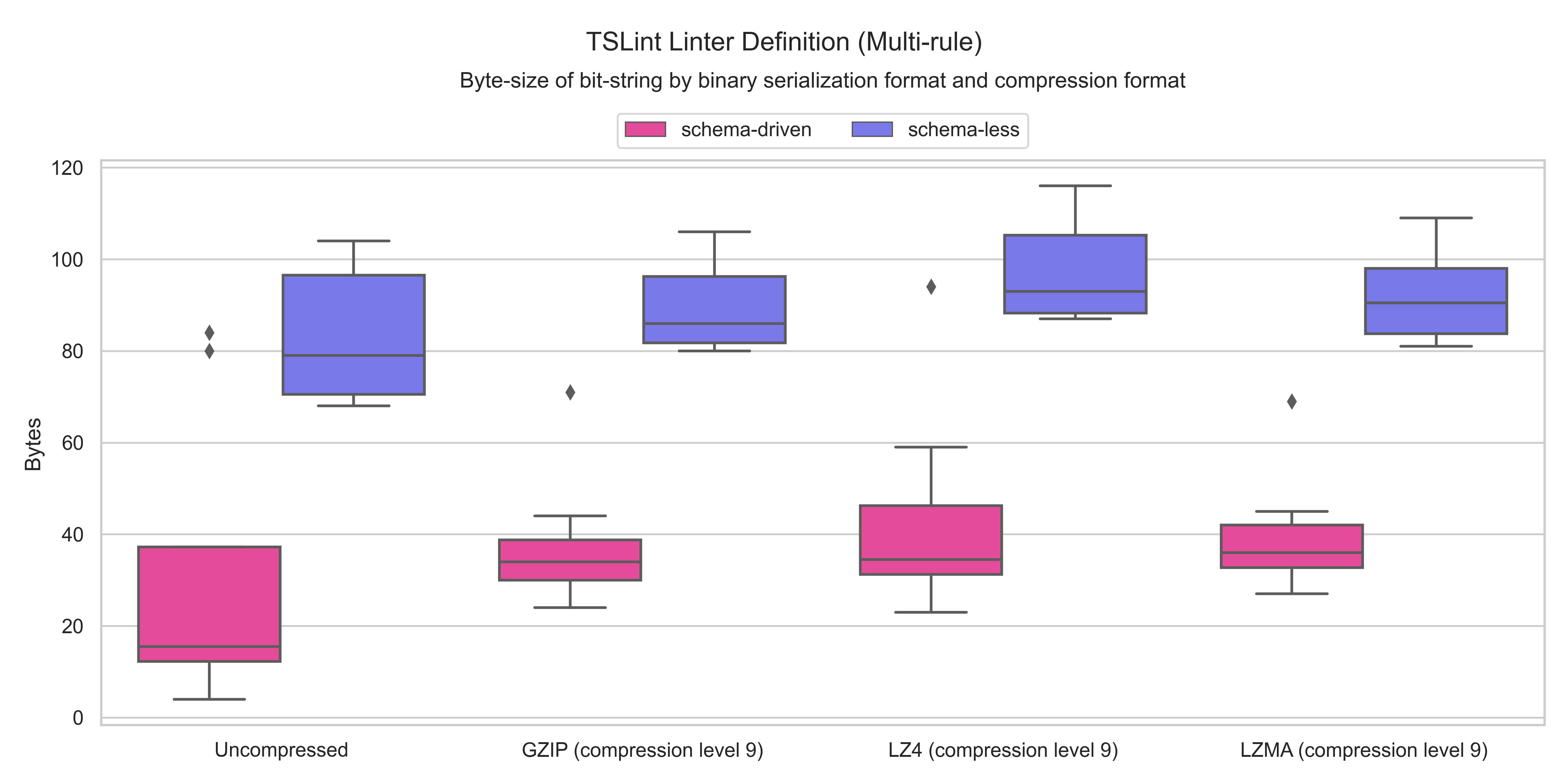}}
\caption{
Box plot of the statistical results in \autoref{table:benchmark-stats-tslintmulti}.
}
\label{fig:benchmark-tslintmulti-boxplot}
\end{figure*}

In \autoref{fig:benchmark-tslintmulti-boxplot}, we observe the medians for
uncompressed schema-driven binary serialization specifications to be smaller in
comparison to uncompressed schema-less binary serialization specifications.  The range
between the upper and lower whiskers of uncompressed schema-driven binary
serialization specifications is smaller than the range between the upper and lower
whiskers of uncompressed schema-less binary serialization specifications.  However,
the inter-quartile range of both both uncompressed schema-driven and
schema-less binary serialization specifications is similar.


In terms of compression, GZIP results in the lower median for both
schema-driven and schema-less binary serialization specifications.  However,
compression is not space-efficient in terms of the median for both
schema-driven and schema-less binary serialization specifications.
Additionally, the use of GZIP, LZ4 and LZMA for schema-driven binary
serialization specifications exhibits upper outliers.  While compression does
not contribute to space-efficiency, it reduces the range between the upper and
lower whiskers and inter-quartile range for both schema-driven and schema-less
binary serialization specifications.  In particular, the compression format
with the smaller range between the upper and lower whiskers for schema-driven
binary serialization specifications is LZMA, the compression formats with the
smaller inter-quartile range for schema-driven binary serialization
specifications are GZIP and LZMA, the compression format with the smaller range
between the upper and lower whiskers for schema-less binary serialization
specifications is GZIP, and the compression formats with the smaller
inter-quartile range for schema-less binary serialization specifications are
GZIP and LZMA.


Overall, \we conclude that uncompressed schema-driven binary serialization
specifications are space-efficient in comparison to uncompressed schema-less
binary serialization specifications and that compression does not contribute to
space-efficiency in comparison to both uncompressed schema-driven and
schema-less binary serialization specifications.

\clearpage

\subsection{CommitLint Configuration (Basic)}
\label{sec:benchmark-commitlintbasic}

\begin{figure*}[ht!]
\frame{\includegraphics[width=\linewidth]{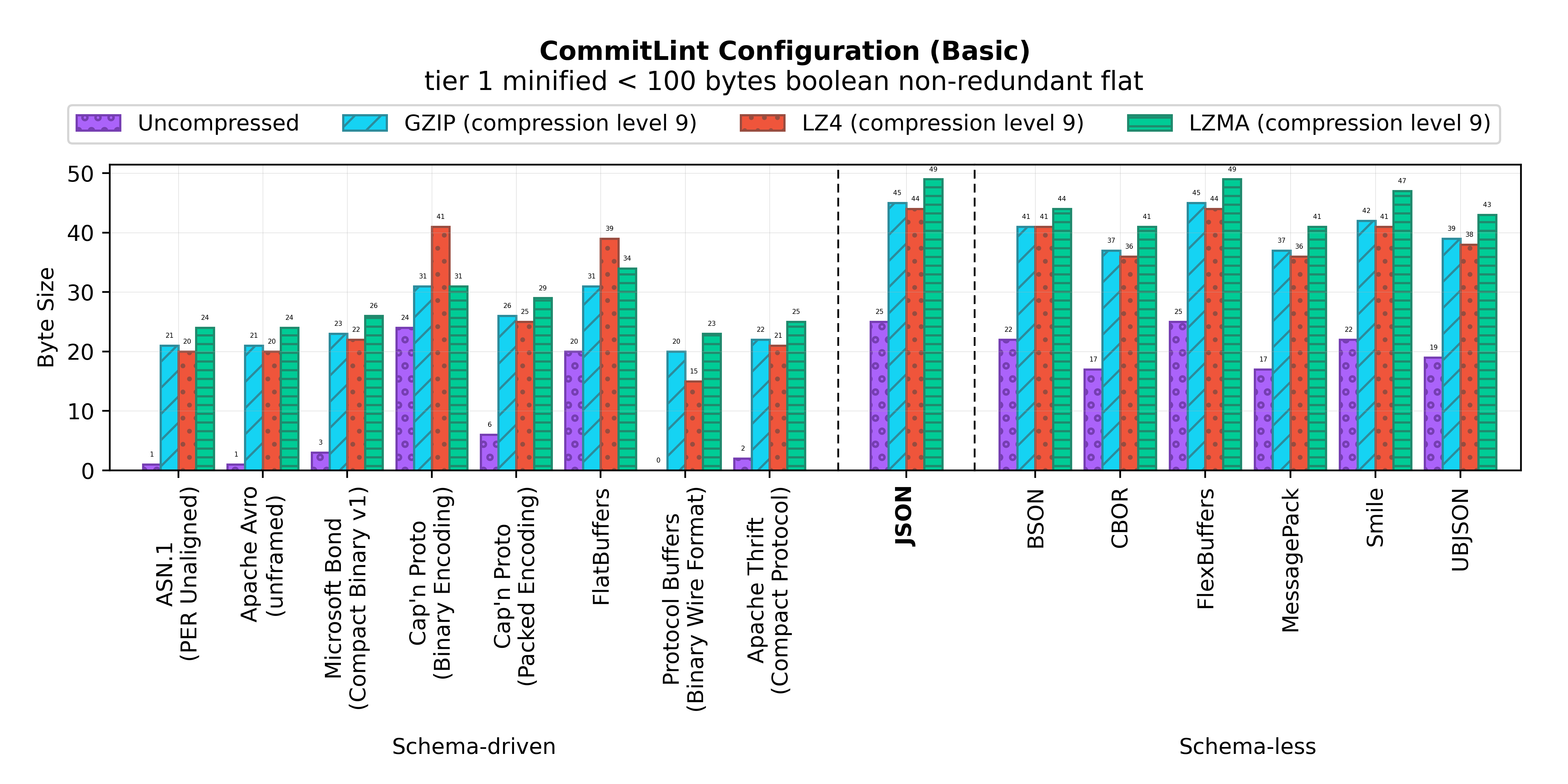}}
\caption{
The benchmark results for the CommitLint Configuration (Basic) test case listed in \autoref{table:benchmark-documents} and \autoref{table:benchmark-documents-1}.
}
\label{fig:benchmark-commitlintbasic}
\end{figure*}

CommitLint \footnote{\url{https://commitlint.js.org/\#/}} is an open-source
command-line tool to enforce version-control commit conventions in software
engineering projects. CommitLint is a community effort under the Conventional
Changelog \footnote{\url{https://github.com/conventional-changelog}}
organization formed by employees from companies including GitHub
\footnote{\url{https://github.com/zeke}} and Google
\footnote{\url{https://github.com/bcoe}}. In
\autoref{fig:benchmark-commitlintbasic}, \we demonstrate a \textbf{Tier 1
minified $<$ 100 bytes boolean non-redundant flat} (Tier 1 BNF from
\autoref{table:json-taxonomy}) JSON document that represents a CommitLint
configuration file which declares that CommitLint must not use its default
commit ignore rules.

The smallest bit-string is produced by Protocol Buffers \cite{protocolbuffers}
(0 bytes), closely followed by both ASN.1 PER Unaligned \cite{asn1-per} and
Apache Avro \cite{avro} (1 byte), and Apache Thrift \cite{slee2007thrift} (2
bytes). Protocol Buffers \cite{protocolbuffers} achieves a zero byte-size as
the input document consists of the boolean value \emph{false}, which Protocol
Buffers represents by not encoding the corresponding field. The binary
serialization specifications that produced the smallest bit-strings are schema-driven
and sequential \cite{viotti2022survey}. Conversely, the larges bit-string is produced by
FlexBuffers \cite{flexbuffers} (152 bytes) followed by UBJSON \cite{ubjson}
(137 bytes) and BSON \cite{bson} (133 bytes).  The binary serialization specifications
that produced the largest bit-strings are schema-less.  With the exception of
FlexBuffers, the binary serialization specifications that produced the largest
bit-strings are also sequential \cite{viotti2022survey}.  In comparison to JSON
\cite{ECMA-404} (25 bytes), binary serialization achieves 0 bytes in the best
case for this input document, the maximal possible size reduction.
Additionally, none of the 14 JSON-compatible binary serialization specifications
listed in \autoref{table:benchmark-specifications-schema-driven} and
\autoref{table:benchmark-specifications-schema-less} result in bit-strings that are
larger than JSON. However, the FlexBuffers \cite{flexbuffers} schema-less
pointer-based serialization specification result in a bit-string that is equal in size
to JSON.

For this Tier 1 BNF document, the best performing schema-less serialization
specifications are CBOR \cite{RFC7049} and MessagePack \cite{messagepack} (17 bytes).
As shown in \autoref{table:benchmark-stats-commitlintbasic}, uncompressed
schema-driven specifications provide smaller \emph{average} and \emph{median}
bit-strings than uncompressed schema-less specifications. However, as highlighted by
the \emph{range} and \emph{standard deviation}, uncompressed schema-driven
specifications exhibit higher size reduction variability depending on the
expressiveness of the schema language (i.e. how the language constructs allow
you to model the data) and the size optimizations devised by its authors. With
the exception of the pointer-based binary serialization specifications Cap'n Proto
Binary Encoding \cite{capnproto} and FlatBuffers \cite{flatbuffers}, the
selection of schema-driven serialization specifications listed in
\autoref{table:benchmark-specifications-schema-driven} produce bit-strings that are
equal to or smaller than their schema-less counterparts listed in
\autoref{table:benchmark-specifications-schema-less}.  The best performing
pointer-based serialization specification is Cap'n Proto Packed Encoding
\cite{capnproto} (6 bytes).

The compression formats listed in
\autoref{sec:benchmark-compression-formats} do not result in positive
gains for any bit-string due to the overhead of encoding the dictionary data
structures and the low redundancy of the input data. The best performing
compression format for JSON is LZ4 (44 bytes).

\begin{table*}[hb!]
\caption{A byte-size statistical analysis of the benchmark results shown in \autoref{fig:benchmark-commitlintbasic} divided by schema-driven and schema-less specifications.}
\label{table:benchmark-stats-commitlintbasic}
\begin{tabularx}{\linewidth}{X|l|l|l|l|l|l|l|l}
\toprule
\multirow{2}{*}{\textbf{Category}} &
\multicolumn{4}{c|}{\textbf{Schema-driven}} &
\multicolumn{4}{c}{\textbf{Schema-less}} \\
\cline{2-9}
& \small\textbf{Average} & \small\textbf{Median} & \small\textbf{Range} & \small\textbf{Std.dev} & \small\textbf{Average} & \small\textbf{Median} & \small\textbf{Range} & \small\textbf{Std.dev} \\
\midrule
Uncompressed & \small{7.1} & \small{2.5} & \small{24} & \small{8.8} & \small{20.3} & \small{20.5} & \small{8} & \small{2.9} \\ \hline
GZIP (compression level 9) & \small{24.4} & \small{22.5} & \small{11} & \small{4.2} & \small{40.2} & \small{40} & \small{8} & \small{2.9} \\ \hline
LZ4 (compression level 9) & \small{25.4} & \small{21.5} & \small{26} & \small{8.8} & \small{39.3} & \small{39.5} & \small{8} & \small{2.9} \\ \hline
LZMA (compression level 9) & \small{27} & \small{25.5} & \small{11} & \small{3.7} & \small{44.2} & \small{43.5} & \small{8} & \small{3.0} \\
\bottomrule
\end{tabularx}
\end{table*}

\begin{table*}[hb!]
\caption{The benchmark raw data results and schemas for the plot in \autoref{fig:benchmark-commitlintbasic}.}
\label{table:benchmark-commitlintbasic}
\begin{tabularx}{\linewidth}{X|l|l|l|l|l}
\toprule
\textbf{Serialization Format} & \textbf{Schema} & \textbf{Uncompressed} & \textbf{GZIP} & \textbf{LZ4} & \textbf{LZMA} \\
\midrule
ASN.1 (PER Unaligned) & \href{https://github.com/jviotti/binary-json-size-benchmark/blob/main/benchmark/commitlintbasic/asn1/schema.asn}{\small{\texttt{schema.asn}}} & 1 & 21 & 20 & 24 \\ \hline
Apache Avro (unframed) & \href{https://github.com/jviotti/binary-json-size-benchmark/blob/main/benchmark/commitlintbasic/avro/schema.json}{\small{\texttt{schema.json}}} & 1 & 21 & 20 & 24 \\ \hline
Microsoft Bond (Compact Binary v1) & \href{https://github.com/jviotti/binary-json-size-benchmark/blob/main/benchmark/commitlintbasic/bond/schema.bond}{\small{\texttt{schema.bond}}} & 3 & 23 & 22 & 26 \\ \hline
Cap'n Proto (Binary Encoding) & \href{https://github.com/jviotti/binary-json-size-benchmark/blob/main/benchmark/commitlintbasic/capnproto/schema.capnp}{\small{\texttt{schema.capnp}}} & 24 & 31 & 41 & 31 \\ \hline
Cap'n Proto (Packed Encoding) & \href{https://github.com/jviotti/binary-json-size-benchmark/blob/main/benchmark/commitlintbasic/capnproto-packed/schema.capnp}{\small{\texttt{schema.capnp}}} & 6 & 26 & 25 & 29 \\ \hline
FlatBuffers & \href{https://github.com/jviotti/binary-json-size-benchmark/blob/main/benchmark/commitlintbasic/flatbuffers/schema.fbs}{\small{\texttt{schema.fbs}}} & 20 & 31 & 39 & 34 \\ \hline
Protocol Buffers (Binary Wire Format) & \href{https://github.com/jviotti/binary-json-size-benchmark/blob/main/benchmark/commitlintbasic/protobuf/schema.proto}{\small{\texttt{schema.proto}}} & 0 & 20 & 15 & 23 \\ \hline
Apache Thrift (Compact Protocol) & \href{https://github.com/jviotti/binary-json-size-benchmark/blob/main/benchmark/commitlintbasic/thrift/schema.thrift}{\small{\texttt{schema.thrift}}} & 2 & 22 & 21 & 25 \\ \hline
\hline \textbf{JSON} & - & 25 & 45 & 44 & 49 \\ \hline \hline
BSON & - & 22 & 41 & 41 & 44 \\ \hline
CBOR & - & 17 & 37 & 36 & 41 \\ \hline
FlexBuffers & - & 25 & 45 & 44 & 49 \\ \hline
MessagePack & - & 17 & 37 & 36 & 41 \\ \hline
Smile & - & 22 & 42 & 41 & 47 \\ \hline
UBJSON & - & 19 & 39 & 38 & 43 \\
\bottomrule
\end{tabularx}
\end{table*}

\begin{figure*}[ht!]
\frame{\includegraphics[width=\linewidth]{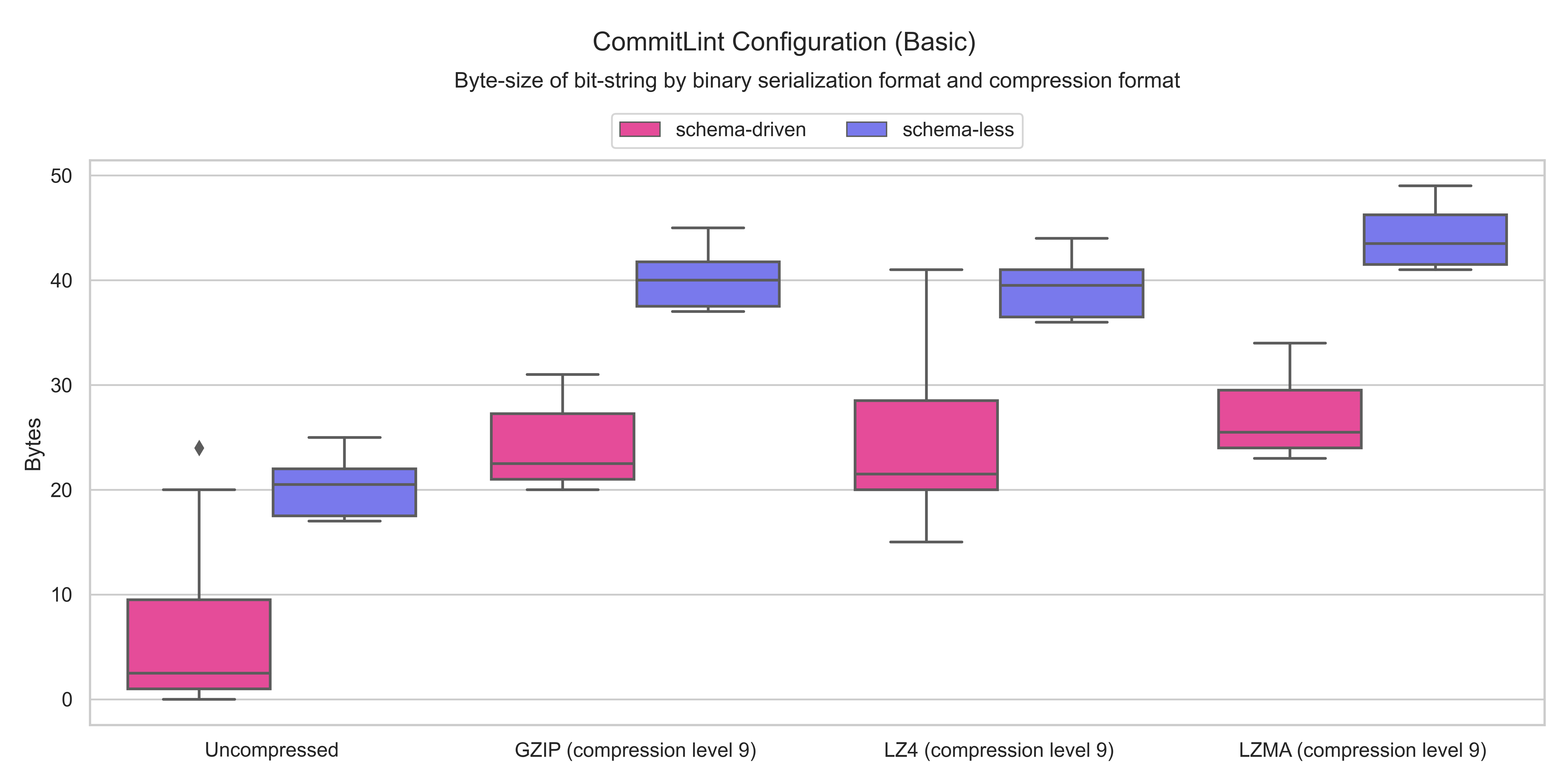}}
\caption{
Box plot of the statistical results in \autoref{table:benchmark-stats-commitlintbasic}.
}
\label{fig:benchmark-commitlintbasic-boxplot}
\end{figure*}

In \autoref{fig:benchmark-commitlintbasic-boxplot}, we observe the medians for
uncompressed schema-driven binary serialization specifications to be smaller in
comparison to uncompressed schema-less binary serialization specifications.  The range
between the upper and lower whiskers and the inter-quartile range of
uncompressed schema-less binary serialization specifications is smaller than the range
between the upper and lower whiskers and the inter-quartile range of
uncompressed schema-driven binary serialization specifications.


In terms of compression, LZ4 results in the lower median for both schema-driven
and schema-less binary serialization specifications.  However, compression is
not space-efficient in terms of the median for both schema-driven and
schema-less binary serialization specifications.  While compression does not
contribute to space-efficiency, it reduces the range between the upper and
lower whiskers and inter-quartile range for schema-driven binary serialization
specifications.  In particular, the compression formats with the smaller range
between the upper and lower whiskers for schema-driven binary serialization
specifications are GZIP and LZMA, the compression format with the smaller
inter-quartile range for schema-driven binary serialization specifications is
LZMA, and the compression format with the smaller inter-quartile range for
schema-less binary serialization specifications is GZIP.


Overall, \we conclude that uncompressed schema-driven binary serialization
specifications are space-efficient in comparison to uncompressed schema-less binary
serialization specifications and that compression does not contribute to
space-efficiency in comparison to both uncompressed schema-driven and
schema-less binary serialization specifications.

\clearpage

\subsection{TSLint Linter Definition (Basic)}
\label{sec:benchmark-tslintbasic}

\begin{figure*}[ht!]
\frame{\includegraphics[width=\linewidth]{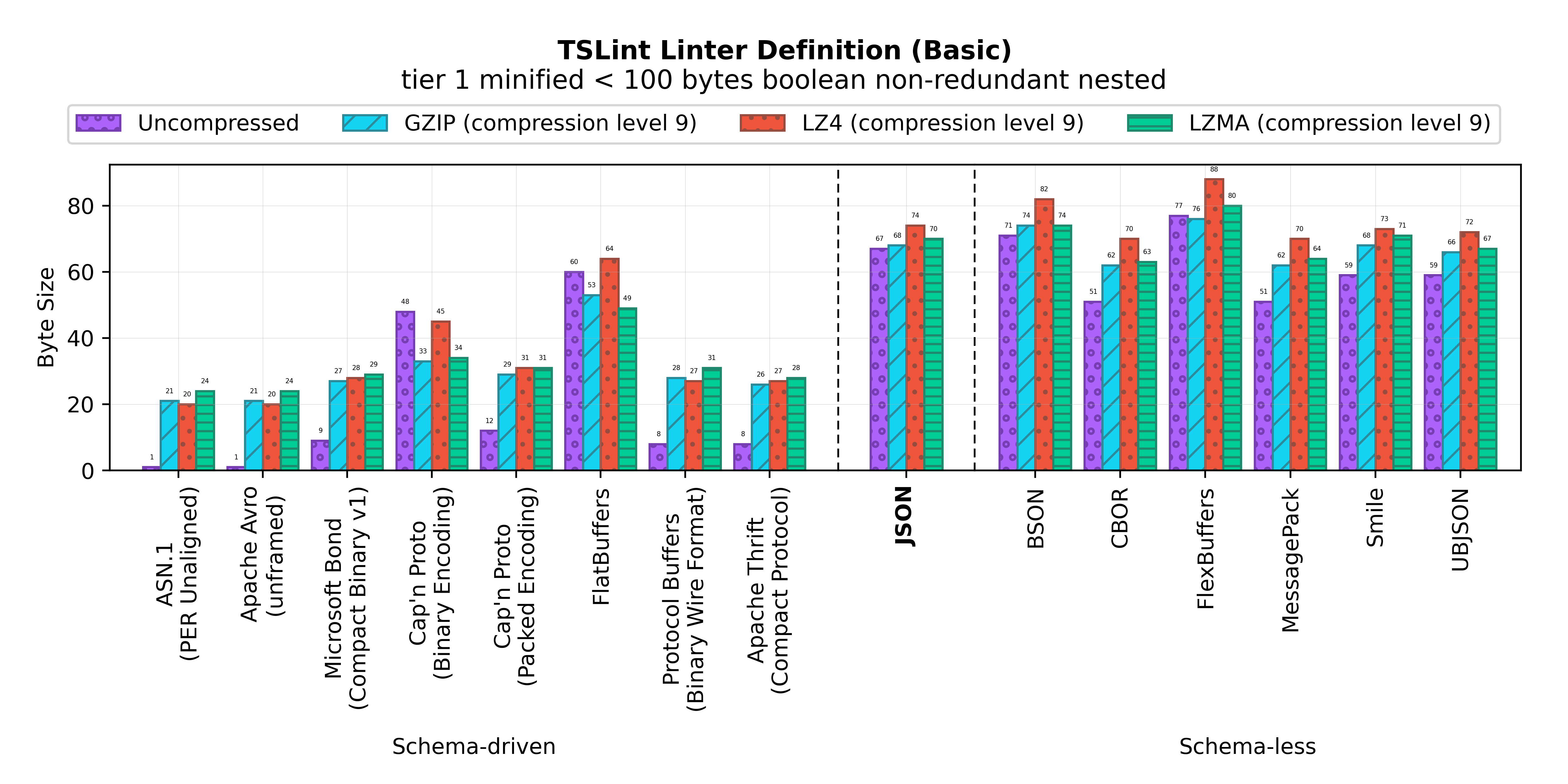}}
\caption{
The benchmark results for the TSLint Linter Definition (Basic) test case listed in \autoref{table:benchmark-documents} and \autoref{table:benchmark-documents-1}.
}
\label{fig:benchmark-tslintbasic}
\end{figure*}

TSLint \footnote{\url{https://palantir.github.io/tslint}} is now an obsolete
open-source linter for the TypeScript
\footnote{\url{https://www.typescriptlang.org}} programming language. TSLint
was created by the Big Data analytics company Palantir
\footnote{\url{https://www.palantir.com}} and was merged with the ESLint
open-source JavaScript linter in 2019
\footnote{\url{https://github.com/palantir/tslint/issues/4534}}. In
\autoref{fig:benchmark-tslintbasic}, \we demonstrate a \textbf{Tier 1 minified
$<$ 100 bytes boolean non-redundant nested} (Tier 1 BNN from
\autoref{table:json-taxonomy}) JSON document that consists of a basic TSLint
configuration that enforces grouped alphabetized imports.

The smallest bit-string is produced by both ASN.1 PER Unaligned \cite{asn1-per}
and Apache Avro \cite{avro} (1 byte), followed by both Protocol Buffers
\cite{protocolbuffers} and Apache Thrift \cite{slee2007thrift} (8 bytes), and
Microsoft Bond \cite{microsoft-bond} (9 bytes). The binary serialization
specifications that produced the smallest bit-strings are schema-driven and sequential
\cite{viotti2022survey}. Conversely, the largest bit-string is produced by FlexBuffers
\cite{flexbuffers} (77 bytes), closely followed by BSON \cite{bson} (71 bytes)
and FlatBuffers \cite{flatbuffers} (60 bytes). With the exception of BSON, the
binary serialization specifications that produced the largest bit-strings are
pointer-based \cite{viotti2022survey}.  In comparison to JSON \cite{ECMA-404} (67 bytes),
binary serialization achieves a \textbf{67x} size reduction in the best case
for this input document. Similar large size reductions are observed in JSON
documents whose content is dominated by \emph{boolean} and \emph{numeric}
values.  However, 2 out of the 14 JSON-compatible binary serialization specifications
listed in \autoref{table:benchmark-specifications-schema-driven} and
\autoref{table:benchmark-specifications-schema-less} result in bit-strings that are
larger than JSON: BSON \cite{bson} and FlexBuffers \cite{flexbuffers}.  These
binary serialization specifications are schema-less.

For this Tier 1 BNN document, the best performing schema-driven serialization
specification achieves a \textbf{51x} size reduction compared to the best performing
schema-less serialization specification: MessagePack \cite{messagepack} (51 bytes).
As shown in \autoref{table:benchmark-stats-tslintbasic}, uncompressed
schema-driven specifications provide smaller \emph{average} and \emph{median}
bit-strings than uncompressed schema-less specifications. However, as highlighted by
the \emph{range} and \emph{standard deviation}, uncompressed schema-driven
specifications exhibit higher size reduction variability depending on the
expressiveness of the schema language (i.e. how the language constructs allow
you to model the data) and the size optimizations devised by its authors.  With
the exception of the pointer-based binary serialization specification FlatBuffers
\cite{flatbuffers}, the selection of schema-driven serialization specifications listed
in \autoref{table:benchmark-specifications-schema-driven} produce bit-strings that are
equal to or smaller than their schema-less counterparts listed in
\autoref{table:benchmark-specifications-schema-less}.  The best performing sequential
serialization specification achieves a \textbf{12x} size reduction compared to the
best performing pointer-based serialization specification: Cap'n Proto Packed Encoding
\cite{capnproto} (12 bytes).

The compression formats listed in
\autoref{sec:benchmark-compression-formats} result in positive gains for
the bit-strings produced by Cap'n Proto Binary Encoding \cite{capnproto},
FlatBuffers \cite{flatbuffers} and FlexBuffers \cite{flexbuffers}. The best
performing uncompressed binary serialization specification achieves a
\textbf{68x} size reduction compared to the best performing compression format
for JSON: GZIP \cite{RFC1952} (68 bytes).

\begin{table*}[hb!]
\caption{A byte-size statistical analysis of the benchmark results shown in \autoref{fig:benchmark-tslintbasic} divided by schema-driven and schema-less specifications.}
\label{table:benchmark-stats-tslintbasic}
\begin{tabularx}{\linewidth}{X|l|l|l|l|l|l|l|l}
\toprule
\multirow{2}{*}{\textbf{Category}} &
\multicolumn{4}{c|}{\textbf{Schema-driven}} &
\multicolumn{4}{c}{\textbf{Schema-less}} \\
\cline{2-9}
& \small\textbf{Average} & \small\textbf{Median} & \small\textbf{Range} & \small\textbf{Std.dev} & \small\textbf{Average} & \small\textbf{Median} & \small\textbf{Range} & \small\textbf{Std.dev} \\
\midrule
Uncompressed & \small{18.4} & \small{8.5} & \small{59} & \small{21.1} & \small{61.3} & \small{59} & \small{26} & \small{9.7} \\ \hline
GZIP (compression level 9) & \small{29.8} & \small{27.5} & \small{32} & \small{9.5} & \small{68} & \small{67} & \small{14} & \small{5.4} \\ \hline
LZ4 (compression level 9) & \small{32.8} & \small{27.5} & \small{44} & \small{13.9} & \small{75.8} & \small{72.5} & \small{18} & \small{6.8} \\ \hline
LZMA (compression level 9) & \small{31.3} & \small{30} & \small{25} & \small{7.4} & \small{69.8} & \small{69} & \small{17} & \small{5.9} \\
\bottomrule
\end{tabularx}
\end{table*}

\begin{table*}[hb!]
\caption{The benchmark raw data results and schemas for the plot in \autoref{fig:benchmark-tslintbasic}.}
\label{table:benchmark-tslintbasic}
\begin{tabularx}{\linewidth}{X|l|l|l|l|l}
\toprule
\textbf{Serialization Format} & \textbf{Schema} & \textbf{Uncompressed} & \textbf{GZIP} & \textbf{LZ4} & \textbf{LZMA} \\
\midrule
ASN.1 (PER Unaligned) & \href{https://github.com/jviotti/binary-json-size-benchmark/blob/main/benchmark/tslintbasic/asn1/schema.asn}{\small{\texttt{schema.asn}}} & 1 & 21 & 20 & 24 \\ \hline
Apache Avro (unframed) & \href{https://github.com/jviotti/binary-json-size-benchmark/blob/main/benchmark/tslintbasic/avro/schema.json}{\small{\texttt{schema.json}}} & 1 & 21 & 20 & 24 \\ \hline
Microsoft Bond (Compact Binary v1) & \href{https://github.com/jviotti/binary-json-size-benchmark/blob/main/benchmark/tslintbasic/bond/schema.bond}{\small{\texttt{schema.bond}}} & 9 & 27 & 28 & 29 \\ \hline
Cap'n Proto (Binary Encoding) & \href{https://github.com/jviotti/binary-json-size-benchmark/blob/main/benchmark/tslintbasic/capnproto/schema.capnp}{\small{\texttt{schema.capnp}}} & 48 & 33 & 45 & 34 \\ \hline
Cap'n Proto (Packed Encoding) & \href{https://github.com/jviotti/binary-json-size-benchmark/blob/main/benchmark/tslintbasic/capnproto-packed/schema.capnp}{\small{\texttt{schema.capnp}}} & 12 & 29 & 31 & 31 \\ \hline
FlatBuffers & \href{https://github.com/jviotti/binary-json-size-benchmark/blob/main/benchmark/tslintbasic/flatbuffers/schema.fbs}{\small{\texttt{schema.fbs}}} & 60 & 53 & 64 & 49 \\ \hline
Protocol Buffers (Binary Wire Format) & \href{https://github.com/jviotti/binary-json-size-benchmark/blob/main/benchmark/tslintbasic/protobuf/schema.proto}{\small{\texttt{schema.proto}}} & 8 & 28 & 27 & 31 \\ \hline
Apache Thrift (Compact Protocol) & \href{https://github.com/jviotti/binary-json-size-benchmark/blob/main/benchmark/tslintbasic/thrift/schema.thrift}{\small{\texttt{schema.thrift}}} & 8 & 26 & 27 & 28 \\ \hline
\hline \textbf{JSON} & - & 67 & 68 & 74 & 70 \\ \hline \hline
BSON & - & 71 & 74 & 82 & 74 \\ \hline
CBOR & - & 51 & 62 & 70 & 63 \\ \hline
FlexBuffers & - & 77 & 76 & 88 & 80 \\ \hline
MessagePack & - & 51 & 62 & 70 & 64 \\ \hline
Smile & - & 59 & 68 & 73 & 71 \\ \hline
UBJSON & - & 59 & 66 & 72 & 67 \\
\bottomrule
\end{tabularx}
\end{table*}

\begin{figure*}[ht!]
\frame{\includegraphics[width=\linewidth]{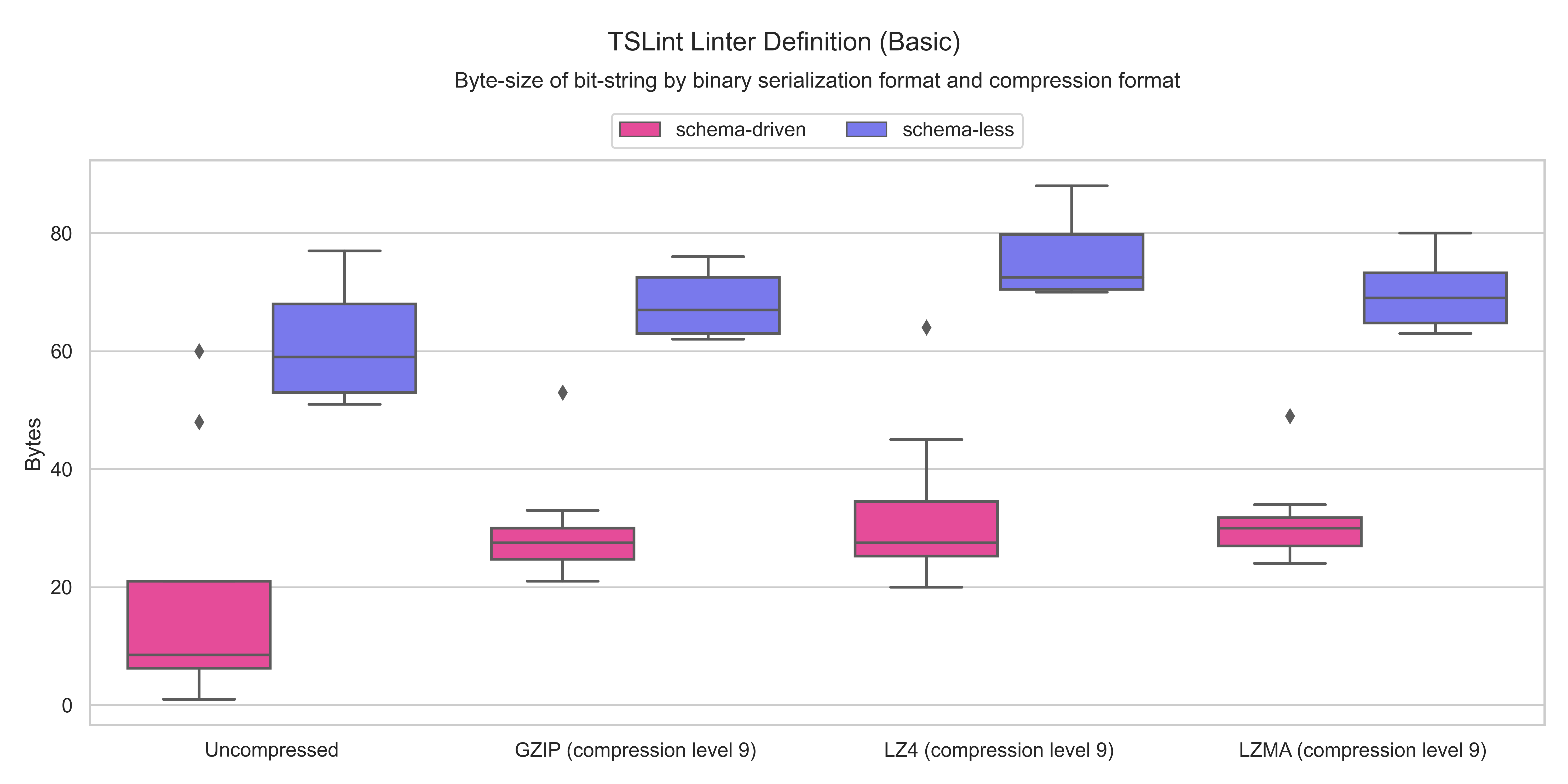}}
\caption{
Box plot of the statistical results in \autoref{table:benchmark-stats-tslintbasic}.
}
\label{fig:benchmark-tslintbasic-boxplot}
\end{figure*}

In \autoref{fig:benchmark-tslintbasic-boxplot}, we observe the medians for
uncompressed schema-driven binary serialization specifications to be smaller in
comparison to uncompressed schema-less binary serialization specifications.  The range
between the upper and lower whiskers of uncompressed schema-driven binary
serialization specifications is smaller than the range between the upper and lower
whiskers of uncompressed schema-less binary serialization specifications.  However,
the inter-quartile range of both both uncompressed schema-driven and
schema-less binary serialization specifications is similar.


In terms of compression, GZIP and LZ4 result in the lower medians for
schema-driven binary serialization specifications while GZIP results in the
lower median for schema-less binary serialization specifications.  However,
compression is not space-efficient in terms of the median for both
schema-driven and schema-less binary serialization specifications.
Additionally, the use of GZIP, LZ4 and LZMA for schema-driven binary
serialization specifications exhibits upper outliers.  While compression does
not contribute to space-efficiency, it reduces the range between the upper and
lower whiskers and inter-quartile range for both schema-driven and schema-less
binary serialization specifications.  In particular, the compression format
with the smaller range between the upper and lower whiskers for schema-driven
binary serialization specifications is LZMA, the compression formats with the
smaller inter-quartile range for schema-driven binary serialization
specifications are GZIP and LZMA, the compression format with the smaller range
between the upper and lower whiskers for schema-less binary serialization
specifications is GZIP, and the compression formats with the smaller
inter-quartile range for schema-less binary serialization specifications are
LZ4 and LZMA.


Overall, \we conclude that uncompressed schema-driven binary serialization
specifications are space-efficient in comparison to uncompressed schema-less binary
serialization specifications and that compression does not contribute to
space-efficiency in comparison to both uncompressed schema-driven and
schema-less binary serialization specifications.

\clearpage

\subsection{GeoJSON Example Document}
\label{sec:benchmark-geojson}

\begin{figure*}[ht!]
\frame{\includegraphics[width=\linewidth]{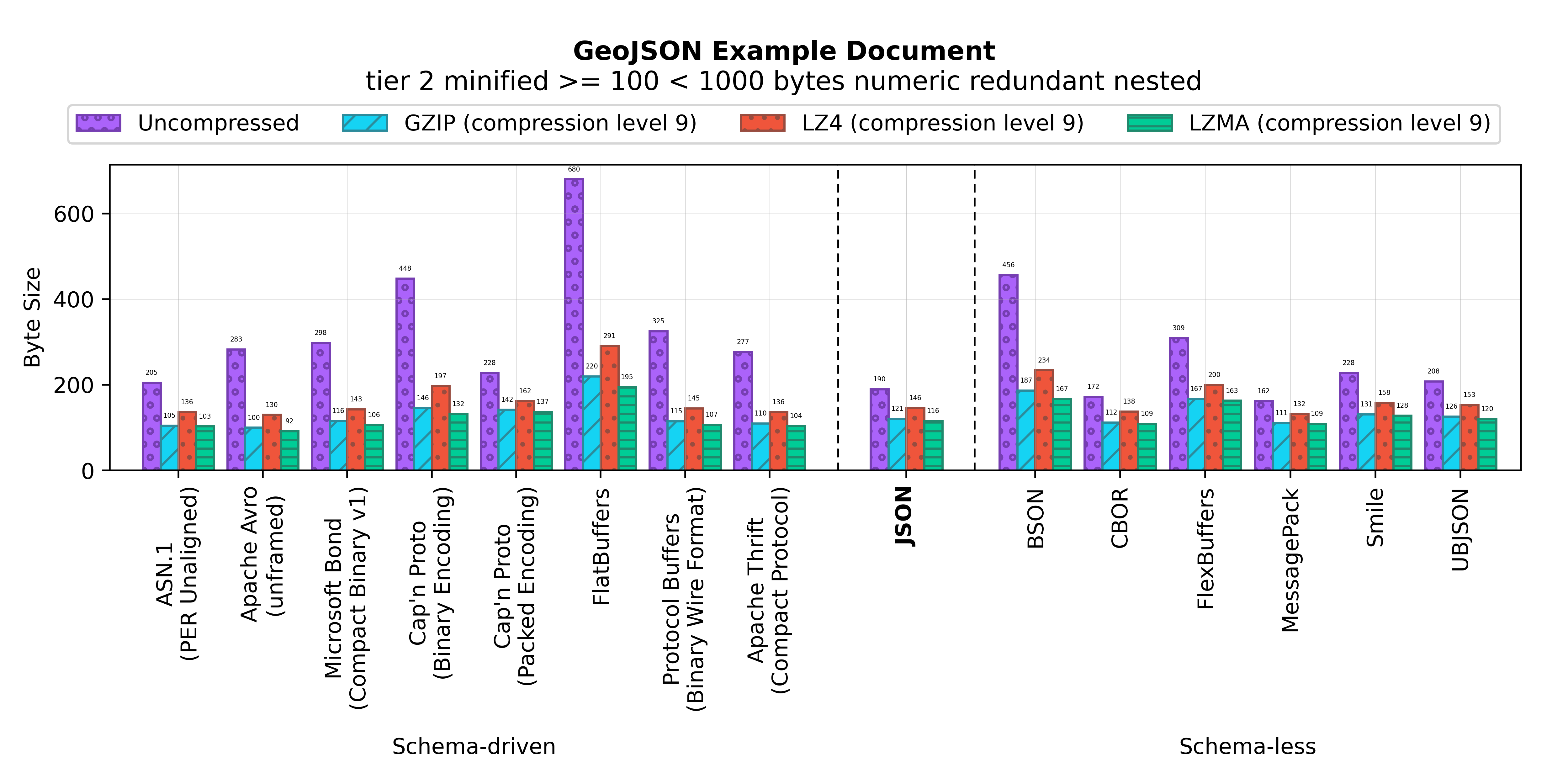}}
\caption{
The benchmark results for the GeoJSON Example Document test case listed in \autoref{table:benchmark-documents} and \autoref{table:benchmark-documents-1}.
}
\label{fig:benchmark-geojson}
\end{figure*}

GeoJSON \cite{RFC7946} is a standard to encode geospatial information using
JSON. GeoJSON is used in industries that have geographical and geospatial use
cases such as engineering, logistics and telecommunications. In
\autoref{fig:benchmark-geojson}, \we demonstrate a \textbf{Tier 2 minified
$\geq$ 100 $<$ 1000 bytes numeric redundant nested} (Tier 2 NRN from
\autoref{table:json-taxonomy}) JSON document that defines an example polygon
using the GeoJSON format.

In this exceptional case, the smallest bit-strings for this input document are
produced by the schema-less sequential specifications MessagePack
\cite{messagepack} (162 bytes) and CBOR \cite{RFC7049} (172 bytes), followed by
the schema-driven sequential specification ASN.1 PER Unaligned \cite{asn1-per}
(205 bytes). In comparison to other input documents, this input document
defines poly-dimensional JSON \cite{ECMA-404} arrays, which the schema-driven
binary serialization specifications from the selection do not encode in a
space-efficient manner. Similarly to other cases, the largest bit-string is
produced by FlatBuffers \cite{flatbuffers} (680 bytes), followed by BSON
\cite{bson} (456 bytes) and Cap'n Proto Binary Encoding \cite{capnproto} (448
bytes). With the exception of BSON, the binary serialization specifications
that produced the largest bit-strings are pointer-based \cite{viotti2022survey}.  In
comparison to JSON \cite{ECMA-404} (190 bytes), binary serialization only
achieves a \textbf{1.1x} size reduction in the best case for this input
document. With the exception of the schema-less sequential MessagePack
\cite{messagepack} and CBOR \cite{RFC7049} binary serialization specifications,
all the other JSON-compatible binary serialization specifications listed in
\autoref{table:benchmark-specifications-schema-driven} and
\autoref{table:benchmark-specifications-schema-less} result in bit-strings that
are larger than JSON.

For this Tier 2 NRN document, the smaller bit-string is produced by a
schema-less specification.  However, the best performing schema-less serialization
specification only achieves a \textbf{1.2x} size reduction compared to the best
performing schema-driven serialization specification: ASN.1 PER Unaligned
\cite{asn1-per} (205 bytes).  As shown in
\autoref{table:benchmark-stats-geojson}, uncompressed schema-less specifications
provide smaller \emph{average} and \emph{median} bit-strings than uncompressed
schema-driven specifications. Additionally, as highlighted by the \emph{range} and
\emph{standard deviation}, uncompressed schema-driven specifications exhibit higher
size reduction variability depending on the expressiveness of the schema
language (i.e. how the language constructs allow you to model the data) and the
size optimizations devised by its authors.  The schema-less and sequential
specifications CBOR \cite{RFC7049} and MessagePack \cite{messagepack} produce
bit-strings that are smaller to all their schema-driven counterparts listed in
\autoref{table:benchmark-specifications-schema-driven}.  The best performing
sequential serialization specification only achieves a \textbf{1.4x} size reduction
compared to the best performing pointer-based serialization specification: Cap'n Proto
Packed Encoding \cite{capnproto} (228 bytes).

The compression formats listed in
\autoref{sec:benchmark-compression-formats} result in positive gains for
all bit-strings. The best performing compression format for JSON, LZMA (116
bytes), achieve a \textbf{1.3x} size reduction compared to the best performing
uncompressed binary serialization specification.

\begin{table*}[hb!]
\caption{A byte-size statistical analysis of the benchmark results shown in \autoref{fig:benchmark-geojson} divided by schema-driven and schema-less specifications.}
\label{table:benchmark-stats-geojson}
\begin{tabularx}{\linewidth}{X|l|l|l|l|l|l|l|l}
\toprule
\multirow{2}{*}{\textbf{Category}} &
\multicolumn{4}{c|}{\textbf{Schema-driven}} &
\multicolumn{4}{c}{\textbf{Schema-less}} \\
\cline{2-9}
& \small\textbf{Average} & \small\textbf{Median} & \small\textbf{Range} & \small\textbf{Std.dev} & \small\textbf{Average} & \small\textbf{Median} & \small\textbf{Range} & \small\textbf{Std.dev} \\
\midrule
Uncompressed & \small{343} & \small{290.5} & \small{475} & \small{144.6} & \small{255.8} & \small{218} & \small{294} & \small{101.5} \\ \hline
GZIP (compression level 9) & \small{131.8} & \small{115.5} & \small{120} & \small{36.8} & \small{139} & \small{128.5} & \small{76} & \small{28.4} \\ \hline
LZ4 (compression level 9) & \small{167.5} & \small{144} & \small{161} & \small{50.8} & \small{169.2} & \small{155.5} & \small{102} & \small{36.3} \\ \hline
LZMA (compression level 9) & \small{122} & \small{106.5} & \small{103} & \small{31.1} & \small{132.7} & \small{124} & \small{58} & \small{23.8} \\
\bottomrule
\end{tabularx}
\end{table*}

\begin{table*}[hb!]
\caption{The benchmark raw data results and schemas for the plot in \autoref{fig:benchmark-geojson}.}
\label{table:benchmark-geojson}
\begin{tabularx}{\linewidth}{X|l|l|l|l|l}
\toprule
\textbf{Serialization Format} & \textbf{Schema} & \textbf{Uncompressed} & \textbf{GZIP} & \textbf{LZ4} & \textbf{LZMA} \\
\midrule
ASN.1 (PER Unaligned) & \href{https://github.com/jviotti/binary-json-size-benchmark/blob/main/benchmark/geojson/asn1/schema.asn}{\small{\texttt{schema.asn}}} & 205 & 105 & 136 & 103 \\ \hline
Apache Avro (unframed) & \href{https://github.com/jviotti/binary-json-size-benchmark/blob/main/benchmark/geojson/avro/schema.json}{\small{\texttt{schema.json}}} & 283 & 100 & 130 & 92 \\ \hline
Microsoft Bond (Compact Binary v1) & \href{https://github.com/jviotti/binary-json-size-benchmark/blob/main/benchmark/geojson/bond/schema.bond}{\small{\texttt{schema.bond}}} & 298 & 116 & 143 & 106 \\ \hline
Cap'n Proto (Binary Encoding) & \href{https://github.com/jviotti/binary-json-size-benchmark/blob/main/benchmark/geojson/capnproto/schema.capnp}{\small{\texttt{schema.capnp}}} & 448 & 146 & 197 & 132 \\ \hline
Cap'n Proto (Packed Encoding) & \href{https://github.com/jviotti/binary-json-size-benchmark/blob/main/benchmark/geojson/capnproto-packed/schema.capnp}{\small{\texttt{schema.capnp}}} & 228 & 142 & 162 & 137 \\ \hline
FlatBuffers & \href{https://github.com/jviotti/binary-json-size-benchmark/blob/main/benchmark/geojson/flatbuffers/schema.fbs}{\small{\texttt{schema.fbs}}} & 680 & 220 & 291 & 195 \\ \hline
Protocol Buffers (Binary Wire Format) & \href{https://github.com/jviotti/binary-json-size-benchmark/blob/main/benchmark/geojson/protobuf/schema.proto}{\small{\texttt{schema.proto}}} & 325 & 115 & 145 & 107 \\ \hline
Apache Thrift (Compact Protocol) & \href{https://github.com/jviotti/binary-json-size-benchmark/blob/main/benchmark/geojson/thrift/schema.thrift}{\small{\texttt{schema.thrift}}} & 277 & 110 & 136 & 104 \\ \hline
\hline \textbf{JSON} & - & 190 & 121 & 146 & 116 \\ \hline \hline
BSON & - & 456 & 187 & 234 & 167 \\ \hline
CBOR & - & 172 & 112 & 138 & 109 \\ \hline
FlexBuffers & - & 309 & 167 & 200 & 163 \\ \hline
MessagePack & - & 162 & 111 & 132 & 109 \\ \hline
Smile & - & 228 & 131 & 158 & 128 \\ \hline
UBJSON & - & 208 & 126 & 153 & 120 \\
\bottomrule
\end{tabularx}
\end{table*}

\begin{figure*}[ht!]
\frame{\includegraphics[width=\linewidth]{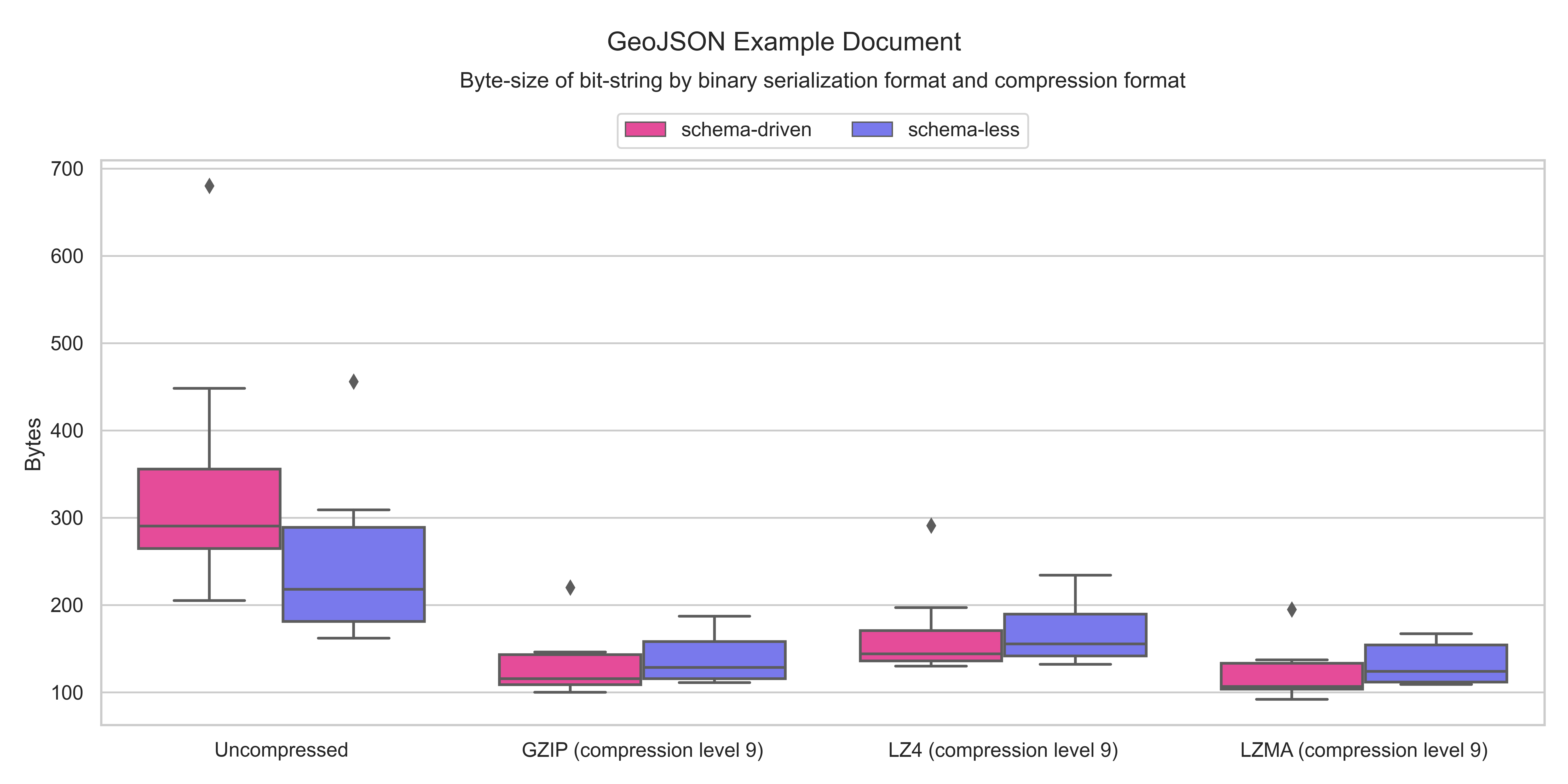}}
\caption{
Box plot of the statistical results in \autoref{table:benchmark-stats-geojson}.
}
\label{fig:benchmark-geojson-boxplot}
\end{figure*}

In \autoref{fig:benchmark-geojson-boxplot}, contrary to other cases, we observe
the medians for uncompressed schema-less binary serialization specifications to be
smaller in comparison to uncompressed schema-driven binary serialization
specifications.  The range between the upper and lower whiskers of uncompressed
schema-less binary serialization specifications is smaller than the range between the
upper and lower whiskers of uncompressed schema-driven binary serialization
specifications.  However, the inter-quartile range of uncompressed schema-less binary
serialization specifications is larger than the inter-quartile range of uncompressed
schema-driven binary serialization specifications. Additionally, their respective
quartiles overlap.


In terms of compression, LZMA results in the lower median for both
schema-driven and schema-less binary serialization specifications.
Additionally, GZIP, LZ4 and LZMA are space-efficient in terms of the median in
comparison to both uncompressed schema-driven and schema-less binary
serialization specifications.  However, the use of GZIP, LZ4 and LZMA for
schema-driven binary serialization specifications exhibits upper outliers.
Nevertheless, compression reduces the range between the upper and lower
whiskers and inter-quartile range for both schema-driven and schema-less binary
serialization specifications.  In particular, the compression formats with the
smaller range between the upper and lower whiskers and the smaller
inter-quartile range for schema-driven binary serialization specifications are
GZIP and LZMA, the compression format with the smaller range between the upper
and lower whiskers for schema-less binary serialization specifications is LZMA,
and the compression formats with the smaller inter-quartile range for
schema-less binary serialization specifications are GZIP and LZMA.


Overall, \we conclude that uncompressed schema-less binary serialization
specifications are space-efficient in comparison to uncompressed schema-driven
binary serialization specifications and that all the considered compression
formats are space-efficient in comparison to uncompressed schema-driven and
schema-less binary serialization specifications.

\clearpage

\subsection{OpenWeatherMap API Example Document}
\label{sec:benchmark-openweathermap}

\begin{figure*}[ht!]
\frame{\includegraphics[width=\linewidth]{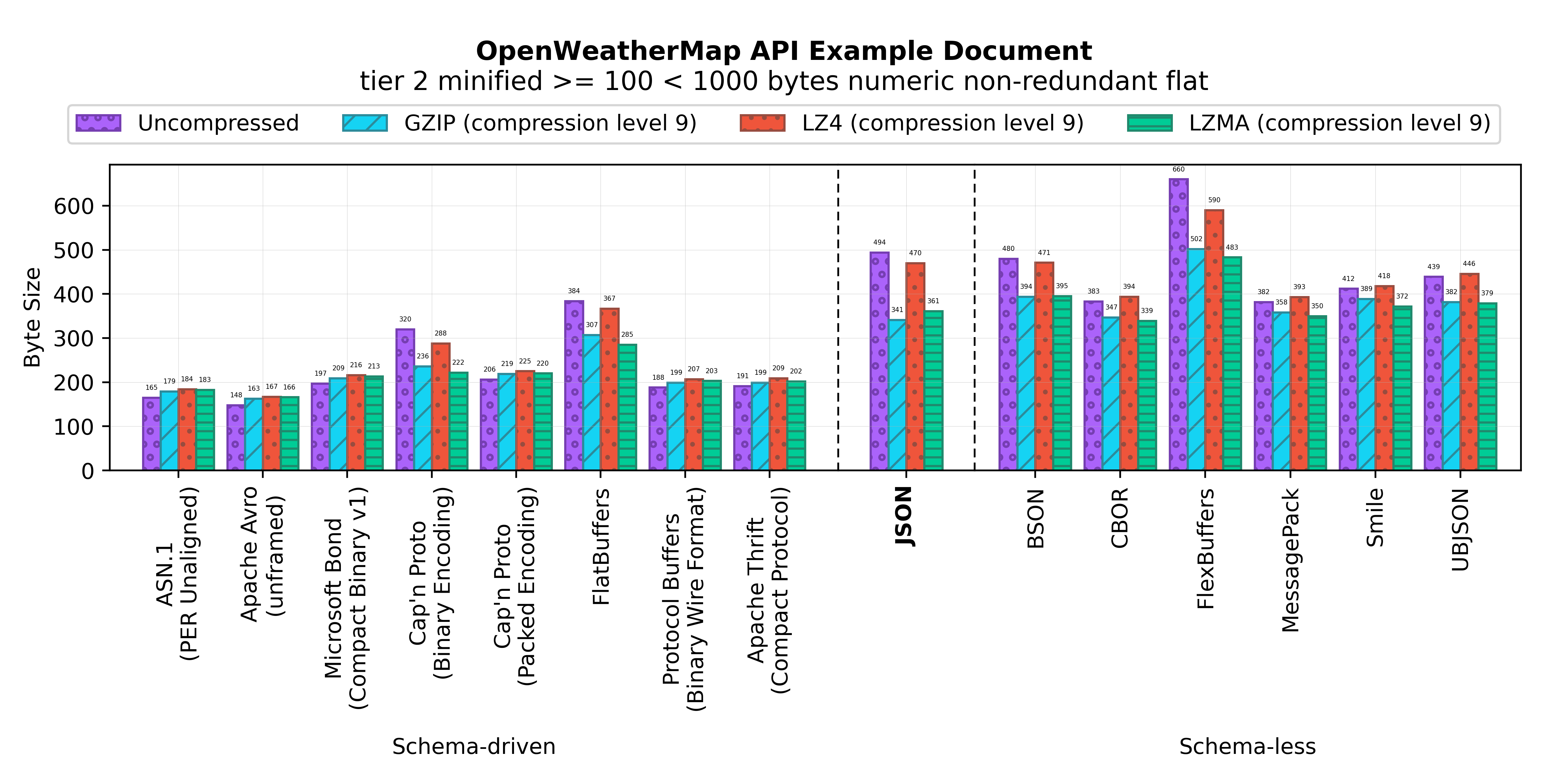}}
\caption{
The benchmark results for the OpenWeatherMap API Example Document test case listed in \autoref{table:benchmark-documents} and \autoref{table:benchmark-documents-1}.
}
\label{fig:benchmark-openweathermap}
\end{figure*}

OpenWeatherMap \footnote{\url{https://openweathermap.org}} is a weather data
and forecast API provider used in industries such as energy, agriculture,
transportation and construction. In \autoref{fig:benchmark-openweathermap}, \we
demonstrate a \textbf{Tier 2 minified $\geq$ 100 $<$ 1000 bytes numeric
non-redundant flat} (Tier 2 NNF from \autoref{table:json-taxonomy}) JSON
document that consists of an HTTP/1.1 \cite{RFC7231} response of the weather
information in Mountain View, California on June 12, 2019 at 2:44:05 PM GMT.

The smallest bit-string is produced by Apache Avro \cite{avro} (148 bytes),
followed by ASN.1 PER Unaligned \cite{asn1-per} (165 bytes) and Protocol
Buffers \cite{protocolbuffers} (188 bytes).  The binary serialization specifications
that produced the smallest bit-strings are schema-driven and sequential
\cite{viotti2022survey}. Conversely, the largest bit-string is produced by FlexBuffers
\cite{flexbuffers} (660 bytes), followed by BSON \cite{bson} (480 bytes) and
UBJSON \cite{ubjson} (439 bytes). The binary serialization specifications that
produced the largest bit-strings are schema-less and with the exception of
FlexBuffers, they are also sequential \cite{viotti2022survey}.  In comparison to JSON
\cite{ECMA-404} (494 bytes), binary serialization achieves a \textbf{3.3x} size
reduction in the best case for this input document.  However, 1 out of the 14
JSON-compatible binary serialization specifications listed in
\autoref{table:benchmark-specifications-schema-driven} and
\autoref{table:benchmark-specifications-schema-less} result in bit-strings that are
larger than JSON: the schema-less and pointer-based FlexBuffers
\cite{flexbuffers} binary serialization specification.

For this Tier 2 NNF document, the best performing schema-driven serialization
specification achieves a \textbf{2.5x} size reduction compared to the best performing
schema-less serialization specification: MessagePack \cite{messagepack} (382 bytes).
As shown in \autoref{table:benchmark-stats-openweathermap}, uncompressed
schema-driven specifications provide smaller \emph{average} and \emph{median}
bit-strings than uncompressed schema-less specifications. Additionally, as highlighted
by the \emph{range} and \emph{standard deviation}, uncompressed schema-less
specifications exhibit higher size reduction variability given that FlexBuffers
\cite{flexbuffers} produces a notably large bit-string. With the exception of
the pointer-based binary serialization specification FlatBuffers \cite{flatbuffers},
the selection of schema-driven serialization specifications listed in
\autoref{table:benchmark-specifications-schema-driven} produce bit-strings that are
equal to or smaller than their schema-less counterparts listed in
\autoref{table:benchmark-specifications-schema-less}.  The best performing sequential
serialization specification only achieves a \textbf{1.3x} size reduction compared to
the best performing pointer-based serialization specification: Cap'n Proto Packed
Encoding \cite{capnproto} (206 bytes).

The compression formats listed in
\autoref{sec:benchmark-compression-formats} result in positive gains for
all bit-strings except the ones produced by ASN.1 PER Unaligned
\cite{asn1-per}, Apache Avro \cite{avro}, Microsoft Bond \cite{microsoft-bond},
Cap'n Proto Packed Encoding \cite{capnproto}, Protocol Buffers
\cite{protocolbuffers} and Apache Thrift \cite{slee2007thrift}. The best
performing uncompressed binary serialization specification achieves a
\textbf{2.3x} size reduction compared to the best performing compression format
for JSON: GZIP \cite{RFC1952} (341 bytes).

\begin{table*}[hb!]
\caption{A byte-size statistical analysis of the benchmark results shown in \autoref{fig:benchmark-openweathermap} divided by schema-driven and schema-less specifications.}
\label{table:benchmark-stats-openweathermap}
\begin{tabularx}{\linewidth}{X|l|l|l|l|l|l|l|l}
\toprule
\multirow{2}{*}{\textbf{Category}} &
\multicolumn{4}{c|}{\textbf{Schema-driven}} &
\multicolumn{4}{c}{\textbf{Schema-less}} \\
\cline{2-9}
& \small\textbf{Average} & \small\textbf{Median} & \small\textbf{Range} & \small\textbf{Std.dev} & \small\textbf{Average} & \small\textbf{Median} & \small\textbf{Range} & \small\textbf{Std.dev} \\
\midrule
Uncompressed & \small{224.9} & \small{194} & \small{236} & \small{77.1} & \small{459.3} & \small{425.5} & \small{278} & \small{95.9} \\ \hline
GZIP (compression level 9) & \small{213.9} & \small{204} & \small{144} & \small{41.0} & \small{395.3} & \small{385.5} & \small{155} & \small{50.5} \\ \hline
LZ4 (compression level 9) & \small{232.9} & \small{212.5} & \small{200} & \small{60.6} & \small{452} & \small{432} & \small{197} & \small{67.6} \\ \hline
LZMA (compression level 9) & \small{211.8} & \small{208} & \small{119} & \small{32.8} & \small{386.3} & \small{375.5} & \small{144} & \small{47.0} \\
\bottomrule
\end{tabularx}
\end{table*}

\begin{table*}[hb!]
\caption{The benchmark raw data results and schemas for the plot in \autoref{fig:benchmark-openweathermap}.}
\label{table:benchmark-openweathermap}
\begin{tabularx}{\linewidth}{X|l|l|l|l|l}
\toprule
\textbf{Serialization Format} & \textbf{Schema} & \textbf{Uncompressed} & \textbf{GZIP} & \textbf{LZ4} & \textbf{LZMA} \\
\midrule
ASN.1 (PER Unaligned) & \href{https://github.com/jviotti/binary-json-size-benchmark/blob/main/benchmark/openweathermap/asn1/schema.asn}{\small{\texttt{schema.asn}}} & 165 & 179 & 184 & 183 \\ \hline
Apache Avro (unframed) & \href{https://github.com/jviotti/binary-json-size-benchmark/blob/main/benchmark/openweathermap/avro/schema.json}{\small{\texttt{schema.json}}} & 148 & 163 & 167 & 166 \\ \hline
Microsoft Bond (Compact Binary v1) & \href{https://github.com/jviotti/binary-json-size-benchmark/blob/main/benchmark/openweathermap/bond/schema.bond}{\small{\texttt{schema.bond}}} & 197 & 209 & 216 & 213 \\ \hline
Cap'n Proto (Binary Encoding) & \href{https://github.com/jviotti/binary-json-size-benchmark/blob/main/benchmark/openweathermap/capnproto/schema.capnp}{\small{\texttt{schema.capnp}}} & 320 & 236 & 288 & 222 \\ \hline
Cap'n Proto (Packed Encoding) & \href{https://github.com/jviotti/binary-json-size-benchmark/blob/main/benchmark/openweathermap/capnproto-packed/schema.capnp}{\small{\texttt{schema.capnp}}} & 206 & 219 & 225 & 220 \\ \hline
FlatBuffers & \href{https://github.com/jviotti/binary-json-size-benchmark/blob/main/benchmark/openweathermap/flatbuffers/schema.fbs}{\small{\texttt{schema.fbs}}} & 384 & 307 & 367 & 285 \\ \hline
Protocol Buffers (Binary Wire Format) & \href{https://github.com/jviotti/binary-json-size-benchmark/blob/main/benchmark/openweathermap/protobuf/schema.proto}{\small{\texttt{schema.proto}}} & 188 & 199 & 207 & 203 \\ \hline
Apache Thrift (Compact Protocol) & \href{https://github.com/jviotti/binary-json-size-benchmark/blob/main/benchmark/openweathermap/thrift/schema.thrift}{\small{\texttt{schema.thrift}}} & 191 & 199 & 209 & 202 \\ \hline
\hline \textbf{JSON} & - & 494 & 341 & 470 & 361 \\ \hline \hline
BSON & - & 480 & 394 & 471 & 395 \\ \hline
CBOR & - & 383 & 347 & 394 & 339 \\ \hline
FlexBuffers & - & 660 & 502 & 590 & 483 \\ \hline
MessagePack & - & 382 & 358 & 393 & 350 \\ \hline
Smile & - & 412 & 389 & 418 & 372 \\ \hline
UBJSON & - & 439 & 382 & 446 & 379 \\
\bottomrule
\end{tabularx}
\end{table*}

\begin{figure*}[ht!]
\frame{\includegraphics[width=\linewidth]{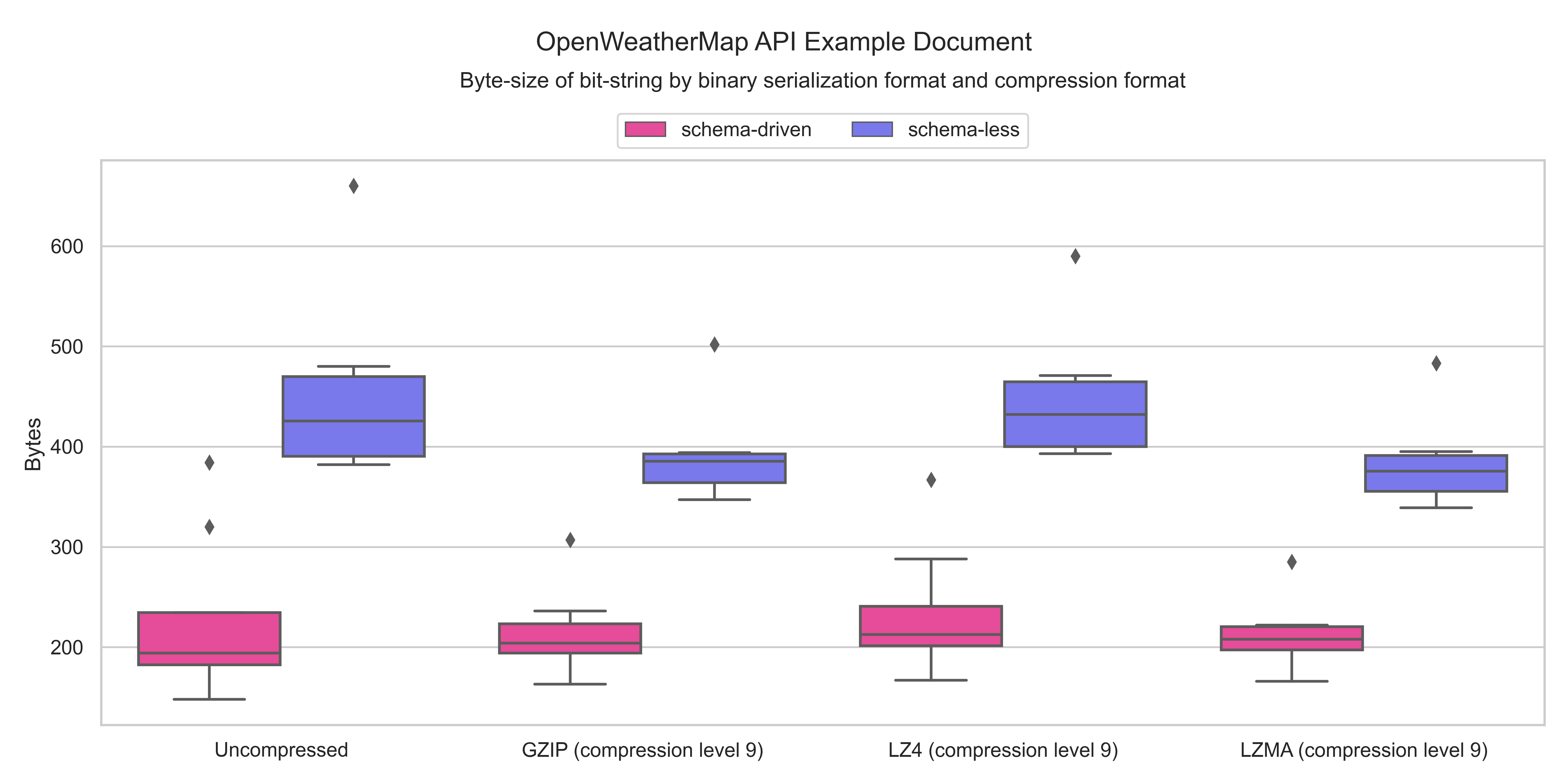}}
\caption{
Box plot of the statistical results in \autoref{table:benchmark-stats-openweathermap}.
}
\label{fig:benchmark-openweathermap-boxplot}
\end{figure*}

In \autoref{fig:benchmark-openweathermap-boxplot}, we observe the medians for
uncompressed schema-driven binary serialization specifications to be smaller in
comparison to uncompressed schema-less binary serialization specifications.  The range
between the upper and lower whiskers of uncompressed schema-driven and
schema-less binary serialization specifications is similar. However, the
inter-quartile range of uncompressed schema-driven binary serialization specifications
is smaller than the inter-quartile range of uncompressed schema-less binary
serialization specifications.


In terms of compression, GZIP results in the lower medians for schema-driven
binary serialization specifications while LZMA results in the lower median for
schema-less binary serialization specifications.  Compression is not
space-efficient in terms of the median in comparison to uncompressed
schema-driven binary serialization specifications. However, GZIP and LZMA are
space-efficient in terms of the median in comparison to uncompressed
schema-less binary serialization specifications. Additionally, the use of GZIP,
LZ4 and LZMA for both schema-driven binary serialization specifications and
schema-less binary serialization specifications exhibits upper outliers.
Nevertheless, compression reduces the range between the upper and lower
whiskers and inter-quartile range for both schema-driven and schema-less binary
serialization specifications.  In particular, the compression format with the
smaller range between the upper and lower whiskers and the smaller
inter-quartile range for schema-driven binary serialization specifications is
LZMA, and the compression format with the smaller range between the upper and
lower whiskers and the smaller inter-quartile range for schema-less binary
serialization specifications is GZIP.


Overall, \we conclude that uncompressed schema-driven binary serialization
specifications are space-efficient in comparison to uncompressed schema-less binary
serialization specifications. Compression does not contribute to space-efficiency in
comparison to schema-driven binary serialization specifications but GZIP and LZMA are
space-efficient in comparison to uncompressed schema-less binary serialization
specifications.

\clearpage

\subsection{OpenWeather Road Risk API Example}
\label{sec:benchmark-openweatherroadrisk}

\begin{figure*}[ht!]
\frame{\includegraphics[width=\linewidth]{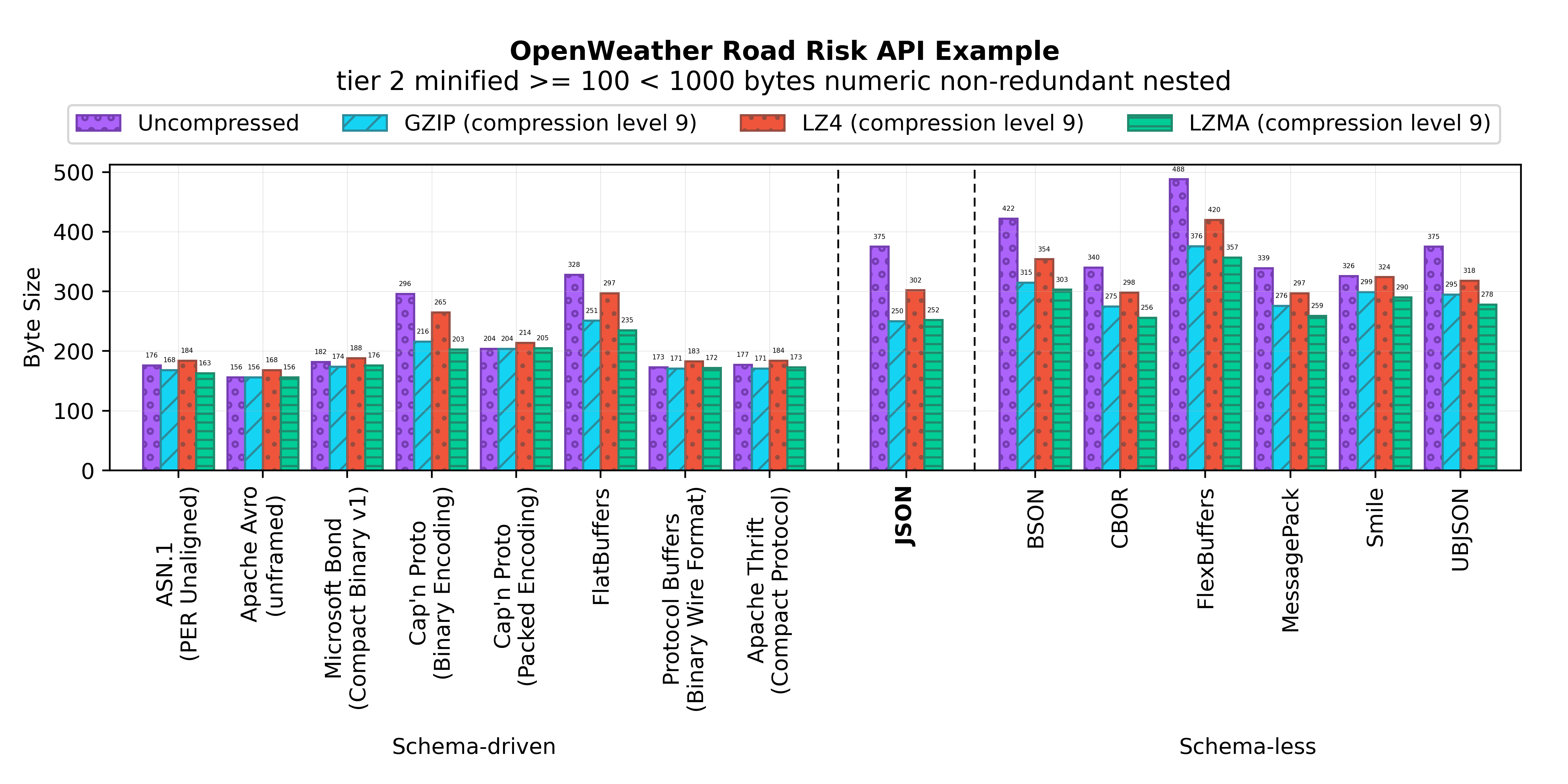}}
\caption{
The benchmark results for the OpenWeather Road Risk API Example test case listed in \autoref{table:benchmark-documents} and \autoref{table:benchmark-documents-1}.
}
\label{fig:benchmark-openweatherroadrisk}
\end{figure*}

OpenWeatherMap \footnote{\url{https://openweathermap.org}} is a weather data
and forecast API provider used in industries such as energy, agriculture,
transportation and construction. In
\autoref{fig:benchmark-openweatherroadrisk}, \we demonstrate a \textbf{Tier 2
minified $\geq$ 100 $<$ 1000 bytes numeric non-redundant nested} (Tier 2 NNN
from \autoref{table:json-taxonomy}) JSON document that consists of an example
HTTP/1.1 \cite{RFC7231} Road Risk API response from the official API
documentation that provides weather data and national alerts along a specific
route.

The smallest bit-string is produced by Apache Avro \cite{avro} (156 bytes),
followed by Protocol Buffers \cite{protocolbuffers} (173 bytes) and ASN.1 PER
Unaligned \cite{asn1-per} (176 bytes). The binary serialization specifications that
produced the smallest bit-strings are schema-driven and sequential \cite{viotti2022survey}.
Conversely, the largest bit-string is produced by FlexBuffers
\cite{flexbuffers} (488 bytes), followed by BSON \cite{bson} (422 bytes) and
UBJSON \cite{ubjson} (375 bytes). The binary serialization specifications that
produced the largest bit-strings are schema-less and with the exception of
FlexBuffers, they are also sequential \cite{viotti2022survey}.  In comparison to JSON
\cite{ECMA-404} (375 bytes), binary serialization achieves a \textbf{2.4x} size
reduction in the best case for this input document.  However, 2 out of the 14
JSON-compatible binary serialization specifications listed in
\autoref{table:benchmark-specifications-schema-driven} and
\autoref{table:benchmark-specifications-schema-less} result in bit-strings that are
larger than JSON: BSON \cite{bson} and FlexBuffers \cite{flexbuffers}.  These
binary serialization specifications are schema-less.  Additionally, the schema-less
and sequential UBJSON \cite{ubjson} serialization specification results in a
bit-string that is equal to JSON in size.

For this Tier 2 NNN document, the best performing schema-driven serialization
specification achieves a \textbf{2x} size reduction compared to the best performing
schema-less serialization specification: Smile \cite{smile} (326 bytes).  As shown in
\autoref{table:benchmark-stats-openweatherroadrisk}, uncompressed schema-driven
specifications provide smaller \emph{average} and \emph{median} bit-strings than
uncompressed schema-less specifications. However, as highlighted by the \emph{range}
and \emph{standard deviation}, uncompressed schema-driven specifications exhibit
higher size reduction variability depending on the expressiveness of the schema
language (i.e. how the language constructs allow you to model the data) and the
size optimizations devised by its authors. With the exception of the
pointer-based binary serialization specification FlatBuffers \cite{flatbuffers}, the
selection of schema-driven serialization specifications listed in
\autoref{table:benchmark-specifications-schema-driven} produce bit-strings that are
equal to or smaller than their schema-less counterparts listed in
\autoref{table:benchmark-specifications-schema-less}.  The best performing sequential
serialization specification only achieves a \textbf{1.3x} size reduction compared to
the best performing pointer-based serialization specification: Cap'n Proto Packed
Encoding \cite{capnproto} (204 bytes).

The compression formats listed in
\autoref{sec:benchmark-compression-formats} result in positive gains for
all bit-strings. The best performing uncompressed binary serialization
specification achieves a \textbf{1.6x} size reduction compared to the best
performing compression format for JSON: GZIP \cite{RFC1952} (250 bytes).

\begin{table*}[hb!]
\caption{A byte-size statistical analysis of the benchmark results shown in \autoref{fig:benchmark-openweatherroadrisk} divided by schema-driven and schema-less specifications.}
\label{table:benchmark-stats-openweatherroadrisk}
\begin{tabularx}{\linewidth}{X|l|l|l|l|l|l|l|l}
\toprule
\multirow{2}{*}{\textbf{Category}} &
\multicolumn{4}{c|}{\textbf{Schema-driven}} &
\multicolumn{4}{c}{\textbf{Schema-less}} \\
\cline{2-9}
& \small\textbf{Average} & \small\textbf{Median} & \small\textbf{Range} & \small\textbf{Std.dev} & \small\textbf{Average} & \small\textbf{Median} & \small\textbf{Range} & \small\textbf{Std.dev} \\
\midrule
Uncompressed & \small{211.5} & \small{179.5} & \small{172} & \small{59.8} & \small{381.7} & \small{357.5} & \small{162} & \small{57.2} \\ \hline
GZIP (compression level 9) & \small{188.9} & \small{172.5} & \small{95} & \small{30.0} & \small{306} & \small{297} & \small{101} & \small{34.2} \\ \hline
LZ4 (compression level 9) & \small{210.4} & \small{186} & \small{129} & \small{43.2} & \small{335.2} & \small{321} & \small{123} & \small{42.4} \\ \hline
LZMA (compression level 9) & \small{185.4} & \small{174.5} & \small{79} & \small{24.9} & \small{290.5} & \small{284} & \small{101} & \small{34.0} \\
\bottomrule
\end{tabularx}
\end{table*}

\begin{table*}[hb!]
\caption{The benchmark raw data results and schemas for the plot in \autoref{fig:benchmark-openweatherroadrisk}.}
\label{table:benchmark-openweatherroadrisk}
\begin{tabularx}{\linewidth}{X|l|l|l|l|l}
\toprule
\textbf{Serialization Format} & \textbf{Schema} & \textbf{Uncompressed} & \textbf{GZIP} & \textbf{LZ4} & \textbf{LZMA} \\
\midrule
ASN.1 (PER Unaligned) & \href{https://github.com/jviotti/binary-json-size-benchmark/blob/main/benchmark/openweatherroadrisk/asn1/schema.asn}{\small{\texttt{schema.asn}}} & 176 & 168 & 184 & 163 \\ \hline
Apache Avro (unframed) & \href{https://github.com/jviotti/binary-json-size-benchmark/blob/main/benchmark/openweatherroadrisk/avro/schema.json}{\small{\texttt{schema.json}}} & 156 & 156 & 168 & 156 \\ \hline
Microsoft Bond (Compact Binary v1) & \href{https://github.com/jviotti/binary-json-size-benchmark/blob/main/benchmark/openweatherroadrisk/bond/schema.bond}{\small{\texttt{schema.bond}}} & 182 & 174 & 188 & 176 \\ \hline
Cap'n Proto (Binary Encoding) & \href{https://github.com/jviotti/binary-json-size-benchmark/blob/main/benchmark/openweatherroadrisk/capnproto/schema.capnp}{\small{\texttt{schema.capnp}}} & 296 & 216 & 265 & 203 \\ \hline
Cap'n Proto (Packed Encoding) & \href{https://github.com/jviotti/binary-json-size-benchmark/blob/main/benchmark/openweatherroadrisk/capnproto-packed/schema.capnp}{\small{\texttt{schema.capnp}}} & 204 & 204 & 214 & 205 \\ \hline
FlatBuffers & \href{https://github.com/jviotti/binary-json-size-benchmark/blob/main/benchmark/openweatherroadrisk/flatbuffers/schema.fbs}{\small{\texttt{schema.fbs}}} & 328 & 251 & 297 & 235 \\ \hline
Protocol Buffers (Binary Wire Format) & \href{https://github.com/jviotti/binary-json-size-benchmark/blob/main/benchmark/openweatherroadrisk/protobuf/schema.proto}{\small{\texttt{schema.proto}}} & 173 & 171 & 183 & 172 \\ \hline
Apache Thrift (Compact Protocol) & \href{https://github.com/jviotti/binary-json-size-benchmark/blob/main/benchmark/openweatherroadrisk/thrift/schema.thrift}{\small{\texttt{schema.thrift}}} & 177 & 171 & 184 & 173 \\ \hline
\hline \textbf{JSON} & - & 375 & 250 & 302 & 252 \\ \hline \hline
BSON & - & 422 & 315 & 354 & 303 \\ \hline
CBOR & - & 340 & 275 & 298 & 256 \\ \hline
FlexBuffers & - & 488 & 376 & 420 & 357 \\ \hline
MessagePack & - & 339 & 276 & 297 & 259 \\ \hline
Smile & - & 326 & 299 & 324 & 290 \\ \hline
UBJSON & - & 375 & 295 & 318 & 278 \\
\bottomrule
\end{tabularx}
\end{table*}

\begin{figure*}[ht!]
\frame{\includegraphics[width=\linewidth]{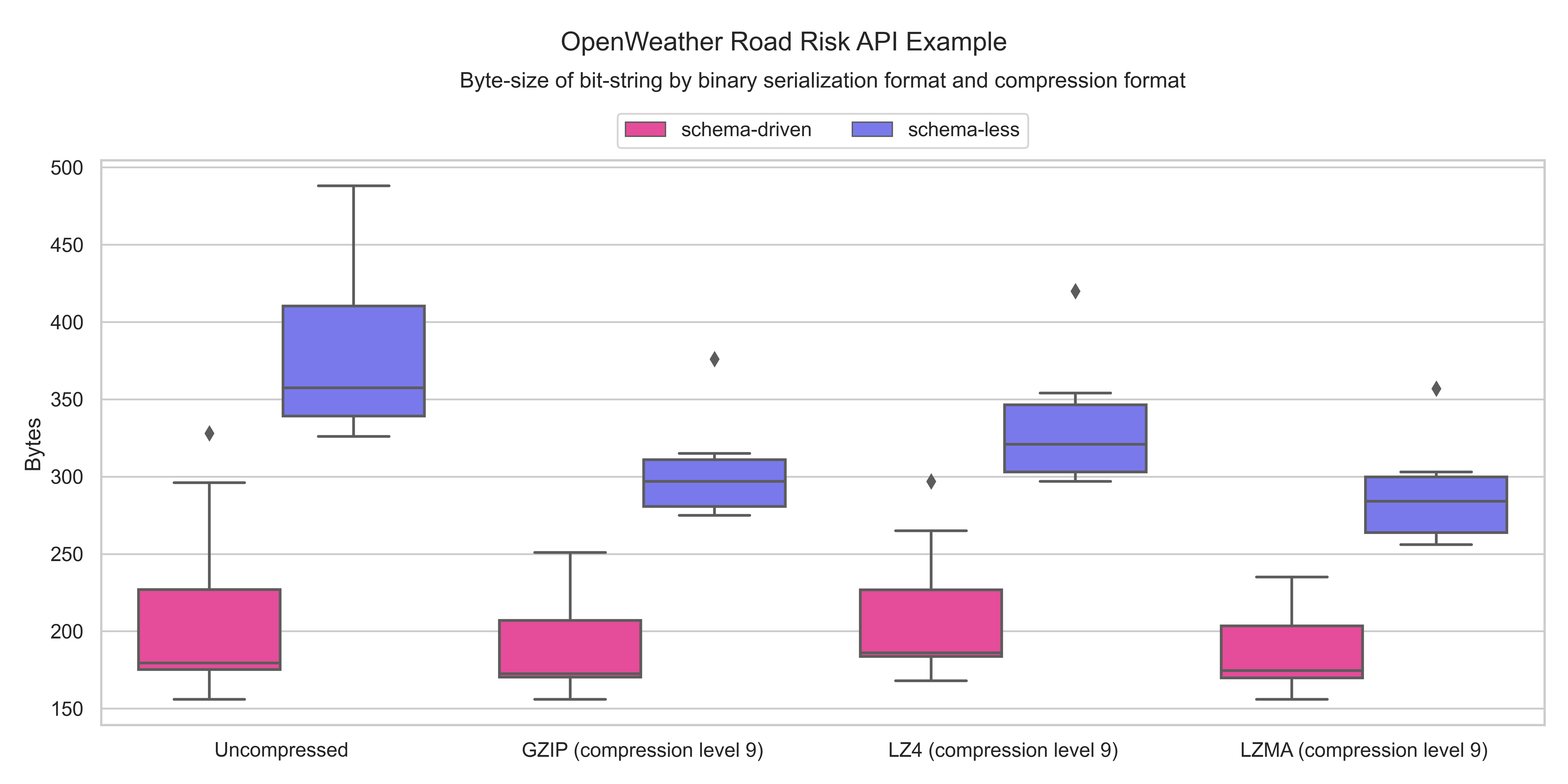}}
\caption{
Box plot of the statistical results in \autoref{table:benchmark-stats-openweatherroadrisk}.
}
\label{fig:benchmark-openweatherroadrisk-boxplot}
\end{figure*}

In \autoref{fig:benchmark-openweatherroadrisk-boxplot}, we observe the medians
for uncompressed schema-driven binary serialization specifications to be smaller in
comparison to uncompressed schema-less binary serialization specifications.  The range
between the upper and lower whiskers and the inter-quartile range of
uncompressed schema-driven binary serialization specifications is smaller than the
range between the upper and lower whiskers and the inter-quartile range of
uncompressed schema-less binary serialization specifications.


In terms of compression, GZIP results in the lower medians for schema-driven
binary serialization specifications while LZMA results in the lower median for
schema-less binary serialization specifications.  Additionally, GZIP and LZMA
are space-efficient in terms of the median in comparison to uncompressed
schema-driven binary serialization specifications and GZIP, LZ4 and LZMA are
space-efficient in terms of the median in comparison to uncompressed
schema-less binary serialization specifications.  However, the use of LZ4 for
schema-driven binary serialization specifications and the use of GZIP, LZ4 and
LZMA for schema-less binary serialization specifications exhibits upper
outliers.  Nevertheless, compression reduces the range between the upper and
lower whiskers and inter-quartile range for both schema-driven and schema-less
binary serialization specifications.  In particular, the compression format
with the smaller range between the upper and lower whiskers for schema-driven
binary serialization specifications is LZMA, the compression formats with the
smaller inter-quartile range for schema-driven binary serialization
specifications are GZIP and LZMA, and the compression format with the smaller
range between the upper and lower whiskers and the smaller inter-quartile range
for schema-less binary serialization specifications is GZIP.


Overall, \we conclude that uncompressed schema-driven binary serialization
specifications are space-efficient in comparison to uncompressed schema-less
binary serialization specifications. GZIP and LZMA are space-efficient in
comparison to uncompressed schema-driven binary serialization specifications
and all the considered compression formats are space-efficient in comparison to
uncompressed schema-less binary serialization specifications.

\clearpage

\subsection{TravisCI Notifications Configuration}
\label{sec:benchmark-travisnotifications}

\begin{figure*}[ht!]
\frame{\includegraphics[width=\linewidth]{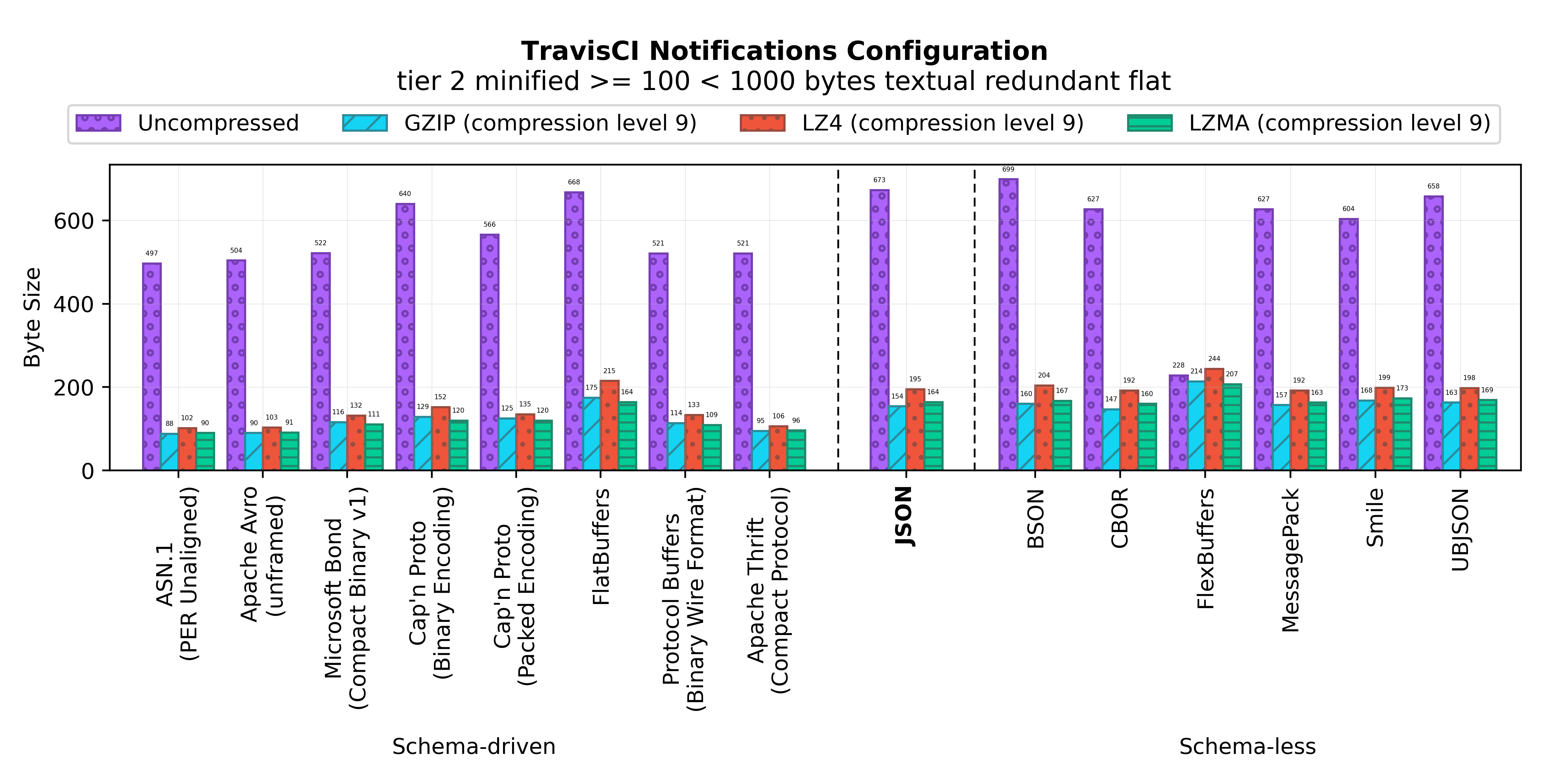}}
\caption{
The benchmark results for the TravisCI Notifications Configuration test case listed in \autoref{table:benchmark-documents} and \autoref{table:benchmark-documents-1}.
}
\label{fig:benchmark-travisnotifications}
\end{figure*}

TravisCI \footnote{\url{https://travis-ci.com}} is a commercial cloud-provider
of continuous integration and deployment pipelines  used by a wide range of
companies in the software development industry such as ZenDesk, BitTorrent, and
Engine Yard. In \autoref{fig:benchmark-travisnotifications}, \we demonstrate a
\textbf{Tier 2 minified $\geq$ 100 $<$ 1000 bytes textual redundant flat} (Tier
2 TRF from \autoref{table:json-taxonomy}) JSON document that consists of an
example pipeline configuration for TravisCI that declares a set of credentials
to post build notifications to various external services.

In this exceptional case, the smallest bit-string for this input document is
produced by the schema-less pointer-based serialization specification FlexBuffers
\cite{flexbuffers} (228 bytes), followed by the schema-driven sequential
specifications ASN.1 PER Unaligned \cite{asn1-per} (497 bytes) and Apache Avro
\cite{avro} (504 bytes). In comparison to other specifications, FlexBuffers detects
and de-duplicates the 7 instances of the same string value from this input
document. Conversely, the largest bit-string is produced by BSON \cite{bson}
(699 bytes), followed by FlatBuffers \cite{flatbuffers} (668 bytes) and UBJSON
\cite{ubjson} (658 bytes). With the exception of FlatBuffers, the binary
serialization specifications that produced the largest bit-strings are schema-less and
sequential \cite{viotti2022survey}.  In comparison to JSON \cite{ECMA-404} (673 bytes), binary
serialization achieves a \textbf{2.9x} size reduction in the best case for this
input document.  However, 1 out of the 14 JSON-compatible binary serialization
specifications listed in \autoref{table:benchmark-specifications-schema-driven} and
\autoref{table:benchmark-specifications-schema-less} result in bit-strings that are
larger than JSON: the schema-less and sequential BSON \cite{bson} binary
serialization specification.

For this Tier 2 TRF document, the smaller bit-string is produced by a
schema-less specification. The best performing schema-less serialization specification
achieves a \textbf{2.1x} size reduction compared to the best performing
schema-driven serialization specification: ASN.1 PER Unaligned \cite{asn1-per} (497
bytes).  As shown in \autoref{table:benchmark-stats-travisnotifications},
uncompressed schema-driven specifications provide smaller \emph{average} and
\emph{median} bit-strings than uncompressed schema-less specifications. Additionally,
as highlighted by the \emph{range} and \emph{standard deviation}, uncompressed
schema-less specifications exhibit higher size reduction variability given that
FlexBuffers \cite{flexbuffers} produces a notably small bit-string.  The
schema-less and pointer-based specification FlexBuffers \cite{flexbuffers} produces
bit-strings that are smaller to all their schema-driven counterparts listed in
\autoref{table:benchmark-specifications-schema-driven}.  The best performing
pointer-based serialization specification achieves a \textbf{2.1x} size reduction
compared to the best performing sequential serialization specification: ASN.1 PER
Unaligned \cite{asn1-per} (497 bytes).

The compression formats listed in
\autoref{sec:benchmark-compression-formats} result in positive gains for
all bit-strings. The best performing compression format for JSON, GZIP
\cite{RFC1952} (154 bytes), achieves a \textbf{1.4x} size reduction compared to
the best performing uncompressed binary serialization specification.

\begin{table*}[hb!]
\caption{A byte-size statistical analysis of the benchmark results shown in \autoref{fig:benchmark-travisnotifications} divided by schema-driven and schema-less specifications.}
\label{table:benchmark-stats-travisnotifications}
\begin{tabularx}{\linewidth}{X|l|l|l|l|l|l|l|l}
\toprule
\multirow{2}{*}{\textbf{Category}} &
\multicolumn{4}{c|}{\textbf{Schema-driven}} &
\multicolumn{4}{c}{\textbf{Schema-less}} \\
\cline{2-9}
& \small\textbf{Average} & \small\textbf{Median} & \small\textbf{Range} & \small\textbf{Std.dev} & \small\textbf{Average} & \small\textbf{Median} & \small\textbf{Range} & \small\textbf{Std.dev} \\
\midrule
Uncompressed & \small{554.9} & \small{521.5} & \small{171} & \small{60.7} & \small{573.8} & \small{627} & \small{471} & \small{157.5} \\ \hline
GZIP (compression level 9) & \small{116.5} & \small{115} & \small{87} & \small{26.5} & \small{168.2} & \small{161.5} & \small{67} & \small{21.5} \\ \hline
LZ4 (compression level 9) & \small{134.8} & \small{132.5} & \small{113} & \small{34.7} & \small{204.8} & \small{198.5} & \small{52} & \small{18.0} \\ \hline
LZMA (compression level 9) & \small{112.6} & \small{110} & \small{74} & \small{22.4} & \small{173.2} & \small{168} & \small{47} & \small{15.7} \\
\bottomrule
\end{tabularx}
\end{table*}

\begin{table*}[hb!]
\caption{The benchmark raw data results and schemas for the plot in \autoref{fig:benchmark-travisnotifications}.}
\label{table:benchmark-travisnotifications}
\begin{tabularx}{\linewidth}{X|l|l|l|l|l}
\toprule
\textbf{Serialization Format} & \textbf{Schema} & \textbf{Uncompressed} & \textbf{GZIP} & \textbf{LZ4} & \textbf{LZMA} \\
\midrule
ASN.1 (PER Unaligned) & \href{https://github.com/jviotti/binary-json-size-benchmark/blob/main/benchmark/travisnotifications/asn1/schema.asn}{\small{\texttt{schema.asn}}} & 497 & 88 & 102 & 90 \\ \hline
Apache Avro (unframed) & \href{https://github.com/jviotti/binary-json-size-benchmark/blob/main/benchmark/travisnotifications/avro/schema.json}{\small{\texttt{schema.json}}} & 504 & 90 & 103 & 91 \\ \hline
Microsoft Bond (Compact Binary v1) & \href{https://github.com/jviotti/binary-json-size-benchmark/blob/main/benchmark/travisnotifications/bond/schema.bond}{\small{\texttt{schema.bond}}} & 522 & 116 & 132 & 111 \\ \hline
Cap'n Proto (Binary Encoding) & \href{https://github.com/jviotti/binary-json-size-benchmark/blob/main/benchmark/travisnotifications/capnproto/schema.capnp}{\small{\texttt{schema.capnp}}} & 640 & 129 & 152 & 120 \\ \hline
Cap'n Proto (Packed Encoding) & \href{https://github.com/jviotti/binary-json-size-benchmark/blob/main/benchmark/travisnotifications/capnproto-packed/schema.capnp}{\small{\texttt{schema.capnp}}} & 566 & 125 & 135 & 120 \\ \hline
FlatBuffers & \href{https://github.com/jviotti/binary-json-size-benchmark/blob/main/benchmark/travisnotifications/flatbuffers/schema.fbs}{\small{\texttt{schema.fbs}}} & 668 & 175 & 215 & 164 \\ \hline
Protocol Buffers (Binary Wire Format) & \href{https://github.com/jviotti/binary-json-size-benchmark/blob/main/benchmark/travisnotifications/protobuf/schema.proto}{\small{\texttt{schema.proto}}} & 521 & 114 & 133 & 109 \\ \hline
Apache Thrift (Compact Protocol) & \href{https://github.com/jviotti/binary-json-size-benchmark/blob/main/benchmark/travisnotifications/thrift/schema.thrift}{\small{\texttt{schema.thrift}}} & 521 & 95 & 106 & 96 \\ \hline
\hline \textbf{JSON} & - & 673 & 154 & 195 & 164 \\ \hline \hline
BSON & - & 699 & 160 & 204 & 167 \\ \hline
CBOR & - & 627 & 147 & 192 & 160 \\ \hline
FlexBuffers & - & 228 & 214 & 244 & 207 \\ \hline
MessagePack & - & 627 & 157 & 192 & 163 \\ \hline
Smile & - & 604 & 168 & 199 & 173 \\ \hline
UBJSON & - & 658 & 163 & 198 & 169 \\
\bottomrule
\end{tabularx}
\end{table*}

\begin{figure*}[ht!]
\frame{\includegraphics[width=\linewidth]{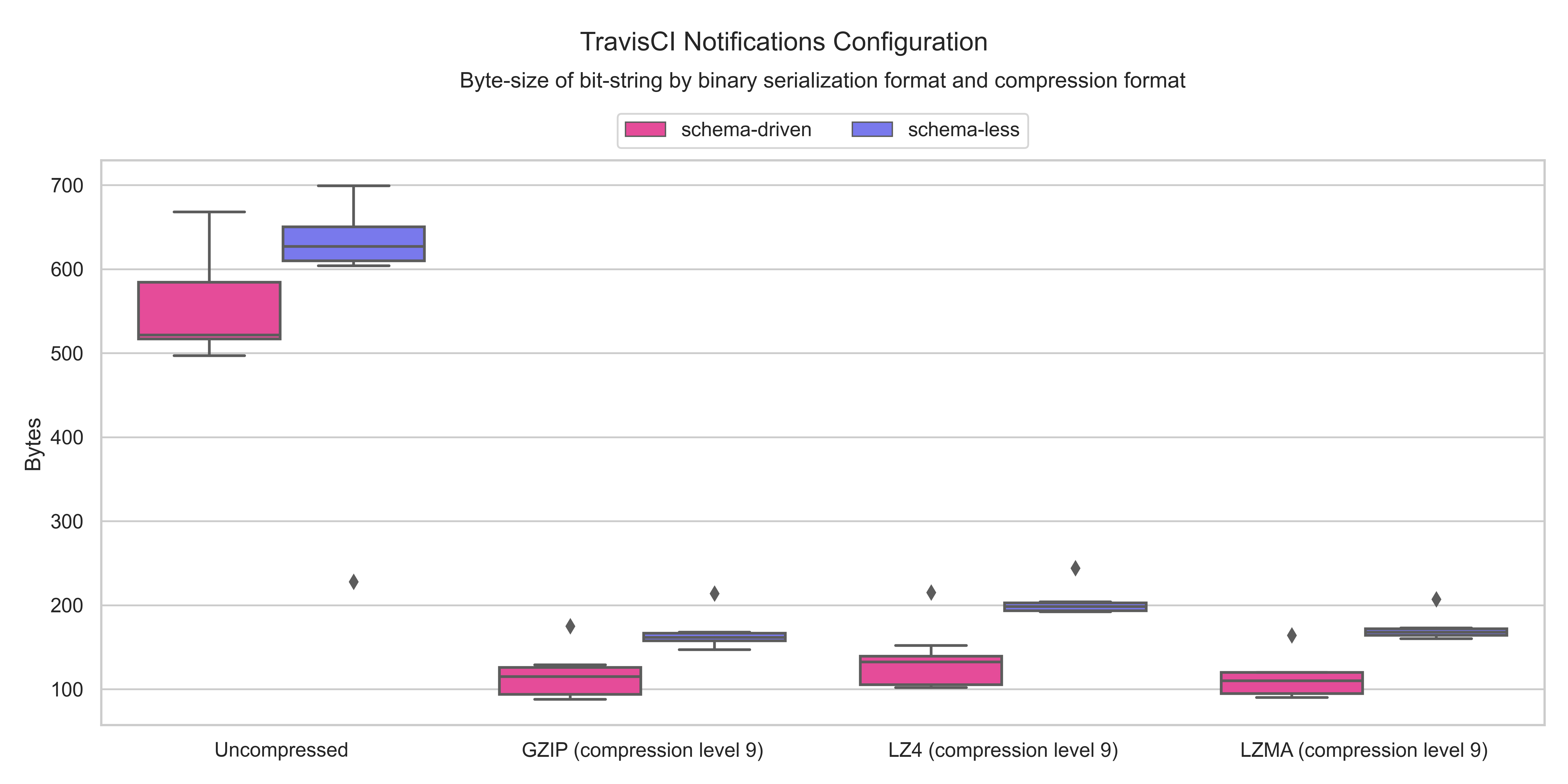}}
\caption{
Box plot of the statistical results in \autoref{table:benchmark-stats-travisnotifications}.
}
\label{fig:benchmark-travisnotifications-boxplot}
\end{figure*}

In \autoref{fig:benchmark-travisnotifications-boxplot}, we observe the medians
for uncompressed schema-driven binary serialization specifications to be smaller in
comparison to uncompressed schema-less binary serialization specifications.  The range
between the upper and lower whiskers and the inter-quartile range of
uncompressed schema-less binary serialization specifications is smaller than the range
between the upper and lower whiskers and the inter-quartile range of
uncompressed schema-driven binary serialization specifications. The low outlier for
uncompressed schema-less binary serialization specification represents the
space-efficiency of FlexBuffers \cite{flexbuffers} given its string
de-duplication features.


In terms of compression, LZMA results in the lower medians for schema-driven
binary serialization specifications while GZIP results in the lower median for
schema-less binary serialization specifications.  Additionally, GZIP, LZ4 and
LZMA are space-efficient in terms of the median in comparison to both
uncompressed schema-driven and schema-less binary serialization specifications.
However, the use of GZIP, LZ4 and LZMA for schema-driven binary serialization
specifications and schema-less binary serialization specifications exhibits
upper outliers.  Nevertheless, compression reduces the range between the upper
and lower whiskers and inter-quartile range for both schema-driven and
schema-less binary serialization specifications.  In particular, the
compression format with the smaller range between the upper and lower whiskers
and the smaller inter-quartile range for schema-driven binary serialization
specifications is LZMA, and the compression formats with the smaller range
between the upper and lower whiskers for schema-less binary serialization
specifications are LZ4 and LZMA.


Overall, \we conclude that uncompressed schema-driven binary serialization
specifications are space-efficient in comparison to uncompressed schema-less
binary serialization specifications and that all the considered compression
formats are space-efficient in comparison to uncompressed schema-driven and
schema-less binary serialization specifications.

\clearpage

\subsection{Entry Point Regulation Manifest}
\label{sec:benchmark-epr}

\begin{figure*}[ht!]
\frame{\includegraphics[width=\linewidth]{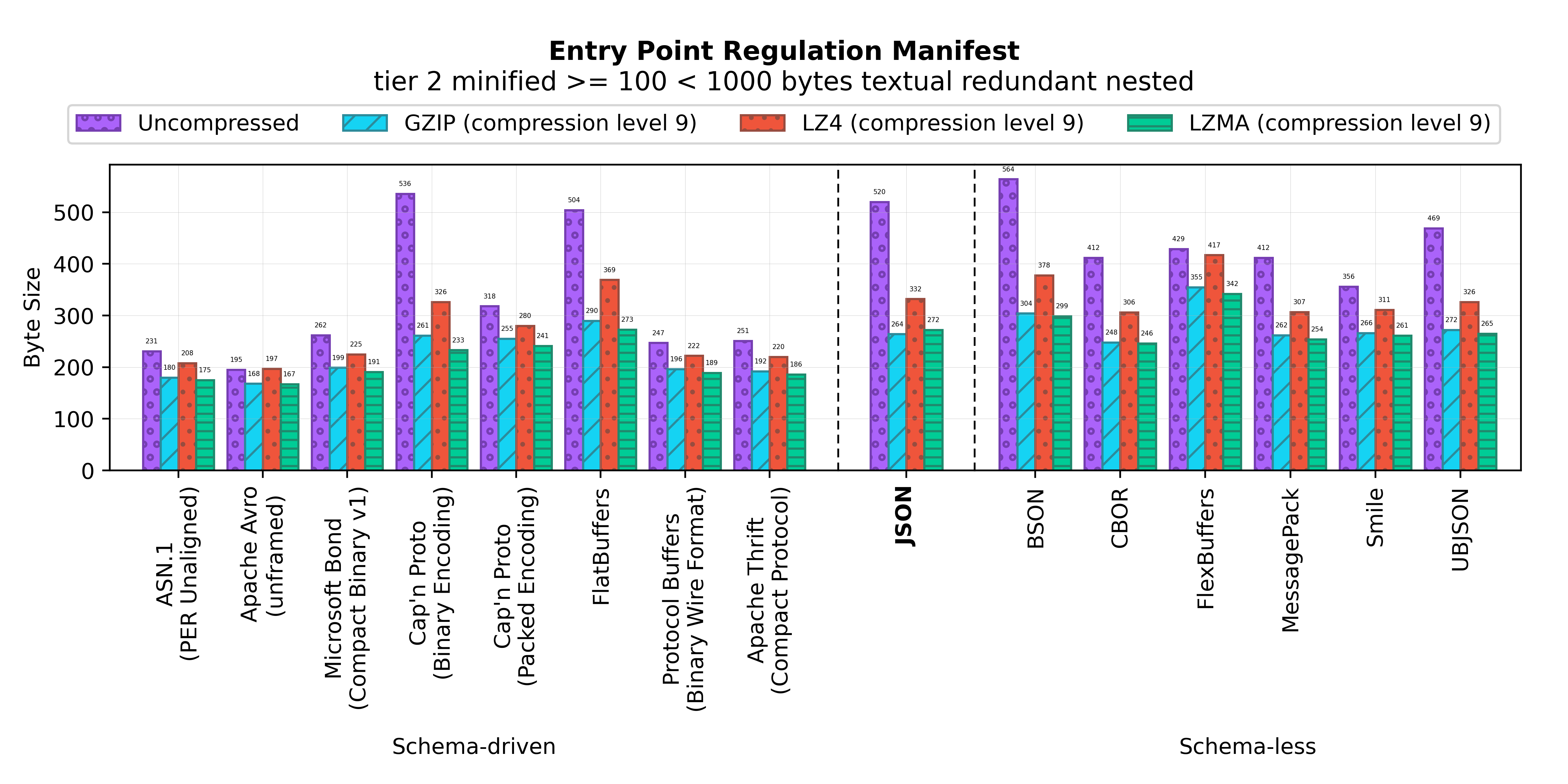}}
\caption{
The benchmark results for the Entry Point Regulation Manifest test case listed in \autoref{table:benchmark-documents} and \autoref{table:benchmark-documents-1}.
}
\label{fig:benchmark-epr}
\end{figure*}

Entry Point Regulation (EPR) \cite{EPR} is a W3C proposal led by Google that
defines a manifest that protects websites against cross-site scripting attacks
by allowing the developer to mark the areas of the application that can be
externally referenced. EPR manifests are used in the web industry. In
\autoref{fig:benchmark-epr}, \we demonstrate a \textbf{Tier 2 minified $\geq$
100 $<$ 1000 bytes textual redundant nested} (Tier 2 TRN from
\autoref{table:json-taxonomy}) JSON document that defines an example EPR policy
for a fictitious website.

The smallest bit-string is produced by Apache Avro \cite{avro} (195 bytes)
followed by ASN.1 PER Unaligned \cite{asn1-per} (231 bytes) and Protocol
Buffers \cite{protocolbuffers} (247 bytes). The binary serialization specifications
that produced the smallest bit-strings are schema-driven and sequential
\cite{viotti2022survey}. Conversely, the largest bit-string is produced by BSON \cite{bson}
(564 bytes) followed by Cap'n Proto Binary Encoding \cite{capnproto} (536
bytes) and FlatBuffers \cite{flatbuffers} (504 bytes).  With the exception of
BSON, the binary serialization specifications that produced the largest bit-strings
are schema-driven and pointer-based \cite{viotti2022survey}.  In comparison to JSON
\cite{ECMA-404} (520 bytes), binary serialization achieves a \textbf{2.6x} size
reduction in the best case for this input document.  However, 2 out of the 14
JSON-compatible binary serialization specifications listed in
\autoref{table:benchmark-specifications-schema-driven} and
\autoref{table:benchmark-specifications-schema-less} result in bit-strings that are
larger than JSON: Cap'n Proto Binary Encoding \cite{capnproto} and BSON
\cite{bson}. These binary serialization specifications are either schema-less or
schema-driven and pointer-based.

For this Tier 2 TRN document, the best performing schema-driven serialization
specification only achieves a \textbf{1.8x} size reduction compared to the best
performing schema-less serialization specification: Smile \cite{smile} (356 bytes).
As shown in \autoref{table:benchmark-stats-epr}, uncompressed schema-driven
specifications provide smaller \emph{average} and \emph{median} bit-strings than
uncompressed schema-less specifications. However, as highlighted by the \emph{range}
and \emph{standard deviation}, uncompressed schema-driven specifications exhibit
higher size reduction variability depending on the expressiveness of the schema
language (i.e. how the language constructs allow you to model the data) and the
size optimizations devised by its authors. With the exception of the
pointer-based binary serialization specifications Cap'n Proto Binary Encoding
\cite{capnproto} and FlatBuffers \cite{flatbuffers}, the selection of
schema-driven serialization specifications listed in
\autoref{table:benchmark-specifications-schema-driven} produce bit-strings that are
equal to or smaller than their schema-less counterparts listed in
\autoref{table:benchmark-specifications-schema-less}.  The best performing sequential
serialization specification only achieves a \textbf{1.6x} size reduction compared to
the best performing pointer-based serialization specification: Cap'n Proto Packed
Encoding \cite{capnproto} (318 bytes).

The compression formats listed in
\autoref{sec:benchmark-compression-formats} result in positive gains for
all bit-strings. The best performing uncompressed binary serialization
specification achieves a \textbf{1.3x} size reduction compared to the best
performing compression format for JSON: GZIP \cite{RFC1952} (264 bytes).

\begin{table*}[hb!]
\caption{A byte-size statistical analysis of the benchmark results shown in \autoref{fig:benchmark-epr} divided by schema-driven and schema-less specifications.}
\label{table:benchmark-stats-epr}
\begin{tabularx}{\linewidth}{X|l|l|l|l|l|l|l|l}
\toprule
\multirow{2}{*}{\textbf{Category}} &
\multicolumn{4}{c|}{\textbf{Schema-driven}} &
\multicolumn{4}{c}{\textbf{Schema-less}} \\
\cline{2-9}
& \small\textbf{Average} & \small\textbf{Median} & \small\textbf{Range} & \small\textbf{Std.dev} & \small\textbf{Average} & \small\textbf{Median} & \small\textbf{Range} & \small\textbf{Std.dev} \\
\midrule
Uncompressed & \small{318} & \small{256.5} & \small{341} & \small{121.2} & \small{440.3} & \small{420.5} & \small{208} & \small{64.5} \\ \hline
GZIP (compression level 9) & \small{217.6} & \small{197.5} & \small{122} & \small{41.6} & \small{284.5} & \small{269} & \small{107} & \small{35.8} \\ \hline
LZ4 (compression level 9) & \small{255.9} & \small{223.5} & \small{172} & \small{58.6} & \small{340.8} & \small{318.5} & \small{111} & \small{42.1} \\ \hline
LZMA (compression level 9) & \small{206.9} & \small{190} & \small{106} & \small{35.1} & \small{277.8} & \small{263} & \small{96} & \small{33.2} \\
\bottomrule
\end{tabularx}
\end{table*}

\begin{table*}[hb!]
\caption{The benchmark raw data results and schemas for the plot in \autoref{fig:benchmark-epr}.}
\label{table:benchmark-epr}
\begin{tabularx}{\linewidth}{X|l|l|l|l|l}
\toprule
\textbf{Serialization Format} & \textbf{Schema} & \textbf{Uncompressed} & \textbf{GZIP} & \textbf{LZ4} & \textbf{LZMA} \\
\midrule
ASN.1 (PER Unaligned) & \href{https://github.com/jviotti/binary-json-size-benchmark/blob/main/benchmark/epr/asn1/schema.asn}{\small{\texttt{schema.asn}}} & 231 & 180 & 208 & 175 \\ \hline
Apache Avro (unframed) & \href{https://github.com/jviotti/binary-json-size-benchmark/blob/main/benchmark/epr/avro/schema.json}{\small{\texttt{schema.json}}} & 195 & 168 & 197 & 167 \\ \hline
Microsoft Bond (Compact Binary v1) & \href{https://github.com/jviotti/binary-json-size-benchmark/blob/main/benchmark/epr/bond/schema.bond}{\small{\texttt{schema.bond}}} & 262 & 199 & 225 & 191 \\ \hline
Cap'n Proto (Binary Encoding) & \href{https://github.com/jviotti/binary-json-size-benchmark/blob/main/benchmark/epr/capnproto/schema.capnp}{\small{\texttt{schema.capnp}}} & 536 & 261 & 326 & 233 \\ \hline
Cap'n Proto (Packed Encoding) & \href{https://github.com/jviotti/binary-json-size-benchmark/blob/main/benchmark/epr/capnproto-packed/schema.capnp}{\small{\texttt{schema.capnp}}} & 318 & 255 & 280 & 241 \\ \hline
FlatBuffers & \href{https://github.com/jviotti/binary-json-size-benchmark/blob/main/benchmark/epr/flatbuffers/schema.fbs}{\small{\texttt{schema.fbs}}} & 504 & 290 & 369 & 273 \\ \hline
Protocol Buffers (Binary Wire Format) & \href{https://github.com/jviotti/binary-json-size-benchmark/blob/main/benchmark/epr/protobuf/schema.proto}{\small{\texttt{schema.proto}}} & 247 & 196 & 222 & 189 \\ \hline
Apache Thrift (Compact Protocol) & \href{https://github.com/jviotti/binary-json-size-benchmark/blob/main/benchmark/epr/thrift/schema.thrift}{\small{\texttt{schema.thrift}}} & 251 & 192 & 220 & 186 \\ \hline
\hline \textbf{JSON} & - & 520 & 264 & 332 & 272 \\ \hline \hline
BSON & - & 564 & 304 & 378 & 299 \\ \hline
CBOR & - & 412 & 248 & 306 & 246 \\ \hline
FlexBuffers & - & 429 & 355 & 417 & 342 \\ \hline
MessagePack & - & 412 & 262 & 307 & 254 \\ \hline
Smile & - & 356 & 266 & 311 & 261 \\ \hline
UBJSON & - & 469 & 272 & 326 & 265 \\
\bottomrule
\end{tabularx}
\end{table*}

\begin{figure*}[ht!]
\frame{\includegraphics[width=\linewidth]{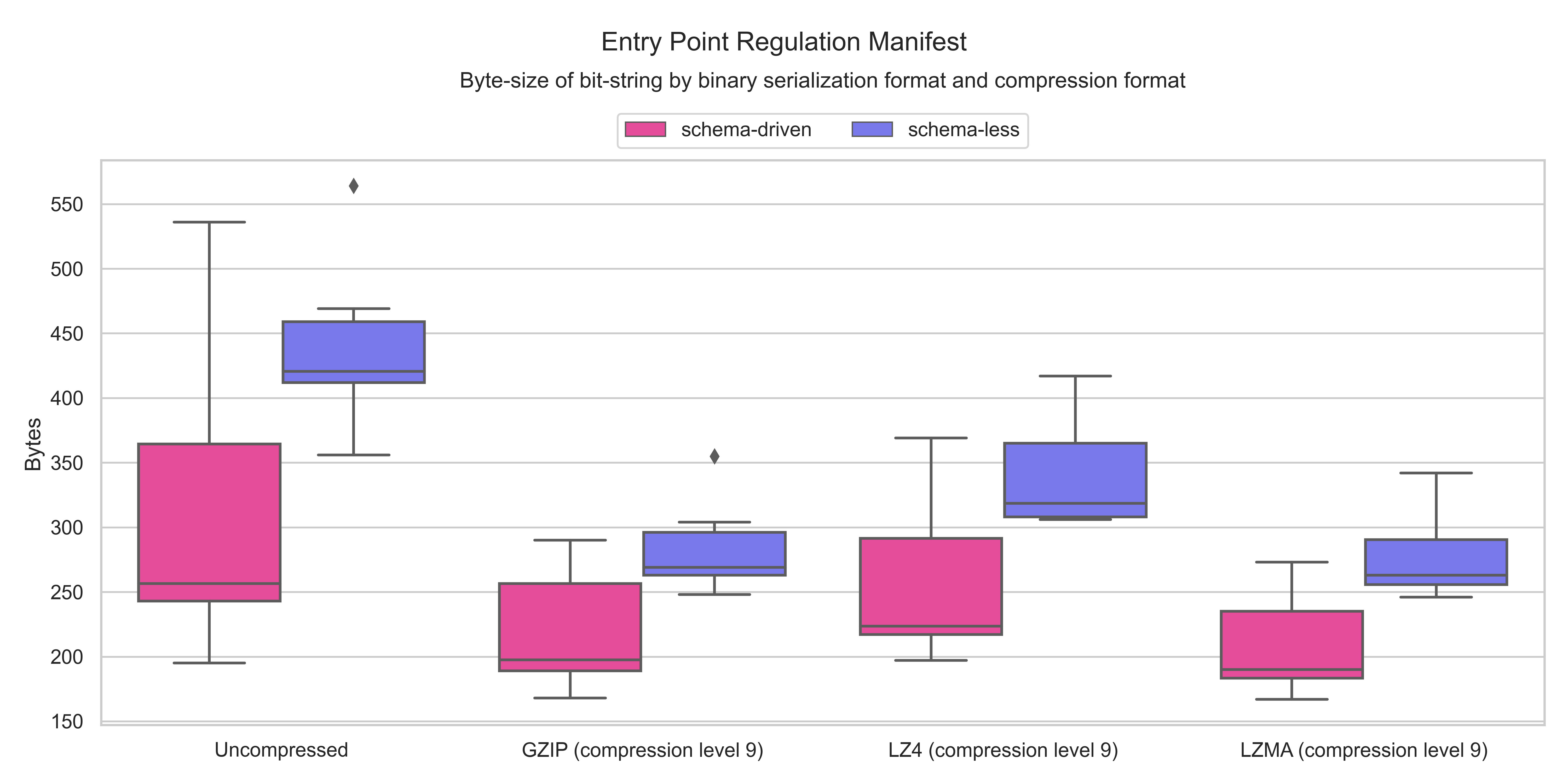}}
\caption{
Box plot of the statistical results in \autoref{table:benchmark-stats-epr}.
}
\label{fig:benchmark-epr-boxplot}
\end{figure*}

In \autoref{fig:benchmark-epr-boxplot}, we observe the medians for uncompressed
schema-driven binary serialization specifications to be smaller in comparison to
uncompressed schema-less binary serialization specifications.  The range between the
upper and lower whiskers and the inter-quartile range of uncompressed
schema-less binary serialization specifications is smaller than the range between the
upper and lower whiskers and the inter-quartile range of uncompressed
schema-driven binary serialization specifications.


In terms of compression, LZMA results in the lower median for both
schema-driven and schema-less binary serialization specifications.
Additionally, GZIP, LZ4 and LZMA are space-efficient in terms of the median in
comparison to both uncompressed schema-driven and schema-less binary
serialization specifications.  However, the use of GZIP for schema-less binary
serialization specifications exhibit upper outliers.  Nevertheless, compression
reduces the range between the upper and lower whiskers and inter-quartile range
for both schema-driven and schema-less binary serialization specifications.  In
particular, the compression format with the smaller range between the upper and
lower whiskers and the smaller inter-quartile range for schema-driven binary
serialization specifications is LZMA, the compression format with the smaller
range between the upper and lower whiskers for schema-less binary serialization
specifications is GZIP, and the compression formats with the smaller
inter-quartile range for schema-less binary serialization specifications are
GZIP and LZMA.


Overall, \we conclude that uncompressed schema-driven binary serialization
specifications are space-efficient in comparison to uncompressed schema-less
binary serialization specifications and that all the considered compression
formats are space-efficient in comparison to uncompressed schema-driven and
schema-less binary serialization specifications.

\clearpage

\subsection{JSON Feed Example Document}
\label{sec:benchmark-jsonfeed}

\begin{figure*}[ht!]
\frame{\includegraphics[width=\linewidth]{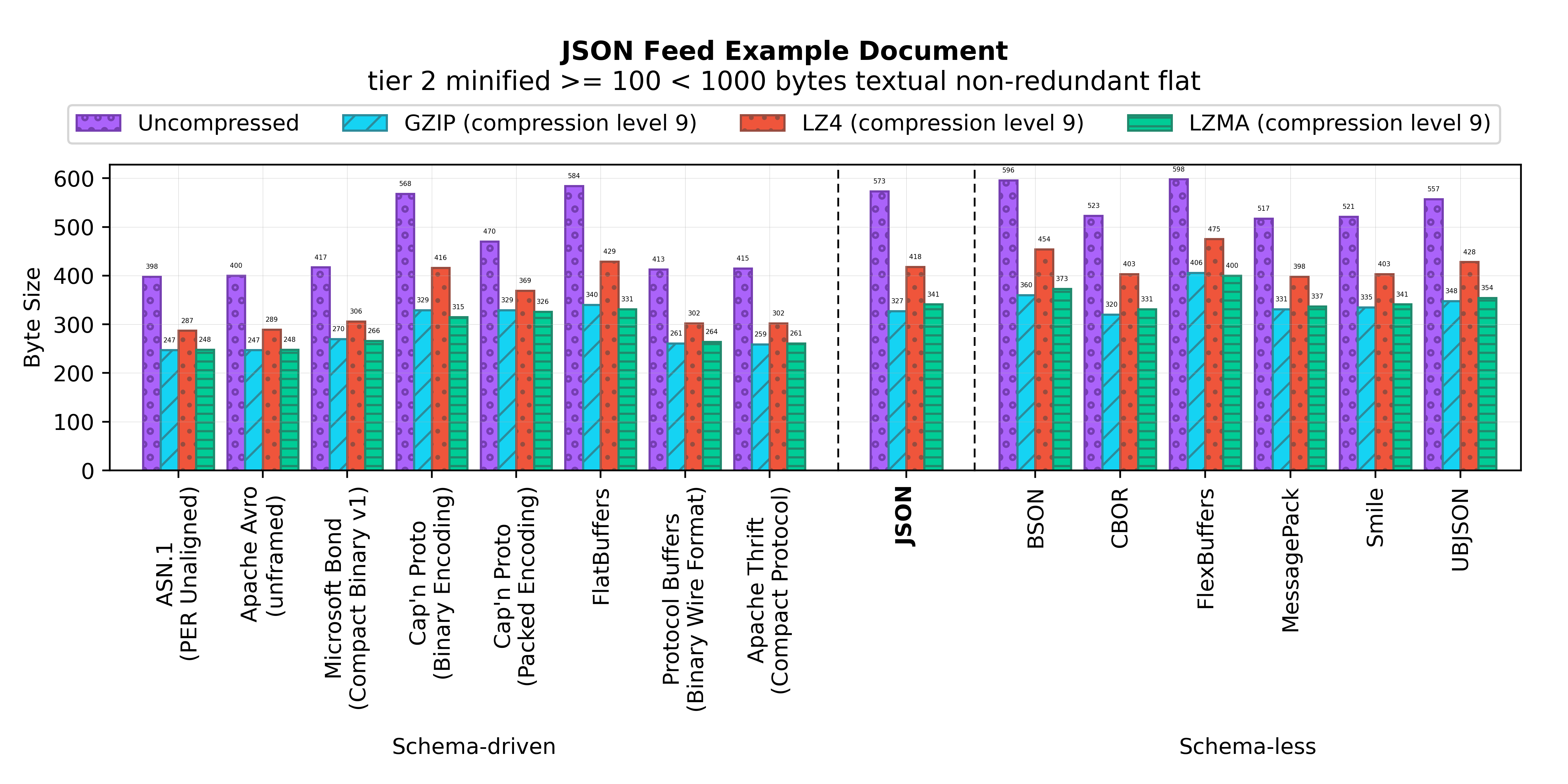}}
\caption{
The benchmark results for the JSON Feed Example Document test case listed in \autoref{table:benchmark-documents} and \autoref{table:benchmark-documents-1}.
}
\label{fig:benchmark-jsonfeed}
\end{figure*}

JSON Feed \cite{jsonfeed} is a specification for a syndication JSON format
similar to RSS \cite{rss} and Atom \cite{RFC4287} used in the publishing
\footnote{\url{https://micro.blog}} and media
\footnote{\url{https://npr.codes/npr-now-supports-json-feed-1c8af29d0ce7}}
industries. In \autoref{fig:benchmark-jsonfeed}, \we demonstrate a \textbf{Tier
2 minified $\geq$ 100 $<$ 1000 bytes textual non-redundant flat} (Tier 2 TNF
from \autoref{table:json-taxonomy}) JSON document that consists of a JSON Feed
manifest for an example website that contains a single blog entry.

The smallest bit-string is produced by ASN.1 PER Unaligned \cite{asn1-per} (398
bytes), closely followed by Apache Avro \cite{avro} (400 bytes) and Protocol
Buffers \cite{protocolbuffers} (413 bytes). The binary serialization specifications
that produced the smallest bit-strings are schema-driven and sequential
\cite{viotti2022survey}. Conversely, the largest bit-string is produced by FlexBuffers
\cite{flexbuffers} (598 bytes), closely followed by BSON \cite{bson} (596
bytes) and FlatBuffers \cite{flatbuffers} (584 bytes). With the exception of
BSON, the binary serialization specifications that produced the largest bit-strings
are pointer-based \cite{viotti2022survey}.  In comparison to JSON \cite{ECMA-404} (573 bytes),
binary serialization only achieves a \textbf{1.4x} size reduction in the best
case for this input document.  Additionally, 3 out of the 14 JSON-compatible
binary serialization specifications listed in
\autoref{table:benchmark-specifications-schema-driven} and
\autoref{table:benchmark-specifications-schema-less} result in bit-strings that are
larger than JSON: FlatBuffers \cite{flatbuffers}, BSON \cite{bson} and
FlexBuffers \cite{flexbuffers}. These binary serialization specifications are either
schema-less or schema-driven and pointer-based.

For this Tier 2 TNF document, the best performing schema-driven serialization
specification only achieves a \textbf{1.2x} size reduction compared to the best
performing schema-less serialization specification: MessagePack \cite{messagepack}
(517 bytes).  As shown in \autoref{table:benchmark-stats-jsonfeed},
uncompressed schema-driven specifications provide smaller \emph{average} and
\emph{median} bit-strings than uncompressed schema-less specifications. However, as
highlighted by the \emph{range} and \emph{standard deviation}, uncompressed
schema-driven specifications exhibit higher size reduction variability depending on
the expressiveness of the schema language (i.e. how the language constructs
allow you to model the data) and the size optimizations devised by its authors.
With the exception of the pointer-based binary serialization specifications Cap'n
Proto Binary Encoding \cite{capnproto} and FlatBuffers \cite{flatbuffers}, the
selection of schema-driven serialization specifications listed in
\autoref{table:benchmark-specifications-schema-driven} produce bit-strings that are
equal to or smaller than their schema-less counterparts listed in
\autoref{table:benchmark-specifications-schema-less}.  The best performing sequential
serialization specification only achieves a \textbf{1.1x} size reduction compared to
the best performing pointer-based serialization specification: Cap'n Proto Packed
Encoding \cite{capnproto} (470 bytes).

The compression formats listed in
\autoref{sec:benchmark-compression-formats} result in positive gains for
all bit-strings. The best performing compression format for JSON, GZIP
\cite{RFC1952} (327 bytes), achieves a \textbf{1.2x} size reduction compared to
the best performing uncompressed binary serialization specification.

\begin{table*}[hb!]
\caption{A byte-size statistical analysis of the benchmark results shown in \autoref{fig:benchmark-jsonfeed} divided by schema-driven and schema-less specifications.}
\label{table:benchmark-stats-jsonfeed}
\begin{tabularx}{\linewidth}{X|l|l|l|l|l|l|l|l}
\toprule
\multirow{2}{*}{\textbf{Category}} &
\multicolumn{4}{c|}{\textbf{Schema-driven}} &
\multicolumn{4}{c}{\textbf{Schema-less}} \\
\cline{2-9}
& \small\textbf{Average} & \small\textbf{Median} & \small\textbf{Range} & \small\textbf{Std.dev} & \small\textbf{Average} & \small\textbf{Median} & \small\textbf{Range} & \small\textbf{Std.dev} \\
\midrule
Uncompressed & \small{458.1} & \small{416} & \small{186} & \small{71.3} & \small{552} & \small{540} & \small{81} & \small{34.4} \\ \hline
GZIP (compression level 9) & \small{285.3} & \small{265.5} & \small{93} & \small{37.5} & \small{350} & \small{341.5} & \small{86} & \small{28.1} \\ \hline
LZ4 (compression level 9) & \small{337.5} & \small{304} & \small{142} & \small{54.7} & \small{426.8} & \small{415.5} & \small{77} & \small{28.9} \\ \hline
LZMA (compression level 9) & \small{282.4} & \small{265} & \small{83} & \small{33.1} & \small{356} & \small{347.5} & \small{69} & \small{23.9} \\
\bottomrule
\end{tabularx}
\end{table*}

\begin{table*}[hb!]
\caption{The benchmark raw data results and schemas for the plot in \autoref{fig:benchmark-jsonfeed}.}
\label{table:benchmark-jsonfeed}
\begin{tabularx}{\linewidth}{X|l|l|l|l|l}
\toprule
\textbf{Serialization Format} & \textbf{Schema} & \textbf{Uncompressed} & \textbf{GZIP} & \textbf{LZ4} & \textbf{LZMA} \\
\midrule
ASN.1 (PER Unaligned) & \href{https://github.com/jviotti/binary-json-size-benchmark/blob/main/benchmark/jsonfeed/asn1/schema.asn}{\small{\texttt{schema.asn}}} & 398 & 247 & 287 & 248 \\ \hline
Apache Avro (unframed) & \href{https://github.com/jviotti/binary-json-size-benchmark/blob/main/benchmark/jsonfeed/avro/schema.json}{\small{\texttt{schema.json}}} & 400 & 247 & 289 & 248 \\ \hline
Microsoft Bond (Compact Binary v1) & \href{https://github.com/jviotti/binary-json-size-benchmark/blob/main/benchmark/jsonfeed/bond/schema.bond}{\small{\texttt{schema.bond}}} & 417 & 270 & 306 & 266 \\ \hline
Cap'n Proto (Binary Encoding) & \href{https://github.com/jviotti/binary-json-size-benchmark/blob/main/benchmark/jsonfeed/capnproto/schema.capnp}{\small{\texttt{schema.capnp}}} & 568 & 329 & 416 & 315 \\ \hline
Cap'n Proto (Packed Encoding) & \href{https://github.com/jviotti/binary-json-size-benchmark/blob/main/benchmark/jsonfeed/capnproto-packed/schema.capnp}{\small{\texttt{schema.capnp}}} & 470 & 329 & 369 & 326 \\ \hline
FlatBuffers & \href{https://github.com/jviotti/binary-json-size-benchmark/blob/main/benchmark/jsonfeed/flatbuffers/schema.fbs}{\small{\texttt{schema.fbs}}} & 584 & 340 & 429 & 331 \\ \hline
Protocol Buffers (Binary Wire Format) & \href{https://github.com/jviotti/binary-json-size-benchmark/blob/main/benchmark/jsonfeed/protobuf/schema.proto}{\small{\texttt{schema.proto}}} & 413 & 261 & 302 & 264 \\ \hline
Apache Thrift (Compact Protocol) & \href{https://github.com/jviotti/binary-json-size-benchmark/blob/main/benchmark/jsonfeed/thrift/schema.thrift}{\small{\texttt{schema.thrift}}} & 415 & 259 & 302 & 261 \\ \hline
\hline \textbf{JSON} & - & 573 & 327 & 418 & 341 \\ \hline \hline
BSON & - & 596 & 360 & 454 & 373 \\ \hline
CBOR & - & 523 & 320 & 403 & 331 \\ \hline
FlexBuffers & - & 598 & 406 & 475 & 400 \\ \hline
MessagePack & - & 517 & 331 & 398 & 337 \\ \hline
Smile & - & 521 & 335 & 403 & 341 \\ \hline
UBJSON & - & 557 & 348 & 428 & 354 \\
\bottomrule
\end{tabularx}
\end{table*}

\begin{figure*}[ht!]
\frame{\includegraphics[width=\linewidth]{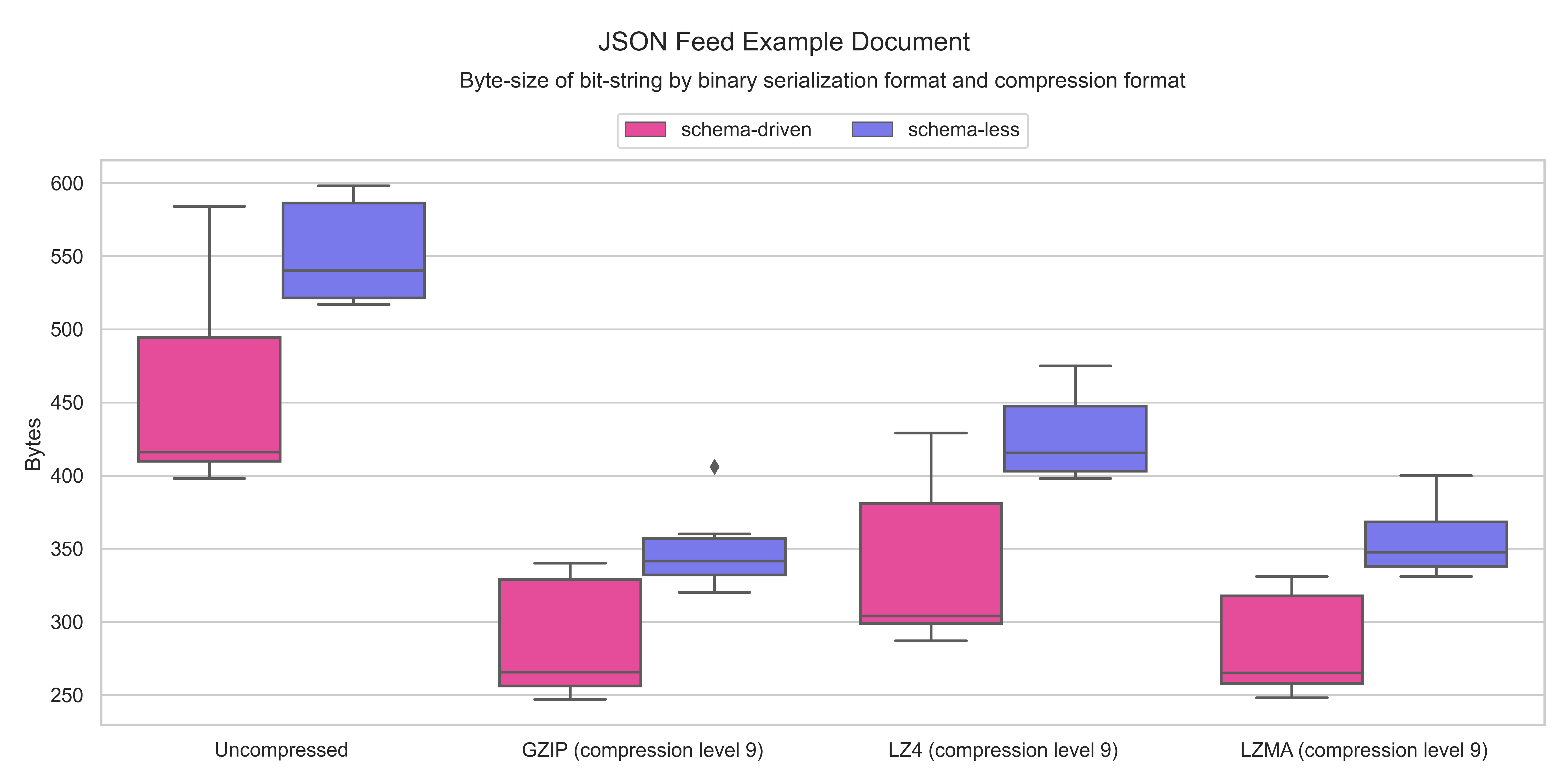}}
\caption{
Box plot of the statistical results in \autoref{table:benchmark-stats-jsonfeed}.
}
\label{fig:benchmark-jsonfeed-boxplot}
\end{figure*}

In \autoref{fig:benchmark-jsonfeed-boxplot}, we observe the medians for
uncompressed schema-driven binary serialization specifications to be smaller in
comparison to uncompressed schema-less binary serialization specifications.  The range
between the upper and lower whiskers and the inter-quartile range of
uncompressed schema-less binary serialization specifications is smaller than the range
between the upper and lower whiskers and the inter-quartile range of
uncompressed schema-driven binary serialization specifications.


In terms of compression, GZIP and LZMA result in the lower medians for
schema-driven binary serialization specifications while GZIP results in the lower
median for schema-less binary serialization specifications.  Additionally, GZIP, LZ4
and LZMA are space-efficient in terms of the median in comparison to both
uncompressed schema-driven and schema-less binary serialization specifications.
However, the use of GZIP for schema-less binary serialization specifications exhibits
upper outliers.  Nevertheless, compression reduces the range between the upper
and lower whiskers and inter-quartile range for both schema-driven and
schema-less binary serialization specifications.  In particular, the compression
format with the smaller range between the upper and lower whiskers and the
smaller inter-quartile range for schema-driven binary serialization specifications is
LZMA, and the compression format with the smaller range between the upper and
lower whiskers and the smaller inter-quartile range for schema-less binary
serialization specifications is GZIP.


Overall, \we conclude that uncompressed schema-driven binary serialization
specifications are space-efficient in comparison to uncompressed schema-less
binary serialization specifications and that all the considered compression
formats are space-efficient in comparison to uncompressed schema-driven and
schema-less binary serialization specifications.

\clearpage

\subsection{GitHub Workflow Definition}
\label{sec:benchmark-githubworkflow}

\begin{figure*}[ht!]
\frame{\includegraphics[width=\linewidth]{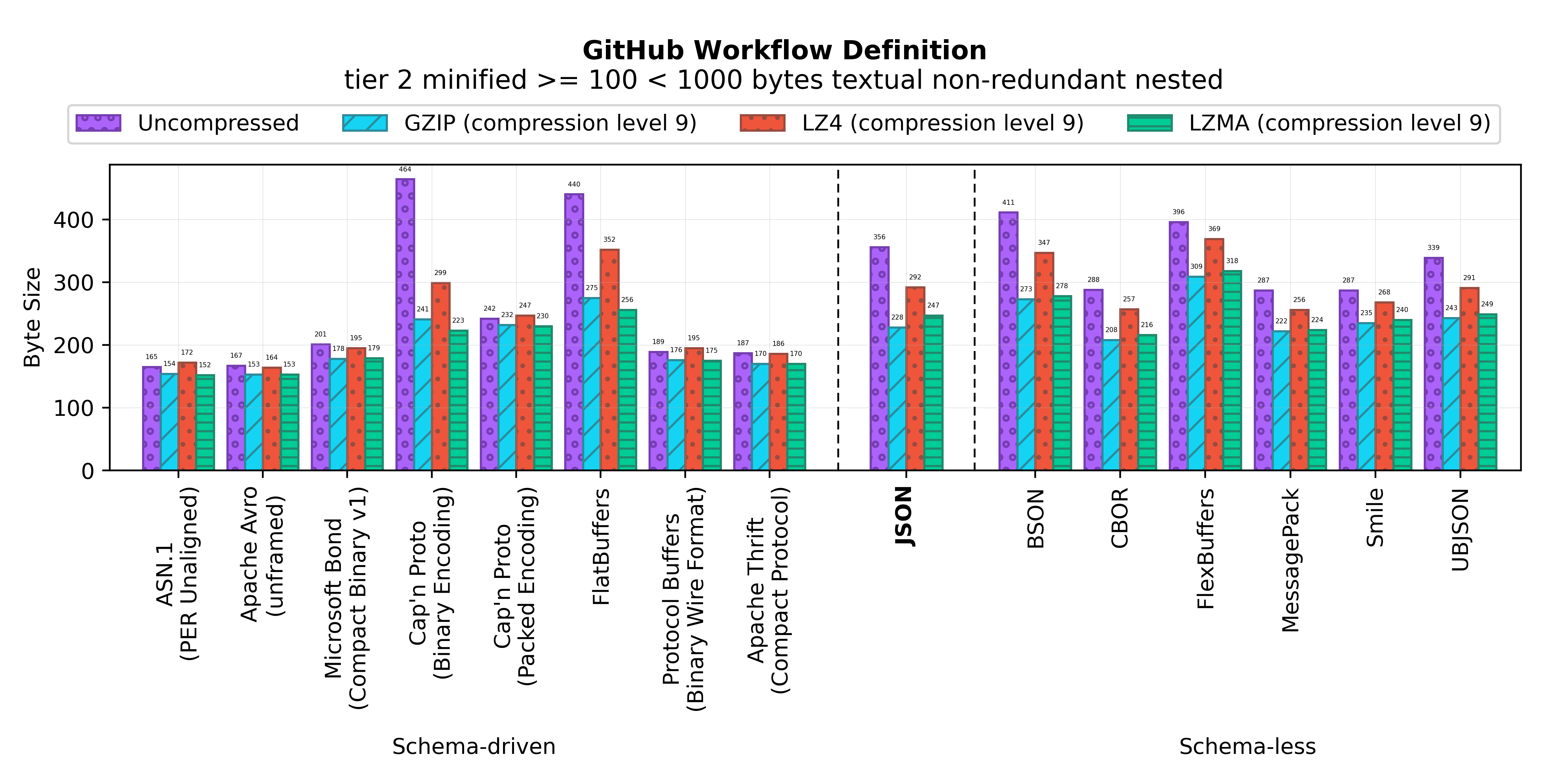}}
\caption{
The benchmark results for the GitHub Workflow Definition test case listed in \autoref{table:benchmark-documents} and \autoref{table:benchmark-documents-1}.
}
\label{fig:benchmark-githubworkflow}
\end{figure*}

The GitHub \footnote{\url{https://github.com}} software hosting provider has an
automation service called GitHub Actions
\footnote{\url{https://github.com/features/actions}} for projects to define
custom workflows. GitHub Actions is used primarily by the open-source software
industry. In \autoref{fig:benchmark-githubworkflow}, \we demonstrate a
\textbf{Tier 2 minified $\geq$ 100 $<$ 1000 bytes textual non-redundant nested}
(Tier 2 TNN from \autoref{table:json-taxonomy}) JSON document that consists of
a simple example workflow definition.

The smallest bit-string is produced by ASN.1 PER Unaligned \cite{asn1-per} (165
bytes), closely followed by Apache Avro \cite{avro} (167 bytes) and Apache
Thrift \cite{slee2007thrift} (187 bytes).  The binary serialization specifications
that produced the smallest bit-strings are schema-driven and sequential
\cite{viotti2022survey}. Conversely, the largest bit-string is produced by Cap'n Proto Binary
Encoding \cite{capnproto} (464 bytes), followed by FlatBuffers
\cite{flatbuffers} (440 bytes) and BSON \cite{bson} (411 bytes).  With the
exception of BSON, the binary serialization specifications that produced the largest
bit-strings are schema-driven and pointer-based \cite{viotti2022survey}.  In comparison to
JSON \cite{ECMA-404} (356 bytes), binary serialization achieves a \textbf{2.1x}
size reduction in the best case for this input document.  However, 4 out of the
14 JSON-compatible binary serialization specifications listed in
\autoref{table:benchmark-specifications-schema-driven} and
\autoref{table:benchmark-specifications-schema-less} result in bit-strings that are
larger than JSON: Cap'n Proto Binary Encoding \cite{capnproto}, FlatBuffers
\cite{flatbuffers}, BSON \cite{bson} and FlexBuffers \cite{flexbuffers}. These
binary serialization specifications are either schema-less or schema-driven and
pointer-based.

For this Tier 2 TNN document, the best performing schema-driven serialization
specification only achieves a \textbf{1.7x} size reduction compared to the best
performing schema-less serialization specification: MessagePack \cite{messagepack} and
Smile \cite{smile} (287 bytes).  As shown in
\autoref{table:benchmark-stats-githubworkflow}, uncompressed schema-driven
specifications provide smaller \emph{average} and \emph{median} bit-strings than
uncompressed schema-less specifications. However, as highlighted by the \emph{range}
and \emph{standard deviation}, uncompressed schema-driven specifications exhibit
higher size reduction variability depending on the expressiveness of the schema
language (i.e. how the language constructs allow you to model the data) and the
size optimizations devised by its authors. With the exception of the
pointer-based binary serialization specifications Cap'n Proto Binary Encoding
\cite{capnproto} and FlatBuffers \cite{flatbuffers}, the selection of
schema-driven serialization specifications listed in
\autoref{table:benchmark-specifications-schema-driven} produce bit-strings that are
equal to or smaller than their schema-less counterparts listed in
\autoref{table:benchmark-specifications-schema-less}.  The best performing sequential
serialization specification only achieves a \textbf{1.4x} size reduction compared to
the best performing pointer-based serialization specification: Cap'n Proto Packed
Encoding \cite{capnproto} (242 bytes).

The compression formats listed in
\autoref{sec:benchmark-compression-formats} result in positive gains for
all bit-strings. The best performing uncompressed binary serialization
specification achieves a \textbf{1.3x} size reduction compared to the best
performing compression format for JSON: GZIP \cite{RFC1952} (228 bytes).

\begin{table*}[hb!]
\caption{A byte-size statistical analysis of the benchmark results shown in \autoref{fig:benchmark-githubworkflow} divided by schema-driven and schema-less specifications.}
\label{table:benchmark-stats-githubworkflow}
\begin{tabularx}{\linewidth}{X|l|l|l|l|l|l|l|l}
\toprule
\multirow{2}{*}{\textbf{Category}} &
\multicolumn{4}{c|}{\textbf{Schema-driven}} &
\multicolumn{4}{c}{\textbf{Schema-less}} \\
\cline{2-9}
& \small\textbf{Average} & \small\textbf{Median} & \small\textbf{Range} & \small\textbf{Std.dev} & \small\textbf{Average} & \small\textbf{Median} & \small\textbf{Range} & \small\textbf{Std.dev} \\
\midrule
Uncompressed & \small{256.9} & \small{195} & \small{299} & \small{115.0} & \small{334.7} & \small{313.5} & \small{124} & \small{52.2} \\ \hline
GZIP (compression level 9) & \small{197.4} & \small{177} & \small{122} & \small{42.7} & \small{248.3} & \small{239} & \small{101} & \small{33.7} \\ \hline
LZ4 (compression level 9) & \small{226.3} & \small{195} & \small{188} & \small{63.2} & \small{298} & \small{279.5} & \small{113} & \small{44.4} \\ \hline
LZMA (compression level 9) & \small{192.3} & \small{177} & \small{104} & \small{36.3} & \small{254.2} & \small{244.5} & \small{102} & \small{34.7} \\
\bottomrule
\end{tabularx}
\end{table*}

\begin{table*}[hb!]
\caption{The benchmark raw data results and schemas for the plot in \autoref{fig:benchmark-githubworkflow}.}
\label{table:benchmark-githubworkflow}
\begin{tabularx}{\linewidth}{X|l|l|l|l|l}
\toprule
\textbf{Serialization Format} & \textbf{Schema} & \textbf{Uncompressed} & \textbf{GZIP} & \textbf{LZ4} & \textbf{LZMA} \\
\midrule
ASN.1 (PER Unaligned) & \href{https://github.com/jviotti/binary-json-size-benchmark/blob/main/benchmark/githubworkflow/asn1/schema.asn}{\small{\texttt{schema.asn}}} & 165 & 154 & 172 & 152 \\ \hline
Apache Avro (unframed) & \href{https://github.com/jviotti/binary-json-size-benchmark/blob/main/benchmark/githubworkflow/avro/schema.json}{\small{\texttt{schema.json}}} & 167 & 153 & 164 & 153 \\ \hline
Microsoft Bond (Compact Binary v1) & \href{https://github.com/jviotti/binary-json-size-benchmark/blob/main/benchmark/githubworkflow/bond/schema.bond}{\small{\texttt{schema.bond}}} & 201 & 178 & 195 & 179 \\ \hline
Cap'n Proto (Binary Encoding) & \href{https://github.com/jviotti/binary-json-size-benchmark/blob/main/benchmark/githubworkflow/capnproto/schema.capnp}{\small{\texttt{schema.capnp}}} & 464 & 241 & 299 & 223 \\ \hline
Cap'n Proto (Packed Encoding) & \href{https://github.com/jviotti/binary-json-size-benchmark/blob/main/benchmark/githubworkflow/capnproto-packed/schema.capnp}{\small{\texttt{schema.capnp}}} & 242 & 232 & 247 & 230 \\ \hline
FlatBuffers & \href{https://github.com/jviotti/binary-json-size-benchmark/blob/main/benchmark/githubworkflow/flatbuffers/schema.fbs}{\small{\texttt{schema.fbs}}} & 440 & 275 & 352 & 256 \\ \hline
Protocol Buffers (Binary Wire Format) & \href{https://github.com/jviotti/binary-json-size-benchmark/blob/main/benchmark/githubworkflow/protobuf/schema.proto}{\small{\texttt{schema.proto}}} & 189 & 176 & 195 & 175 \\ \hline
Apache Thrift (Compact Protocol) & \href{https://github.com/jviotti/binary-json-size-benchmark/blob/main/benchmark/githubworkflow/thrift/schema.thrift}{\small{\texttt{schema.thrift}}} & 187 & 170 & 186 & 170 \\ \hline
\hline \textbf{JSON} & - & 356 & 228 & 292 & 247 \\ \hline \hline
BSON & - & 411 & 273 & 347 & 278 \\ \hline
CBOR & - & 288 & 208 & 257 & 216 \\ \hline
FlexBuffers & - & 396 & 309 & 369 & 318 \\ \hline
MessagePack & - & 287 & 222 & 256 & 224 \\ \hline
Smile & - & 287 & 235 & 268 & 240 \\ \hline
UBJSON & - & 339 & 243 & 291 & 249 \\
\bottomrule
\end{tabularx}
\end{table*}

\begin{figure*}[ht!]
\frame{\includegraphics[width=\linewidth]{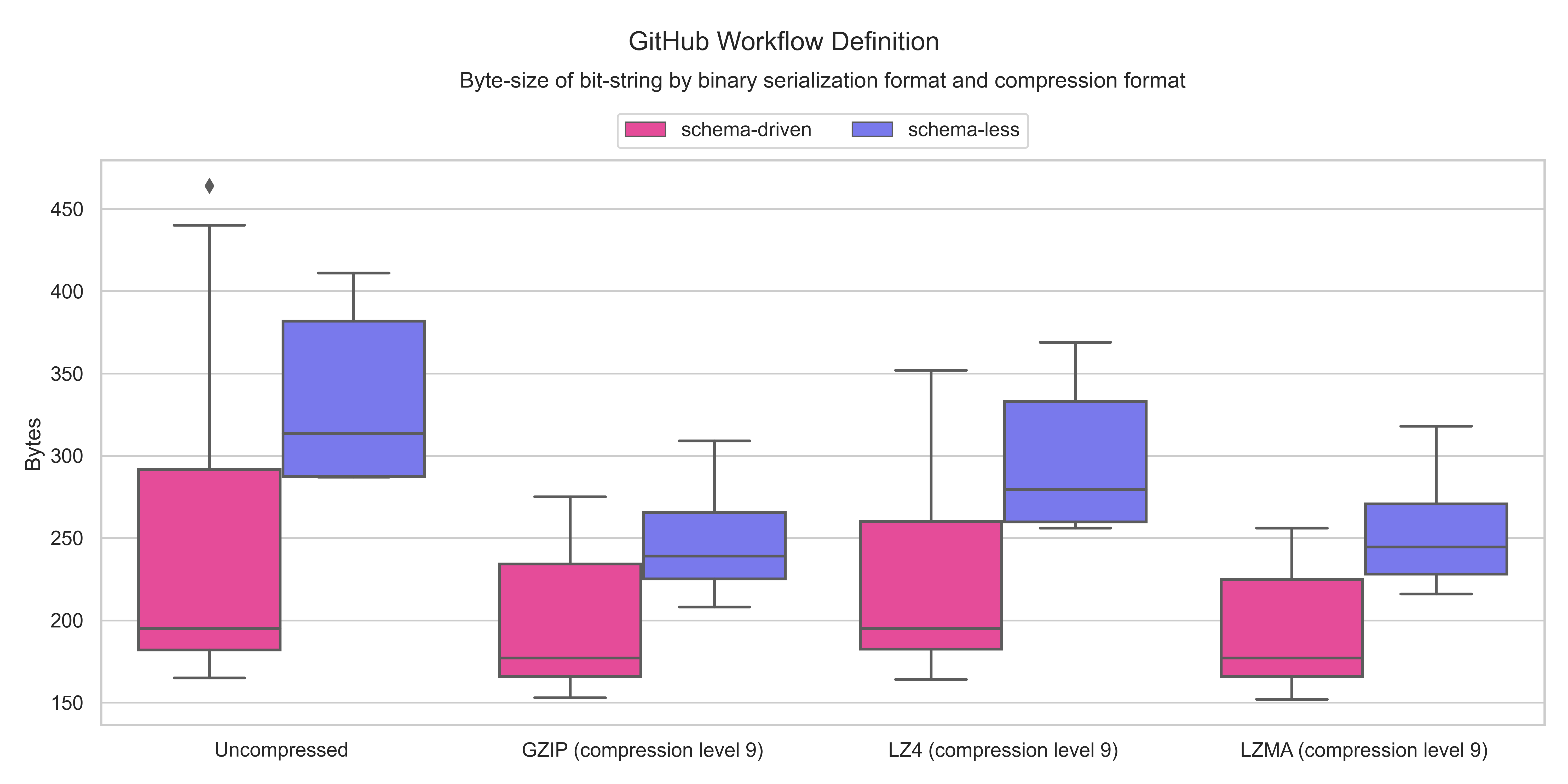}}
\caption{
Box plot of the statistical results in \autoref{table:benchmark-stats-githubworkflow}.
}
\label{fig:benchmark-githubworkflow-boxplot}
\end{figure*}

In \autoref{fig:benchmark-githubworkflow-boxplot}, we observe the medians for
uncompressed schema-driven binary serialization specifications to be smaller in
comparison to uncompressed schema-less binary serialization specifications.  The range
between the upper and lower whiskers and the inter-quartile range of
uncompressed schema-less binary serialization specifications is smaller than the range
between the upper and lower whiskers and the inter-quartile range of
uncompressed schema-driven binary serialization specifications. However, their
respective quartiles overlap.


In terms of compression, GZIP and LZMA result in the lower medians for
schema-driven binary serialization specifications while GZIP results in the
lower median for schema-less binary serialization specifications.
Additionally, GZIP and LZMA are space-efficient in terms of the median in
comparison to uncompressed schema-driven binary serialization specifications
and GZIP, LZ4 and LZMA are space-efficient in terms of the median in comparison
to uncompressed schema-less binary serialization specifications.  Also,
compression reduces the range between the upper and lower whiskers and
inter-quartile range for both schema-driven and schema-less binary
serialization specifications.  In particular, the compression format with the
smaller range between the upper and lower whiskers and the smaller
inter-quartile range for schema-driven binary serialization specifications is
LZMA, the compression formats with the smaller range between the upper and
lower whiskers for schema-less binary serialization specifications are GZIP and
LZMA, and the compression format with the smaller inter-quartile range for
schema-less binary serialization specifications is GZIP.


Overall, \we conclude that uncompressed schema-driven binary serialization
specifications are space-efficient in comparison to uncompressed schema-less
binary serialization specifications. GZIP and LZMA are space-efficient in
comparison to uncompressed schema-driven binary serialization specifications
and all the considered compression formats are space-efficient in comparison to
uncompressed schema-less binary serialization specifications.

\clearpage

\subsection{GitHub FUNDING Sponsorship Definition (Empty)}
\label{sec:benchmark-githubfundingblank}

\begin{figure*}[ht!]
\frame{\includegraphics[width=\linewidth]{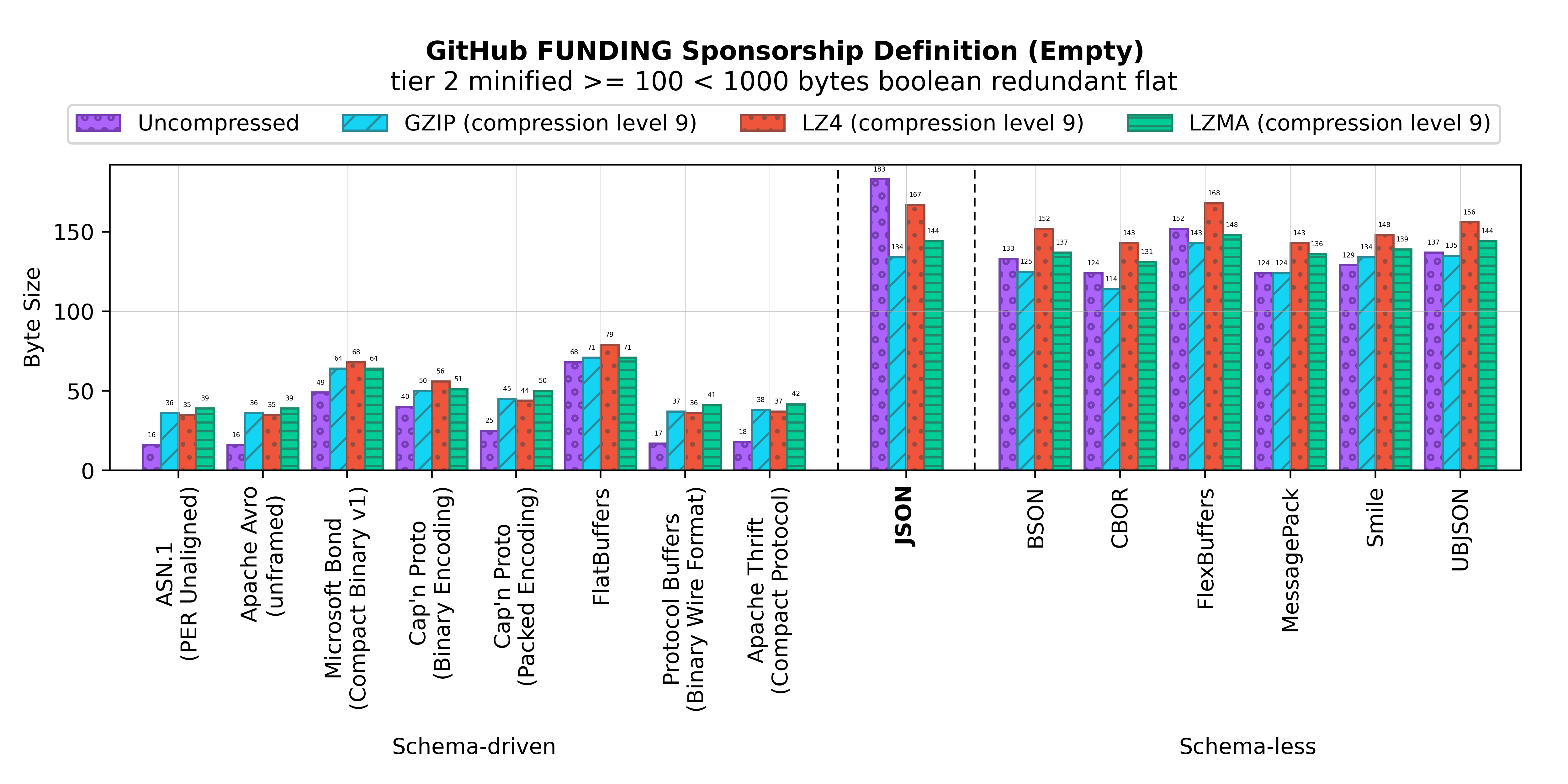}}
\caption{
The benchmark results for the GitHub FUNDING Sponsorship Definition (Empty) test case listed in \autoref{table:benchmark-documents} and \autoref{table:benchmark-documents-1}.
}
\label{fig:benchmark-githubfundingblank}
\end{figure*}

The GitHub \footnote{\url{https://github.com}} software hosting provider
defines a \texttt{FUNDING}
\footnote{\url{https://docs.github.com/en/github/administering-a-repository/managing-repository-settings/displaying-a-sponsor-button-in-your-repository}}
file format to declare the funding platforms that an open-source project
supports. The \texttt{FUNDING} file format is used by the open-source software
industry. In \autoref{fig:benchmark-githubfundingblank}, \we demonstrate a
\textbf{Tier 2 minified $\geq$ 100 $<$ 1000 bytes boolean redundant flat} (Tier
2 BRF from \autoref{table:json-taxonomy}) JSON document that consists of a
definition that does not declare any supported funding platforms.

The smallest bit-string is produced by both ASN.1 PER Unaligned \cite{asn1-per}
and Apache Avro \cite{avro} (16 bytes), closely followed by Protocol Buffers
\cite{protocolbuffers} (17 bytes) and Apache Thrift \cite{slee2007thrift} (18
bytes). The binary serialization specifications that produced the smallest bit-strings
are schema-driven and sequential \cite{viotti2022survey}.  Conversely, the largest bit-string
is produced by FlexBuffers \cite{flexbuffers} (152 bytes), followed by UBJSON
\cite{ubjson} (137 bytes) and BSON \cite{bson} (133 bytes). The binary
serialization specifications that produced the largest bit-strings are schema-less and
with the exception of FlexBuffers, they are also sequential \cite{viotti2022survey}.  In
comparison to JSON \cite{ECMA-404} (183 bytes), binary serialization achieves a
\textbf{11.4x} size reduction in the best case for this input document.
Similar large size reductions are observed in JSON documents whose content is
dominated by \emph{boolean} and \emph{numeric} values.  None of the 14
JSON-compatible binary serialization specifications listed in
\autoref{table:benchmark-specifications-schema-driven} and
\autoref{table:benchmark-specifications-schema-less} result in bit-strings that are
larger than JSON.

For this Tier 2 BRF document, the best performing schema-driven serialization
specification achieves a \textbf{7.7x} size reduction compared to the best performing
schema-less serialization specification: CBOR \cite{RFC7049} and MessagePack
\cite{messagepack} (124 bytes).  As shown in
\autoref{table:benchmark-stats-githubfundingblank}, uncompressed schema-driven
specifications provide smaller \emph{average} and \emph{median} bit-strings than
uncompressed schema-less specifications. However, as highlighted by the \emph{range}
and \emph{standard deviation}, uncompressed schema-driven specifications exhibit
higher size reduction variability depending on the expressiveness of the schema
language (i.e. how the language constructs allow you to model the data) and the
size optimizations devised by its authors.  The entire selection of
schema-driven serialization specifications listed in
\autoref{table:benchmark-specifications-schema-driven} produce bit-strings that are
equal to or smaller than their schema-less counterparts listed in
\autoref{table:benchmark-specifications-schema-less}.  The best performing sequential
serialization specification only achieves a \textbf{1.5x} size reduction compared to
the best performing pointer-based serialization specification: Cap'n Proto Packed
Encoding \cite{capnproto} (25 bytes).

The compression formats listed in
\autoref{sec:benchmark-compression-formats} result in positive gains for
the bit-strings produced by JSON \cite{ECMA-404}, BSON \cite{bson}, CBOR
\cite{RFC7049}, FlexBuffers \cite{flexbuffers} and UBJSON \cite{ubjson}. The
best performing uncompressed binary serialization specification achieves a
\textbf{8.3x} size reduction compared to the best performing compression format
for JSON: GZIP \cite{RFC1952} (134 bytes).

\begin{table*}[hb!]
\caption{A byte-size statistical analysis of the benchmark results shown in \autoref{fig:benchmark-githubfundingblank} divided by schema-driven and schema-less specifications.}
\label{table:benchmark-stats-githubfundingblank}
\begin{tabularx}{\linewidth}{X|l|l|l|l|l|l|l|l}
\toprule
\multirow{2}{*}{\textbf{Category}} &
\multicolumn{4}{c|}{\textbf{Schema-driven}} &
\multicolumn{4}{c}{\textbf{Schema-less}} \\
\cline{2-9}
& \small\textbf{Average} & \small\textbf{Median} & \small\textbf{Range} & \small\textbf{Std.dev} & \small\textbf{Average} & \small\textbf{Median} & \small\textbf{Range} & \small\textbf{Std.dev} \\
\midrule
Uncompressed & \small{31.1} & \small{21.5} & \small{52} & \small{18.1} & \small{133.2} & \small{131} & \small{28} & \small{9.6} \\ \hline
GZIP (compression level 9) & \small{47.1} & \small{41.5} & \small{35} & \small{12.8} & \small{129.2} & \small{129.5} & \small{29} & \small{9.3} \\ \hline
LZ4 (compression level 9) & \small{48.8} & \small{40.5} & \small{44} & \small{16.0} & \small{151.7} & \small{150} & \small{25} & \small{8.7} \\ \hline
LZMA (compression level 9) & \small{49.6} & \small{46} & \small{32} & \small{11.3} & \small{139.2} & \small{138} & \small{17} & \small{5.5} \\
\bottomrule
\end{tabularx}
\end{table*}

\begin{table*}[hb!]
\caption{The benchmark raw data results and schemas for the plot in \autoref{fig:benchmark-githubfundingblank}.}
\label{table:benchmark-githubfundingblank}
\begin{tabularx}{\linewidth}{X|l|l|l|l|l}
\toprule
\textbf{Serialization Format} & \textbf{Schema} & \textbf{Uncompressed} & \textbf{GZIP} & \textbf{LZ4} & \textbf{LZMA} \\
\midrule
ASN.1 (PER Unaligned) & \href{https://github.com/jviotti/binary-json-size-benchmark/blob/main/benchmark/githubfundingblank/asn1/schema.asn}{\small{\texttt{schema.asn}}} & 16 & 36 & 35 & 39 \\ \hline
Apache Avro (unframed) & \href{https://github.com/jviotti/binary-json-size-benchmark/blob/main/benchmark/githubfundingblank/avro/schema.json}{\small{\texttt{schema.json}}} & 16 & 36 & 35 & 39 \\ \hline
Microsoft Bond (Compact Binary v1) & \href{https://github.com/jviotti/binary-json-size-benchmark/blob/main/benchmark/githubfundingblank/bond/schema.bond}{\small{\texttt{schema.bond}}} & 49 & 64 & 68 & 64 \\ \hline
Cap'n Proto (Binary Encoding) & \href{https://github.com/jviotti/binary-json-size-benchmark/blob/main/benchmark/githubfundingblank/capnproto/schema.capnp}{\small{\texttt{schema.capnp}}} & 40 & 50 & 56 & 51 \\ \hline
Cap'n Proto (Packed Encoding) & \href{https://github.com/jviotti/binary-json-size-benchmark/blob/main/benchmark/githubfundingblank/capnproto-packed/schema.capnp}{\small{\texttt{schema.capnp}}} & 25 & 45 & 44 & 50 \\ \hline
FlatBuffers & \href{https://github.com/jviotti/binary-json-size-benchmark/blob/main/benchmark/githubfundingblank/flatbuffers/schema.fbs}{\small{\texttt{schema.fbs}}} & 68 & 71 & 79 & 71 \\ \hline
Protocol Buffers (Binary Wire Format) & \href{https://github.com/jviotti/binary-json-size-benchmark/blob/main/benchmark/githubfundingblank/protobuf/schema.proto}{\small{\texttt{schema.proto}}} & 17 & 37 & 36 & 41 \\ \hline
Apache Thrift (Compact Protocol) & \href{https://github.com/jviotti/binary-json-size-benchmark/blob/main/benchmark/githubfundingblank/thrift/schema.thrift}{\small{\texttt{schema.thrift}}} & 18 & 38 & 37 & 42 \\ \hline
\hline \textbf{JSON} & - & 183 & 134 & 167 & 144 \\ \hline \hline
BSON & - & 133 & 125 & 152 & 137 \\ \hline
CBOR & - & 124 & 114 & 143 & 131 \\ \hline
FlexBuffers & - & 152 & 143 & 168 & 148 \\ \hline
MessagePack & - & 124 & 124 & 143 & 136 \\ \hline
Smile & - & 129 & 134 & 148 & 139 \\ \hline
UBJSON & - & 137 & 135 & 156 & 144 \\
\bottomrule
\end{tabularx}
\end{table*}

\begin{figure*}[ht!]
\frame{\includegraphics[width=\linewidth]{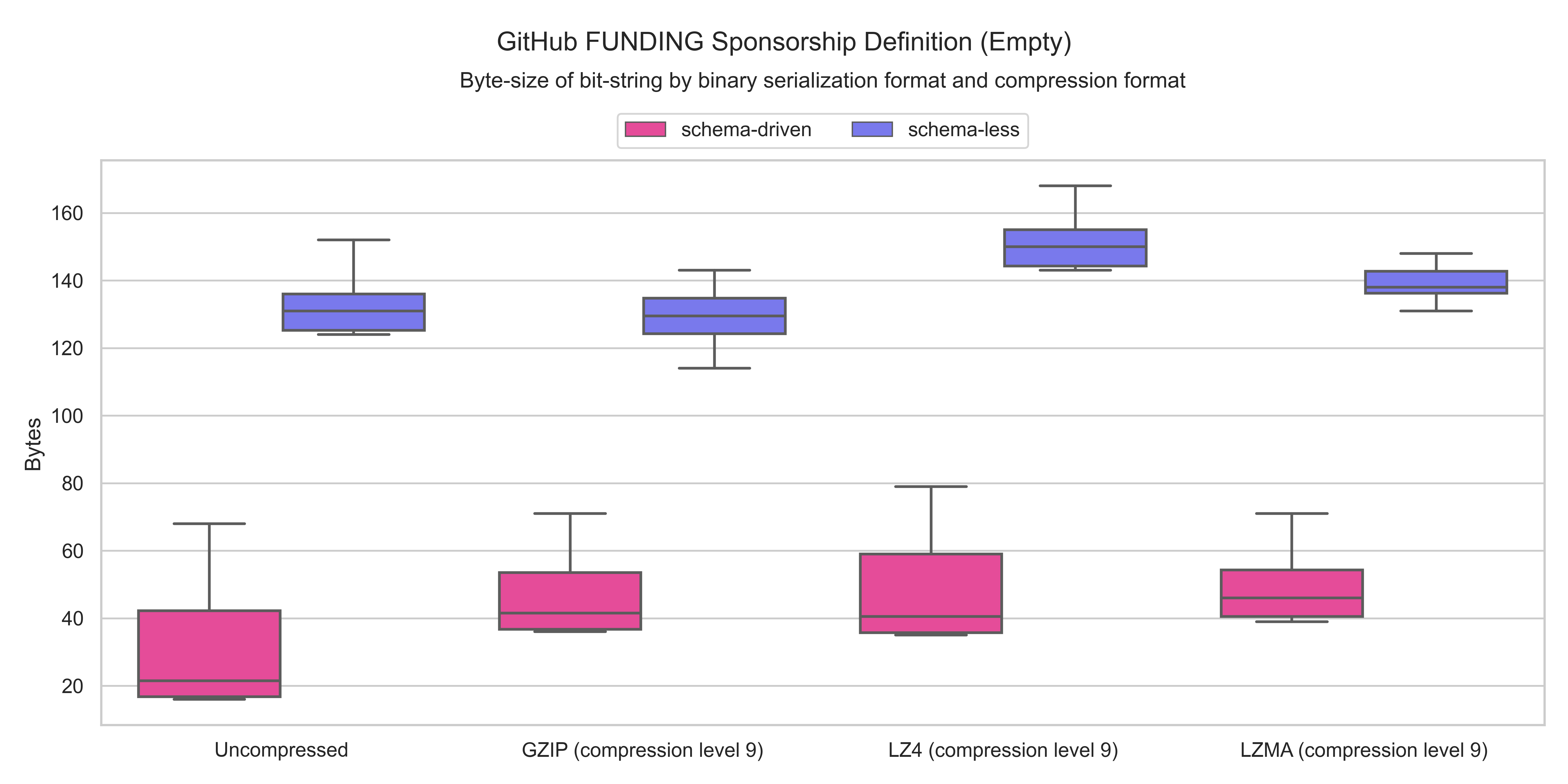}}
\caption{
Box plot of the statistical results in \autoref{table:benchmark-stats-githubfundingblank}.
}
\label{fig:benchmark-githubfundingblank-boxplot}
\end{figure*}

In \autoref{fig:benchmark-githubfundingblank-boxplot}, we observe the medians
for uncompressed schema-driven binary serialization specifications to be smaller in
comparison to uncompressed schema-less binary serialization specifications.  The range
between the upper and lower whiskers and the inter-quartile range of
uncompressed schema-less binary serialization specifications is smaller than the range
between the upper and lower whiskers and the inter-quartile range of
uncompressed schema-driven binary serialization specifications.


In terms of compression, LZ4 results in the lower median for schema-driven
binary serialization specifications while GZIP results in the lower median for
schema-less binary serialization specifications.  Compression is not
space-efficient in terms of the median in comparison to uncompressed
schema-driven binary serialization specifications. However, GZIP is
space-efficient in terms of the median in comparison to uncompressed
schema-less binary serialization specifications.  Nevertheless, compression
reduces the range between the upper and lower whiskers and inter-quartile range
for both schema-driven and schema-less binary serialization specifications.  In
particular, the compression format with the smaller range between the upper and
lower whiskers and the smaller inter-quartile range for both schema-driven and
schema-less binary serialization specifications is LZMA.


Overall, \we conclude that uncompressed schema-driven binary serialization
specifications are space-efficient in comparison to uncompressed schema-less binary
serialization specifications. Compression does not contribute to space-efficiency in
comparison to schema-driven binary serialization specifications but GZIP is
space-efficient in comparison to schema-less binary serialization specifications.

\clearpage

\subsection{ECMAScript Module Loader Definition}
\label{sec:benchmark-esmrc}

\begin{figure*}[ht!]
\frame{\includegraphics[width=\linewidth]{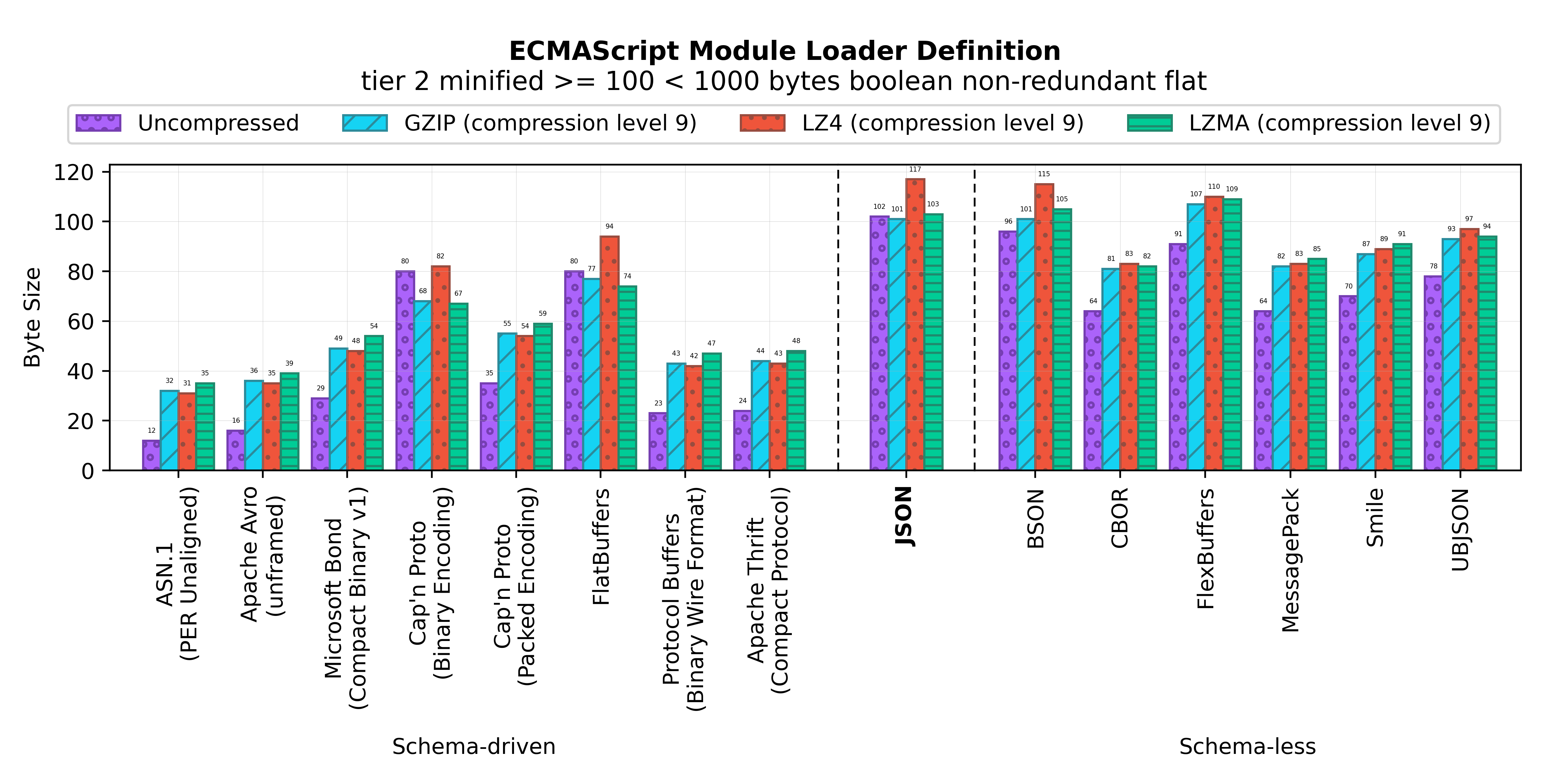}}
\caption{
The benchmark results for the ECMAScript Module Loader Definition test case listed in \autoref{table:benchmark-documents} and \autoref{table:benchmark-documents-1}.
}
\label{fig:benchmark-esmrc}
\end{figure*}

\texttt{esm} \footnote{\url{https://github.com/standard-things/esm}} is an
open-source ECMAScript \cite{ECMA-262} module loader for the Node.js
\footnote{\url{https://nodejs.org}} JavaScript runtime that allows developers
to use the modern \texttt{import} module syntax on older runtime versions.
\texttt{esm} is used in the web industry. In \autoref{fig:benchmark-esmrc}, \we
demonstrate a \textbf{Tier 2 minified $\geq$ 100 $<$ 1000 bytes boolean
non-redundant flat} (Tier 2 BNF from \autoref{table:json-taxonomy}) JSON
document that defines an example \texttt{esm} configuration.

The smallest bit-string is produced by ASN.1 PER Unaligned \cite{asn1-per} (12
bytes), followed by Apache Avro \cite{avro} (16 bytes) and Protocol Buffers
\cite{protocolbuffers} (23 bytes). The binary serialization specifications that
produced the smallest bit-strings are schema-driven and sequential \cite{viotti2022survey}.
Conversely, the largest bit-string is produced by BSON \cite{bson} (96 bytes),
followed by FlexBuffers \cite{flexbuffers} (91 bytes) and FlatBuffers
\cite{flatbuffers} (80 bytes). With the exception of BSON, the binary
serialization specifications that produced the largest bit-strings are pointer-based
\cite{viotti2022survey}.  In comparison to JSON \cite{ECMA-404} (102 bytes), binary
serialization achieves a \textbf{8.5x} size reduction in the best case for this
input document.  None of the 14 JSON-compatible binary serialization specifications
listed in \autoref{table:benchmark-specifications-schema-driven} and
\autoref{table:benchmark-specifications-schema-less} result in bit-strings that are
larger than JSON.

For this Tier 2 BNF document, the best performing schema-driven serialization
specification achieves a \textbf{5.3x} size reduction compared to the best performing
schema-less serialization specification: CBOR \cite{RFC7049} and MessagePack
\cite{messagepack} (64 bytes).  As shown in
\autoref{table:benchmark-stats-esmrc}, uncompressed schema-driven specifications
provide smaller \emph{average} and \emph{median} bit-strings than uncompressed
schema-less specifications. However, as highlighted by the \emph{range} and
\emph{standard deviation}, uncompressed schema-driven specifications exhibit higher
size reduction variability depending on the expressiveness of the schema
language (i.e. how the language constructs allow you to model the data) and the
size optimizations devised by its authors. With the exception of the
pointer-based binary serialization specifications Cap'n Proto Binary Encoding
\cite{capnproto} and FlatBuffers \cite{flatbuffers}, the selection of
schema-driven serialization specifications listed in
\autoref{table:benchmark-specifications-schema-driven} produce bit-strings that are
equal to or smaller than their schema-less counterparts listed in
\autoref{table:benchmark-specifications-schema-less}.  The best performing sequential
serialization specification achieves a \textbf{2.9x} size reduction compared to the
best performing pointer-based serialization specification: Cap'n Proto Packed Encoding
\cite{capnproto} (35 bytes).

The compression formats listed in
\autoref{sec:benchmark-compression-formats} result in positive gains for
the bit-strings produced by Cap'n Proto Binary Encoding \cite{capnproto},
FlatBuffers \cite{flatbuffers} and JSON \cite{ECMA-404}. The best performing
uncompressed binary serialization specification achieves a \textbf{8.4x} size
reduction compared to the best performing compression format for JSON: GZIP
\cite{RFC1952} (101 bytes).

\begin{table*}[hb!]
\caption{A byte-size statistical analysis of the benchmark results shown in \autoref{fig:benchmark-esmrc} divided by schema-driven and schema-less specifications.}
\label{table:benchmark-stats-esmrc}
\begin{tabularx}{\linewidth}{X|l|l|l|l|l|l|l|l}
\toprule
\multirow{2}{*}{\textbf{Category}} &
\multicolumn{4}{c|}{\textbf{Schema-driven}} &
\multicolumn{4}{c}{\textbf{Schema-less}} \\
\cline{2-9}
& \small\textbf{Average} & \small\textbf{Median} & \small\textbf{Range} & \small\textbf{Std.dev} & \small\textbf{Average} & \small\textbf{Median} & \small\textbf{Range} & \small\textbf{Std.dev} \\
\midrule
Uncompressed & \small{37.4} & \small{26.5} & \small{68} & \small{25.5} & \small{77.2} & \small{74} & \small{32} & \small{12.5} \\ \hline
GZIP (compression level 9) & \small{50.5} & \small{46.5} & \small{45} & \small{14.5} & \small{91.8} & \small{90} & \small{26} & \small{9.6} \\ \hline
LZ4 (compression level 9) & \small{53.6} & \small{45.5} & \small{63} & \small{21.1} & \small{96.2} & \small{93} & \small{32} & \small{12.5} \\ \hline
LZMA (compression level 9) & \small{52.9} & \small{51} & \small{39} & \small{12.5} & \small{94.3} & \small{92.5} & \small{27} & \small{9.8} \\
\bottomrule
\end{tabularx}
\end{table*}

\begin{table*}[hb!]
\caption{The benchmark raw data results and schemas for the plot in \autoref{fig:benchmark-esmrc}.}
\label{table:benchmark-esmrc}
\begin{tabularx}{\linewidth}{X|l|l|l|l|l}
\toprule
\textbf{Serialization Format} & \textbf{Schema} & \textbf{Uncompressed} & \textbf{GZIP} & \textbf{LZ4} & \textbf{LZMA} \\
\midrule
ASN.1 (PER Unaligned) & \href{https://github.com/jviotti/binary-json-size-benchmark/blob/main/benchmark/esmrc/asn1/schema.asn}{\small{\texttt{schema.asn}}} & 12 & 32 & 31 & 35 \\ \hline
Apache Avro (unframed) & \href{https://github.com/jviotti/binary-json-size-benchmark/blob/main/benchmark/esmrc/avro/schema.json}{\small{\texttt{schema.json}}} & 16 & 36 & 35 & 39 \\ \hline
Microsoft Bond (Compact Binary v1) & \href{https://github.com/jviotti/binary-json-size-benchmark/blob/main/benchmark/esmrc/bond/schema.bond}{\small{\texttt{schema.bond}}} & 29 & 49 & 48 & 54 \\ \hline
Cap'n Proto (Binary Encoding) & \href{https://github.com/jviotti/binary-json-size-benchmark/blob/main/benchmark/esmrc/capnproto/schema.capnp}{\small{\texttt{schema.capnp}}} & 80 & 68 & 82 & 67 \\ \hline
Cap'n Proto (Packed Encoding) & \href{https://github.com/jviotti/binary-json-size-benchmark/blob/main/benchmark/esmrc/capnproto-packed/schema.capnp}{\small{\texttt{schema.capnp}}} & 35 & 55 & 54 & 59 \\ \hline
FlatBuffers & \href{https://github.com/jviotti/binary-json-size-benchmark/blob/main/benchmark/esmrc/flatbuffers/schema.fbs}{\small{\texttt{schema.fbs}}} & 80 & 77 & 94 & 74 \\ \hline
Protocol Buffers (Binary Wire Format) & \href{https://github.com/jviotti/binary-json-size-benchmark/blob/main/benchmark/esmrc/protobuf/schema.proto}{\small{\texttt{schema.proto}}} & 23 & 43 & 42 & 47 \\ \hline
Apache Thrift (Compact Protocol) & \href{https://github.com/jviotti/binary-json-size-benchmark/blob/main/benchmark/esmrc/thrift/schema.thrift}{\small{\texttt{schema.thrift}}} & 24 & 44 & 43 & 48 \\ \hline
\hline \textbf{JSON} & - & 102 & 101 & 117 & 103 \\ \hline \hline
BSON & - & 96 & 101 & 115 & 105 \\ \hline
CBOR & - & 64 & 81 & 83 & 82 \\ \hline
FlexBuffers & - & 91 & 107 & 110 & 109 \\ \hline
MessagePack & - & 64 & 82 & 83 & 85 \\ \hline
Smile & - & 70 & 87 & 89 & 91 \\ \hline
UBJSON & - & 78 & 93 & 97 & 94 \\
\bottomrule
\end{tabularx}
\end{table*}

\begin{figure*}[ht!]
\frame{\includegraphics[width=\linewidth]{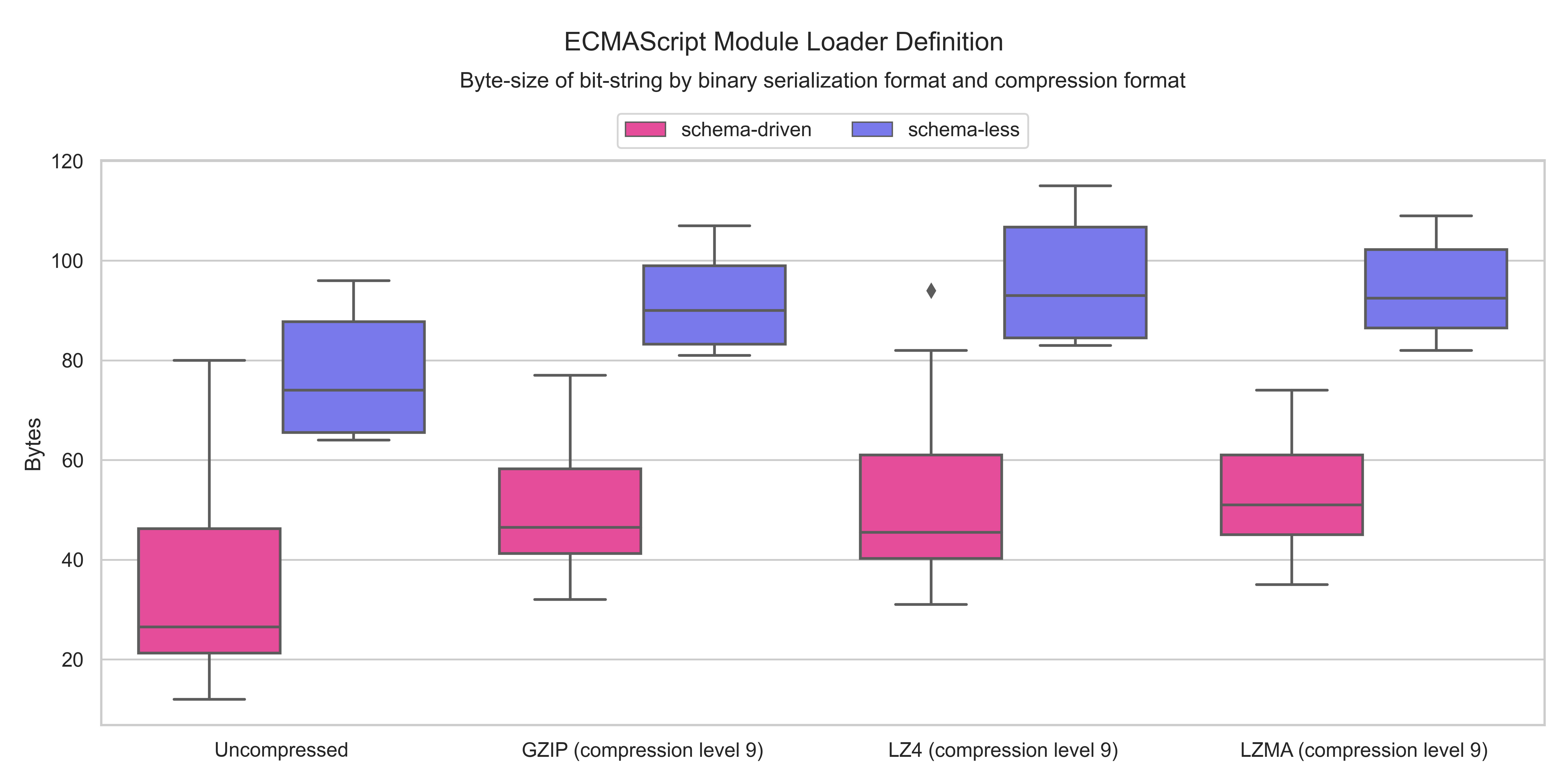}}
\caption{
Box plot of the statistical results in \autoref{table:benchmark-stats-esmrc}.
}
\label{fig:benchmark-esmrc-boxplot}
\end{figure*}

In \autoref{fig:benchmark-esmrc-boxplot}, we observe the medians for
uncompressed schema-driven binary serialization specifications to be smaller in
comparison to uncompressed schema-less binary serialization specifications.  The range
between the upper and lower whiskers and the inter-quartile range of
uncompressed schema-less binary serialization specifications is smaller than the range
between the upper and lower whiskers and the inter-quartile range of
uncompressed schema-driven binary serialization specifications.


In terms of compression, LZ4 results in the lower median for schema-driven
binary serialization specifications while GZIP results in the lower median for
schema-less binary serialization specifications.  However, compression is not
space-efficient in terms of the median for both schema-driven and schema-less
binary serialization specifications.  Additionally, the use of LZ4 for
schema-driven binary serialization specifications exhibits upper outliers.
While compression does not contribute to space-efficiency, it reduces the range
between the upper and lower whiskers and inter-quartile range for both
schema-driven and schema-less binary serialization specifications.  In
particular, the compression format with the smaller range between the upper and
lower whiskers and the smaller inter-quartile range for schema-driven binary
serialization specifications is LZMA, the compression format with the smaller
range between the upper and lower whiskers for schema-less binary serialization
specifications is GZIP, and the compression formats with the smaller
inter-quartile range for schema-less binary serialization specifications are
GZIP and LZMA.


Overall, \we conclude that uncompressed schema-driven binary serialization
specifications are space-efficient in comparison to uncompressed schema-less binary
serialization specifications and that compression does not contribute to
space-efficiency in comparison to both uncompressed schema-driven and
schema-less binary serialization specifications.

\clearpage

\subsection{ESLint Configuration Document}
\label{sec:benchmark-eslintrc}

\begin{figure*}[ht!]
\frame{\includegraphics[width=\linewidth]{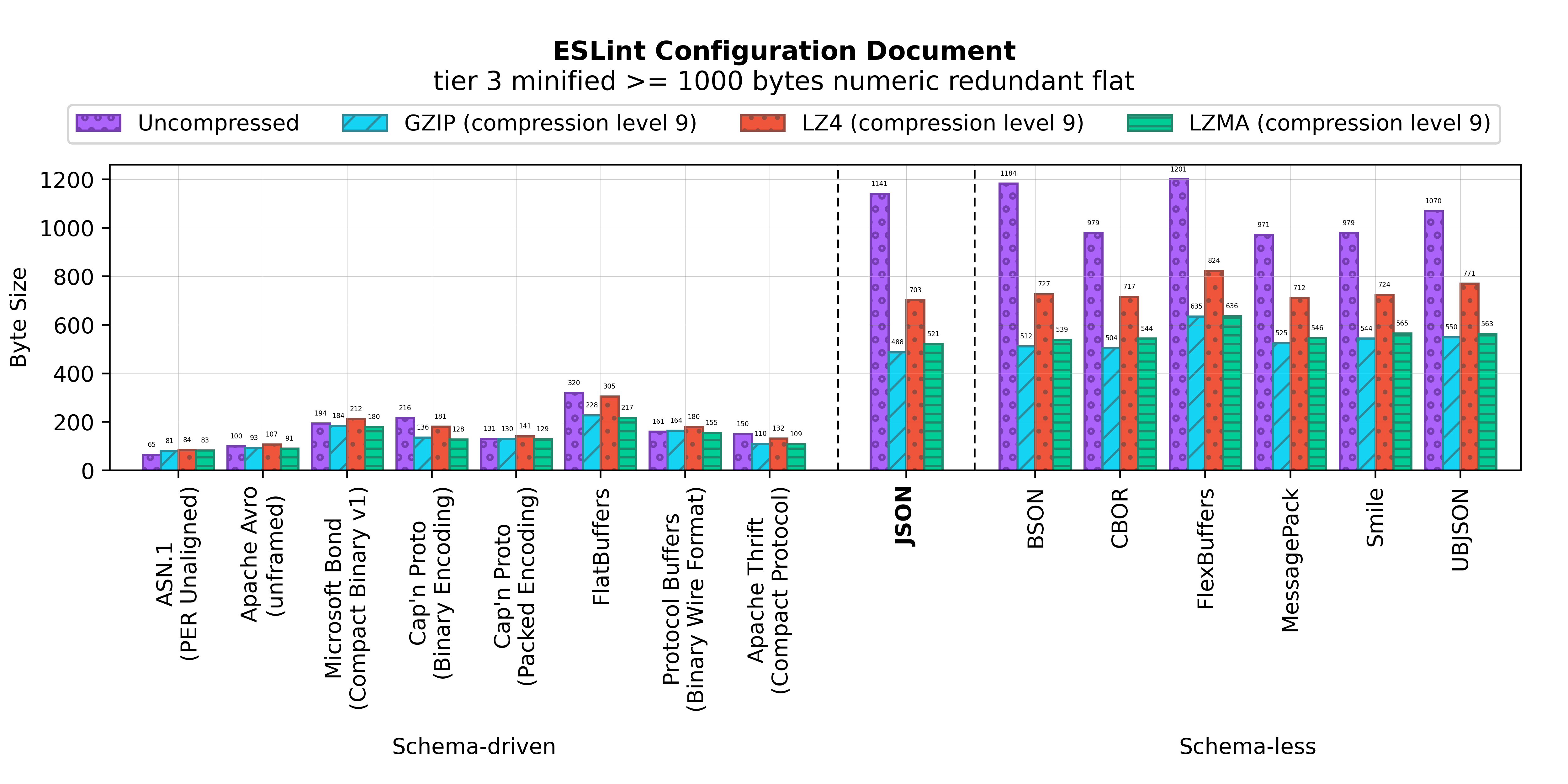}}
\caption{
The benchmark results for the ESLint Configuration Document test case listed in \autoref{table:benchmark-documents} and \autoref{table:benchmark-documents-1}.
}
\label{fig:benchmark-eslintrc}
\end{figure*}

ESLint \footnote{\url{https://eslint.org}} is a popular open-source extensible
linter for the JavaScript \cite{ECMA-262} programming language used by a wide
range of companies in the software development industry such as Google,
Salesforce, and Airbnb. In \autoref{fig:benchmark-eslintrc}, \we demonstrate a
\textbf{Tier 3 minified $\geq$ 1000 bytes numeric redundant flat} (Tier 3 NRF
from \autoref{table:json-taxonomy}) JSON document that defines a browser and
Node.js linter configuration that defines general-purposes and
\emph{React.js}-specific \footnote{\url{https://reactjs.org}} linting rules.

The smallest bit-string is produced by ASN.1 PER Unaligned \cite{asn1-per} (65
bytes) followed by Apache Avro \cite{avro} (100 bytes) and Cap'n Proto Packed
Encoding \cite{capnproto} (131 bytes). These serialization specifications are
schema-driven and with the exception of Cap'n Proto Packed Encoding, which
occupies the third place, they are also sequential \cite{viotti2022survey}.  Conversely, the
largest bit-string is produced by FlexBuffers \cite{flexbuffers} (1201 bytes)
followed by BSON \cite{bson} (1184 bytes) and UBJSON \cite{ubjson} (1070
bytes). The binary serialization specifications that produced the largest bit-strings
are schema-less and with the exception of FlexBuffers, they are also sequential
\cite{viotti2022survey}.  In comparison to JSON \cite{ECMA-404} (1141 bytes), binary
serialization achieves a \textbf{17.5x} size reduction in the best case for
this input document.  Similar large size reductions are observed in JSON
documents whose content is dominated by \emph{boolean} and \emph{numeric}
values.  However, 2 out of the 14 JSON-compatible binary serialization specifications
listed in \autoref{table:benchmark-specifications-schema-driven} and
\autoref{table:benchmark-specifications-schema-less} result in bit-strings that are
larger than JSON: BSON \cite{bson} and FlexBuffers \cite{flexbuffers}. These
binary serialization specifications are schema-less.

For this Tier 3 NRF document, the best performing schema-driven serialization
specification achieves a \textbf{14.9x} size reduction compared to the best performing
schema-less serialization specification: MessagePack \cite{messagepack} (971 bytes).
As shown in \autoref{table:benchmark-stats-eslintrc}, uncompressed
schema-driven specifications provide smaller \emph{average} and \emph{median}
bit-strings than uncompressed schema-less specifications. The entire selection of
schema-driven serialization specifications listed in
\autoref{table:benchmark-specifications-schema-driven} produce bit-strings that are
equal to or smaller than their schema-less counterparts listed in
\autoref{table:benchmark-specifications-schema-less}.  The best performing sequential
serialization specification achieves a \textbf{2x} size reduction compared to the best
performing pointer-based serialization specification: Cap'n Proto Packed Encoding
\cite{capnproto} (131 bytes).

The compression formats listed in
\autoref{sec:benchmark-compression-formats} result in positive gains for
all the bit-strings except the one produced by ASN.1 PER Unaligned
\cite{asn1-per}. The best performing uncompressed binary serialization
specification achieves a \textbf{7.5x} size reduction compared to the best
performing compression format for JSON: GZIP \cite{RFC1952} (488 bytes).

\begin{table*}[hb!]
\caption{A byte-size statistical analysis of the benchmark results shown in \autoref{fig:benchmark-eslintrc} divided by schema-driven and schema-less specifications.}
\label{table:benchmark-stats-eslintrc}
\begin{tabularx}{\linewidth}{X|l|l|l|l|l|l|l|l}
\toprule
\multirow{2}{*}{\textbf{Category}} &
\multicolumn{4}{c|}{\textbf{Schema-driven}} &
\multicolumn{4}{c}{\textbf{Schema-less}} \\
\cline{2-9}
& \small\textbf{Average} & \small\textbf{Median} & \small\textbf{Range} & \small\textbf{Std.dev} & \small\textbf{Average} & \small\textbf{Median} & \small\textbf{Range} & \small\textbf{Std.dev} \\
\midrule
Uncompressed & \small{167.1} & \small{155.5} & \small{255} & \small{73.4} & \small{1064} & \small{1024.5} & \small{230} & \small{96.9} \\ \hline
GZIP (compression level 9) & \small{140.8} & \small{133} & \small{147} & \small{46.0} & \small{545} & \small{534.5} & \small{131} & \small{43.4} \\ \hline
LZ4 (compression level 9) & \small{167.8} & \small{160.5} & \small{221} & \small{65.0} & \small{745.8} & \small{725.5} & \small{112} & \small{39.9} \\ \hline
LZMA (compression level 9) & \small{136.5} & \small{128.5} & \small{134} & \small{42.6} & \small{565.5} & \small{554.5} & \small{97} & \small{33.0} \\
\bottomrule
\end{tabularx}
\end{table*}

\begin{table*}[hb!]
\caption{The benchmark raw data results and schemas for the plot in \autoref{fig:benchmark-eslintrc}.}
\label{table:benchmark-eslintrc}
\begin{tabularx}{\linewidth}{X|l|l|l|l|l}
\toprule
\textbf{Serialization Format} & \textbf{Schema} & \textbf{Uncompressed} & \textbf{GZIP} & \textbf{LZ4} & \textbf{LZMA} \\
\midrule
ASN.1 (PER Unaligned) & \href{https://github.com/jviotti/binary-json-size-benchmark/blob/main/benchmark/eslintrc/asn1/schema.asn}{\small{\texttt{schema.asn}}} & 65 & 81 & 84 & 83 \\ \hline
Apache Avro (unframed) & \href{https://github.com/jviotti/binary-json-size-benchmark/blob/main/benchmark/eslintrc/avro/schema.json}{\small{\texttt{schema.json}}} & 100 & 93 & 107 & 91 \\ \hline
Microsoft Bond (Compact Binary v1) & \href{https://github.com/jviotti/binary-json-size-benchmark/blob/main/benchmark/eslintrc/bond/schema.bond}{\small{\texttt{schema.bond}}} & 194 & 184 & 212 & 180 \\ \hline
Cap'n Proto (Binary Encoding) & \href{https://github.com/jviotti/binary-json-size-benchmark/blob/main/benchmark/eslintrc/capnproto/schema.capnp}{\small{\texttt{schema.capnp}}} & 216 & 136 & 181 & 128 \\ \hline
Cap'n Proto (Packed Encoding) & \href{https://github.com/jviotti/binary-json-size-benchmark/blob/main/benchmark/eslintrc/capnproto-packed/schema.capnp}{\small{\texttt{schema.capnp}}} & 131 & 130 & 141 & 129 \\ \hline
FlatBuffers & \href{https://github.com/jviotti/binary-json-size-benchmark/blob/main/benchmark/eslintrc/flatbuffers/schema.fbs}{\small{\texttt{schema.fbs}}} & 320 & 228 & 305 & 217 \\ \hline
Protocol Buffers (Binary Wire Format) & \href{https://github.com/jviotti/binary-json-size-benchmark/blob/main/benchmark/eslintrc/protobuf/schema.proto}{\small{\texttt{schema.proto}}} & 161 & 164 & 180 & 155 \\ \hline
Apache Thrift (Compact Protocol) & \href{https://github.com/jviotti/binary-json-size-benchmark/blob/main/benchmark/eslintrc/thrift/schema.thrift}{\small{\texttt{schema.thrift}}} & 150 & 110 & 132 & 109 \\ \hline
\hline \textbf{JSON} & - & 1141 & 488 & 703 & 521 \\ \hline \hline
BSON & - & 1184 & 512 & 727 & 539 \\ \hline
CBOR & - & 979 & 504 & 717 & 544 \\ \hline
FlexBuffers & - & 1201 & 635 & 824 & 636 \\ \hline
MessagePack & - & 971 & 525 & 712 & 546 \\ \hline
Smile & - & 979 & 544 & 724 & 565 \\ \hline
UBJSON & - & 1070 & 550 & 771 & 563 \\
\bottomrule
\end{tabularx}
\end{table*}

\begin{figure*}[ht!]
\frame{\includegraphics[width=\linewidth]{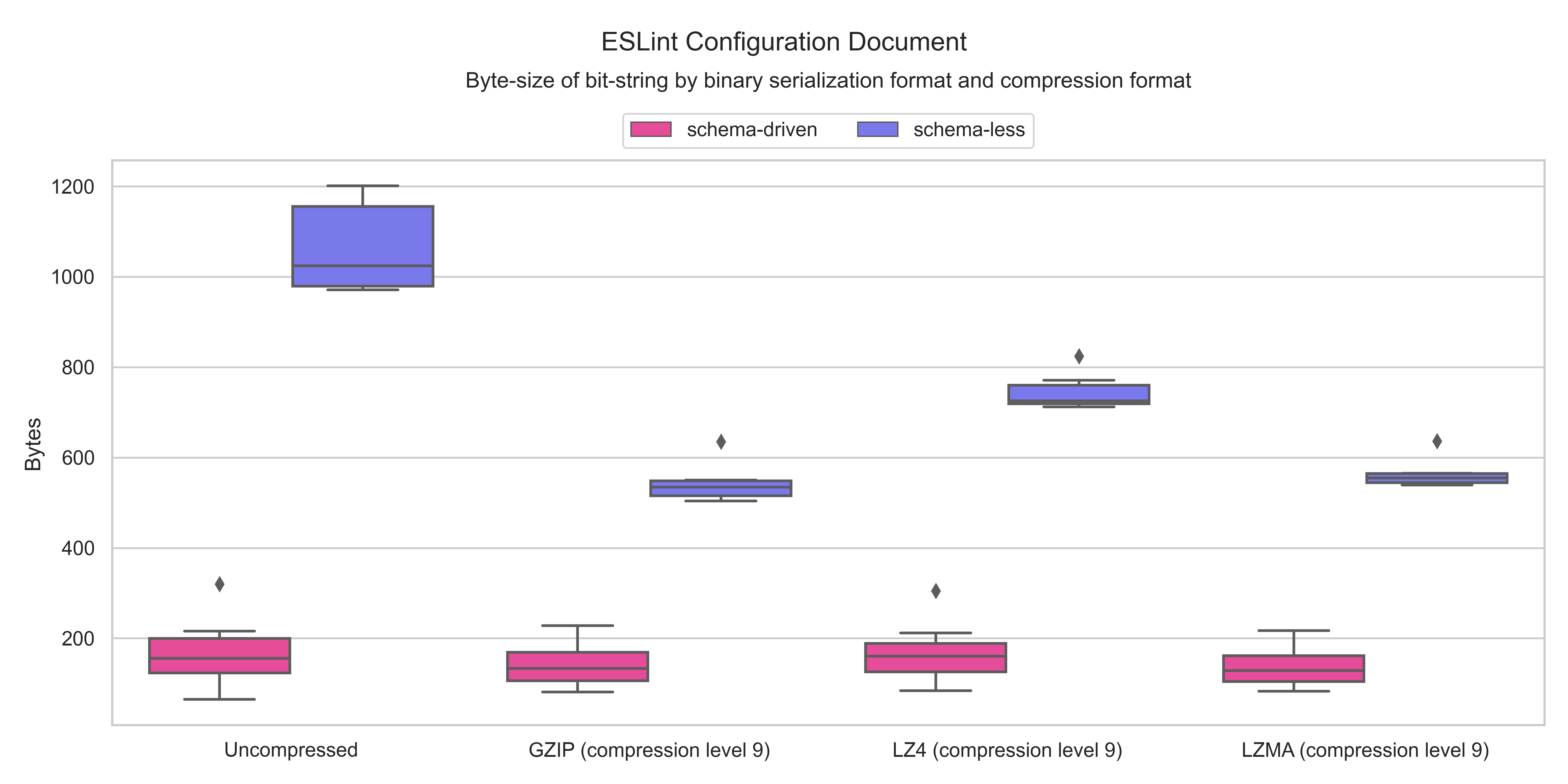}}
\caption{
Box plot of the statistical results in \autoref{table:benchmark-stats-eslintrc}.
}
\label{fig:benchmark-eslintrc-boxplot}
\end{figure*}

In \autoref{fig:benchmark-eslintrc-boxplot}, we observe the medians for
uncompressed schema-driven binary serialization specifications to be smaller in
comparison to uncompressed schema-less binary serialization specifications.  The range
between the upper and lower whiskers and the inter-quartile range of
uncompressed schema-driven binary serialization specifications is smaller than the
range between the upper and lower whiskers and the inter-quartile range of
uncompressed schema-less binary serialization specifications.


In terms of compression, LZMA results in the lower median for schema-driven
binary serialization specifications while GZIP results in the lower median for
schema-less binary serialization specifications.  Additionally, GZIP and LZMA are
space-efficient in terms of the median in comparison to uncompressed
schema-driven binary serialization specifications and GZIP, LZ4 and LZMA are
space-efficient in terms of the median in comparison to uncompressed
schema-less binary serialization specifications.  However, the use of LZ4 for
schema-driven binary serialization specifications and the use of GZIP, LZ4 and LZMA
for schema-less binary serialization specifications exhibit upper outliers.
Nevertheless, compression reduces the range between the upper and lower
whiskers and inter-quartile range for both schema-driven and schema-less binary
serialization specifications.  In particular, the compression format with the smaller
range between the upper and lower whiskers for schema-driven binary
serialization specifications is LZ4, and the compression format with the smaller range
between the upper and lower whiskers and the smaller inter-quartile range for
schema-less binary serialization specifications is LZMA.


Overall, \we conclude that uncompressed schema-driven binary serialization
specifications are space-efficient in comparison to uncompressed schema-less binary
serialization specifications. GZIP and LZMA are space-efficient in comparison to
uncompressed schema-driven binary serialization specifications and all the considered
compression formats are space-efficient in comparison to uncompressed
schema-less binary serialization specifications.

\clearpage

\subsection{NPM Package.json Linter Configuration Manifest}
\label{sec:benchmark-packagejsonlintrc}

\begin{figure*}[ht!]
\frame{\includegraphics[width=\linewidth]{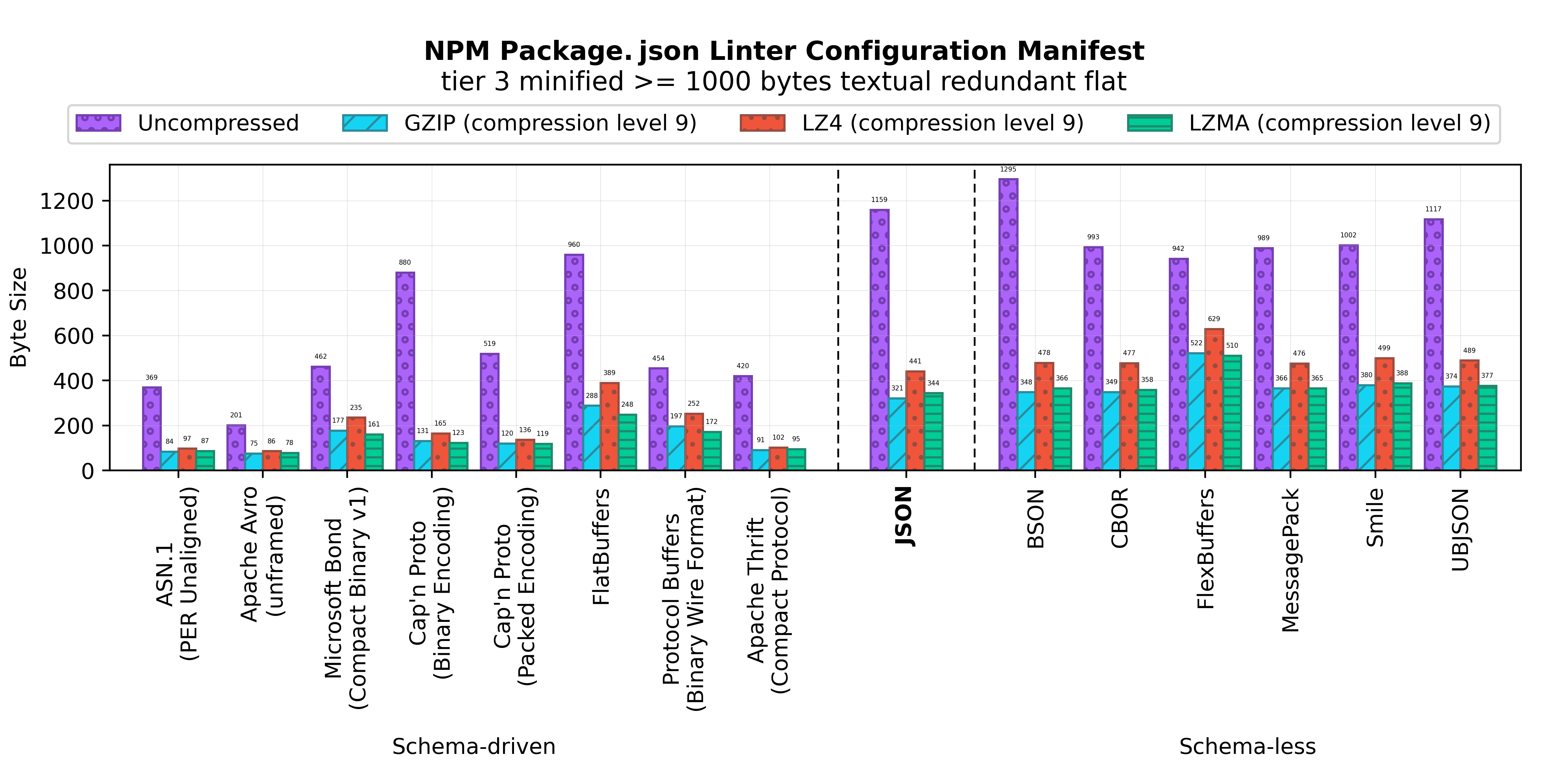}}
\caption{
The benchmark results for the NPM Package.json Linter Configuration Manifest test case listed in \autoref{table:benchmark-documents} and \autoref{table:benchmark-documents-1}.
}
\label{fig:benchmark-packagejsonlintrc}
\end{figure*}

Node.js Package Manager (NPM) \footnote{\url{https://www.npmjs.com}} is an
open-source package manager for Node.js \footnote{\url{https://nodejs.org}}, a
JavaScript \cite{ECMA-262} runtime targetted at the web development industry.
\texttt{npm-package-json-lint}
\footnote{\url{https://npmpackagejsonlint.org/en/}} is an open-source tool to
enforce a set of configurable rules for a Node.js Package Manager (NPM)
\footnote{\url{https://www.npmjs.com}} configuration manifest. In
\autoref{fig:benchmark-packagejsonlintrc}, \we demonstrate a \textbf{Tier 3
minified $\geq$ 1000 bytes textual redundant flat} (Tier 3 TRF from
\autoref{table:json-taxonomy}) JSON document that consists of an example
\texttt{npm-package-json-lint} configuration.

The smallest bit-string is produced by Apache Avro \cite{avro} (201 bytes),
followed by ASN.1 PER Unaligned \cite{asn1-per} (369 bytes) and Apache Thrift
\cite{slee2007thrift} (420 bytes).  The binary serialization specifications that
produced the smallest bit-strings are schema-driven and sequential \cite{viotti2022survey}.
Conversely, the largest bit-string is produced by BSON \cite{bson} (1295
bytes), followed by UBJSON \cite{ubjson} (1117 bytes) and Smile \cite{smile}
(1002 bytes). The binary serialization specifications that produced the largest
bit-strings are schema-less and sequential.  In comparison to JSON
\cite{ECMA-404} (1159 bytes), binary serialization achieves a \textbf{5.7x}
size reduction in the best case for this input document.  However, 1 out of the
14 JSON-compatible binary serialization specifications listed in
\autoref{table:benchmark-specifications-schema-driven} and
\autoref{table:benchmark-specifications-schema-less} result in bit-strings that are
larger than JSON: the schema-less and sequential BSON \cite{bson} binary
serialization specification.

For this Tier 3 TRF document, the best performing schema-driven serialization
specification achieves a \textbf{4.6x} size reduction compared to the best performing
schema-less serialization specification: FlexBuffers \cite{flexbuffers} (942 bytes).
As shown in \autoref{table:benchmark-stats-packagejsonlintrc}, uncompressed
schema-driven specifications provide smaller \emph{average} and \emph{median}
bit-strings than uncompressed schema-less specifications. However, as highlighted by
the \emph{range} and \emph{standard deviation}, uncompressed schema-driven
specifications exhibit higher size reduction variability depending on the
expressiveness of the schema language (i.e. how the language constructs allow
you to model the data) and the size optimizations devised by its authors.  With
the exception of the pointer-based binary serialization specification FlatBuffers
\cite{flatbuffers}, the selection of schema-driven serialization specifications listed
in \autoref{table:benchmark-specifications-schema-driven} produce bit-strings that are
equal to or smaller than their schema-less counterparts listed in
\autoref{table:benchmark-specifications-schema-less}.  The best performing sequential
serialization specification achieves a \textbf{2.5x} size reduction compared to the
best performing pointer-based serialization specification: Cap'n Proto Packed Encoding
\cite{capnproto} (519 bytes).

The compression formats listed in
\autoref{sec:benchmark-compression-formats} result in positive gains for
all bit-strings. The best performing uncompressed binary serialization
specification achieves a \textbf{1.5x} size reduction compared to the best
performing compression format for JSON: GZIP \cite{RFC1952} (321 bytes).

\begin{table*}[hb!]
\caption{A byte-size statistical analysis of the benchmark results shown in \autoref{fig:benchmark-packagejsonlintrc} divided by schema-driven and schema-less specifications.}
\label{table:benchmark-stats-packagejsonlintrc}
\begin{tabularx}{\linewidth}{X|l|l|l|l|l|l|l|l}
\toprule
\multirow{2}{*}{\textbf{Category}} &
\multicolumn{4}{c|}{\textbf{Schema-driven}} &
\multicolumn{4}{c}{\textbf{Schema-less}} \\
\cline{2-9}
& \small\textbf{Average} & \small\textbf{Median} & \small\textbf{Range} & \small\textbf{Std.dev} & \small\textbf{Average} & \small\textbf{Median} & \small\textbf{Range} & \small\textbf{Std.dev} \\
\midrule
Uncompressed & \small{533.1} & \small{458} & \small{759} & \small{240.9} & \small{1056.3} & \small{997.5} & \small{353} & \small{119.2} \\ \hline
GZIP (compression level 9) & \small{145.4} & \small{125.5} & \small{213} & \small{67.6} & \small{389.8} & \small{370} & \small{174} & \small{60.3} \\ \hline
LZ4 (compression level 9) & \small{182.8} & \small{150.5} & \small{303} & \small{97.3} & \small{508} & \small{483.5} & \small{153} & \small{54.7} \\ \hline
LZMA (compression level 9) & \small{135.4} & \small{121} & \small{170} & \small{52.9} & \small{394} & \small{371.5} & \small{152} & \small{52.8} \\
\bottomrule
\end{tabularx}
\end{table*}

\begin{table*}[hb!]
\caption{The benchmark raw data results and schemas for the plot in \autoref{fig:benchmark-packagejsonlintrc}.}
\label{table:benchmark-packagejsonlintrc}
\begin{tabularx}{\linewidth}{X|l|l|l|l|l}
\toprule
\textbf{Serialization Format} & \textbf{Schema} & \textbf{Uncompressed} & \textbf{GZIP} & \textbf{LZ4} & \textbf{LZMA} \\
\midrule
ASN.1 (PER Unaligned) & \href{https://github.com/jviotti/binary-json-size-benchmark/blob/main/benchmark/packagejsonlintrc/asn1/schema.asn}{\small{\texttt{schema.asn}}} & 369 & 84 & 97 & 87 \\ \hline
Apache Avro (unframed) & \href{https://github.com/jviotti/binary-json-size-benchmark/blob/main/benchmark/packagejsonlintrc/avro/schema.json}{\small{\texttt{schema.json}}} & 201 & 75 & 86 & 78 \\ \hline
Microsoft Bond (Compact Binary v1) & \href{https://github.com/jviotti/binary-json-size-benchmark/blob/main/benchmark/packagejsonlintrc/bond/schema.bond}{\small{\texttt{schema.bond}}} & 462 & 177 & 235 & 161 \\ \hline
Cap'n Proto (Binary Encoding) & \href{https://github.com/jviotti/binary-json-size-benchmark/blob/main/benchmark/packagejsonlintrc/capnproto/schema.capnp}{\small{\texttt{schema.capnp}}} & 880 & 131 & 165 & 123 \\ \hline
Cap'n Proto (Packed Encoding) & \href{https://github.com/jviotti/binary-json-size-benchmark/blob/main/benchmark/packagejsonlintrc/capnproto-packed/schema.capnp}{\small{\texttt{schema.capnp}}} & 519 & 120 & 136 & 119 \\ \hline
FlatBuffers & \href{https://github.com/jviotti/binary-json-size-benchmark/blob/main/benchmark/packagejsonlintrc/flatbuffers/schema.fbs}{\small{\texttt{schema.fbs}}} & 960 & 288 & 389 & 248 \\ \hline
Protocol Buffers (Binary Wire Format) & \href{https://github.com/jviotti/binary-json-size-benchmark/blob/main/benchmark/packagejsonlintrc/protobuf/schema.proto}{\small{\texttt{schema.proto}}} & 454 & 197 & 252 & 172 \\ \hline
Apache Thrift (Compact Protocol) & \href{https://github.com/jviotti/binary-json-size-benchmark/blob/main/benchmark/packagejsonlintrc/thrift/schema.thrift}{\small{\texttt{schema.thrift}}} & 420 & 91 & 102 & 95 \\ \hline
\hline \textbf{JSON} & - & 1159 & 321 & 441 & 344 \\ \hline \hline
BSON & - & 1295 & 348 & 478 & 366 \\ \hline
CBOR & - & 993 & 349 & 477 & 358 \\ \hline
FlexBuffers & - & 942 & 522 & 629 & 510 \\ \hline
MessagePack & - & 989 & 366 & 476 & 365 \\ \hline
Smile & - & 1002 & 380 & 499 & 388 \\ \hline
UBJSON & - & 1117 & 374 & 489 & 377 \\
\bottomrule
\end{tabularx}
\end{table*}

\begin{figure*}[ht!]
\frame{\includegraphics[width=\linewidth]{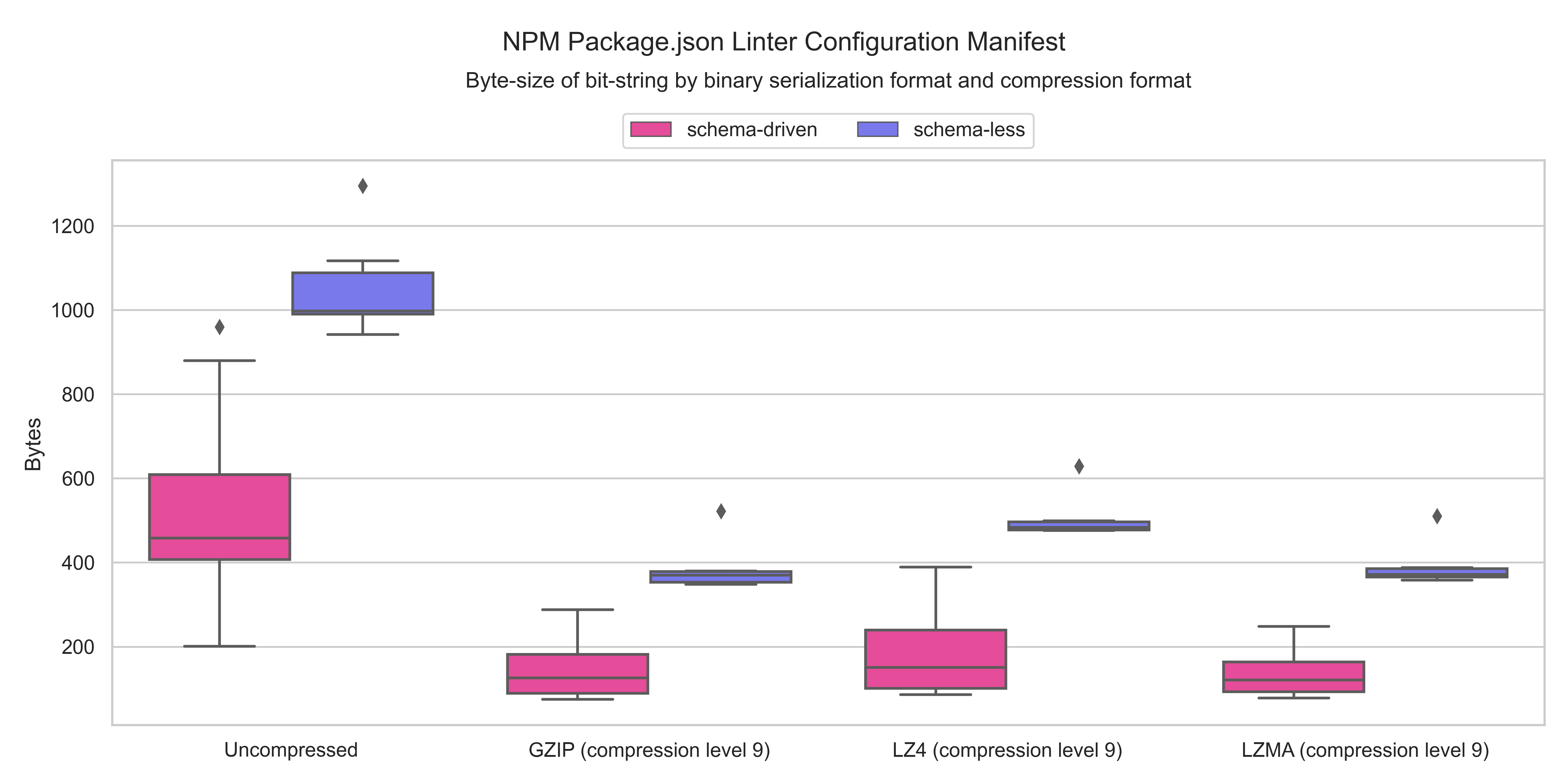}}
\caption{
Box plot of the statistical results in \autoref{table:benchmark-stats-packagejsonlintrc}.
}
\label{fig:benchmark-packagejsonlintrc-boxplot}
\end{figure*}

In \autoref{fig:benchmark-packagejsonlintrc-boxplot}, we observe the medians
for uncompressed schema-driven binary serialization specifications to be smaller in
comparison to uncompressed schema-less binary serialization specifications.  The range
between the upper and lower whiskers and the inter-quartile range of
uncompressed schema-less binary serialization specifications is smaller than the range
between the upper and lower whiskers and the inter-quartile range of
uncompressed schema-driven binary serialization specifications.


In terms of compression, LZMA results in the lower medians for schema-driven
binary serialization specifications while GZIP and LZMA result in the lower
median for schema-less binary serialization specifications.  Additionally,
GZIP, LZ4 and LZMA are space-efficient in terms of the median in comparison to
both uncompressed schema-driven and schema-less binary serialization
specifications.  However, the use of GZIP, LZ4 and LZMA for schema-less binary
serialization specifications exhibits upper outliers.  Nevertheless,
compression reduces the range between the upper and lower whiskers and
inter-quartile range for both schema-driven and schema-less binary
serialization specifications.  In particular, the compression format with the
smaller range between the upper and lower whiskers and the smaller
inter-quartile range for schema-driven binary serialization specifications is
LZMA.


Overall, \we conclude that uncompressed schema-driven binary serialization
specifications are space-efficient in comparison to uncompressed schema-less
binary serialization specifications and that all the considered compression
formats are space-efficient in comparison to uncompressed schema-driven and
schema-less binary serialization specifications.

\clearpage

\subsection{.NET Core Project}
\label{sec:benchmark-netcoreproject}

\begin{figure*}[ht!]
\frame{\includegraphics[width=\linewidth]{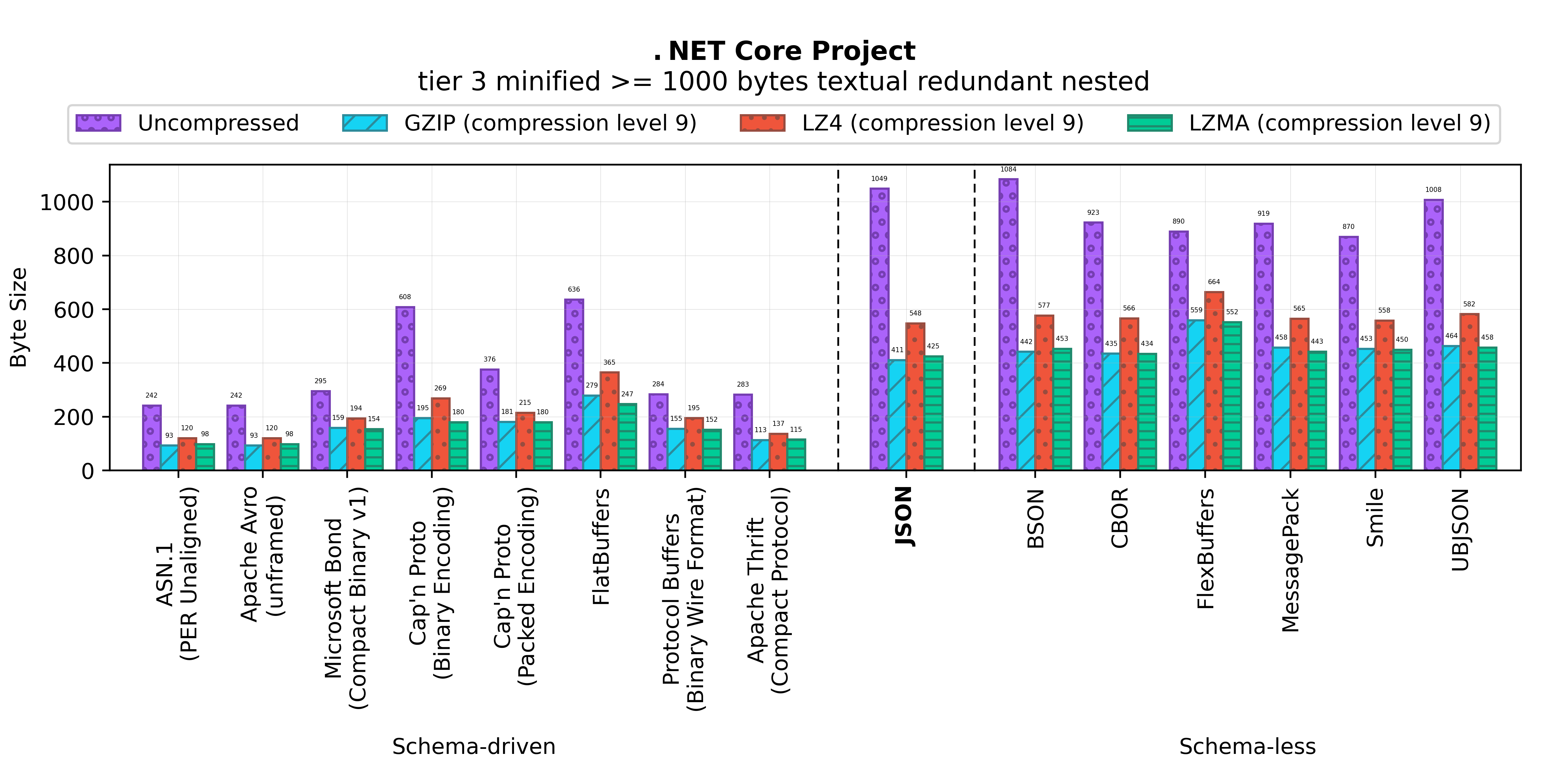}}
\caption{
The benchmark results for the .NET Core Project test case listed in \autoref{table:benchmark-documents} and \autoref{table:benchmark-documents-1}.
}
\label{fig:benchmark-netcoreproject}
\end{figure*}

The ASP.NET \footnote{\url{https://dotnet.microsoft.com/apps/aspnet}} Microsoft
web-application framework defined a now-obsolete JSON-based project manifest
called \texttt{project.json}
\footnote{\url{http://web.archive.org/web/20150322033428/https://github.com/aspnet/Home/wiki/Project.json-file}}
used in the web industry. In \autoref{fig:benchmark-netcoreproject}, \we
demonstrate a \textbf{Tier 3 minified $\geq$ 1000 bytes textual redundant
nested} (Tier 3 TRN from \autoref{table:json-taxonomy}) JSON document that
consists of a detailed an example \texttt{project.json} manifest that lists
several dependencies.

The smallest bit-string is produced by both ASN.1 PER Unaligned \cite{asn1-per}
and Apache Avro \cite{avro} (242 bytes), followed by Apache Thrift
\cite{slee2007thrift} (283 bytes) and Protocol Buffers \cite{protocolbuffers}
(284 bytes). The binary serialization specifications that produced the smallest
bit-strings are schema-driven and sequential \cite{viotti2022survey}.  Conversely, the largest
bit-string is produced by BSON \cite{bson} (1084 bytes), followed by UBJSON
\cite{ubjson} (1008 bytes) and CBOR \cite{RFC7049} (923 bytes).  The binary
serialization specifications that produced the largest bit-strings are schema-less and
sequential \cite{viotti2022survey}.  In comparison to JSON \cite{ECMA-404} (1049 bytes),
binary serialization achieves a \textbf{4.3x} size reduction in the best case
for this input document.  However, 1 out of the 14 JSON-compatible binary
serialization specifications listed in \autoref{table:benchmark-specifications-schema-driven}
and \autoref{table:benchmark-specifications-schema-less} result in bit-strings that
are larger than JSON: the schema-less and sequential BSON \cite{bson} binary
serialization specification.

For this Tier 3 TRN document, the best performing schema-driven serialization
specification achieves a \textbf{3.5x} size reduction compared to the best performing
schema-less serialization specification: Smile \cite{smile} (870 bytes).  As shown in
\autoref{table:benchmark-stats-netcoreproject}, uncompressed schema-driven
specifications provide smaller \emph{average} and \emph{median} bit-strings than
uncompressed schema-less specifications. However, as highlighted by the \emph{range}
and \emph{standard deviation}, uncompressed schema-driven specifications exhibit
higher size reduction variability depending on the expressiveness of the schema
language (i.e. how the language constructs allow you to model the data) and the
size optimizations devised by its authors. The entire selection of
schema-driven serialization specifications listed in
\autoref{table:benchmark-specifications-schema-driven} produce bit-strings that are
equal to or smaller than their schema-less counterparts listed in
\autoref{table:benchmark-specifications-schema-less}.  The best performing sequential
serialization specification only achieves a \textbf{1.5x} size reduction compared to
the best performing pointer-based serialization specification: Cap'n Proto Packed
Encoding \cite{capnproto} (376 bytes).

The compression formats listed in
\autoref{sec:benchmark-compression-formats} result in positive gains for
all bit-strings. The best performing uncompressed binary serialization
specification achieves a \textbf{1.6x} size reduction compared to the best
performing compression format for JSON: GZIP \cite{RFC1952} (411 bytes).

\begin{table*}[hb!]
\caption{A byte-size statistical analysis of the benchmark results shown in \autoref{fig:benchmark-netcoreproject} divided by schema-driven and schema-less specifications.}
\label{table:benchmark-stats-netcoreproject}
\begin{tabularx}{\linewidth}{X|l|l|l|l|l|l|l|l}
\toprule
\multirow{2}{*}{\textbf{Category}} &
\multicolumn{4}{c|}{\textbf{Schema-driven}} &
\multicolumn{4}{c}{\textbf{Schema-less}} \\
\cline{2-9}
& \small\textbf{Average} & \small\textbf{Median} & \small\textbf{Range} & \small\textbf{Std.dev} & \small\textbf{Average} & \small\textbf{Median} & \small\textbf{Range} & \small\textbf{Std.dev} \\
\midrule
Uncompressed & \small{370.8} & \small{289.5} & \small{394} & \small{150.3} & \small{949} & \small{921} & \small{214} & \small{74.2} \\ \hline
GZIP (compression level 9) & \small{158.5} & \small{157} & \small{186} & \small{58.1} & \small{468.5} & \small{455.5} & \small{124} & \small{41.6} \\ \hline
LZ4 (compression level 9) & \small{201.9} & \small{194.5} & \small{245} & \small{78.3} & \small{585.3} & \small{571.5} & \small{106} & \small{36.1} \\ \hline
LZMA (compression level 9) & \small{153} & \small{153} & \small{149} & \small{47.2} & \small{465} & \small{451.5} & \small{118} & \small{39.6} \\
\bottomrule
\end{tabularx}
\end{table*}

\begin{table*}[hb!]
\caption{The benchmark raw data results and schemas for the plot in \autoref{fig:benchmark-netcoreproject}.}
\label{table:benchmark-netcoreproject}
\begin{tabularx}{\linewidth}{X|l|l|l|l|l}
\toprule
\textbf{Serialization Format} & \textbf{Schema} & \textbf{Uncompressed} & \textbf{GZIP} & \textbf{LZ4} & \textbf{LZMA} \\
\midrule
ASN.1 (PER Unaligned) & \href{https://github.com/jviotti/binary-json-size-benchmark/blob/main/benchmark/netcoreproject/asn1/schema.asn}{\small{\texttt{schema.asn}}} & 242 & 93 & 120 & 98 \\ \hline
Apache Avro (unframed) & \href{https://github.com/jviotti/binary-json-size-benchmark/blob/main/benchmark/netcoreproject/avro/schema.json}{\small{\texttt{schema.json}}} & 242 & 93 & 120 & 98 \\ \hline
Microsoft Bond (Compact Binary v1) & \href{https://github.com/jviotti/binary-json-size-benchmark/blob/main/benchmark/netcoreproject/bond/schema.bond}{\small{\texttt{schema.bond}}} & 295 & 159 & 194 & 154 \\ \hline
Cap'n Proto (Binary Encoding) & \href{https://github.com/jviotti/binary-json-size-benchmark/blob/main/benchmark/netcoreproject/capnproto/schema.capnp}{\small{\texttt{schema.capnp}}} & 608 & 195 & 269 & 180 \\ \hline
Cap'n Proto (Packed Encoding) & \href{https://github.com/jviotti/binary-json-size-benchmark/blob/main/benchmark/netcoreproject/capnproto-packed/schema.capnp}{\small{\texttt{schema.capnp}}} & 376 & 181 & 215 & 180 \\ \hline
FlatBuffers & \href{https://github.com/jviotti/binary-json-size-benchmark/blob/main/benchmark/netcoreproject/flatbuffers/schema.fbs}{\small{\texttt{schema.fbs}}} & 636 & 279 & 365 & 247 \\ \hline
Protocol Buffers (Binary Wire Format) & \href{https://github.com/jviotti/binary-json-size-benchmark/blob/main/benchmark/netcoreproject/protobuf/schema.proto}{\small{\texttt{schema.proto}}} & 284 & 155 & 195 & 152 \\ \hline
Apache Thrift (Compact Protocol) & \href{https://github.com/jviotti/binary-json-size-benchmark/blob/main/benchmark/netcoreproject/thrift/schema.thrift}{\small{\texttt{schema.thrift}}} & 283 & 113 & 137 & 115 \\ \hline
\hline \textbf{JSON} & - & 1049 & 411 & 548 & 425 \\ \hline \hline
BSON & - & 1084 & 442 & 577 & 453 \\ \hline
CBOR & - & 923 & 435 & 566 & 434 \\ \hline
FlexBuffers & - & 890 & 559 & 664 & 552 \\ \hline
MessagePack & - & 919 & 458 & 565 & 443 \\ \hline
Smile & - & 870 & 453 & 558 & 450 \\ \hline
UBJSON & - & 1008 & 464 & 582 & 458 \\
\bottomrule
\end{tabularx}
\end{table*}

\begin{figure*}[ht!]
\frame{\includegraphics[width=\linewidth]{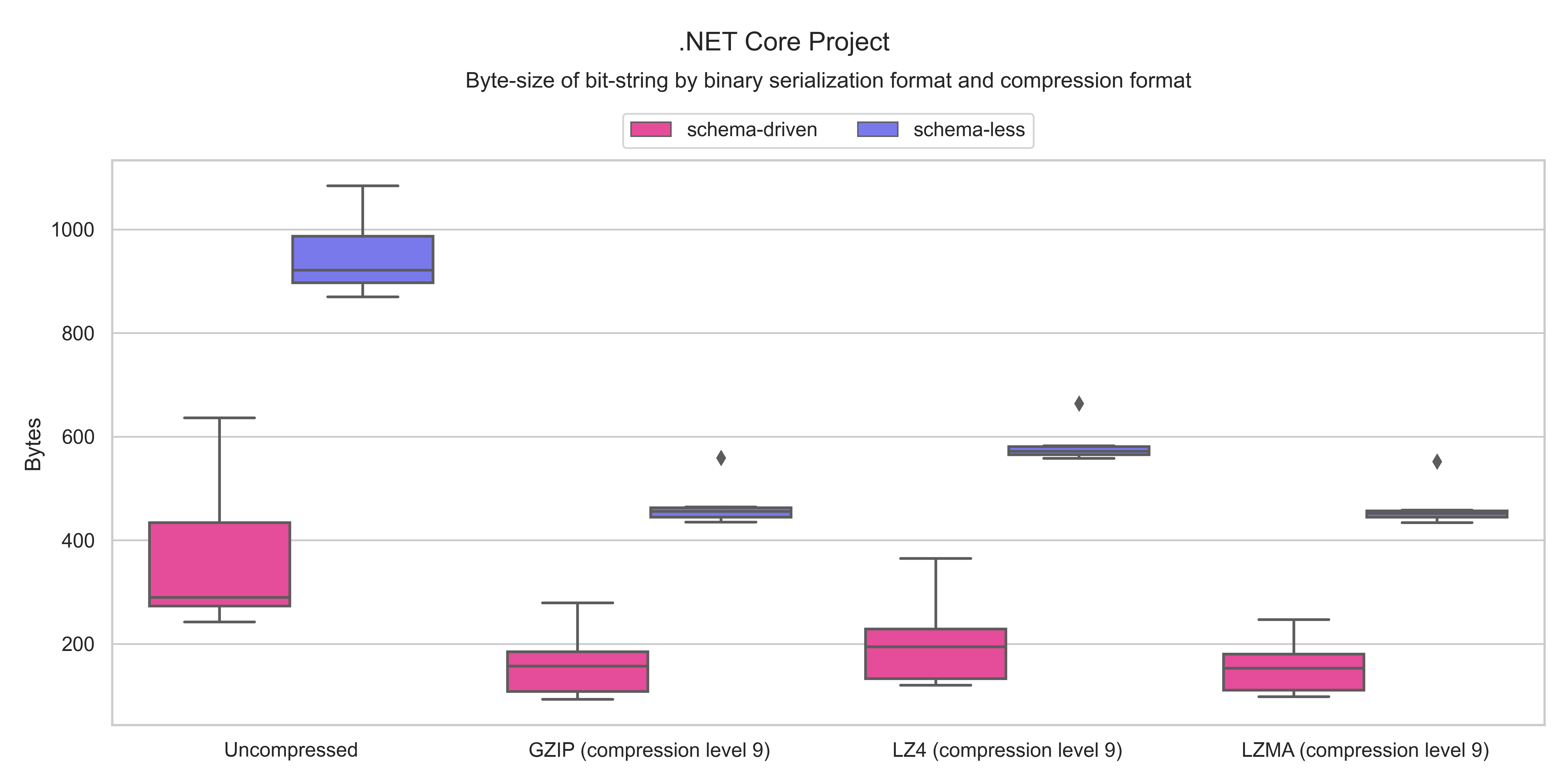}}
\caption{
Box plot of the statistical results in \autoref{table:benchmark-stats-netcoreproject}.
}
\label{fig:benchmark-netcoreproject-boxplot}
\end{figure*}

In \autoref{fig:benchmark-netcoreproject-boxplot}, we observe the medians for
uncompressed schema-driven binary serialization specifications to be smaller in
comparison to uncompressed schema-less binary serialization specifications.  The range
between the upper and lower whiskers and the inter-quartile range of
uncompressed schema-less binary serialization specifications is smaller than the range
between the upper and lower whiskers and the inter-quartile range of
uncompressed schema-driven binary serialization specifications.


In terms of compression, LZMA results in the lower median for both
schema-driven and schema-less binary serialization specifications.
Additionally, GZIP, LZ4 and LZMA are space-efficient in terms of the median in
comparison to both uncompressed schema-driven and schema-less binary
serialization specifications.  However, the use of GZIP, LZ4 and LZMA for
schema-less binary serialization specifications exhibits upper outliers.
Nevertheless, compression reduces the range between the upper and lower
whiskers and inter-quartile range for both schema-driven and schema-less binary
serialization specifications.  In particular, the compression format with the
smaller range between the upper and lower whiskers for schema-driven binary
serialization specifications is LZMA, and the compression formats with the
smaller inter-quartile range for schema-driven binary serialization
specifications are GZIP and LZMA.


Overall, \we conclude that uncompressed schema-driven binary serialization
specifications are space-efficient in comparison to uncompressed schema-less
binary serialization specifications and that all the considered compression
formats are space-efficient in comparison to uncompressed schema-driven and
schema-less binary serialization specifications.

\clearpage

\subsection{NPM Package.json Example Manifest}
\label{sec:benchmark-packagejson}

\begin{figure*}[ht!]
\frame{\includegraphics[width=\linewidth]{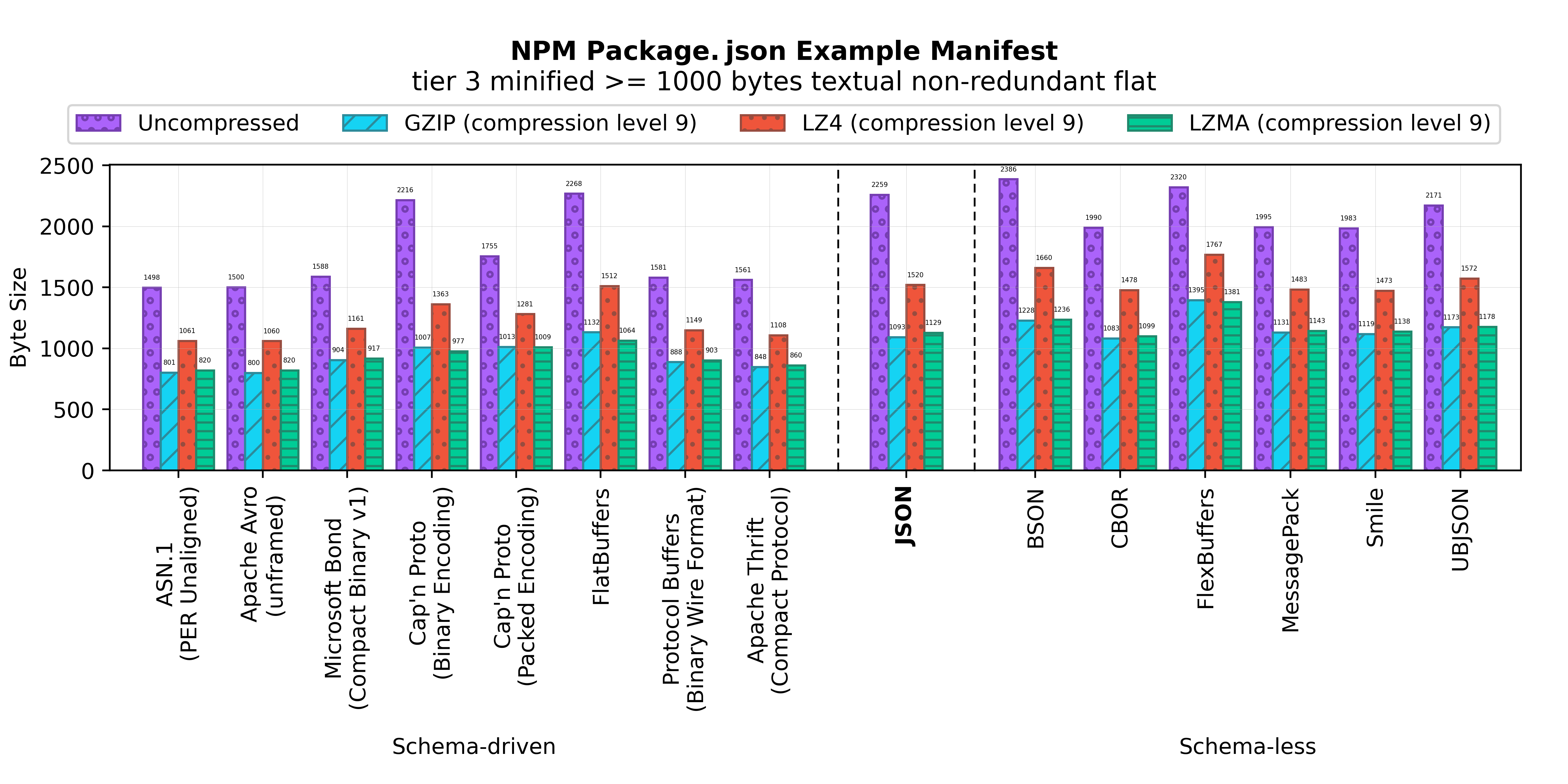}}
\caption{
The benchmark results for the NPM Package.json Example Manifest test case listed in \autoref{table:benchmark-documents} and \autoref{table:benchmark-documents-1}.
}
\label{fig:benchmark-packagejson}
\end{figure*}

Node.js Package Manager (NPM) \footnote{\url{https://www.npmjs.com}} is an
open-source package manager for Node.js \footnote{\url{https://nodejs.org}}, a
JavaScript \cite{ECMA-262} runtime targetted at the web development industry. A
package that is published to NPM is declared using a JSON file called
\texttt{package.json}
\footnote{\url{https://docs.npmjs.com/cli/v6/configuring-npm/package-json}}.
In \autoref{fig:benchmark-packagejson}, \we demonstrate a \textbf{Tier 3
minified $\geq$ 1000 bytes textual non-redundant flat} (Tier 3 TNF from
\autoref{table:json-taxonomy}) JSON document that consists of a
\texttt{package.json} manifest that declares a particular version of the
Grunt.js \footnote{\url{https://gruntjs.com}} task runner.

The smallest bit-string is produced by ASN.1 PER Unaligned \cite{asn1-per}
(1498 bytes), closely followed by Apache Avro \cite{avro} (1500 bytes) and
Apache Thrift \cite{slee2007thrift} (1561 bytes). The binary serialization
specifications that produced the smallest bit-strings are schema-driven and sequential
\cite{viotti2022survey}. Conversely, the largest bit-string is produced by BSON \cite{bson}
(2386 bytes), followed by FlexBuffers \cite{flexbuffers} (2320 bytes) and
FlatBuffers \cite{flatbuffers} (2268 bytes). With the exception of BSON, the
binary serialization specifications that produced the largest bit-strings are
pointer-based \cite{viotti2022survey}.  In comparison to JSON \cite{ECMA-404} (2259 bytes),
binary serialization only achieves a \textbf{1.5x} size reduction in the best
case for this input document.  Additionally, 3 out of the 14 JSON-compatible
binary serialization specifications listed in
\autoref{table:benchmark-specifications-schema-driven} and
\autoref{table:benchmark-specifications-schema-less} result in bit-strings that are
larger than JSON: FlatBuffers \cite{flatbuffers}, BSON \cite{bson} and
FlexBuffers \cite{flexbuffers}. These binary serialization specifications are either
schema-less or schema-driven and pointer-based.

For this Tier 3 TNF document, the best performing schema-driven serialization
specification only achieves a \textbf{1.3x} size reduction compared to the best
performing schema-less serialization specification: Smile \cite{smile} (1983 bytes).
As shown in \autoref{table:benchmark-stats-packagejson}, uncompressed
schema-driven specifications provide smaller \emph{average} and \emph{median}
bit-strings than uncompressed schema-less specifications. However, as highlighted by
the \emph{range} and \emph{standard deviation}, uncompressed schema-driven
specifications exhibit higher size reduction variability depending on the
expressiveness of the schema language (i.e. how the language constructs allow
you to model the data) and the size optimizations devised by its authors. With
the exception of the pointer-based binary serialization specifications Cap'n Proto
Binary Encoding \cite{capnproto} and FlatBuffers \cite{flatbuffers}, the
selection of schema-driven serialization specifications listed in
\autoref{table:benchmark-specifications-schema-driven} produce bit-strings that are
equal to or smaller than their schema-less counterparts listed in
\autoref{table:benchmark-specifications-schema-less}.  The best performing sequential
serialization specification only achieves a \textbf{1.1x} size reduction compared to
the best performing pointer-based serialization specification: Cap'n Proto Packed
Encoding \cite{capnproto} (1755 bytes).

The compression formats listed in
\autoref{sec:benchmark-compression-formats} result in positive gains for
all bit-strings. The best performing compression format for JSON, GZIP
\cite{RFC1952} (1093 bytes), achieves a \textbf{1.3x} size reduction compared
to the best performing uncompressed binary serialization specification.

\begin{table*}[hb!]
\caption{A byte-size statistical analysis of the benchmark results shown in \autoref{fig:benchmark-packagejson} divided by schema-driven and schema-less specifications.}
\label{table:benchmark-stats-packagejson}
\begin{tabularx}{\linewidth}{X|l|l|l|l|l|l|l|l}
\toprule
\multirow{2}{*}{\textbf{Category}} &
\multicolumn{4}{c|}{\textbf{Schema-driven}} &
\multicolumn{4}{c}{\textbf{Schema-less}} \\
\cline{2-9}
& \small\textbf{Average} & \small\textbf{Median} & \small\textbf{Range} & \small\textbf{Std.dev} & \small\textbf{Average} & \small\textbf{Median} & \small\textbf{Range} & \small\textbf{Std.dev} \\
\midrule
Uncompressed & \small{1745.9} & \small{1584.5} & \small{770} & \small{296.2} & \small{2140.8} & \small{2083} & \small{403} & \small{164.3} \\ \hline
GZIP (compression level 9) & \small{924.1} & \small{896} & \small{332} & \small{109.6} & \small{1188.2} & \small{1152} & \small{312} & \small{103.0} \\ \hline
LZ4 (compression level 9) & \small{1211.9} & \small{1155} & \small{452} & \small{150.4} & \small{1572.2} & \small{1527.5} & \small{294} & \small{109.8} \\ \hline
LZMA (compression level 9) & \small{921.3} & \small{910} & \small{244} & \small{83.5} & \small{1195.8} & \small{1160.5} & \small{282} & \small{92.9} \\
\bottomrule
\end{tabularx}
\end{table*}

\begin{table*}[hb!]
\caption{The benchmark raw data results and schemas for the plot in \autoref{fig:benchmark-packagejson}.}
\label{table:benchmark-packagejson}
\begin{tabularx}{\linewidth}{X|l|l|l|l|l}
\toprule
\textbf{Serialization Format} & \textbf{Schema} & \textbf{Uncompressed} & \textbf{GZIP} & \textbf{LZ4} & \textbf{LZMA} \\
\midrule
ASN.1 (PER Unaligned) & \href{https://github.com/jviotti/binary-json-size-benchmark/blob/main/benchmark/packagejson/asn1/schema.asn}{\small{\texttt{schema.asn}}} & 1498 & 801 & 1061 & 820 \\ \hline
Apache Avro (unframed) & \href{https://github.com/jviotti/binary-json-size-benchmark/blob/main/benchmark/packagejson/avro/schema.json}{\small{\texttt{schema.json}}} & 1500 & 800 & 1060 & 820 \\ \hline
Microsoft Bond (Compact Binary v1) & \href{https://github.com/jviotti/binary-json-size-benchmark/blob/main/benchmark/packagejson/bond/schema.bond}{\small{\texttt{schema.bond}}} & 1588 & 904 & 1161 & 917 \\ \hline
Cap'n Proto (Binary Encoding) & \href{https://github.com/jviotti/binary-json-size-benchmark/blob/main/benchmark/packagejson/capnproto/schema.capnp}{\small{\texttt{schema.capnp}}} & 2216 & 1007 & 1363 & 977 \\ \hline
Cap'n Proto (Packed Encoding) & \href{https://github.com/jviotti/binary-json-size-benchmark/blob/main/benchmark/packagejson/capnproto-packed/schema.capnp}{\small{\texttt{schema.capnp}}} & 1755 & 1013 & 1281 & 1009 \\ \hline
FlatBuffers & \href{https://github.com/jviotti/binary-json-size-benchmark/blob/main/benchmark/packagejson/flatbuffers/schema.fbs}{\small{\texttt{schema.fbs}}} & 2268 & 1132 & 1512 & 1064 \\ \hline
Protocol Buffers (Binary Wire Format) & \href{https://github.com/jviotti/binary-json-size-benchmark/blob/main/benchmark/packagejson/protobuf/schema.proto}{\small{\texttt{schema.proto}}} & 1581 & 888 & 1149 & 903 \\ \hline
Apache Thrift (Compact Protocol) & \href{https://github.com/jviotti/binary-json-size-benchmark/blob/main/benchmark/packagejson/thrift/schema.thrift}{\small{\texttt{schema.thrift}}} & 1561 & 848 & 1108 & 860 \\ \hline
\hline \textbf{JSON} & - & 2259 & 1093 & 1520 & 1129 \\ \hline \hline
BSON & - & 2386 & 1228 & 1660 & 1236 \\ \hline
CBOR & - & 1990 & 1083 & 1478 & 1099 \\ \hline
FlexBuffers & - & 2320 & 1395 & 1767 & 1381 \\ \hline
MessagePack & - & 1995 & 1131 & 1483 & 1143 \\ \hline
Smile & - & 1983 & 1119 & 1473 & 1138 \\ \hline
UBJSON & - & 2171 & 1173 & 1572 & 1178 \\
\bottomrule
\end{tabularx}
\end{table*}

\begin{figure*}[ht!]
\frame{\includegraphics[width=\linewidth]{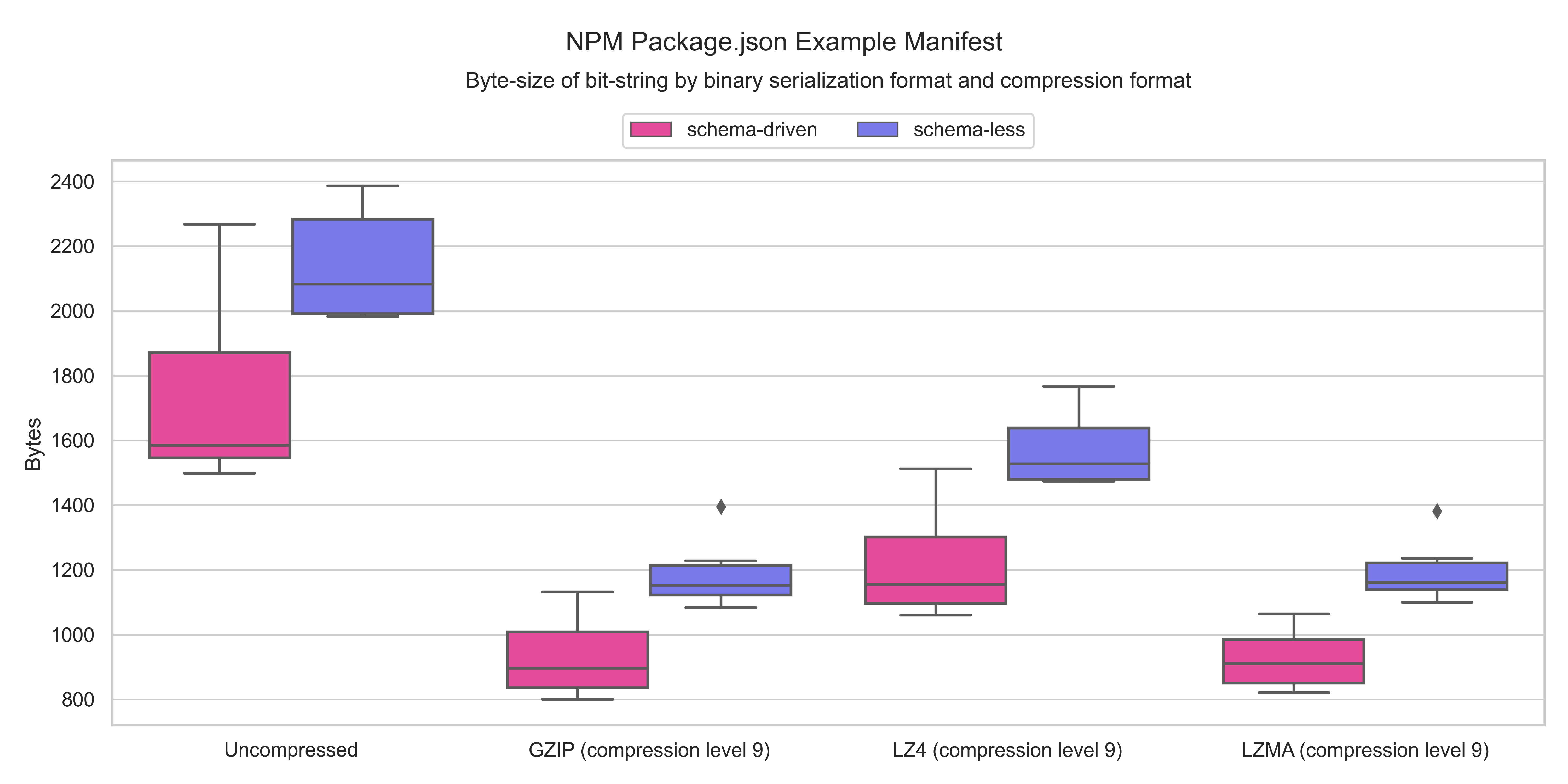}}
\caption{
Box plot of the statistical results in \autoref{table:benchmark-stats-packagejson}.
}
\label{fig:benchmark-packagejson-boxplot}
\end{figure*}

In \autoref{fig:benchmark-packagejson-boxplot}, we observe the medians for
uncompressed schema-driven binary serialization specifications to be smaller in
comparison to uncompressed schema-less binary serialization specifications.  The range
between the upper and lower whiskers and the inter-quartile range of
uncompressed schema-less binary serialization specifications is smaller than the range
between the upper and lower whiskers and the inter-quartile range of
uncompressed schema-driven binary serialization specifications.


In terms of compression, GZIP results in the lower median for both
schema-driven and schema-less binary serialization specifications.  Additionally,
GZIP, LZ4 and LZMA are space-efficient in terms of the median in comparison to
both uncompressed schema-driven and schema-less binary serialization specifications.
However, the use of GZIP and LZMA for schema-less binary serialization specifications
exhibits upper outliers.  Nevertheless, compression reduces the range between
the upper and lower whiskers and inter-quartile range for both schema-driven
and schema-less binary serialization specifications.  In particular, the compression
format with the smaller range between the upper and lower whiskers and the
smaller inter-quartile range for schema-driven binary serialization specifications is
LZMA, and the compression formats with the smaller range between the upper and
lower whiskers and the smaller inter-quartile range for schema-less binary
serialization specifications are GZIP and LZMA.


Overall, \we conclude that uncompressed schema-driven binary serialization
specifications are space-efficient in comparison to uncompressed schema-less
binary serialization specifications and that all the considered compression
formats are space-efficient in comparison to uncompressed schema-driven and
schema-less binary serialization specifications.

\clearpage

\subsection{JSON Resume Example}
\label{sec:benchmark-jsonresume}

\begin{figure*}[ht!]
\frame{\includegraphics[width=\linewidth]{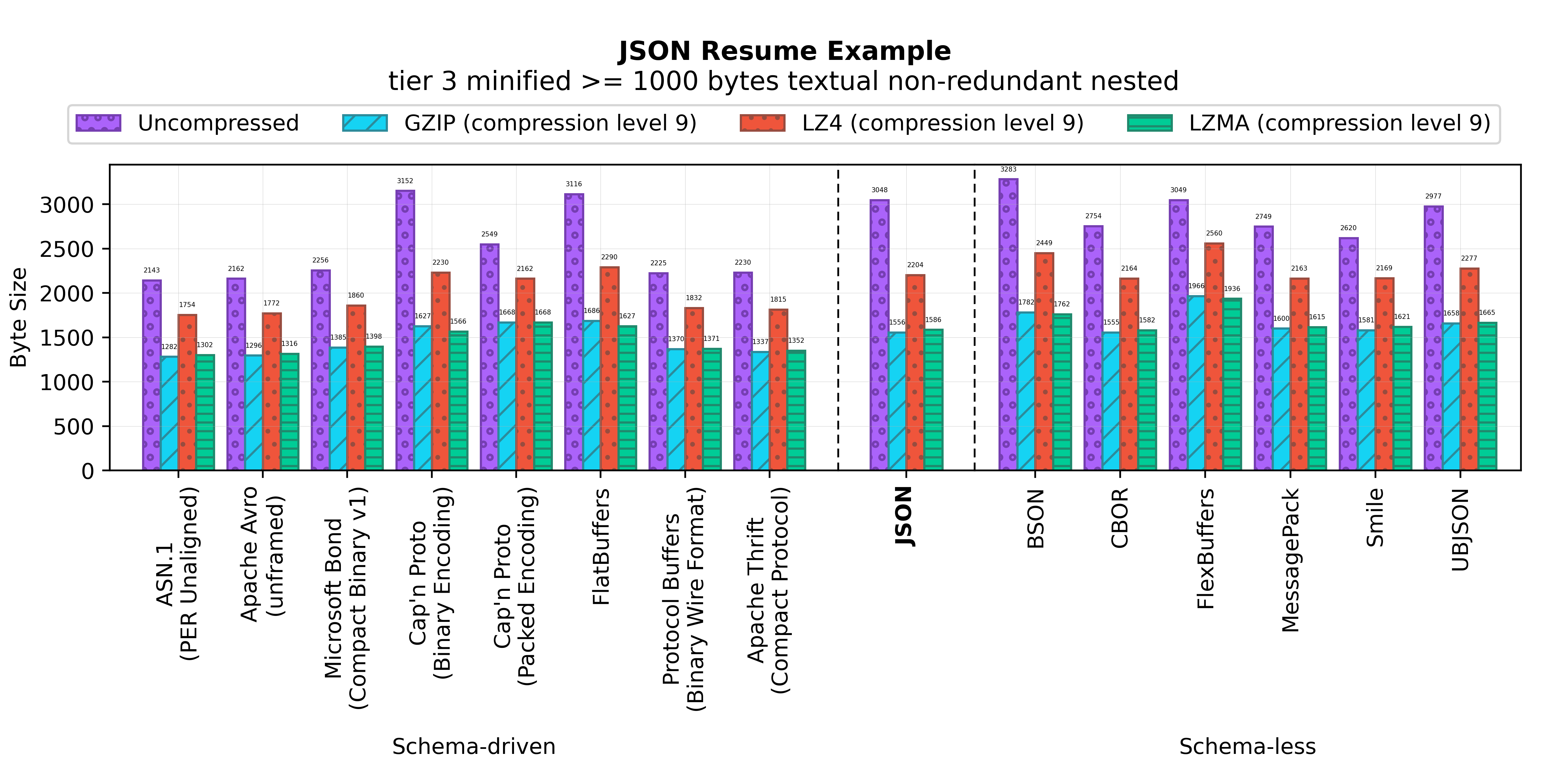}}
\caption{
The benchmark results for the JSON Resume Example test case listed in \autoref{table:benchmark-documents} and \autoref{table:benchmark-documents-1}.
}
\label{fig:benchmark-jsonresume}
\end{figure*}

JSON Resume \footnote{\url{https://jsonresume.org}} is a community-driven
proposal for a JSON-based file format that declares and renders themable
resumes used in the recruitment industry. In
\autoref{fig:benchmark-jsonresume}, \we demonstrate a \textbf{Tier 3 minified
$\geq$ 1000 bytes textual non-redundant nested} (Tier 3 TNN from
\autoref{table:json-taxonomy}) JSON document that consists of a detailed
example resume for a fictitious software programmer.

The smallest bit-string is produced by ASN.1 PER Unaligned \cite{asn1-per}
(2143 bytes), closely followed by Apache Avro \cite{avro} (2162 bytes) and
Protocol Buffers \cite{protocolbuffers} (2225 bytes). The binary serialization
specifications that produced the smallest bit-strings are schema-driven and sequential
\cite{viotti2022survey}. Conversely, the largest bit-string is produced by BSON \cite{bson}
(3283 bytes), followed by Cap'n Proto Binary Encoding \cite{capnproto} (3152
bytes) and FlatBuffers \cite{flatbuffers} (3116 bytes).  With the exception of
BSON, the binary serialization specifications that produced the largest bit-strings
are schema-driven and pointer-based \cite{viotti2022survey}.  In comparison to JSON
\cite{ECMA-404} (3048 bytes), binary serialization only achieves a
\textbf{1.4x} size reduction in the best case for this input document.
Additionally, 4 out of the 14 JSON-compatible binary serialization specifications
listed in \autoref{table:benchmark-specifications-schema-driven} and
\autoref{table:benchmark-specifications-schema-less} result in bit-strings that are
larger than JSON: Cap'n Proto Binary Encoding \cite{capnproto}, FlatBuffers
\cite{flatbuffers}, BSON \cite{bson} and FlexBuffers \cite{flexbuffers}. These
binary serialization specifications are either schema-less or schema-driven and
pointer-based.

For this Tier 3 TNN document, the best performing schema-driven serialization
specification only achieves a \textbf{1.2x} size reduction compared to the best
performing schema-less serialization specification: Smile \cite{smile} (2620 bytes).
As shown in \autoref{table:benchmark-stats-jsonresume}, uncompressed
schema-driven specifications provide smaller \emph{average} and \emph{median}
bit-strings than uncompressed schema-less specifications. However, as highlighted by
the \emph{range} and \emph{standard deviation}, uncompressed schema-driven
specifications exhibit higher size reduction variability depending on the
expressiveness of the schema language (i.e. how the language constructs allow
you to model the data) and the size optimizations devised by its authors. With
the exception of the pointer-based binary serialization specifications Cap'n Proto
Binary Encoding \cite{capnproto} and FlatBuffers \cite{flatbuffers}, the
selection of schema-driven serialization specifications listed in
\autoref{table:benchmark-specifications-schema-driven} produce bit-strings that are
equal to or smaller than their schema-less counterparts listed in
\autoref{table:benchmark-specifications-schema-less}.  The best performing sequential
serialization specification only achieves a \textbf{1.1x} size reduction compared to
the best performing pointer-based serialization specification: Cap'n Proto Packed
Encoding \cite{capnproto} (2549 bytes).

The compression formats listed in
\autoref{sec:benchmark-compression-formats} result in positive gains for
all bit-strings. The best performing compression format for JSON, GZIP
\cite{RFC1952} (1556 bytes), achieves a \textbf{1.3x} size reduction compared
to the best performing uncompressed binary serialization specification.

\begin{table*}[hb!]
\caption{A byte-size statistical analysis of the benchmark results shown in \autoref{fig:benchmark-jsonresume} divided by schema-driven and schema-less specifications.}
\label{table:benchmark-stats-jsonresume}
\begin{tabularx}{\linewidth}{X|l|l|l|l|l|l|l|l}
\toprule
\multirow{2}{*}{\textbf{Category}} &
\multicolumn{4}{c|}{\textbf{Schema-driven}} &
\multicolumn{4}{c}{\textbf{Schema-less}} \\
\cline{2-9}
& \small\textbf{Average} & \small\textbf{Median} & \small\textbf{Range} & \small\textbf{Std.dev} & \small\textbf{Average} & \small\textbf{Median} & \small\textbf{Range} & \small\textbf{Std.dev} \\
\midrule
Uncompressed & \small{2479.1} & \small{2243} & \small{1009} & \small{395.8} & \small{2905.3} & \small{2865.5} & \small{663} & \small{222.5} \\ \hline
GZIP (compression level 9) & \small{1456.4} & \small{1377.5} & \small{404} & \small{161.8} & \small{1690.3} & \small{1629} & \small{411} & \small{143.7} \\ \hline
LZ4 (compression level 9) & \small{1964.4} & \small{1846} & \small{536} & \small{208.5} & \small{2297} & \small{2223} & \small{397} & \small{155.3} \\ \hline
LZMA (compression level 9) & \small{1450} & \small{1384.5} & \small{366} & \small{137.3} & \small{1696.8} & \small{1643} & \small{354} & \small{121.2} \\
\bottomrule
\end{tabularx}
\end{table*}

\begin{table*}[hb!]
\caption{The benchmark raw data results and schemas for the plot in \autoref{fig:benchmark-jsonresume}.}
\label{table:benchmark-jsonresume}
\begin{tabularx}{\linewidth}{X|l|l|l|l|l}
\toprule
\textbf{Serialization Format} & \textbf{Schema} & \textbf{Uncompressed} & \textbf{GZIP} & \textbf{LZ4} & \textbf{LZMA} \\
\midrule
ASN.1 (PER Unaligned) & \href{https://github.com/jviotti/binary-json-size-benchmark/blob/main/benchmark/jsonresume/asn1/schema.asn}{\small{\texttt{schema.asn}}} & 2143 & 1282 & 1754 & 1302 \\ \hline
Apache Avro (unframed) & \href{https://github.com/jviotti/binary-json-size-benchmark/blob/main/benchmark/jsonresume/avro/schema.json}{\small{\texttt{schema.json}}} & 2162 & 1296 & 1772 & 1316 \\ \hline
Microsoft Bond (Compact Binary v1) & \href{https://github.com/jviotti/binary-json-size-benchmark/blob/main/benchmark/jsonresume/bond/schema.bond}{\small{\texttt{schema.bond}}} & 2256 & 1385 & 1860 & 1398 \\ \hline
Cap'n Proto (Binary Encoding) & \href{https://github.com/jviotti/binary-json-size-benchmark/blob/main/benchmark/jsonresume/capnproto/schema.capnp}{\small{\texttt{schema.capnp}}} & 3152 & 1627 & 2230 & 1566 \\ \hline
Cap'n Proto (Packed Encoding) & \href{https://github.com/jviotti/binary-json-size-benchmark/blob/main/benchmark/jsonresume/capnproto-packed/schema.capnp}{\small{\texttt{schema.capnp}}} & 2549 & 1668 & 2162 & 1668 \\ \hline
FlatBuffers & \href{https://github.com/jviotti/binary-json-size-benchmark/blob/main/benchmark/jsonresume/flatbuffers/schema.fbs}{\small{\texttt{schema.fbs}}} & 3116 & 1686 & 2290 & 1627 \\ \hline
Protocol Buffers (Binary Wire Format) & \href{https://github.com/jviotti/binary-json-size-benchmark/blob/main/benchmark/jsonresume/protobuf/schema.proto}{\small{\texttt{schema.proto}}} & 2225 & 1370 & 1832 & 1371 \\ \hline
Apache Thrift (Compact Protocol) & \href{https://github.com/jviotti/binary-json-size-benchmark/blob/main/benchmark/jsonresume/thrift/schema.thrift}{\small{\texttt{schema.thrift}}} & 2230 & 1337 & 1815 & 1352 \\ \hline
\hline \textbf{JSON} & - & 3048 & 1556 & 2204 & 1586 \\ \hline \hline
BSON & - & 3283 & 1782 & 2449 & 1762 \\ \hline
CBOR & - & 2754 & 1555 & 2164 & 1582 \\ \hline
FlexBuffers & - & 3049 & 1966 & 2560 & 1936 \\ \hline
MessagePack & - & 2749 & 1600 & 2163 & 1615 \\ \hline
Smile & - & 2620 & 1581 & 2169 & 1621 \\ \hline
UBJSON & - & 2977 & 1658 & 2277 & 1665 \\
\bottomrule
\end{tabularx}
\end{table*}

\begin{figure*}[ht!]
\frame{\includegraphics[width=\linewidth]{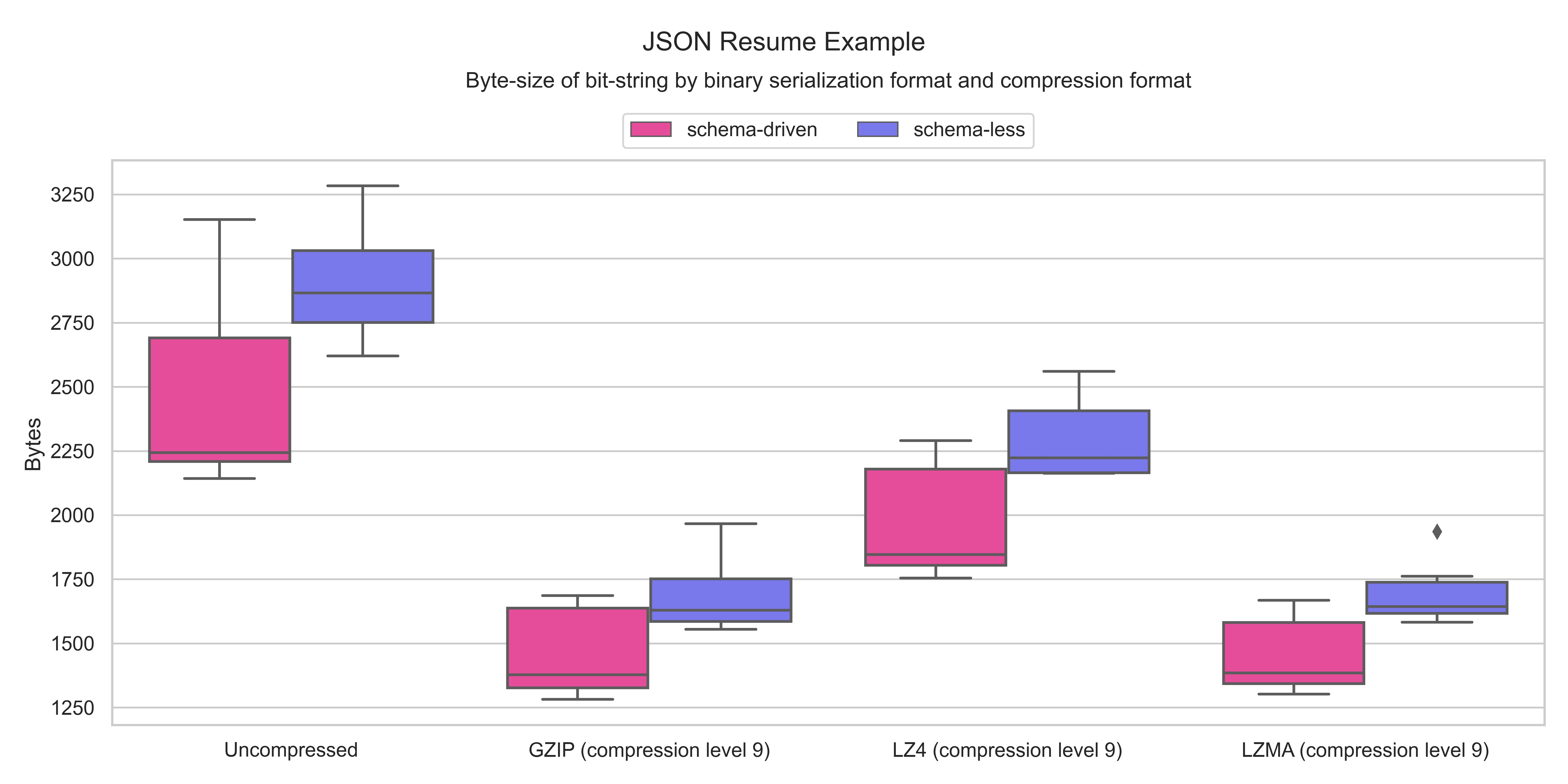}}
\caption{
Box plot of the statistical results in \autoref{table:benchmark-stats-jsonresume}.
}
\label{fig:benchmark-jsonresume-boxplot}
\end{figure*}

In \autoref{fig:benchmark-jsonresume-boxplot}, we observe the medians for
uncompressed schema-driven binary serialization specifications to be smaller in
comparison to uncompressed schema-less binary serialization specifications.  The range
between the upper and lower whiskers and the inter-quartile range of
uncompressed schema-less binary serialization specifications is smaller than the range
between the upper and lower whiskers and the inter-quartile range of
uncompressed schema-driven binary serialization specifications.


In terms of compression, GZIP results in the lower median for both
schema-driven and schema-less binary serialization specifications.  Additionally,
GZIP, LZ4 and LZMA are space-efficient in terms of the median in comparison to
both uncompressed schema-driven and schema-less binary serialization specifications.
However, the use of LZMA for schema-less binary serialization specifications exhibits
upper outliers.  Nevertheless, compression reduces the range between the upper
and lower whiskers and inter-quartile range for both schema-driven and
schema-less binary serialization specifications.  In particular, the compression
format with the smaller range between the upper and lower whiskers and the
smaller inter-quartile range for both schema-driven and schema-less binary
serialization specifications is LZMA.


Overall, \we conclude that uncompressed schema-driven binary serialization
specifications are space-efficient in comparison to uncompressed schema-less
binary serialization specifications and that all the considered compression
formats are space-efficient in comparison to uncompressed schema-driven and
schema-less binary serialization specifications.

\clearpage

\subsection{Nightwatch.js Test Framework Configuration}
\label{sec:benchmark-nightwatch}

\begin{figure*}[ht!]
\frame{\includegraphics[width=\linewidth]{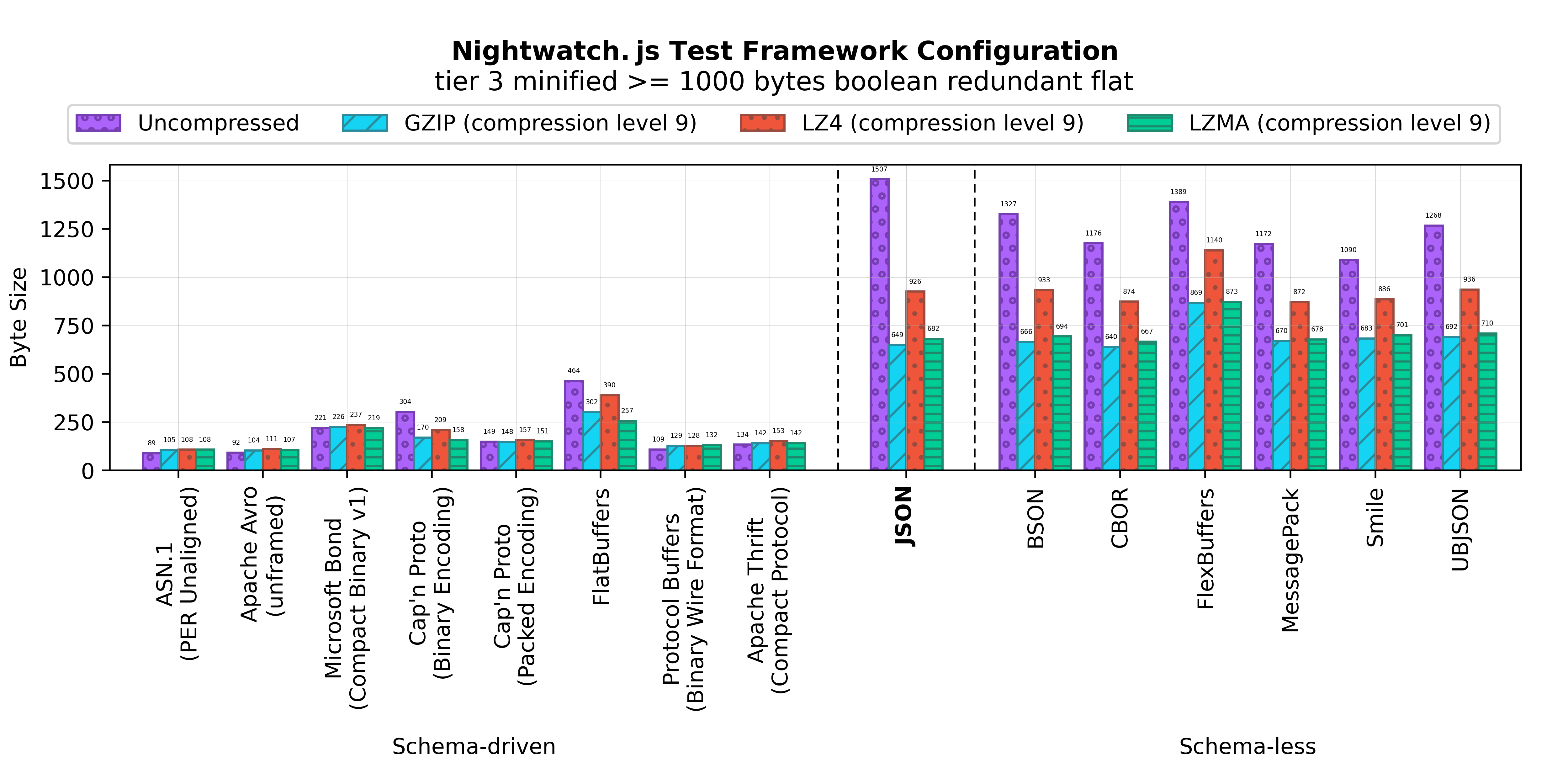}}
\caption{
The benchmark results for the Nightwatch.js Test Framework Configuration test case listed in \autoref{table:benchmark-documents} and \autoref{table:benchmark-documents-1}.
}
\label{fig:benchmark-nightwatch}
\end{figure*}

Nightwatch.js \footnote{\url{https://nightwatchjs.org}} is an open-source
browser automation solution used in the software testing industry. In
\autoref{fig:benchmark-nightwatch}, \we demonstrate a \textbf{Tier 3 minified
$\geq$ 1000 bytes boolean redundant flat} (Tier 3 BRF from
\autoref{table:json-taxonomy}) JSON document that consists of a Nightwatch.js
configuration file that defines a set of general-purpose WebDriver
\cite{webdriver} and Selenium \footnote{\url{https://www.selenium.dev}}
options.

The smallest bit-string is produced by ASN.1 PER Unaligned \cite{asn1-per} (89
bytes), followed by Apache Avro \cite{avro} (92 bytes) and Protocol Buffers
\cite{protocolbuffers} (109 bytes). The binary serialization specifications that
produced the smallest bit-strings are schema-driven and sequential \cite{viotti2022survey}.
Conversely, the largest bit-string is produced by FlexBuffers
\cite{flexbuffers} (1389 bytes), followed by BSON \cite{bson} (1327 bytes) and
UBJSON \cite{ubjson} (1268 bytes).  The binary serialization specifications that
produced the largest bit-strings are schema-less and with the exception of
FlexBuffers, they are also sequential \cite{viotti2022survey}.  In comparison to JSON
\cite{ECMA-404} (1507 bytes), binary serialization achieves a \textbf{16.9x}
size reduction in the best case for this input document.  Similar large size
reductions are observed in JSON documents whose content is dominated by
\emph{boolean} and \emph{numeric} values.  None of the 14 JSON-compatible
binary serialization specifications listed in
\autoref{table:benchmark-specifications-schema-driven} and
\autoref{table:benchmark-specifications-schema-less} result in bit-strings that are
larger than JSON.

For this Tier 3 BRF document, the best performing schema-driven serialization
specification achieves a \textbf{12.2x} size reduction compared to the best performing
schema-less serialization specification: Smile \cite{smile} (1090 bytes).  As shown in
\autoref{table:benchmark-stats-nightwatch}, uncompressed schema-driven specifications
provide smaller \emph{average} and \emph{median} bit-strings than uncompressed
schema-less specifications. However, as highlighted by the \emph{range} and
\emph{standard deviation}, uncompressed schema-driven specifications exhibit higher
size reduction variability depending on the expressiveness of the schema
language (i.e. how the language constructs allow you to model the data) and the
size optimizations devised by its authors. The entire selection of
schema-driven serialization specifications listed in
\autoref{table:benchmark-specifications-schema-driven} produce bit-strings that are
equal to or smaller than their schema-less counterparts listed in
\autoref{table:benchmark-specifications-schema-less}.  The best performing sequential
serialization specification only achieves a \textbf{1.6x} size reduction compared to
the best performing pointer-based serialization specification: Cap'n Proto Packed
Encoding \cite{capnproto} (149 bytes).

The compression formats listed in
\autoref{sec:benchmark-compression-formats} result in positive gains for
all bit-strings except the ones produced by ASN.1 PER Unaligned
\cite{asn1-per}, Apache Avro \cite{avro}, Microsoft Bond \cite{microsoft-bond},
Protocol Buffers \cite{protocolbuffers} and Apache Thrift
\cite{slee2007thrift}. The best performing uncompressed binary serialization
specification achieves a \textbf{7.2x} size reduction compared to the best
performing compression format for JSON: GZIP \cite{RFC1952} (649 bytes).

\begin{table*}[hb!]
\caption{A byte-size statistical analysis of the benchmark results shown in \autoref{fig:benchmark-nightwatch} divided by schema-driven and schema-less specifications.}
\label{table:benchmark-stats-nightwatch}
\begin{tabularx}{\linewidth}{X|l|l|l|l|l|l|l|l}
\toprule
\multirow{2}{*}{\textbf{Category}} &
\multicolumn{4}{c|}{\textbf{Schema-driven}} &
\multicolumn{4}{c}{\textbf{Schema-less}} \\
\cline{2-9}
& \small\textbf{Average} & \small\textbf{Median} & \small\textbf{Range} & \small\textbf{Std.dev} & \small\textbf{Average} & \small\textbf{Median} & \small\textbf{Range} & \small\textbf{Std.dev} \\
\midrule
Uncompressed & \small{195.3} & \small{141.5} & \small{375} & \small{122.5} & \small{1237} & \small{1222} & \small{299} & \small{101.4} \\ \hline
GZIP (compression level 9) & \small{165.8} & \small{145} & \small{198} & \small{63.2} & \small{703.3} & \small{676.5} & \small{229} & \small{75.8} \\ \hline
LZ4 (compression level 9) & \small{186.6} & \small{155} & \small{282} & \small{87.9} & \small{940.2} & \small{909.5} & \small{268} & \small{93.1} \\ \hline
LZMA (compression level 9) & \small{159.3} & \small{146.5} & \small{150} & \small{49.5} & \small{720.5} & \small{697.5} & \small{206} & \small{69.7} \\
\bottomrule
\end{tabularx}
\end{table*}

\begin{table*}[hb!]
\caption{The benchmark raw data results and schemas for the plot in \autoref{fig:benchmark-nightwatch}.}
\label{table:benchmark-nightwatch}
\begin{tabularx}{\linewidth}{X|l|l|l|l|l}
\toprule
\textbf{Serialization Format} & \textbf{Schema} & \textbf{Uncompressed} & \textbf{GZIP} & \textbf{LZ4} & \textbf{LZMA} \\
\midrule
ASN.1 (PER Unaligned) & \href{https://github.com/jviotti/binary-json-size-benchmark/blob/main/benchmark/nightwatch/asn1/schema.asn}{\small{\texttt{schema.asn}}} & 89 & 105 & 108 & 108 \\ \hline
Apache Avro (unframed) & \href{https://github.com/jviotti/binary-json-size-benchmark/blob/main/benchmark/nightwatch/avro/schema.json}{\small{\texttt{schema.json}}} & 92 & 104 & 111 & 107 \\ \hline
Microsoft Bond (Compact Binary v1) & \href{https://github.com/jviotti/binary-json-size-benchmark/blob/main/benchmark/nightwatch/bond/schema.bond}{\small{\texttt{schema.bond}}} & 221 & 226 & 237 & 219 \\ \hline
Cap'n Proto (Binary Encoding) & \href{https://github.com/jviotti/binary-json-size-benchmark/blob/main/benchmark/nightwatch/capnproto/schema.capnp}{\small{\texttt{schema.capnp}}} & 304 & 170 & 209 & 158 \\ \hline
Cap'n Proto (Packed Encoding) & \href{https://github.com/jviotti/binary-json-size-benchmark/blob/main/benchmark/nightwatch/capnproto-packed/schema.capnp}{\small{\texttt{schema.capnp}}} & 149 & 148 & 157 & 151 \\ \hline
FlatBuffers & \href{https://github.com/jviotti/binary-json-size-benchmark/blob/main/benchmark/nightwatch/flatbuffers/schema.fbs}{\small{\texttt{schema.fbs}}} & 464 & 302 & 390 & 257 \\ \hline
Protocol Buffers (Binary Wire Format) & \href{https://github.com/jviotti/binary-json-size-benchmark/blob/main/benchmark/nightwatch/protobuf/schema.proto}{\small{\texttt{schema.proto}}} & 109 & 129 & 128 & 132 \\ \hline
Apache Thrift (Compact Protocol) & \href{https://github.com/jviotti/binary-json-size-benchmark/blob/main/benchmark/nightwatch/thrift/schema.thrift}{\small{\texttt{schema.thrift}}} & 134 & 142 & 153 & 142 \\ \hline
\hline \textbf{JSON} & - & 1507 & 649 & 926 & 682 \\ \hline \hline
BSON & - & 1327 & 666 & 933 & 694 \\ \hline
CBOR & - & 1176 & 640 & 874 & 667 \\ \hline
FlexBuffers & - & 1389 & 869 & 1140 & 873 \\ \hline
MessagePack & - & 1172 & 670 & 872 & 678 \\ \hline
Smile & - & 1090 & 683 & 886 & 701 \\ \hline
UBJSON & - & 1268 & 692 & 936 & 710 \\
\bottomrule
\end{tabularx}
\end{table*}

\begin{figure*}[ht!]
\frame{\includegraphics[width=\linewidth]{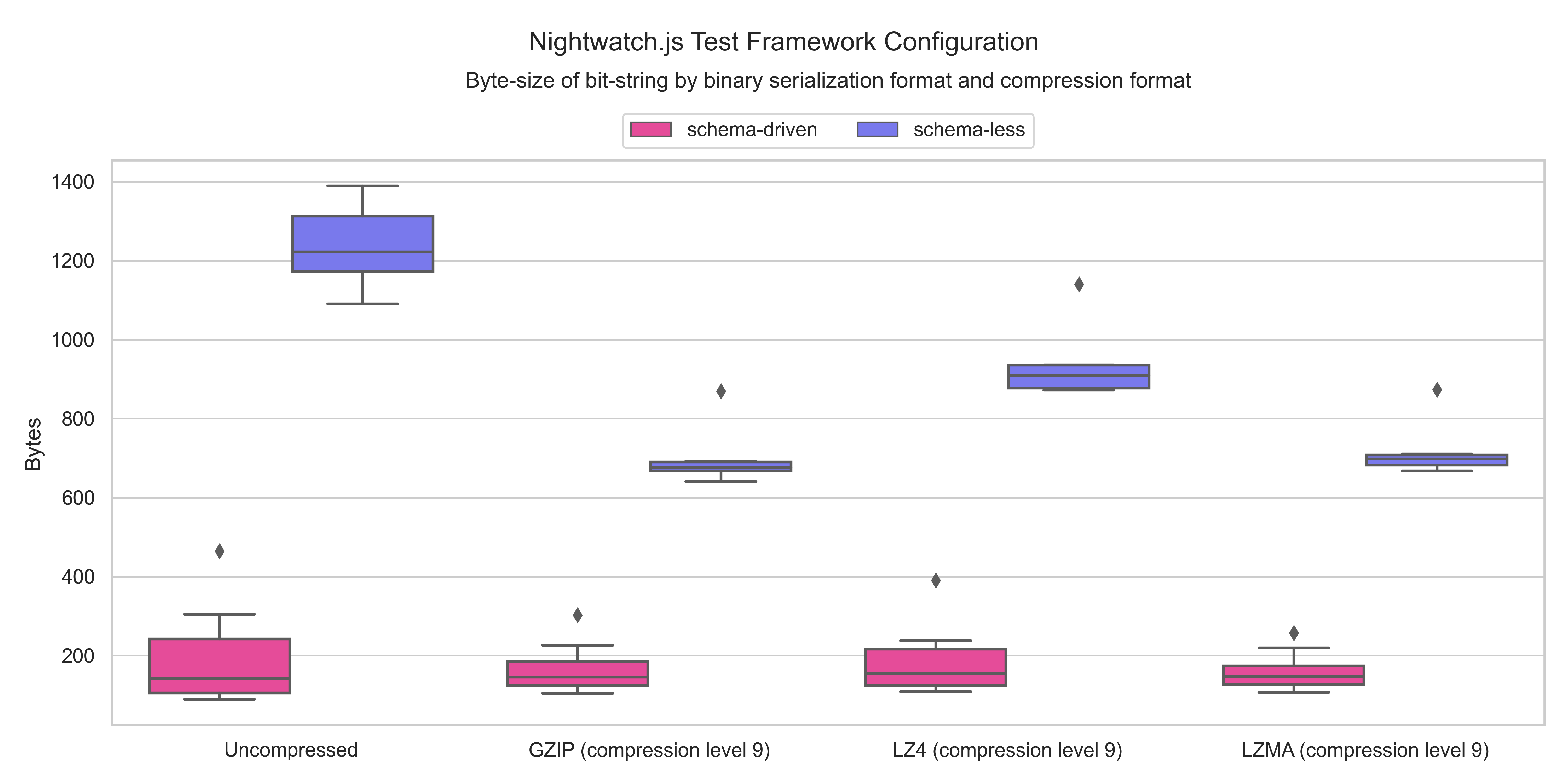}}
\caption{
Box plot of the statistical results in \autoref{table:benchmark-stats-nightwatch}.
}
\label{fig:benchmark-nightwatch-boxplot}
\end{figure*}

In \autoref{fig:benchmark-nightwatch-boxplot}, we observe the medians for
uncompressed schema-driven binary serialization specifications to be smaller in
comparison to uncompressed schema-less binary serialization specifications.  The range
between the upper and lower whiskers of uncompressed schema-driven binary
serialization specifications is smaller than the range between the upper and lower
whiskers of uncompressed schema-less binary serialization specifications.  However,
the inter-quartile range of both both uncompressed schema-driven and
schema-less binary serialization specifications is similar.


In terms of compression, GZIP and LZMA result in the lower medians for
schema-driven binary serialization specifications while GZIP results in the
lower median for schema-less binary serialization specifications.  Compression
is not space-efficient in terms of the median in comparison to uncompressed
schema-driven binary serialization specifications. However, GZIP, LZ4 and LZMA
are space-efficient in terms of the median in comparison to uncompressed
schema-less binary serialization specifications.  Additionally, the use of
GZIP, LZ4 and LZMA for both schema-driven binary serialization specifications
and schema-less binary serialization specifications exhibits upper outliers.
Nevertheless, compression reduces the range between the upper and lower
whiskers and inter-quartile range for both schema-driven and schema-less binary
serialization specifications.  In particular, the compression format with the
smaller inter-quartile range for schema-driven binary serialization
specifications is LZMA, the compression format with the smaller range between
the upper and lower whiskers for schema-less binary serialization
specifications is LZMA, and the compression format with the smaller
inter-quartile range for schema-less binary serialization specifications is
GZIP.


Overall, \we conclude that uncompressed schema-driven binary serialization
specifications are space-efficient in comparison to uncompressed schema-less binary
serialization specifications. Compression does not contribute to space-efficiency in
comparison to schema-driven binary serialization specifications but all the considered
compression formats are space-efficient in comparison to schema-less binary
serialization specifications.

\section{Reproducibility}
\label{sec:benchmark-reproducibility}

To encourage reproducibility for the benchmarking datasets, \we follow every
reproducibility level introduced by \cite{sam_harrison_2021_4761867}. \We found
that the implemented benchmark software matches the definition of Level 3, the
highest-level of reproducibility, as justified in the following sections.

\begin{itemize}

\item \textbf{Automation.} The benchmark software, from the generation of the
serialized bit-strings to the generation of the plots using Matplotlib
\footnote{\url{https://matplotlib.org}}, is automated through a GNU Make
\footnote{\url{https://www.gnu.org/software/make/}} declarative and
parallelizable build definition.

\item \textbf{Testing.} The POSIX shell and Python scripts distributed with the
benchmark are automatically linted using the \emph{shellcheck}
\footnote{\url{https://www.shellcheck.net}} and \emph{flake8}
\footnote{\url{https://flake8.pycqa.org/}} open-source tools, respectively. The
serialization and deserialization procedures of the benchmark are automatically
tested as explained in \autoref{sec:benchmark-testing}.

\item \textbf{Supported Environments.} The benchmark software is ensured to
  work on the macOS (Intel processors) and GNU/Linux operating systems. \We do
    not make any effort to support the Microsoft Windows operating system, but
    \we expect the benchmark software to run on an \emph{msys2}
    \footnote{\url{https://www.msys2.org}} or \emph{Windows Subsystem for
    Linux} \footnote{\url{https://docs.microsoft.com/en-us/windows/wsl/}}
    environment with minor changes at most. The benchmark is exclusively
    concerned with the byte-size of the bit-strings produced by the binary
    serialization specifications.  Therefore, the CPU, memory, and network
    bandwidth characteristics of the test machine do not affect the results of
    the benchmark. No further conditions apart from the exact software versions
    of the dependencies included and required by the project are necessary to
    replicate the results.

\item \textbf{Documentation and Readability.} The \texttt{README} file
  \footnote{\url{https://github.com/jviotti/binary-json-size-benchmark\#running-locally}}
    in the repository contains precise instructions for running the benchmark
    locally and generate the data files and plots.  The project documentation
    includes a detailed list of the system dependencies that are required to
    succesfully execute every part of the benchmark and a detailed list of the
    required binary serialization specifications, implementations, versions,
    and encodings. The benchmark source code is compact and easy to understand
    and navigate due to the declarative rule definition nature of GNU Make.

\item \textbf{DOI.} The version of the benchmark software described in this
  study is archived with a DOI \cite{juan_cruz_viotti_2022_5829569}. The DOI
    includes the source code for reproducing the benchmark and the presented
    results.

\item \textbf{Dependencies.} The benchmark software is implemented using
  established open-source software with the exception of the ASN-1Step
    \footnote{\url{https://www.oss.com/asn1/products/asn-1step/asn-1step.html}}
    command-line tool, which is a proprietary implementation of ASN.1
    \cite{asn1} distributed by OSS Nokalva with a 30 days free trial. Every
    binary serialization specification implementation used in the benchmark
    with the exception of ASN-1Step is pinned to its specific version to ensure
    reproducibility.  As explained in the online documentation, the benchmark
    software expects the ASN-1Step command-line tool version 10.0.2 to be
    installed and globally-accessible in the system in order to benchmark the
    ASN.1 PER Unaligned \cite{asn1-per} binary serialization specification.

\item \textbf{Version Control.} The benchmark repository utilises the
  \emph{git} \footnote{\url{https://git-scm.com}} version control system and
    its publicly hosted on GitHub
    \footnote{\url{https://github.com/jviotti/binary-json-size-benchmark}} as
    recommended by \cite{peng2011reproducible}.

\item \textbf{Continuous Integration.} The GitHub repository hosting the
  benchmark software is setup with the GitHub Actions
  \footnote{\url{https://github.com/features/actions}} continuous integration
  provided to re-run the benchmark automatically on new commits using a
  GNU/Linux Ubuntu 20.04 LTS cloud worker. This process prevents changes to the
  benchmark software from introducing regressions and new software errors. \We
  make use of this process to validate GitHub internal and external pull
  requests before merging them into the trunk.

\item \textbf{Availability.} The benchmark software and results are publicly
  available and governed by the \emph{Apache License 2.0}
  \footnote{\url{https://www.apache.org/licenses/LICENSE-2.0.html}} open-source
  software license. The results of the benchmark are also published as a
  website hosted at \url{https://www.jviotti.com/binary-json-size-benchmark/}
  using the GitHub Pages free static hosting provider. The website provides
  direct links to the JSON \cite{ECMA-404} documents being encoded by the
  benchmark and direct links to the schema definitions used in every case. Both
  the JSON documents and the schema definitions are hosted in the benchmark
  GitHub repository to ensure their availability even if the original sources
  do not exist anymore.

\item \textbf{Continuity.} \We plan to continue extending the benchmark
  software in the future to test new versions of the current selection of
  binary serialization specifications and to include new JSON-compatible binary
  serialization specifications. \We hope for this project to become a
  collaborative effort to measure the space-efficiency of every new
  JSON-compatible serialization specifications and \Weare comitted to accepting
  open-source contributions.

\end{itemize}

\section{Conclusions}
\label{sec:benchmark-conclusions}

\subsection{Q1: How do JSON-compatible schema-less binary serialization specifications
compare to JSON in terms of space-efficiency?}
\label{sec:benchmark-conclusion-json-vs-schema-less}

\begin{figure*}[ht!]
  \frame{\includegraphics[width=\linewidth]{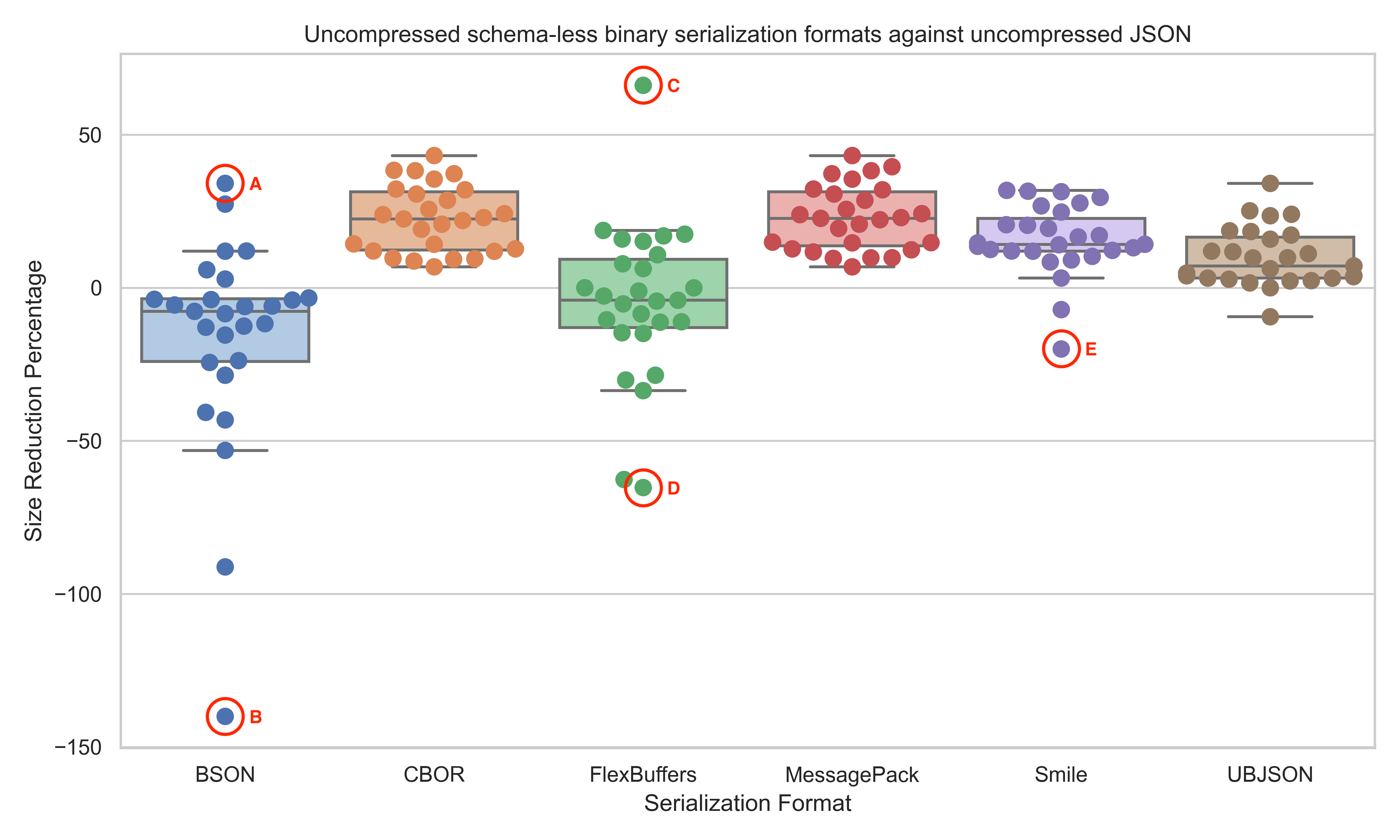}}
  \caption{A box plot that demonstrates the size reduction (in percentages) of
  the selection of schema-less binary serialization specifications listed in
  \autoref{table:benchmark-specifications-schema-less} in comparison to
  uncompressed JSON \cite{ECMA-404} given the input data listed in
  \autoref{table:benchmark-documents} and
  \autoref{table:benchmark-documents-1}.}
\label{fig:benchmark-json-vs-schema-less} \end{figure*}

\autoref{table:json-vs-schema-less} demonstrates that the median size reduction
of the selection of schema-less binary serialization specifications listed in
\autoref{table:benchmark-specifications-schema-less} is 9.1\% and the average
size reductions of the selection of schema-less binary serialization
specifications listed in \autoref{table:benchmark-specifications-schema-less}
is 8.2\% for the selection of input data set described in
\autoref{table:benchmark-documents} and \autoref{table:benchmark-documents-1}.
In comparison to JSON \cite{ECMA-404}, FlexBuffers \cite{flexbuffers} and BSON
\cite{bson} often result in larger bit-strings. In comparison to JSON, both
CBOR \cite{RFC7049} and MessagePack \cite{messagepack} are strictly superior in
terms of space-efficiency.  In both cases, the median and average size
reductions ranged between 22.4\% and 22.8\% for the selection of input data.
Compared to the other schema-less binary serialization specifications,
MessagePack \cite{messagepack} tends to provide the best size reductions in the
\emph{Tier 1 Minified $<$ 100 bytes} and \emph{Tier 2 Minified $\geq$ 100 $<$
1000 bytes} categories while Smile \cite{smile} tends to provide the best size
reductions for \emph{Tier 3 Minified $\geq$ 1000 bytes} JSON documents. As a
notable positive exception shown in
\autoref{fig:benchmark-json-vs-schema-less}, FlexBuffers \cite{flexbuffers}
outperforms the rest of the schema-less binary serialization specifications in
two cases: the \emph{Tier 2 Minified $\geq$ 100 $<$ 1000 bytes, textual,
redundant, and flat} (Tier 2 TRF) JSON document from
\autoref{sec:benchmark-travisnotifications} (C) and the \emph{Tier 3 Minified
$\geq$ 1000 bytes, textual, redundant, and flat} (Tier 3 TRF) JSON document
from \autoref{sec:benchmark-packagejsonlintrc} given its automatic string
deduplication features \cite{viotti2022survey}.
\autoref{fig:benchmark-json-vs-schema-less} shows that CBOR \cite{RFC7049} and
MessagePack \cite{messagepack} tend to outperform the other schema-less binary
serialization specifications in terms of space-efficiency while producing
stable results with no noticeable outliers.  In comparison, BSON \cite{bson} (A
and B) and FlexBuffers \cite{flexbuffers} (C and D) produce noticeable outliers
at both sides of the spectrum while remaining less space-efficient than the
rest of the schema-less binary serialization specifications in most cases. Like
BSON \cite{bson} (B), Smile \cite{smile} produces a negative outlier (E) for
the Tier 2 NRN case from \autoref{sec:benchmark-geojson}.

\textbf{Summary.} There exists schema-less binary serialization specifications
that are space-efficient in comparison to JSON \cite{ECMA-404}. Based on \our
findings, \we conclude that using MessagePack \cite{messagepack} on \emph{Tier
1 Minified $<$ 100 bytes} and \emph{Tier 2 Minified $\geq$ 100 $<$ 1000 bytes}
JSON documents, Smile \cite{smile} on \emph{Tier 3 Minified $\geq$ 1000 bytes}
JSON documents, and FlexBuffers \cite{flexbuffers} on JSON documents with
high-redundancy of \emph{textual} values increases space-efficiency.

\begin{table}[hb!]

\caption{A summary of the size reduction results in comparison to JSON
  \cite{ECMA-404} of the selection of schema-less binary serialization
  specifications listed in \autoref{table:benchmark-specifications-schema-less}
  against the input data listed in \autoref{table:benchmark-documents} and
  \autoref{table:benchmark-documents-1}. See
  \autoref{fig:benchmark-json-vs-schema-less} for a visual representation of
  this data.}

\label{table:json-vs-schema-less}
\begin{tabularx}{\linewidth}{X|l|l|l|l|l|l}
\toprule
  \multirow{2}{\benchmarkconclusionrow}{\textbf{Serialization Specification}} & \multicolumn{5}{c|}{\textbf{Size Reductions in Comparison To JSON}} & \multirow{2}{*}{\textbf{Negative Cases}} \\ \cline{2-6}
& \textbf{Maximum} & \textbf{Minimum} & \textbf{Range} & \textbf{Median} & \textbf{Average} & \\
\midrule
BSON        & 34.1\% & -140.0\% & 174.1 & -7.7\% & -16.8\% & 21 / 27 (77.7\%) \\ \hline
CBOR        & 43.2\% &    6.8\% &  36.3 & 22.5\% &  22.4\% & 0 / 27 (0\%) \\ \hline
FlexBuffers & 66.1\% &  -65.3\% & 131.4 & -4.1\% &  -4.9\% & 16 / 27 (59.2\%) \\ \hline
MessagePack & 43.2\% &    6.8\% &  36.3 & 22.7\% &  22.8\% & 0 / 27 (0\%) \\ \hline
Smile       & 31.8\% &  -20.0\% &  51.8 & 14.2\% &  15.5\% & 2 / 27 (7.4\%) \\ \hline
UBJSON      & 34.1\% &   -9.5\% &  43.6 &  7.1\% &   9.9\% & 1 / 27 (3.7\%) \\ \hline
\textbf{Averages} & \textbf{42.1\%} & \textbf{-36.8\%} & \textbf{78.9} & \textbf{9.1\%} & \textbf{8.2\%} & \textbf{24.6\%} \\
\bottomrule
\end{tabularx}
\end{table}

\subsection{Q2: How do JSON-compatible schema-driven binary serialization
specifications compare to JSON and JSON-compatible schema-less binary serialization
specifications in terms of space-efficiency?}

\begin{figure*}[ht!]
  \frame{\includegraphics[width=\linewidth]{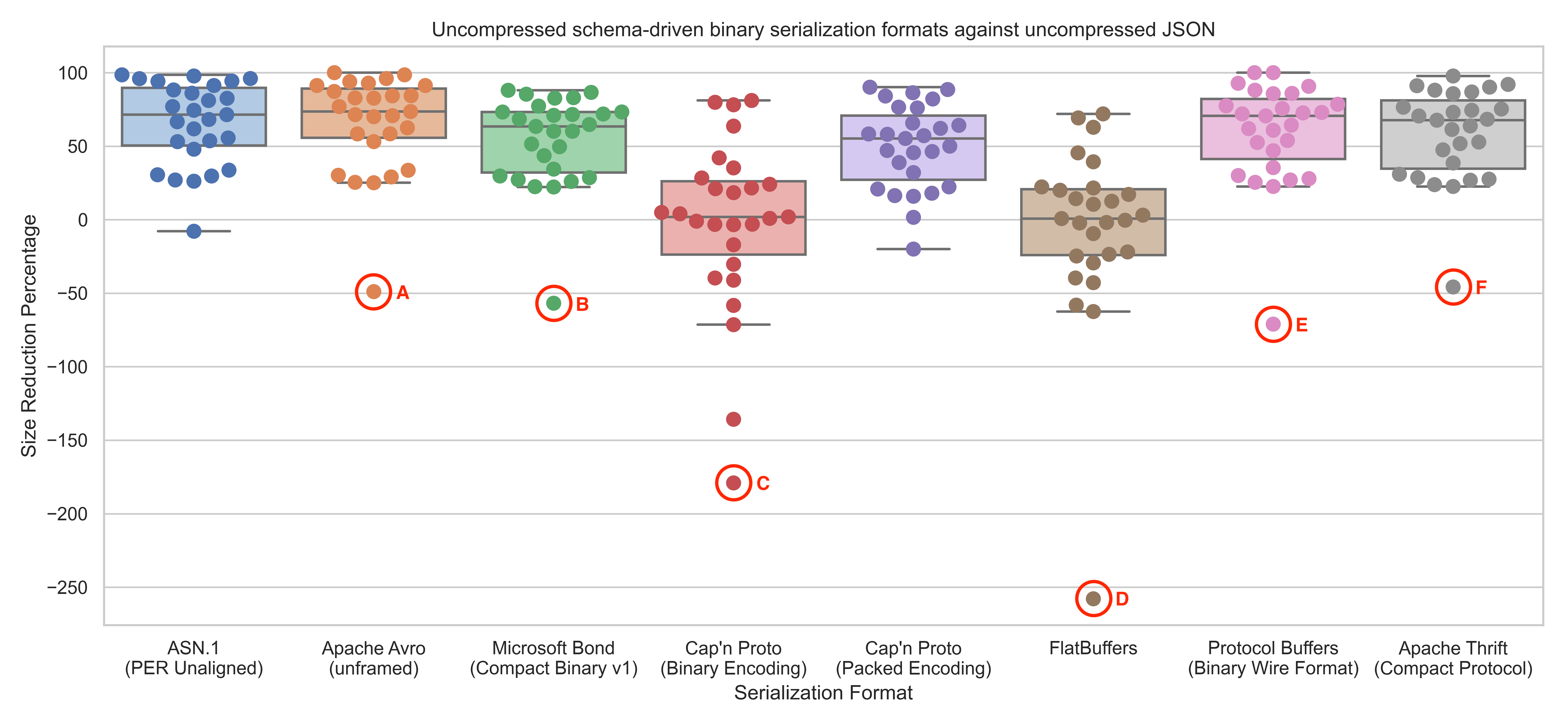}}
  \caption{A box plot that demonstrates the size reduction (in percentages) of
  the selection of schema-driven binary serialization specifications listed in
  \autoref{table:benchmark-specifications-schema-driven} in comparison to
  uncompressed JSON \cite{ECMA-404} given the input data listed in
  \autoref{table:benchmark-documents} and
  \autoref{table:benchmark-documents-1}.}
\label{fig:benchmark-json-vs-schema-driven} \end{figure*}

As illustrated in \autoref{table:json-vs-schema-driven}, the median size
reduction of the selection of schema-driven binary serialization specifications
listed in \autoref{table:benchmark-specifications-schema-driven} is more than
five times higher that the schema-less binary serialization specification size
reductions listed in \autoref{table:json-vs-schema-less} and the average size
reduction of the selection of schema-driven binary serialization specifications
listed in \autoref{table:benchmark-specifications-schema-driven} is more than
five times higher that the schema-less binary serialization specification size
reductions listed in \autoref{table:json-vs-schema-less} for the given input
data described in \autoref{table:benchmark-documents} and
\autoref{table:benchmark-documents-1}. FlatBuffers \cite{flatbuffers} and Cap'n
Proto \cite{capnproto} (unpacked) tend to be less space-efficient than the
selection of schema-less binary serialization specifications and are surpassed
by the rest of the schema-driven binary serialization specifications listed in
\autoref{table:benchmark-specifications-schema-driven} for most cases. On the
other side, ASN.1 PER Unaligned \cite{asn1-per} and Apache Avro (unframed)
\cite{avro} are the most space-efficient schema-driven binary serialization
specifications in 23 out of the 27 cases listed in
\autoref{table:benchmark-documents}.  Most of the schema-driven binary
serialization specifications \we considered are strictly superior to JSON
\cite{ECMA-404} and to the schema-less binary serialization specifications
listed in \autoref{table:benchmark-specifications-schema-less} in terms of
message size with a common exception: ASN.1 PER Unaligned \cite{asn1-per},
Apache Avro (unframed) \cite{avro}, Microsoft Bond (Compact Binary v1)
\cite{microsoft-bond}, Protocol Buffers \cite{protocolbuffers}, Apache Thrift
(Compact Protocol) \cite{slee2007thrift}, and Cap'n Proto \cite{capnproto}
(packed) perform less space-efficiently than JSON \cite{ECMA-404} and the
schema-less binay serialization specifications listed in
\autoref{table:benchmark-specifications-schema-less} in the \emph{Tier 2
Minified $\geq$ 100 $<$ 1000 bytes, numeric, redundant, and nested} (Tier 2
NRN) GeoJSON \cite{RFC7946} document from \autoref{sec:benchmark-geojson}. With
the exception of ASN.1 PER Unaligned \cite{asn1-per} and Cap'n Proto Packed
Encoding \cite{capnproto}, the selection of schema-driven binary serialization
specifications result in negative outliers for the Tier 2 NRN case as shown in
\autoref{fig:benchmark-json-vs-schema-driven} (A, B, C, D, E and F). Compared
to the other JSON documents from the input data set, this JSON document
consists of highly nested arrays and almost no object keys. Leaving that
exception aside, \we found that in general, the schema-driven binary
serialization specifications listed
\autoref{table:benchmark-specifications-schema-driven} provide the highest
space-efficiency improvements in comparison to JSON \cite{ECMA-404} on
\emph{boolean} documents and tend to provide the least space-efficient
improvements on \emph{textual} JSON documents.
\autoref{fig:benchmark-json-vs-schema-driven} shows that schema-driven binary
serialization specifications, in particular ASN.1 PER Unaligned
\cite{asn1-per}, Apache Avro \cite{avro}, Protocol Buffers
\cite{protocolbuffers} and Apache Thrift \cite{slee2007thrift}, result in high
size reductions in comparison to JSON.  However, every considered schema-driven
binary serialization specification results in at least one negative
space-efficiency exception.

\textbf{Summary.} The schema-driven binary serialization specifications listed
in \autoref{table:benchmark-specifications-schema-driven} tend to be more
space-efficient than the schema-less binary serialization specifications listed
in \autoref{table:benchmark-specifications-schema-less} and JSON
\cite{ECMA-404} in most cases. Based on \our findings, \we conclude that ASN.1
PER Unaligned \cite{asn1-per} and Apache Avro (unpacked) \cite{avro} are
space-efficient in comparison to schema-less binary serialization
specifications in almost all cases as they provide over 70\% median size
reductions and over 65\% average size reductions in comparison to JSON
\cite{ECMA-404}.

\begin{table}[hb!]

\caption{A summary of the size reduction results in comparison to JSON
  \cite{ECMA-404} of the selection of schema-driven binary serialization
  specifications listed in
  \autoref{table:benchmark-specifications-schema-driven} against the input data
  listed in \autoref{table:benchmark-documents} and
  \autoref{table:benchmark-documents-1}. See
  \autoref{fig:benchmark-json-vs-schema-driven} for a visual representation of
  this data.}

\label{table:json-vs-schema-driven}
\begin{tabularx}{\linewidth}{X|l|l|l|l|l|l}
\toprule
  \multirow{2}{\benchmarkconclusionrow}{\textbf{Serialization Specification}} & \multicolumn{5}{c|}{\textbf{Size Reductions in Comparison To JSON}} & \multirow{2}{*}{\textbf{Negative Cases}} \\ \cline{2-6}
& \textbf{Maximum} & \textbf{Minimum} & \textbf{Range} & \textbf{Median} & \textbf{Average} & \\
\midrule
ASN.1 (PER Unaligned)              & 98.5\% & -7.9\% & 106.4 & 71.4\% & 65.7\% & 1 / 27 (3.7\%) \\ \hline
Apache Avro (unframed)             & 100\%  & -48.9\% & 148.9 & 73.5\% & 65.7\% & 1 / 27 (3.7\%) \\ \hline
Microsoft Bond (Compact Binary v1) & 88\%  & -56.8\% & 144.8 & 63.4\% & 54\% & 1 / 27 (3.7\%) \\ \hline
Cap'n Proto                        & 81.1\%  & -179.1\% & 260.1 & 1.9\% & -2.9\% & 12 / 27 (44.4\%) \\ \hline
Cap'n Proto (packed)               & 90.1\%  & -20\% & 110.1 & 55.2\% & 49.6\% & 1 / 27 (3.7\%) \\ \hline
FlatBuffers                        & 72\%  & -257.9\% & 329.8 & 0.7\% & -6.1\% & 13 / 27 (48.1\%) \\ \hline
Protocol Buffers                   & 100\%  & -71.1\% & 171.1 & 70.6\% & 59.3\% & 1 / 27 (3.7\%) \\ \hline
Apache Thrift (Compact Protocol)   & 97.7\%  & -45.8\% & 143.5 & 67.6\% & 58.1\% & 1 / 27 (3.7\%) \\ \hline
\textbf{Averages} & \textbf{90.9\%} & \textbf{-85.9\%} & \textbf{176.9} & \textbf{50.6\%} & \textbf{42.9\%} & \textbf{14.3\%} \\
\bottomrule
\end{tabularx}
\end{table}

\subsection{Q3: How do JSON-compatible sequential binary serialization specifications
compare to JSON-compatible pointer-based binary serialization specifications in terms
of space-efficiency?}

\begin{figure}[hb!]
  \frame{\includegraphics[width=\linewidth]{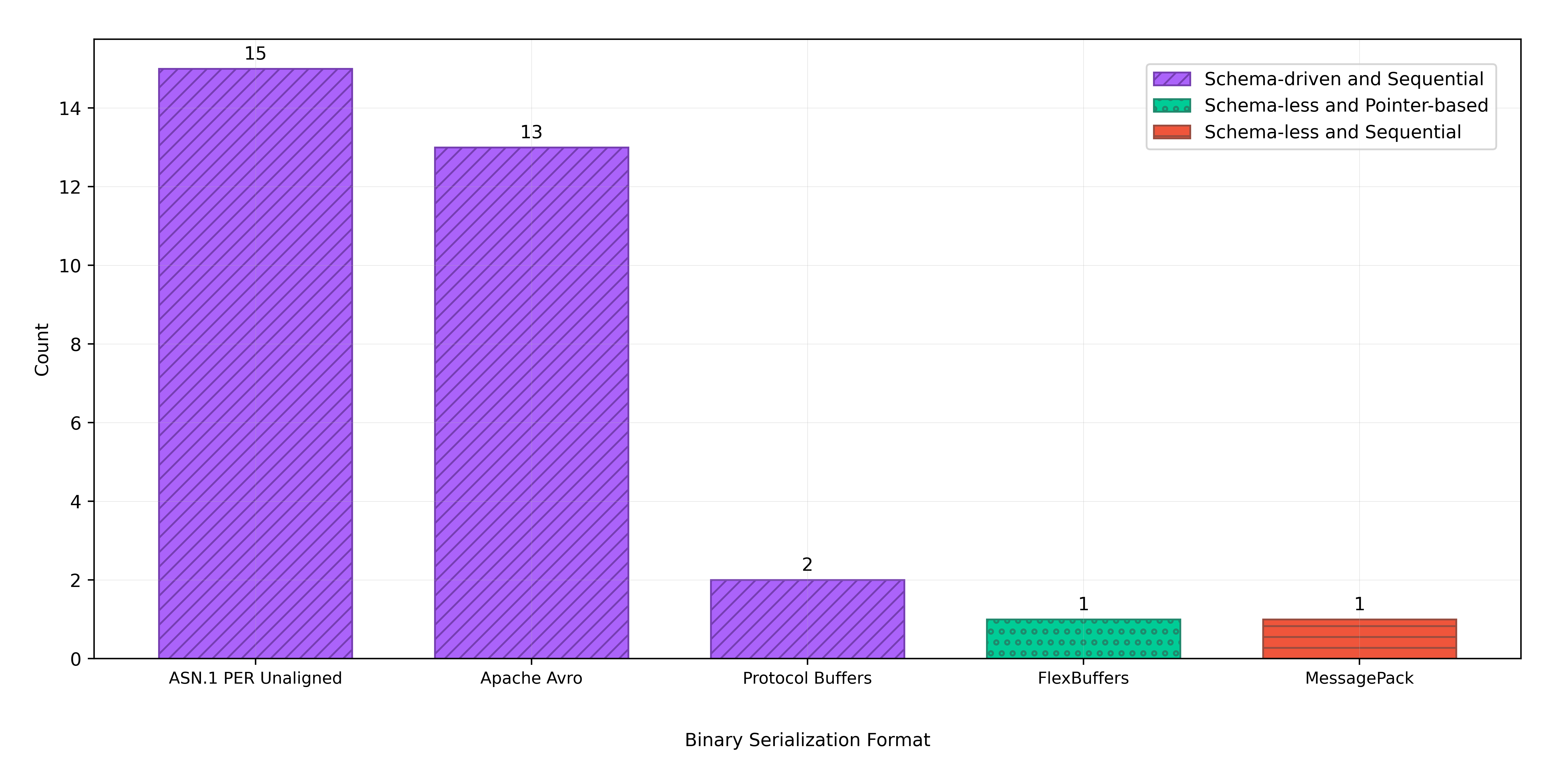}} \caption{The
  binary serialization specifications that resulted in the highest size
  reductions for each JSON \cite{ECMA-404} document for the input data listed
  in \autoref{table:benchmark-documents} and
  \autoref{table:benchmark-documents-1}, broken down by type. Schema-driven
  sequential binary serialization specifications, in particular ASN.1 PER
  Unaligned \cite{asn1-per} and Apache Avro \cite{avro}, resulted in the
  highest size reductions in most cases.} \label{fig:benchmark-rank}
\end{figure}

In terms of the schema-less binary serialization specifications listed in
\autoref{table:benchmark-specifications-schema-less},
\autoref{table:json-vs-schema-less} illustrates that in comparison to JSON
\cite{ECMA-404}, FlexBuffers \cite{flexbuffers} results in negative median and
average size reductions, a characteristic only otherwise applicable to BSON
\cite{bson}. Leaving BSON aside, FlexBuffers only results in more
space-efficient messages than a strict subset of the sequential schema-less
binary serialization specifications in three cases: \emph{Tier 1 TRN}
(\autoref{sec:benchmark-commitlint}), \emph{Tier 2 TRN}
(\autoref{sec:benchmark-epr}) and \emph{Tier 3 TRN}
(\autoref{sec:benchmark-netcoreproject}). Furthemore, FlexBuffers
\cite{flexbuffers} is comparatively more space-efficient that all the other
schema-less binary serialization specifications listed in
\autoref{table:benchmark-specifications-schema-less} for the \emph{Tier 2 TRF} JSON
document from \autoref{sec:benchmark-travisnotifications} and the \emph{Tier 3
TRF} JSON document from \autoref{sec:benchmark-packagejsonlintrc}. However, as
explained in \autoref{sec:benchmark-conclusion-json-vs-schema-less}, this is
due to FlexBuffers automatic string deduplication feature, which is orthogonal
to whether a binary serialization specification is sequential or pointer-based.

\We refer to the schema-driven binary serialization specifications listed in
\autoref{table:benchmark-specifications-schema-driven}.
\autoref{table:json-vs-schema-driven} illustrates that the selection of
sequential schema-driven binary serialization specifications are strictly
superior to FlatBuffers \cite{flatbuffers} in terms of space reductions.
Similarly, Cap'n Proto \cite{capnproto} (unpacked) provides a more
space-efficient bit-string than a single sequential schema-driven binary
serialization specification, Microsoft Bond \cite{microsoft-bond} (Compact
Binary v1), in a single case: \emph{Tier 2 BRF}
(\autoref{sec:benchmark-githubfundingblank}). However, Cap'n Proto
\cite{capnproto} (packed) results in more space-efficient messages than a
strict subset of the sequential schema-driven binary serialization
specifications in six cases: \emph{Tier 1 NNF}
(\autoref{sec:benchmark-circleciblank}), \emph{Tier 1 BRF}
(\autoref{sec:benchmark-sapcloudsdkpipeline}), \emph{Tier 2 NRN}
(\autoref{sec:benchmark-geojson}), \emph{Tier 2 BRF}
(\autoref{sec:benchmark-githubfundingblank}), \emph{Tier 3 NRF}
(\autoref{sec:benchmark-eslintrc}), and \emph{Tier 3 BRF}
(\autoref{sec:benchmark-nightwatch}); but never surpasses the entire set of
sequential schema-driven binary serialization specifications for any JSON
document from the input data set listed in \autoref{table:benchmark-documents}
and \autoref{table:benchmark-documents-1}.

\textbf{Summary.} Based on \our findings, sequential binary serialization
specifications are typically more space-efficient than pointer-based binary
serialization specifications, independent of whether they are schema-less or
schema-driven.

\subsection{Q4: How does compressed JSON compares to uncompressed and
compressed JSON-compatible binary serialization specifications?}


\subsubsection{Data Compression}

\We found that data compression tends to yield negative results on \emph{Tier 1
Minified $<$ 100 bytes} JSON documents. As an extreme, LZMA resulted in a
negative 171.4\% size reduction for \autoref{sec:benchmark-circleciblank}. The
entire selection of data compression formats produced negative results for all
the \emph{Tier 1 Minified $<$ 100 bytes} JSON documents \we considered except
for \autoref{sec:benchmark-tslintmulti}, for which LZ4 produced a negative
result but GZIP \cite{RFC1952} and LZMA resulted in a 8.2\% and 6.1\%
reduction, respectively, and \autoref{sec:benchmark-commitlint}, for which all
data compression formats produced positive results ranging from 10.4\% in the
case of LZ4 to 16.7\% in the case of  GZIP \cite{RFC1952}.  Leaving \emph{Tier
1 Minified $<$ 100 bytes} JSON documents aside, all the data compression
formats \we selected offered better average and median compression ratios on
\emph{textual} JSON documents as seen in \autoref{table:json-compressed}. Out
of the selection of data compression formats, GZIP \cite{RFC1952} performed
better in terms of the average and median size reduction in all \emph{Tier 2
Minified $\geq$ 100 $<$ 1000 bytes} and \emph{Tier 3 Minified $\geq$ 1000
bytes} categories.

\begin{table}[hb!]

\caption{The average and median size reduction of using the selection of data
  compression formats on the \emph{Tier 2 Minified $\geq$ 100 $<$ 1000 bytes}
  and \emph{Tier 3 Minified $\geq$ 1000 bytes} input JSON documents.  GZIP
  \cite{RFC1952} resulted in higher compression ratios for all categories.}

\label{table:json-compressed}
\begin{tabularx}{\linewidth}{X|l|l|l|l|l|l}
\toprule
\multirow{2}{*}{\textbf{Compression Format}} &
\multicolumn{2}{c|}{\textbf{Numeric}} & \multicolumn{2}{c|}{\textbf{Textual}} & \multicolumn{2}{c}{\textbf{Boolean}} \\
\cline{2-7}
& \emph{Average} & \emph{Median} & \emph{Average} & \emph{Median} & \emph{Average} & \emph{Median} \\ \midrule
GZIP (compression level 9) & 39\% & 33.3\% & 54\% & 49.2\% & 28\% & 26.8\% \\ \hline
LZ4 (compression level 9) & 21\% & 19.5\% & 40\% & 32.7\% & 20\% & 8.7\% \\ \hline
LZMA (compression level 9) & 38\% & 32.8\% & 52\% & 48\% & 25\% & 21.3\% \\
\bottomrule
\end{tabularx}
\end{table}


\subsubsection{Schema-less Binary Serialization Specifications}

\begin{figure*}[ht!]
  \frame{\includegraphics[width=\linewidth]{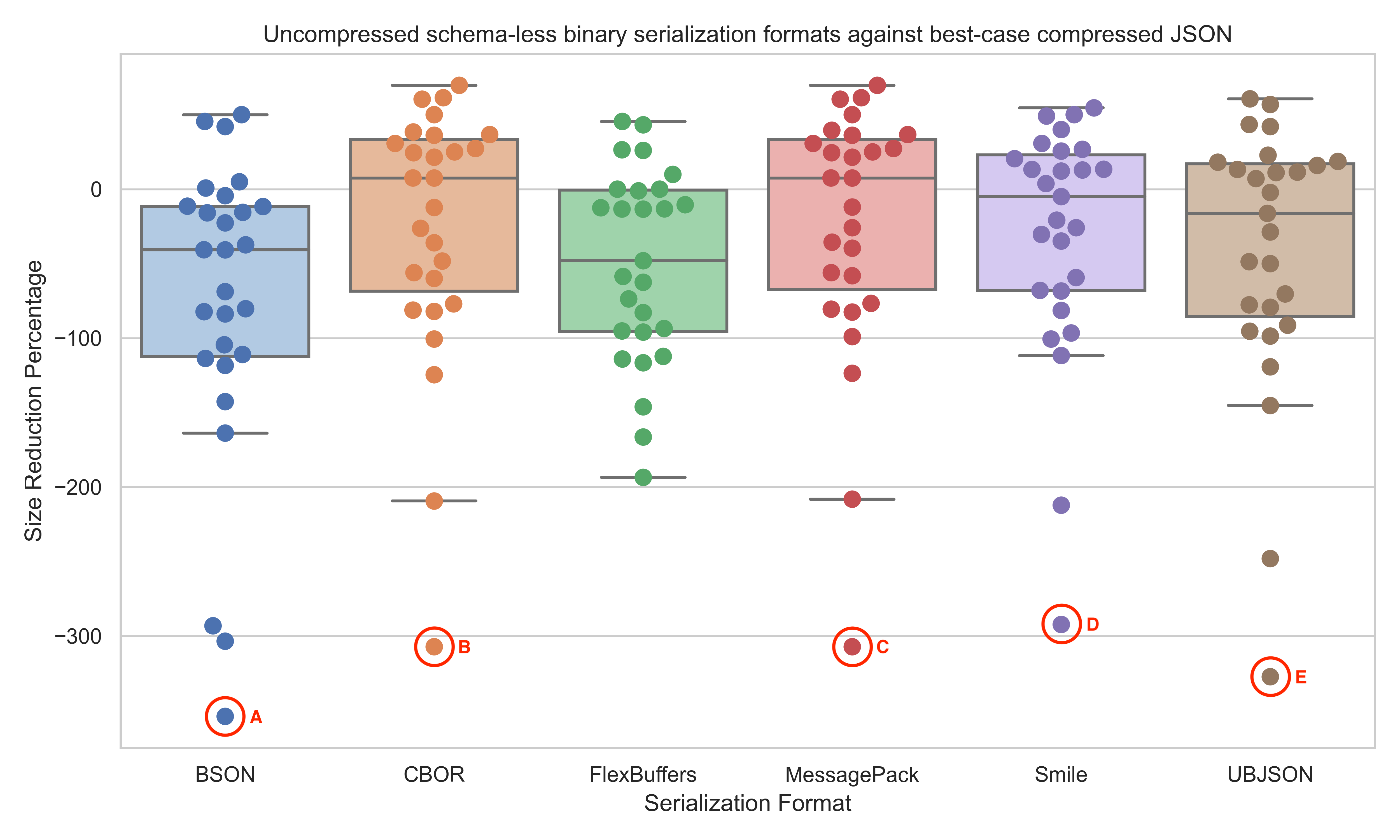}}
  \caption{A box plot that demonstrates the size reduction (in percentages) of
  the selection of schema-less binary serialization specifications listed in
  \autoref{table:benchmark-specifications-schema-less} in comparison to the
  best-case compressed JSON \cite{ECMA-404} given the compression formats
  listed in \autoref{table:benchmark-compression-formats} and the input data
listed in \autoref{table:benchmark-documents} and
\autoref{table:benchmark-documents-1}.}
\label{fig:benchmark-compressed-json-vs-schema-less} \end{figure*}

\autoref{table:compressed-json-vs-schema-less} summarizes the size reductions
provided by schema-less binary serialization specifications in comparison to compresed
JSON \cite{ECMA-404}. Leaving BSON \cite{bson} and FlexBuffers
\cite{flexbuffers} aside, schema-less binary serialization specifications typically
provide space-efficient results in \emph{Tier 1 Minified $<$ 100 bytes} JSON
documents, as these usually resulted in negative compression ratios. However,
compressed JSON provides space-efficient results in 15 out of the 27 listed in
\autoref{fig:schemastore-taxonomy}. In comparison to compressed JSON, no
schema-less binary serialization provides both a positive median and average
size reduction. As shown in
\autoref{fig:benchmark-compressed-json-vs-schema-less}, the selection of
schema-less binary serialization specifications listed in
\autoref{table:benchmark-specifications-schema-less}, with the exception of
FlexBuffers \cite{flexbuffers}, result in negative outliers for the Tier 2 TRF
case from \autoref{sec:benchmark-travisnotifications} (A, B, C, D, E).

\begin{figure*}[ht!]
  \frame{\includegraphics[width=\linewidth]{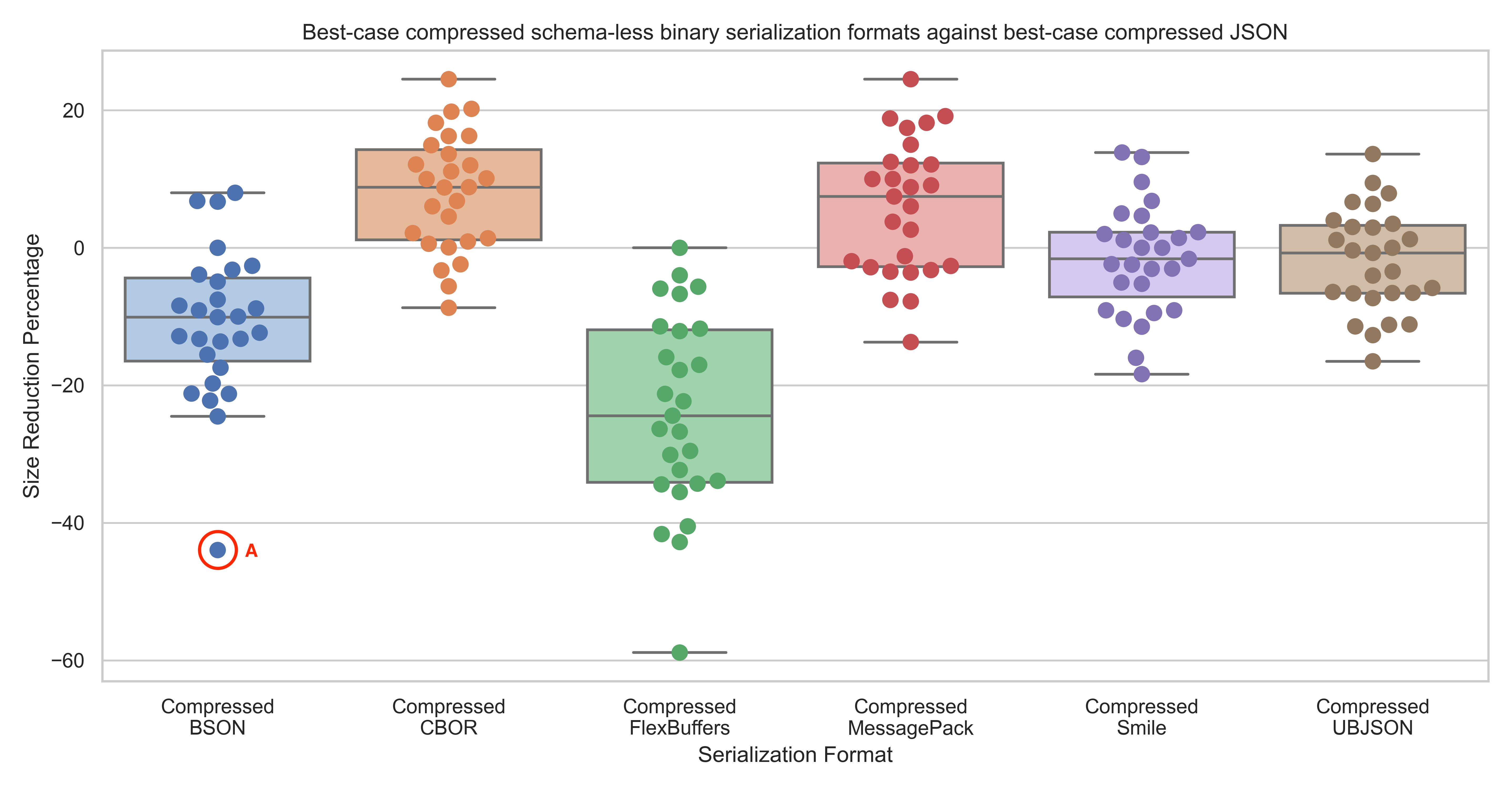}}
  \caption{A box plot that demonstrates the size reduction (in percentages) of
  the selection of schema-less binary serialization specifications listed in
  \autoref{table:benchmark-specifications-schema-less} in their best-case
  compressed forms given the compression formats listed in
  \autoref{table:benchmark-compression-formats} in comparison to the best-case
  compressed JSON \cite{ECMA-404} given the compression formats listed in
  \autoref{table:benchmark-compression-formats} and the input data listed in
  \autoref{table:benchmark-documents} and
  \autoref{table:benchmark-documents-1}.}
\label{fig:benchmark-compressed-json-vs-compressed-schema-less} \end{figure*}

As summarized in \autoref{table:compressed-json-vs-compressed-schema-less},
compressing the bit-strings produced by schema-less binary serialization
specifications results in 22 out 90 instances that are space-efficient in comparison
to compressed JSON on \emph{Tier 2 Minified $\geq$ 100 $<$ 1000 bytes} and
\emph{Tier 3 Minified $\geq$ 1000 bytes} JSON documents but reduces the
advantages that uncompressed schema-less binary serialization specifications have over
compressed JSON on \emph{Tier 1 Minified $<$ 100 bytes} JSON documents. In
comparison to compressed JSON, compressed CBOR \cite{RFC7049} is strictly equal
or superior than the rest of the compressed schema-less binary serialization
specifications in all but a single case: \emph{Tier 1 NRN} from
\autoref{sec:benchmark-jsonereversesort}, providing the highest median (8.8\%)
and highest average (8.1\%) size reductions. As a notable outlier shown in
\autoref{fig:benchmark-compressed-json-vs-compressed-schema-less}, best-case
compressed BSON \cite{bson} results in a negative size reduction of 44\% in
comparison to compressed JSON \cite{ECMA-404} for the Tier 2 NRN case from
\autoref{sec:benchmark-geojson}.

\begin{table}[hb!]

\caption{A summary of the size reduction results in comparison to the best case
  scenarios of compressed JSON \cite{ECMA-404} given the compression formats
  listed in \autoref{table:benchmark-compression-formats} of the selection of
  schema-less binary serialization specifications listed in
  \autoref{table:benchmark-specifications-schema-less} against the input data
  listed in \autoref{table:benchmark-documents} and
  \autoref{table:benchmark-documents-1}. See
  \autoref{fig:benchmark-compressed-json-vs-schema-less} for a visual
  representation of this data.}

\label{table:compressed-json-vs-schema-less}
\begin{tabularx}{\linewidth}{X|l|l|l|l|l|l}
\toprule
  \multirow{2}{\benchmarkconclusionrow}{\textbf{Serialization Specification}} & \multicolumn{5}{c|}{\textbf{Size Reductions in Comparison To Compressed JSON}} & \multirow{2}{*}{\textbf{Negative Cases}} \\ \cline{2-6}
& \textbf{Maximum} & \textbf{Minimum} & \textbf{Range} & \textbf{Median} & \textbf{Average} & \\
\midrule
BSON        & 50.0\% & -353.9\% & 403.9 & -40.8\% & -76.9\% & 22 / 27 (81.4\%) \\ \hline
CBOR        & 69.7\% &    -307.1\% &  376.8 & 7.5\% &  -26.8\% & 13 / 27 (48.1\%) \\ \hline
FlexBuffers & 45.5\% &  -193.5\% & 238.9 & -48.1\% &  -50.8\% & 20 / 27 (74\%) \\ \hline
MessagePack & 69.7\% &    -307.1\% &  376.8 & 7.5\% &  -26.2\% & 13 / 27 (48.1\%) \\ \hline
Smile       & 54.5\% &  -292.2\% &  346.8 & -5\% &  -31.7\% & 14 / 27 (51.8\%) \\ \hline
UBJSON      & 60.6\% &   -327.3\% &  387.9 &  -16.3\% &   -43.6\% & 15 / 27 (55.5\%) \\ \hline
\textbf{Averages} & \textbf{58.3\%} & \textbf{-296.9\%} & \textbf{355.2} & \textbf{-15.9\%} & \textbf{-42.7\%} & \textbf{59.8\%} \\
\bottomrule
\end{tabularx}
\end{table}

\begin{table}[hb!]

\caption{A summary of the size reduction results of the best case scenarios of
  compressed schema-less binary serialization specifications listed in
  \autoref{table:benchmark-specifications-schema-less} in comparison to the
  best case scenarios of compressed JSON \cite{ECMA-404} given the compression
  formats listed in \autoref{table:benchmark-compression-formats} and the the
  input data listed in \autoref{table:benchmark-documents} and
  \autoref{table:benchmark-documents-1}.  See
  \autoref{fig:benchmark-compressed-json-vs-compressed-schema-less} for a
  visual representation of this data.}

\label{table:compressed-json-vs-compressed-schema-less}
\begin{tabularx}{\linewidth}{X|l|l|l|l|l|l}
\toprule
\multirow{2}{\benchmarkconclusionrow}{\textbf{Serialization Specification}} & \multicolumn{5}{c|}{\textbf{Size Reductions in Comparison To Compressed JSON}} & \multirow{2}{*}{\textbf{Negative Cases}} \\ \cline{2-6}
& \textbf{Maximum} & \textbf{Minimum} & \textbf{Range} & \textbf{Median} & \textbf{Average} & \\
\midrule
Compressed BSON        & 8\%    & -44\% & 52 & -10.1\% & -11\% & 23 / 27 (85.1\%) \\ \hline
Compressed CBOR        & 24.5\% &    -8.7\% &  33.3 & 8.8\% &  8.1\% & 4 / 27 (14.8\%) \\ \hline
Compressed FlexBuffers & 0\% &  -58.9\% & 58.9 & -24.4\% &  -23.8\% & 27 / 27 (100\%) \\ \hline
Compressed MessagePack & 24.5\% &    -13.7\% &  38.2 & 7.5\% &  5.9\% & 10 / 27 (37\%) \\ \hline
Compressed Smile       & 13.9\% &  -18.4\% &  32.2 & -1.6\% &  -1.6\% & 14 / 27 (51.8\%) \\ \hline
Compressed UBJSON      & 13.6\% &   -16.5\% &  30.1 &  -0.7\% &   -1.9\% & 15 / 27 (55.5\%) \\ \hline
\textbf{Averages} & \textbf{14.1\%} & \textbf{-26.7\%} & \textbf{40.8} & \textbf{-3.4\%} & \textbf{-4.1\%} & \textbf{57.3\%} \\
\bottomrule
\end{tabularx}
\end{table}

\subsubsection{Schema-driven Binary Serialization Specifications}

\begin{figure*}[ht!]
  \frame{\includegraphics[width=\linewidth]{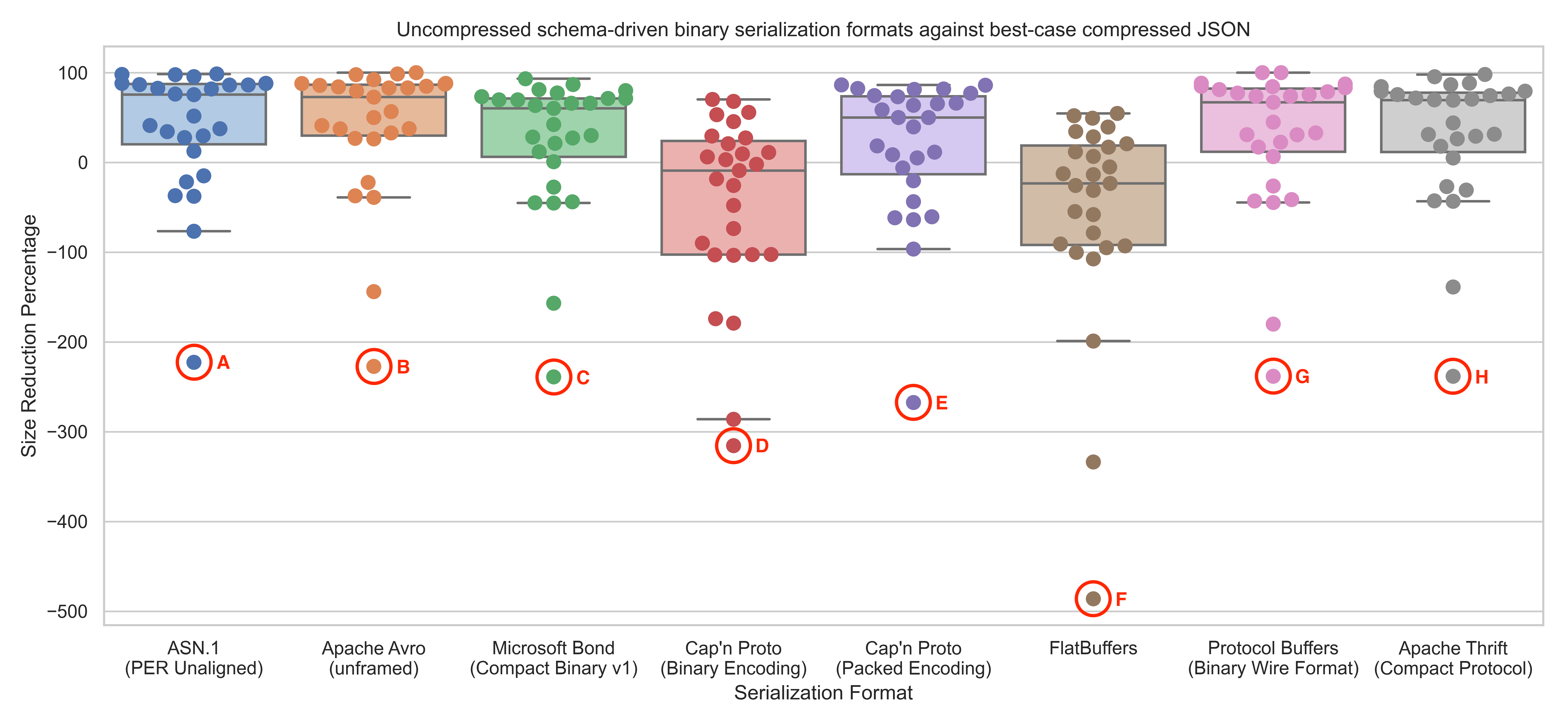}}
  \caption{A box plot that demonstrates the size reduction (in percentages) of
  the selection of schema-driven binary serialization specifications listed in
  \autoref{table:benchmark-specifications-schema-driven} in comparison to the
  best-case compressed JSON \cite{ECMA-404} given the compression formats
  listed in \autoref{table:benchmark-compression-formats} and the input data
listed in \autoref{table:benchmark-documents} and
\autoref{table:benchmark-documents-1}.}
\label{fig:benchmark-compressed-json-vs-schema-driven} \end{figure*}

As shown in \autoref{table:compressed-json-vs-schema-driven}, schema-driven
binary serialization specifications provide positive median and average size
reductions in comparison to compressed JSON \cite{ECMA-404}. However,
schema-driven binary serialization specifications tend to produce negative results in
comparison to compressed JSON mostly on \emph{Tier 2 Minified $\geq$ 100 $<$
1000 bytes} \emph{textual} (22 out of 32 cases) and \emph{Tier 3 Minified
$\geq$ 1000 bytes} \emph{textual} (25 out of 32) JSON documents. Even when
taking compression into account, both ASN.1 PER Unaligned \cite{asn1-per} and
Apache Avro (unpacked) \cite{avro} continue to provide over 70\% median size
reductions and almost 40\% average size reductions. As shown in
\autoref{fig:benchmark-compressed-json-vs-schema-driven}, the entire selection
of schema-driven binary serialization specifications listed in
\autoref{table:benchmark-specifications-schema-driven} results in negative outliers
for the Tier 2 TRF case from \autoref{sec:benchmark-travisnotifications} (A, B,
C, D, E, G and H) and the Tier 2 NRN case from \autoref{sec:benchmark-geojson}
(F).

\begin{figure*}[ht!]
  \frame{\includegraphics[width=\linewidth]{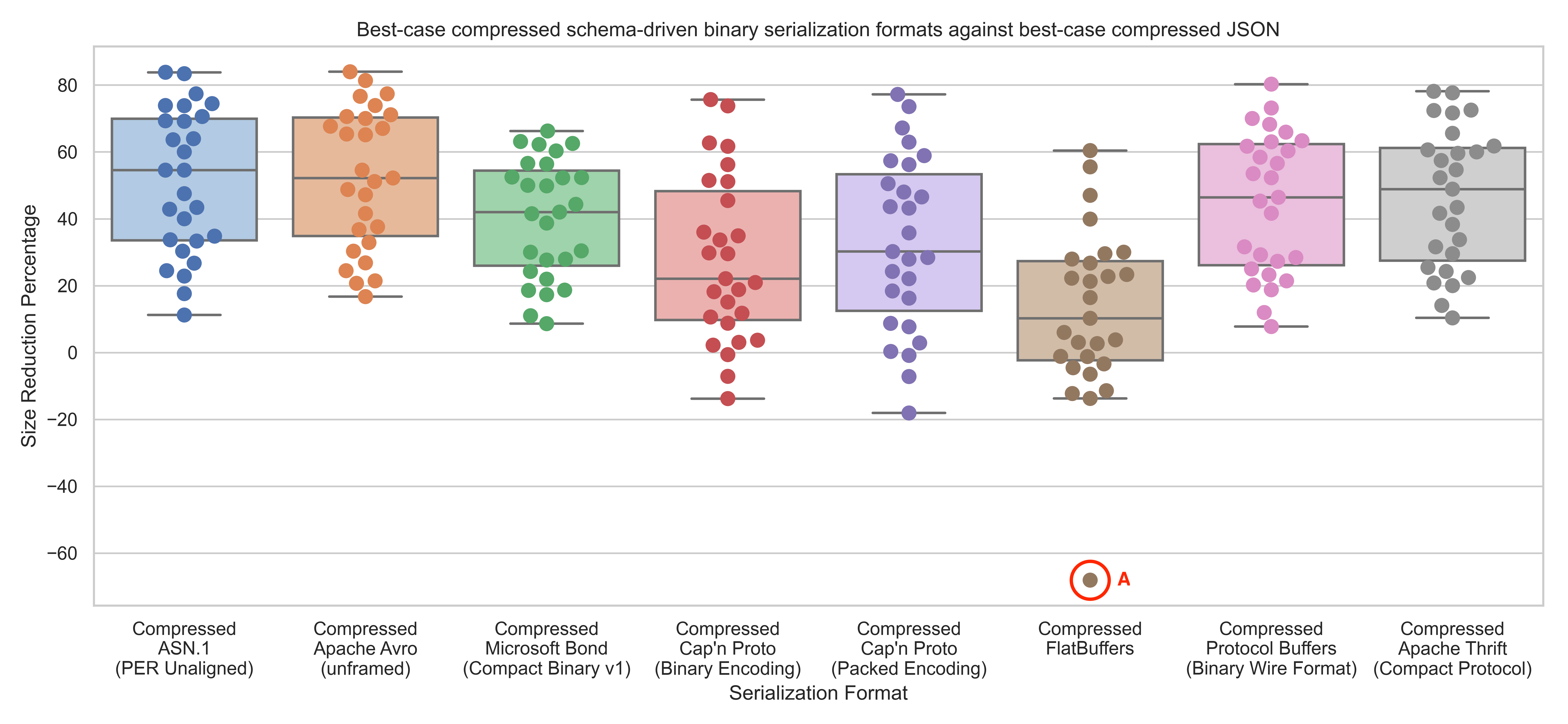}}
  \caption{A box plot that demonstrates the size reduction (in percentages) of
  the selection of schema-driven binary serialization specifications listed in
  \autoref{table:benchmark-specifications-schema-driven} in their best-case
  compressed forms given the compression formats listed in
  \autoref{table:benchmark-compression-formats} in comparison to the best-case
  compressed JSON \cite{ECMA-404} given the compression formats listed in
  \autoref{table:benchmark-compression-formats} and the input data listed in
  \autoref{table:benchmark-documents} and
  \autoref{table:benchmark-documents-1}.}
\label{fig:benchmark-compressed-json-vs-compressed-schema-driven} \end{figure*}

Compressing the bit-strings produced by schema-driven binary serialization
specifications shows that compressed \emph{sequential} schema-driven binary
serialization specifications are strictly superior than compressed JSON
\cite{ECMA-404} as shown in
\autoref{table:compressed-json-vs-compressed-schema-driven}. On the higher end,
both ASN.1 PER Unaligned \cite{asn1-per} and Apache Avro \cite{avro} provide
median and average size reductions of over 50\% in comparison to compressed
JSON, with a minimum size reduction of over 11\% in the Tier 2 NRN case from
\autoref{sec:benchmark-geojson} for which all the schema-driven binary
serialization specifications previously resulted in negative size reductions in
comparison to uncompressed JSON. As a notable exception shown in
\autoref{fig:benchmark-compressed-json-vs-compressed-schema-driven}, best-case
compressed FlatBuffers \cite{flatbuffers} results in a negative size reduction
of 68.1\% (A) in comparison to compressed JSON \cite{ECMA-404} for the Tier 2
NRN case.

\begin{table}[hb!]

\caption{A summary of the size reduction results in comparison to the best case
  scenarios of compressed JSON \cite{ECMA-404} given the compression formats
  listed in \autoref{table:benchmark-compression-formats} of the selection of
  schema-diven binary serialization specifications listed in
  \autoref{table:benchmark-specifications-schema-driven} against the input data
  listed in \autoref{table:benchmark-documents} and
  \autoref{table:benchmark-documents-1}. See
  \autoref{fig:benchmark-compressed-json-vs-schema-driven} for a visual
  representation of this data.}

\label{table:compressed-json-vs-schema-driven}
\begin{tabularx}{\linewidth}{X|l|l|l|l|l|l}
\toprule
\multirow{2}{\benchmarkconclusionrow}{\textbf{Serialization Specification}} & \multicolumn{5}{c|}{\textbf{Size Reductions in Comparison To Compressed JSON}} & \multirow{2}{*}{\textbf{Negative Cases}} \\ \cline{2-6}
& \textbf{Maximum} & \textbf{Minimum} & \textbf{Range} & \textbf{Median} & \textbf{Average} & \\
\midrule
ASN.1 (PER Unaligned)              & 98.5\% & -222.7\% & 321.3 & 75.5\% & 39\% & 6 / 27 (22.2\%) \\ \hline
Apache Avro (unframed)             & 100\%  & -227.3\% & 327.3 & 72.7\% & 39.4\% & 5 / 27 (18.5\%) \\ \hline
Microsoft Bond (Compact Binary v1) & 93.2\%  & -239\% & 332.1 & 60.2\% & 23.4\% & 6 / 27 (22.2\%) \\ \hline
Cap'n Proto                        & 70.1\%  & -315.6\% & 385.7 & -9.1\% & -45.7\% & 15 / 27 (55.5\%) \\ \hline
Cap'n Proto (packed)               & 86.4\%  & -267.5\% & 353.9 & 50\% & 17\% & 8 / 27 (29.6\%) \\ \hline
FlatBuffers                        & 54.5\%  & -486.2\% & 540.8 & -23.4\% & -55.4\% & 17 / 27 (62.9\%) \\ \hline
Protocol Buffers                   & 100\%  & -238.3\% & 338.3 & 67\% & 28.4\% & 6 / 27 (22.2\%) \\ \hline
Apache Thrift (Compact Protocol)   & 98\%  & -238.3\% & 336.3 & 69.3\% & 29\% & 6 / 27 (22.2\%) \\ \hline
\textbf{Averages} & \textbf{87.6\%} & \textbf{-279.4\%} & \textbf{367} & \textbf{45.3\%} & \textbf{9.4\%} & \textbf{31.9\%} \\
\bottomrule
\end{tabularx}
\end{table}

\begin{table}[hb!]

\caption{A summary of the size reduction results of the best case scenarios of
  compressed schema-driven binary serialization specifications listed in
  \autoref{table:benchmark-specifications-schema-driven} in comparison to the
  best case scenarios of compressed JSON \cite{ECMA-404} given the compression
  formats listed in \autoref{table:benchmark-compression-formats} and the the
  input data listed in \autoref{table:benchmark-documents} and
  \autoref{table:benchmark-documents-1}. See
  \autoref{fig:benchmark-compressed-json-vs-compressed-schema-driven} for a
  visual representation of this data.}

\label{table:compressed-json-vs-compressed-schema-driven}
\begin{tabularx}{\linewidth}{X|l|l|l|l|l|l}
\toprule
\multirow{2}{\benchmarkconclusionrow}{\textbf{Serialization Specification}} & \multicolumn{5}{c|}{\textbf{Size Reductions in Comparison To Compressed JSON}} & \multirow{2}{*}{\textbf{Negative Cases}} \\ \cline{2-6}
  & \textbf{Maximum} & \textbf{Minimum} & \textbf{Range} & \textbf{Median} & \textbf{Average} & \\
\midrule
Compressed ASN.1 (PER Unaligned)              & 83.8\% & 11.2\% & 72.6 & 54.5\% & 51.2\%    & 0 / 27 (0\%) \\ \hline
Compressed Apache Avro (unframed)             & 84\%  & 16.7\% & 67.3 & 52.2\% & 52.3\%     & 0 / 27 (0\%) \\ \hline
Compressed Microsoft Bond (Compact Binary v1) & 66.3\%  & 8.6\% & 57.6 & 42\% & 40.3\%      & 0 / 27 (0\%) \\ \hline
Compressed Cap'n Proto                        & 75.7\%  & -13.8\% & 89.4 & 22.1\% & 28\%    & 3 / 27 (11.1\%) \\ \hline
Compressed Cap'n Proto (packed)               & 77.2\%  & -18.1\% & 95.3 & 30.2\% & 32.7\%  & 3 / 27 (11.1\%) \\ \hline
Compressed FlatBuffers                        & 60.4\%  & -68.1\% & 128.5 & 10.2\% & 12.1\% & 9 / 27 (33.3\%) \\ \hline
Compressed Protocol Buffers                   & 80.3\%  & 7.8\% & 72.5 & 46.4\% & 44.6\%    & 0 / 27 (0\%) \\ \hline
Compressed Apache Thrift (Compact Protocol)   & 78.1\%  & 10.3\% & 67.8 & 48.9\% & 46.2\%   & 0 / 27 (0\%) \\ \hline
\textbf{Averages} & \textbf{75.7\%} & \textbf{-5.7\%} & \textbf{81.4} & \textbf{38.3\%} & \textbf{38.4\%} & \textbf{6.9\%} \\
\bottomrule
\end{tabularx}
\end{table}

\textbf{Summary.} In comparison to compressed JSON, both compressed and
uncompressed schema-less binary serialization specifications result in negative
median and average size reductions. However, both compressed and uncompressed
schema-driven binary serialization specifications result in positive median and
average reduction. Furthermore, compressed sequential schema-driven binary
serialization specifications are strictly superior to compressed JSON in all
the cases from the input data.

\subsection{JSON Compatibility}
Implementing the benchmark and writing schemas for the set of schema-driven
serialization specification revealed that some of the considered schema-driven
serialization specifications are not strictly compatible with JSON
\cite{ECMA-404} and required transformations in order for the input data to be
accepted by the implementations or the respective schema definition languages
when encoding JSON documents from the input data set listed in
\autoref{table:benchmark-documents} and \autoref{table:benchmark-documents-1}.
These transformations can be inspected in the benchmark public GitHub
repository
\footnote{\url{https://github.com/jviotti/binary-json-size-benchmark}}. These
transformations are divided into the following categories:

\begin{itemize}

\item \textbf{Keys.} The schema definition languages provided by ASN.1
  \cite{asn1}, Microsoft Bond \cite{microsoft-bond}, Cap'n Proto
    \cite{capnproto}, FlatBuffers \cite{flatbuffers}, Protocol Buffers
    \cite{protocolbuffers}, and Apache Thrift \cite{slee2007thrift} disallow
    property names that include hyphens, slashes, dollar signs, parenthesis,
    and periods. Also, ASN.1 \cite{asn1} disallows property names that start
    with an underscore and Cap'n Proto \cite{capnproto} disallows property
    names include underscores and capitalised property names. Furthermore,
    Protocol Buffers \cite{protocolbuffers} and Apache Thrift
    \cite{slee2007thrift} disallow property names that equal the reserved
    keywords \emph{async}, \emph{extends}, \emph{in}, and \emph{with}. To
    handle these cases, the disallowed properties are renamed to a close
    variation that the schema language permits.

\item \textbf{Values.} Protocol Buffers \cite{protocolbuffers} defines the
  \emph{null} type as an enumeration consisting of a single constant: zero
    \footnote{\url{https://github.com/protocolbuffers/protobuf/blob/master/src/google/protobuf/struct.proto}}.
    FlatBuffers \cite{flatbuffers} does not support a \emph{null} type.  When
    using FlatBuffers \cite{flatbuffers}, \we represent this type with an
    enumeration consisting of a single constant in the same manner as Protocol
    Buffers \cite{protocolbuffers}. In both cases, \we transform any JSON
    \cite{ECMA-404} \emph{null} value into zero.

  \item \textbf{Structural.} Neither Microsoft Bond \cite{microsoft-bond},
    Cap'n Proto \cite{capnproto}, FlatBuffers \cite{flatbuffers}, Protocol
    Buffers \cite{protocolbuffers}, nor Apache Thrift \cite{slee2007thrift}
    support encoding a JSON document that consists of a top level array. In
    these cases, \we move the array into a wrapper structure.  FlatBuffers
    \cite{flatbuffers} and Protocol Buffers \cite{protocolbuffers} also do not
    support nested arrays. In these cases, \we introduce wrapper structures at
    every array nesting level. Finally, ASN.1 \cite{asn1}, Microsoft Bond
    \cite{microsoft-bond}, Cap'n Proto \cite{capnproto}, FlatBuffers
    \cite{flatbuffers}, Protocol Buffers \cite{protocolbuffers}, and Apache
    Thrift \cite{slee2007thrift} do not support heterogenous arrays of
    non-composite types. In these cases, \we convert the heterogenous arrays
    into arrays of union structures. Microsoft Bond \cite{microsoft-bond} does
    not support union types and in this case \we introduce a structure
    consisting of optional fields.  Additionally, the use of unions in
    FlatBuffers \cite{flatbuffers} requires the introduction of an additional
    textual property to signify the union choice. In order not to put this
    specification at a disadvantage, \we encode the fixed-length heterogenous
    array as tables where their property names correspond to the array indexes.

\end{itemize}

The type of transformations that were necessary for each JSON document from the
input data defined in \autoref{table:benchmark-documents} and
\autoref{table:benchmark-documents-1} are listed in
\autoref{table:benchmark-compatibility}. In summary, every schema-less binary
serialization specifications listed in
\autoref{table:benchmark-specifications-schema-less} is compatible with the
input data set. In terms of schema-driven specifications, only Apache Avro
\cite{avro} is strictly compatible with the input data set.

\begin{table}[hb!]

\caption{A summary of the transformations needed to serialize the input data
  JSON documents listed in \autoref{table:benchmark-documents} and
  \autoref{table:benchmark-documents-1} using the set of binary serialization
  specifications listed in
  \autoref{table:benchmark-specifications-schema-driven} and
  \autoref{table:benchmark-specifications-schema-less}. The JSON documents from
  \autoref{table:benchmark-documents} and \autoref{table:benchmark-documents-1}
  that are not present in this table did not require any type of
  transformation. Each letter signifies the type of required transformation as
  defined in this section. The letter K stands for \emph{Keys}, the letter V
  stands for \emph{Values}, and the letter S stands for \emph{Structural}.}

\label{table:benchmark-compatibility}
\begin{tabularx}{\linewidth}{X|l|l|l|l|l|l|l|l|l|l|l|l|l}
\toprule
\textbf{Input Data} &

\textbf{\rotatebox[origin=c]{90}{ASN.1}} &
\textbf{\rotatebox[origin=c]{90}{Apache Avro}} &
\textbf{\rotatebox[origin=c]{90}{Microsoft Bond}} &
\textbf{\rotatebox[origin=c]{90}{BSON}} &
\textbf{\rotatebox[origin=c]{90}{Cap'n Proto}} &
\textbf{\rotatebox[origin=c]{90}{CBOR}} &
\textbf{\rotatebox[origin=c]{90}{FlatBuffers}} &
\textbf{\rotatebox[origin=c]{90}{FlexBuffers}} &
\textbf{\rotatebox[origin=c]{90}{MessagePack}} &
\textbf{\rotatebox[origin=c]{90}{Protocol Buffers}} &
\textbf{\rotatebox[origin=c]{90}{Smile}} &
\textbf{\rotatebox[origin=c]{90}{Apache Thrift}} &
\textbf{\rotatebox[origin=c]{90}{UBJSON}} \\ \midrule

Tier 1 NRF & K   &  & K   &  & K   &  & K   &  &  & K   &  & K   & \\ \midrule
Tier 1 NRN & K   &  & K   &  & K   &  & K   &  &  & K   &  & K   & \\ \midrule
Tier 1 TRF & K   &  & K   &  & K   &  & K   &  &  & K   &  & K   & \\ \midrule
Tier 1 TRN & K+S &  & K+S &  & K+S &  & K+S &  &  & K+S &  & K+S & \\ \midrule
Tier 1 TNF &     &  &     &  &     &  &     &  &  & K   &  & K   & \\ \midrule
Tier 1 BRF &     &  &     &  &     &  & V   &  &  & V   &  &     & \\ \midrule
Tier 1 BRN & K   &  & K   &  & K   &  & K   &  &  & K   &  & K   & \\ \midrule
Tier 1 BNN & K   &  & K   &  & K   &  & K   &  &  & K   &  & K   & \\ \midrule
                                                                 
Tier 2 NRN &     &  &     &  &     &  & S   &  &  & S   &  &     & \\ \midrule
Tier 2 NNF &     &  &     &  & K   &  &     &  &  &     &  &     & \\ \midrule
Tier 2 NNN &     &  & S   &  & K+S &  & S   &  &  & S   &  & S   & \\ \midrule
Tier 2 TNF &     &  &     &  & K   &  &     &  &  &     &  &     & \\ \midrule
Tier 2 TNN & K   &  & K   &  & K   &  & K   &  &  & K   &  & K   & \\ \midrule
Tier 2 BRF &     &  &     &  & K   &  & V   &  &  & V   &  &     & \\ \midrule
                                                                 
Tier 3 NRF & K+S &  & K+S &  & K+S &  & K+S &  &  & K+S &  & K+S & \\ \midrule
Tier 3 TRF & K+S &  & K+S &  & K+S &  & K+S &  &  & K+S &  & K+S & \\ \midrule
Tier 3 TRN & K   &  & K   &  & K   &  & K   &  &  & K   &  & K   & \\ \midrule
Tier 3 BRF &     &  &     &  & K   &  & V   &  &  & V   &  &     & \\ \midrule
Tier 3 TNF & K   &  & K   &  & K   &  & K   &  &  & K   &  & K   & \\ 

\bottomrule
\end{tabularx}
\end{table}

\section{Future Work}
In this paper, \we present the results of a comprehensive benchmark of 13
JSON-compatible schema-driven and schema-less binary serialization
specifications across 27 real-world JSON documents test cases across
industries.

\Our findings provide a number of conclusions. When \we investigated how
JSON-compatible schema-less binary serialization specifications compare to JSON
in terms of space-efficiency, \we found that using MessagePack
\cite{messagepack} on \emph{Tier 1 Minified $<$ 100 bytes} and \emph{Tier 2
Minified $\geq$ 100 $<$ 1000 bytes} JSON documents, Smile \cite{smile} on
\emph{Tier 3 Minified $\geq$ 1000 bytes} JSON documents, and FlexBuffers
\cite{flexbuffers} on JSON documents with high-redundancy of \emph{textual}
values increases space-efficiency. When \we investigated how JSON-compatible
schema-driven binary serialization specifications compare to JSON and
JSON-compatible schema-less binary serialization specifications in terms of
space-efficiency, \we found that ASN.1 PER Unaligned \cite{asn1-per} and Apache
Avro (unpacked) \cite{avro} are space-efficient in comparison to schema-less
binary serialization specifications in almost all cases. When \we investigated
how JSON-compatible sequential binary serialization specifications to compare
to JSON-compatible pointer-based binary serialization specifications in terms
of space-efficiency, \we found that sequential binary serialization
specifications are typically more space-efficient than pointer-based binary
serialization specifications, independent of whether they are schema-less or
schema-driven. When \we investigated how compressed JSON compares to
uncompressed and compressed JSON-compatible binary serialization
specifications, \we found that in comparison to compressed JSON, both
compressed and uncompressed schema-less binary serialization specifications
result in negative median and average size reductions.  However, both
compressed and uncompressed schema-driven binary serialization specifications
result in positive median and average reduction.  Furthermore, compressed
sequential schema-driven binary serialization specifications are strictly
superior to compressed JSON in all the cases from the input data.

Based on \our findings, \we believe there is room to augment the input data set
to include JSON documents that match the 9 missing taxonomy categories
described in \autoref{sec:benchmark-methodology-input-data} and to increase the
sample proportionality. \We hope to encourage contributions to \our open-source
space-efficiency benchmark automation software for general improvements and
support for new JSON-compatible binary serialization specifications. Using \our
learnings, \we hope to propose a new JSON-compatible binary serialization
specification with better space-efficiency characteristics.

\clearpage
\begin{ack}

Thanks to OSS Nokalva \footnote{\url{https://www.ossnokalva.com}} for offering
  access to and help for using their proprietary ASN-1Step ASN.1 \cite{asn1}
  implementation.

\end{ack}

\bibliographystyle{ACM-Reference-Format}
\bibliography{arxiv}

\end{document}